\documentclass{aa}  

\usepackage{graphicx}
\usepackage{natbib}
\usepackage{caption}
\usepackage{xcolor} 
\usepackage{amsmath}
\usepackage{hyperref}
\bibpunct{(}{)}{;}{a}{}{,} 
\usepackage[varg]{txfonts}
\usepackage[pdftex]{pict2e}

\newcommand\gaia{\textit{Gaia}}

\makeatletter
\renewcommand*\aa@pageof{, page \thepage{} of \pageref*{LastPage}}
\makeatother

\begin{document}

   \title{\gaia\ Data Release 3: Spectroscopic binary-star orbital solutions}
    
   \subtitle{The SB1 processing chain}

   \author{E.\,Gosset\inst{1,\thanks{Research Director, F.R.S.-FNRS (Belgium)}}
               \and Y.\,Damerdji\inst{2,}\inst{1}
               \and T.\,Morel\inst{1}
               \and L.\,Delchambre\inst{1}
               \and J.-L.\,Halbwachs\inst{3}
              \and G.\,Sadowski\inst{4}
              \and D.\,Pourbaix\inst{4,\dag \thanks{\dag\ Deceased 2021}}
              \and A.\,Sozzetti\inst{5}
              \and P.\,Panuzzo\inst{6}
              \and F.\,Arenou\inst{6}
              }

   \institute{Institut d'Astrophysique et de G\'eophysique, 
              Universit\'e de Li\`ege, 19c, All\'ee du 6 Ao\^ut, B-4000 Li\`ege, Belgium
              \and CRAAG-Centre de Recherche en Astronomie, Astrophysique et G\'eophysique, 
              Route de l'Observatoire, Bp 63 Bouzareah, DZ-16340 Algiers, Algeria
              \and Universit\'e de Strasbourg, CNRS, Observatoire Astronomique de Strasbourg,
              UMR 7550, 11 rue de l'Universit\'e, F-67000 Strasbourg, France
              \and Institut d'Astronomie et d'Astrophysique, Universit\'e Libre de Bruxelles 
              CP 226, Boulevard du Triomphe, B-1050 Brussels, Belgium
              \and INAF - Osservatorio Astrofisico di Torino, via Osservatorio 20,
              I-10025 Pino Torinese (TO), Italy
              \and GEPI, Observatoire de Paris, Universit\'e PSL, CNRS, 5 Place Jules Janssen,
              F-92190 Meudon, France}

   \date{Received 3 May 2024; accepted 8 October 2024}

   \authorrunning{Gosset et al.}
   \titlerunning{\gaia\ Data Release 3: spectroscopic binary-star orbital solutions}

  \abstract
   {The \gaia\ satellite constitutes one of ESA's cornerstone
   missions. Being primarily
   an astrometric space experiment measuring positions,
   proper motions, and parallaxes for a huge
   number of stars, it also performs photometric
   and spectrophotometric observations. \gaia\ 
   operates a medium-dispersion spectrometer, known as Radial Velocity
   Spectrometer (RVS), which
   provides spectra and radial velocity (RV) time series.}
   {The paper is focussed on the analysis of the
   RV time series.
   We fit orbital and trend models,
   restricting our study to  objects of
   spectral types F-G-K that are brighter than a magnitude of 12, 
   presenting only one single spectrum (SB1).} 
   {Suitable time series were processed and analysed on an 
   object-per-object basis, providing orbital or trend solutions.
   The results of the various fits were further filtered
   internally on the basis of several quality measures
   to discard spurious solutions.
   The objects with solely a spectroscopic solution were classified in one of the
   three classes: {\tt{SB1}} (eccentric model), {\tt{SB1C}} (circular model), 
   or {\tt{TrendSB1}} (mere trend model).}
   {We detail the methods used in this work and describe the derived
   parameters and results. After a description of the models considered
   and the related quality tests of the fit, we detail
   the internal filtering process aimed at rejecting
   bad solutions. We also present a full 
   validation of the pipeline.
   A description of the current content of the catalogue is also provided.}
   {We present the {\tt{SB1}}, {\tt{SB1C}}, and 
   {\tt{TrendSB1}} spectroscopic solutions contained in 
   the SB subcatalogue, part of the DR3 catalogue. 
   We deliver some 181\,327 orbital solutions
   in  class {\tt{SB1}}, 202 in  class {\tt{SB1C}}, and
   56\,808 in the associated class {\tt{TrendSB1}}. This is a
   first release and the delivered SB subcatalogue could
   be further tuned and refined. However, the majority
   of the entries are correct. Thus, this data set constitutes
   by far the largest set of spectroscopic orbital solutions
   to be computed.}

   \maketitle
%
%
\section{Introduction}\label{sec:spectroSB1_introduction}
The \gaia\ satellite constitutes one of the cornerstone missions of the
European Space Agency (ESA). This astrometric space experiment is intended
to represent a new breakthrough, compared to the previous pioneering one,
{\sc{Hipparcos}}
\citep{2009aaat.book.....P}.
The \gaia\ satellite has two telescopes, observing at 106\fdg 5 of each 
other, and three associated scientific instruments that are continuously scanning 
the celestial sphere thanks to its six-hour rotation
(around an axis perpendicular to the two telescope's lines of sight), 
as well as the precession of the spinning axis around the direction of a 
fictitious, nominal Sun (compensating for the orbit of the satellite 
around the Sun-Earth Lagrangian point L2), with a 63-day period, and thanks to 
the one-year orbit of the satellite around the Sun, while it accompanies the Earth
in its revolution \citep{2016A&A...595A...1G}.
Multiple visits are necessary to reach a convenient 
signal-to-noise (S/N) ratio, but also
to gather information on binarity (or higher order multiplicity)
and variability.

The \gaia\ satellite is measuring the positions and transverse proper 
motions for some 1.8 billion stars, as well as precise parallaxes
and thus distances for a very large number of objects. 
The detectors are working in 
a Time Delay Integration mode (TDI). The satellite
was launched on 19 December 2013, and has been scanning the sky
since the end of the commissioning period in July 2014. The detailed
description of the project and of the payload can be found in
\citet{2016A&A...595A...1G},
published alongside the first Data Release (DR1).
The description of DR1 and of the related data products is provided in
\citet{2016A&A...595A...2G}
whereas  DR2 was detailed in
\citet{2018A&A...616A...1G}.
The present paper is to be associated with the DR3
\citep{2023A&A...674A...1G}.
In addition to  astrometry performed with the 
Astrometric Field (AF) instrument
\citep{2018A&A...616A...2L, 2021A&A...649A...2L}, 
the satellite
also produces photometry from the same instrument as well as
spectrophotometry provided by another channel: the BP/RP
spectrophotometer purveyor of very low-resolution spectra as well as 
of photometric colours
\citep{2018A&A...616A...3R, 2021A&A...649A...3R, 2018A&A...616A...4E}.
The two above-mentioned channels are supposed to observe stellar objects
down to a  magnitude of about 20 in the  $G$ band. 

In addition to astrometry and spectrophotometry, the satellite also incorporates
a third channel: a medium-resolution spectrometer to measure the 
radial velocities (RVs) of stellar objects, known as 
the Radial Velocity Spectrometer (RVS). 
This channel and the
related instrument are described in great detail in
\citet{2018A&A...616A...5C}, which also includes 
a comprehensive review of the genesis of the RVS instrument.
The RVS brings the third component of the space velocity 
(to join with the transverse proper motion on the sky). 
It also provides, for the brightest stars, information
on the nature of the object ($T_{\mathrm{eff}}$, $\log g$, 
chemical abundances, chromospheric activity, etc.).
Stars brighter than a magnitude of
$G_{\mathrm{RVS}}^{\mathrm{int}}$~=~2.76
produce saturated pixels and, thus, are not considered here
\citep[for the definition of the $G_{\mathrm{RVS}}$ magnitudes, see][]{2023A&A...674A...6S}.

The scanning law for RVS is intended to typically produce 40 transits
over the nominal five-year mission.
The focal plane of the RVS is covered by 12 CCDs (four rows of three in a row). 
The RVS instrument acquires, in TDI mode, spectra in the range 845-872\,nm,
including the well-known \ion{Ca}{ii} triplet (in late-type stars). With a 
resolving power of about $R \sim 11\,500$
\citep[see Sects. 8.2 and 11 of][]{2018A&A...616A...5C}, 
it obtains three spectra per transit, each corresponding to a maximum
exposure time of 4.42\,s (they are analysed individually and 
concomitantly). These spectra are spread over about 1100 pixels
\citep{2018A&A...616A...5C}, but the part used for RV determination
is restricted to the range 846-870\,nm 
\citep{2023A&A...674A...6S}. 
Very bright objects benefit from 2D spectra
whose across scan component is 10 pixels. 

DR2 was the first release concerning the spectrometer channel and some 280
million spectra were treated and resulted in the publication of median
(over transits) RVs for over 7 million objects (brighter than mag.\,12,
being single-line spectra and corresponding to assumed constant RVs) in the range
of effective temperatures $\left[3550, 6900\right]$\,K.
This catalogue of single-line, non-exotic stars has already been
widely used to study the structure of the Galaxy
\citep{2018A&A...616A..11G,
2018Natur.561..360A,
2018MNRAS.479L.108K,
2018A&A...619A..72R,
2019A&A...626A..41M,
2019A&A...625A...5K,
2019MNRAS.490..157M,
2019MNRAS.484.3291T,
2019A&A...621A..48L,
2020A&A...634A..66L,
2020A&A...634L...8K,
2020MNRAS.494.6001N}.

The data coming out of the RVS are treated by ground-based spectroscopic processing.
The spectroscopic processing is ensured by the Coordination Unit 6 (CU6) of
the Data Processing and Analysis Consortium
\citep[DPAC;][]{2016A&A...595A...1G}. The first
description of this processing was given in
\citet{2018A&A...616A...6S} for DR2
and the properties and validation of the RV set were described in 
\citet{2019A&A...622A.205K}.
The stars exhibiting a double-line spectrum (also known as the composite spectrum) 
were rejected from the process \citep[as described in][]{2018A&A...616A...6S, dr3-dpacp-161}. 
The spectroscopic processing pipeline is made of a series of modules, each taking
into account a series of tasks. The first modules are working per transit and per
trending epochs. The flowchart is illustrated in Sect. 2.3 of 
\citet{2018A&A...616A...6S}.
The single transit analysis (STA) refers to the workflow that actually performs
the measurements of the RV per transit.
Three submodules are measuring the RVs of all the selected convenient stars
(single spectrum ones).
They are based on the idea to derive the RV by cross-correlating
the observed spectrum with a synthetic spectrum built in {\tt{generate\_template}}
\citep{2018A&A...616A...6S}. 
The incentive in making use of several methods is explained in the early investigation
of the problem, with conclusions detailed in
\citet[][]{2014A&A...562A..97D}
\citep[see also Sect. 7 of][]{2018A&A...616A...6S}.
The STA measures the RVs per object and per transit. Thereafter,
the multiple transit analysis (MTA) workflow performs some statistical
analysis tests based on the RVs from all objects at all transits. It computes (per star)
a probability of constancy that is used to spot out variable objects.
It also computes the median of the RVs over all transits for all the bright objects
\citep{2018A&A...616A...6S}. 
The constancy probability is not used to define
the list of assumed constant objects that are considered to compute their median RV.
The updated version of the spectroscopic chain for DR3 is described in
\citet{CU6documentation,2023A&A...674A...5K}.
In DR3, MTA further computes mean spectra for assumed constant objects
(Seabroke et al., in preparation).

In the context of DR3, 
the RV values corresponding to variable objects
are transmitted to the CU4 (object processing) and in particular to the
non-single star (NSS) part of the object processing that analyses these data
in order to derive possible orbital solutions. Three channels running in parallel
deal with the astrometric data \citep{2023A&A...674A...9H, 2023A&A...674A..10H},  
photometric data 
\citep[][see Sect. 7.6 by Siopis \& Sadowski]{dr3-dpacp-179,pourbaixetaldoc}, 
and the spectroscopic data
\citep[the present paper, and][]{dr3-dpacp-161}.
The detailed description of the CU4 spectroscopic processing chain 
for SB1 (spectroscopic binaries with only one spectrum visible;
hereafter labelled NSS-SB1)
orbital solutions or trend ones is the subject of the present paper
whereas \citet{dr3-dpacp-161} describe the SB2 channel. 
These DR3 products are preliminary results;
this is the first time that orbital solutions issued from \gaia\ have been delivered.
The results of the three channels can be
further processed by combining their independent outputs. 
This is the task of the combiner 
\citep[][see Sect. 7.7 by 
Gavras \& Arenou]{2023A&A...674A..34G, pourbaixetaldoc}. 
The scientific conception of the SB1 chain 
described here was elaborated on by Yassine
Damerdji, Eric Gosset and Thierry Morel;
most of the codes were benchmarked in MATLAB and finally written in
Java by YD with some help from TM; the validation tests (including the
critical building of the ground-based data catalogues used for comparison) 
were mainly performed by TM. The whole work was carried out in the highly 
favourable and boosting context of the NSS team.

Section\,\ref{sec:spectroSB1_input} describes the nature 
of the data that are entering the process, whereas
Sect.\,\ref{sec:spectroSB1_model} describes the simple 
mathematical formalism used and lists the models
that are considered. 
Section\,\ref{sec:spectroSB1_periodsearch} explicits the 
importance of the period search 
and the related difficulties. 
Section\,\ref{sec:spectroSB1_method} briefly summarises 
the main trials we did to  arrive at the
method described in Sect.\,\ref{sec:spectroSB1_procchain}, where we 
present the chain as adopted for the DR3 (NSS-SB1 main processing). In 
Sect.\,\ref{sec:spectroSB1_postproc}, we introduce the notion of post-processing.
Section\,\ref{sec:spectroSB1_results} illustrates the results. 
Section\,\ref{sec:spectroSB1_validation} concerns
the validation of the results of the chain. 
Additional considerations in Sect.\,\ref{sec:spectroSB1_add_cons} explain the
details of the post-filtering and provide additional information in
the light of the conclusions from the previous section.
The catalogue description, the caveats
and the precautions necessary in using the results of 
the NSS-SB1 processing are given in 
Sect.\,\ref{sec:spectroSB1_catalog}. 
Section\,\ref{sec:spectroSB1_conclusion} presents our conclusions. 
The contents of the present paper are more detailed and contain 
material that supplants the information given in 
the \gaia\ archives
\citep[see Sects. 7.4 and 7.5 in][]
{pourbaixetaldoc}\footnote{{{\url{https://gea.esac.esa.int/archive/documentation/GDR3/}}}}
at the time of the release.
The main core of the paper is accompanied by appendices.
Appendices D to J are only available on Zenodo
(see \url{https://zenodo.org/records/13990211}).
\section{Description of the input data}
\label{sec:spectroSB1_input}
The DR3 data correspond to the first
34 months of the nominal mission 
(25 July 2014 - 28 May 2017, almost 1038\,d, 2.84\,y).
The instrumental spectra obtained
at each RVS transit are converted to proper physical spectra (reduced) by the
pipeline developed by the CU6 (spectroscopic processing), see
\citet{CU6documentation, 2023A&A...674A...5K}.
Per transit, one spectrum is acquired by each of the three CCDs
in a row of the RVS \citep{2018A&A...616A...5C}. 
The RVs are measured individually on the three spectra.
Three methods are used to derive the radial velocities
\citep{2018A&A...616A...6S}, each of them
providing three velocities (one per CCD) and a combined RV value
(over the three CCDs).
The three combined RVs issued from the three methods 
are transmitted to the {\emph{Integrator}} that provides a 
single combined value (per transit).
Only this single combined value is considered here
but the other intermediate RVs are used as a sanity check 
to validate the final combined value
\citep[see Sects.\,6.4.8.2 and 6.4.8.5 of][]{CU6documentation}. 
The RVs are measured by cross-correlating
the observed reduced and calibrated spectra with a
theoretical/synthetic spectrum computed on the basis
of various stellar atmosphere models. 
These so-called templates are identified through
a set of three atmospheric parameters
(effective temperature, surface gravity,
metallicity). The selection of the template
used to measure the RVs is based
on these three parameters only; the alpha element content is not 
considered as an independent parameter.
The assignation of the
set of parameters to the object treated is
performed on the basis of three databases.
They are ordered here in the order of preference.
The first possible origin is an internal compilation
of ground-based catalogues. The second one is from an 
early run of the {\emph{Gaia General Stellar
Parametrizer from Spectroscopy}}
\citep[GSP-Spec,][]{2023A&A...674A..29R}. The third one
is from an early run of the 
{\emph{Gaia General Stellar
Parametrizer from Photometry}}
\citep[GSP-Phot,][]{2023A&A...674A..27A}. 
Finally, if none of these
sources provides the necessary parameters, the latter were
estimated as previously done for DR2
\citep[{\emph{DetermineAP}}: Sect. 6.5 of][]{2018A&A...616A...6S}. 
This process {\emph{DetermineAP}} cross-correlates the observed
spectrum with each template from a limited subset of 28 selected
ones, and this for every transit. The template giving per transit 
the highest correlation is retained, whatever is the velocity.
The final adopted template is the one that is the major contributor
over all the transits.
More details can be found in
Sect. 3.6.1 of \citet{2023A&A...674A...5K}.
The repartition among the various origins for the data set
treated here is the following:
ground-based catalogues (14.4\,\%), GSP-Spec (33.4\,\%), GSP-Phot (45.6\,\%)
and {\emph{DetermineAP}} (6.6\,\%).
The RV measurements from transit to transit
are performed with the same template. 
Consequently, a template mismatch is expected
to essentially introduce a systematic RV shift
(for single-line objects).

The measurements of the RVs were performed
under the supervision of the STA 
Development Unit. The list of good RVs
corresponding to a particular star are then analysed 
by the MTA 
to detect and separate 
variable RV sets from constant RV sets. 
For each star, the median of the RVs 
is computed and appears in the main catalogue. 
The stars
with variable RVs were further considered by the global pipeline and forwarded
to the CU4 (Object processing: NSS). For DR3, the adopted cutting threshold
corresponds to a probability of variability of 0.99
({\tt{rv\_chisq\_pvalue}} $\le \, 0.01)$.
The composite spectra and SB2 objects were already  
treated  separately at the level of the
CU6 processing \citep{dr3-dpacp-161} for the
determination of both RVs and are
then dispatched to CU4/NSS in order to search for the
corresponding orbital solutions
\citep{dr3-dpacp-161}.

The non-single star (NSS) processing is intended to perform the reduction
of the corresponding data from the point of view of multiple star studies.
In the present paper, we  consider the sole spectroscopic channel for the
SB1 objects. 

The input data were drawn from a list of RVs 
(and their 1$\sigma$-uncertainties) that were corrected
to make reference to the barycentre of the solar system. 
It should be pointed out that for DR3
the rms error on the barycentric correction 
($\sigma \, \lesssim \, 0.05$ km\,s$^{-1}$) 
was not propagated to the uncertainty on the
individual RVs. 

The RVs are accompanied
by a time of observations expressed in barycentric Julian days (BJD).
The number of 
points of the time series corresponds to the number of transits for which the
RV has been properly measured with success. The data could present 
several weaknesses such that outliers and wrong RVs due, for instance, to template
mismatch errors (for the latter case, most likely the same systematic for
all values within a time series). 
These problems and additional selection rules
are further documented in \citet{2023A&A...674A...5K} and will
not be further detailed here. 
It should however be noticed that some objects have been rejected at the level of the
input data filtering in CU6 because they correspond to 
a poorly performing {\em{Astrometric Global Iterative Solution (AGIS)}} 
solution or to a large error on the along-scan field angle $\eta$
\citep[see its definition in][]{2018A&A...616A...6S}. 
The threshold has been chosen
at $\sigma_\eta \, = \, 200 \,$mas, equivalent to about 
29 km\,s$^{-1}$. Other filtering
applied to the input CU6 data is described
in Sect. 4 of \citet{2023A&A...674A...5K}.
In the course of the CU6 STA/MTA validation, a bias on output RVs was noticed 
for stars fainter than 
$G_{\mathrm{RVS}}^{\mathrm{int}} \, = \, 11$
\citep[being 0\,km\,s$^{-1}$ at magnitude 11 and reaching 75\,m\,s$^{-1}$ at 
magnitude 12,][]{2023A&A...674A...5K}. 
The correction was not applied for the catalogue, 
nor for the transmission of the RVs to the NSS processing. 
The effect is quite small and should only induce 
a global shift of all the RVs per object along with, as a consequence, 
a small bias on the systemic velocities (see Sect.\,\ref{sec:spectroSB1_model}). 

The current astrometric solution for DR3 has been released as part
of the EDR3 and contains, among others, positions for a large
number of objects. The astrometric solution (AGIS) is based on a 
treatment assuming the objects are single, unresolved stars 
and thus having an image characterised by the instrumental
line-spread function (for the 1D observations in the
astrometric channel) or point-spread function (for the 2D
observations), as described in
\citet{2021A&A...649A...2L}. Some objects do not behave as expected
according to this assumption. The pathological behaviour could be due
to anomalies, but a large contribution to this problem originates
in unresolved or poorly resolved double (or multiple) objects.
The separation threshold is considered to be around 0\farcs 4-0\farcs 5
\citep[see][]{2021A&A...649A...5F, 2023A&A...674A..25H}.
The double objects not following the behaviour for single stars
are associated with a bad adjustment to the global astrometric
solution and are thus accompanied by a large
renormalised unit weighted error ({\tt{ruwe}}).
\citet{2021A&A...649A...2L} also associated 
an excess noise (impacting the position) to these objects, 
which is delivered in the
catalogue under the name 
{\tt{astrometric\_excess\_noise}} and which corresponds
to the additional dispersion necessary to accept the
concerned object in the global solution.
Objects with a {\tt{ruwe}} larger than 1.4 are certainly
problematic, but smaller values already have an impact on the RVS
calibration. Since the RVS is a slitless spectrometer, the calibration
in wavelength of the spectra is requesting an epoch position of the
object that is accurate enough. Therefore, for the objects that are
deviating from the single star behaviour, we could expect that
the wavelength calibration is not correct, and thus 
the measured transit RVs are somewhat biased and this creates
artefacts in the time series.
No correction of this effect is offered for DR3.
As a first approach, we preferred not to discard these objects 
on the basis of the {\tt{ruwe}}. In order to take into account
this problem, it was decided to inflate the uncertainty on the RV
value by introducing (quadratically) an additional component
linked to the astrometry. This term is derived as being
$\sigma_{\mathrm{add}}$ (km\,s$^{-1}$) = 
\mbox{0.146 $\times$ {\tt{astrometric\_excess\_noise}} (mas)} where the factor
0.146 is related to the dispersion of the spectrometer
(transformation of mas into km\,s$^{-1}$). 
Typically, the values of {\tt{astrometric\_excess\_noise}} 
are below 10 mas and, thus,
$\sigma_{\mathrm{add}}$ is below 1.46 km\,s$^{-1}$.

Some selections were also made on the objects at the moment 
of their entry in the NSS-SB1 processing.  
In the case of DR3, these kinds
of stars are restricted to the range of effective temperatures
3875 K to 8125 K (mainly ((M))-K-G-F-(A) stars). No emission-line object
was  selected and, in principle, only stars 
without spectral peculiarities were permitted.
The objects with a {\tt{rv\_renormalised\_gof}} value greater than 4 were 
selected \citep{2023A&A...674A...5K} to enter the NSS-SB1
processing because they were considered as clearly non-constant.
The measurements of the individual (per transit) RVs
are restricted to the range --1000 to +1000 km\,s$^{-1}$.
A global view of the selections applied beforehand is
available in Table\,\ref{tab:appcatalogsel}.
This measurement of the individual RVs on individual spectra is only possible
for bright objects. 
The distribution in magnitude of the treated objects
is provided in Fig.\,\ref{fig:EGhistogrvsmag} in the validation
Sect.\,\ref{sssec:spectroSB1_validation_intern_SB1}. 
A threshold on $G_{\mathrm{RVS}}^{\mathrm{int}}$ was selected at 12 mag.
Fainter objects are considered to produce individual spectra for which
the S/N ratio does not enable the derivation of a sound radial velocity.
This strongly reduces the amount of selected objects. The stellar RVs
are not processed by the NSS spectroscopic channel if the number of transits 
(data points) is less than 10 (see details in Sect.\,\ref{sssec:spectroSB1_processing_ingest}).

The nature of the measurement itself (cross-correlation with templates)
implies that it is not possible to discriminate between a change in RV
due to a global shift of the line (as expected for SB1 objects) and a 
shift due to a line-profile variation (as expected e.g.\ for some intrinsic variables). 
Although possibilities to get rid of this problem exist,
they are not implemented for DR3 and this certainly constitutes a limitation 
of the present orbital-solution pipeline. 
In particular, fake SB1 could persist in the sample corresponding to never-deblending SB2.

When the SB2 chain \citep{dr3-dpacp-161} analyses an SB2 time series and is not able 
to derive the corresponding SB2 orbital solution, the object can be redirected 
to the SB1 chain.
An SB1 time series is built with the RVs characterised by the smallest uncertainties.
The object is then
considered as an SB1 and is forwarded to the present chain. This is a second
possibility of entering the SB1 channel.
These objects are flagged 
\citep[for more details, see][and also 
Appendix\,\ref{sec:appI}]{dr3-dpacp-161}.
\section{The underlying orbital models}\label{sec:spectroSB1_model}
Four kinds of orbital solutions are considered here: they are labelled
SB1, SB1C, TrendSB1, and StochasticSB1. They are described here below.
The single-line RVs treated here and consequently the deduced motion
is intended to refer to the sole object that is visible in the spectra.
The model will be fitted through a classical least-square procedure and the
selected objective function is the $\chi^2$-function except when otherwise
stated. The uncertainties on the parameters are calculated in the classical way
on the basis of the diagonal elements of the inverse 
of the curvature matrix \citep[i.e.\ the 
variance-covariance matrix; see][]{2003drea.book.....B}.
Despite the fact that we use the Greek symbol $\sigma$ for all
(i.e.\ concerning input data as well as output parameters) uncertainties, 
it should be pointed out that all of them are of course estimators.
They are always expressed as 1$\sigma$ values.
\subsection{Orbital models of type SB1}\label{ssec:spectroSB1_modelSB1}
The general eccentric Keplerian model is, for the spectroscopic channel,
expressed in terms of the Campbell coefficients 
\citep[see the appendices in][as the seminal reference]{2023A&A...674A...9H, 1960pdss.book.....B}.
The model RV is given by
\begin{equation}
RV(t) \, = \, \gamma \, + \, K \left[ \cos(v(t)+\omega) \, + \, e \, \cos \omega \, \right]
\label{Equa_RV_Eccentric}
,\end{equation}
where $v(t)$ is the true anomaly, which is deduced from the 
eccentric anomaly $E(t)$ by
\begin{equation}
\cos{v(t)} \, = \, \frac{\cos E(t) - e}{1-e \, \cos{E(t)}}
\label{COSV}
,\end{equation}
\begin{equation}
\sin{v(t)} \, = \, \frac{\sqrt{1-e^2} \, \sin E(t)}{1-e \, \cos{E(t)}}
\label{SINV}
,\end{equation}
which in turn is expressed as a function of the mean anomaly $M(t)$
\begin{equation}
M(t) = E(t) - e \, \sin E(t) \, = \, \frac{2\pi}{P} (t-T_{\mathrm{0}}) \, .
\label{ECCENTRICANOMALIES}
\end{equation}
Here, $T_{\mathrm{0}}$ is the time of passage at periastron, while $\omega$ is the argument
of periastron measured in the sense of the orbit with its origin
at the ascending node. In this model, the eccentricity $e$ enters as a strongly
non-linear parameter. This is also the case for $T_0$. The parameters to be determined are
$P$, $\gamma$, $K$, $e$, $\omega$, $T_{\mathrm{0}}$ (respectively, 
the period, the centre-of-mass velocity, the semi-amplitude, the eccentricity, 
the argument of periastron and the time of passage at periastron).
The first step of the computation is devoted to the determination of the
period. Some significance tests are made on the very existence of the period.
If the period is not significant, the model is fitting noise and
is considered as being invalid. This is a statistical decision. Owing to the insufficient
span of time covered for the DR3, no apsidal motion is considered; $\omega$ is considered
constant in time.
\subsection{Orbital models of type SB1C}\label{ssec:spectroSB1_modelSB1C}
Another particular kind of orbit is the circular one, i.e. $e = 0$.
In this case, the model is much more simple and is expressed by the following
equation
\begin{equation}
RV(t) \, = \, \gamma \, + \, K \left[\cos(\frac{2\pi}{P} \, (t-T_{\mathrm{0}})) \right]
,\end{equation}
where here $T_{\mathrm{0}}$ is the time of maximum velocity and the $\omega$ is
fixed at zero corresponding to the maximum velocity.
The free parameters to determine are restricted to 
$P$, $\gamma$, $K$, $T_{\mathrm{0}}$. The model is linear except, of course, for the
determination of the period. Concerning the
$T_{\mathrm{0}}$, the model can be linearised
by using the following equivalent equation
\begin{equation}
RV(t) \, = \, \gamma \, + \, A \left[\cos(\frac{2\pi}{P} \, t) \right]
+ \, B \left[\sin(\frac{2\pi}{P} \, t) \right]
\, \, \, .
\label{EQCIRCULAREXPANDED}
\end{equation}
We then computed
\begin{equation}
K = \sqrt{A^2 + B^2} 
\end{equation}
and 
\begin{equation}
T_0 \, = \, \frac{P}{2 \pi} \times \, \mathrm{atan2}(-B,A) 
,\end{equation}
where this $T_0$ refers to a cyclic variable and
is assigned the value closest to $t\, = \, 0$.
The significance level of the period is here also
determined.
\subsection{Models of type TrendSB1}\label{ssec:spectroSB1_modelTrendSB1}
It can happen that the RV evolution as a function of time
just exhibits a trend. This could be a transient behaviour
(e.g.\ outburst) just
observed by chance or correspond to a small 
piece of slowly accelerated motion (e.g. of a stellar
atmosphere).
It could also correspond to an orbital motion with a period well longer 
than the span of observing time covered by the
satellite. This span of time is 34 months for the DR3. However, it could be
slightly shorter for some objects because it combines with the gaps in the
scanning law. The present modelisation is restricted to 
polynomials of one to four degrees and presenting only one extremum on the interval.
The degree of the polynomial is chosen on the basis
of an F-test for nested models.
In practice, it turned out that only the one- and
two-degree polynomials were necessary. No higher degree fit was required. 
The formalism used for the fit is the same as for Hipparcos
\citep{1997ESASP1200.....E}.
The TrendSB1 solution is compared to the orbital
solutions using a Fisher Snedecor F-test 
on the comparison of the respective $\chi^2$s. 
It should be made clear that this comparison
could be an approximation.
\subsection{Models of type StochasticSB1}\label{ssec:spectroSB1_modelStochSB1}
When the fit of a TrendSB1 solution is not possible nor the fit of an SB1 or SB1C
orbital solution corresponding to a significant period, the solution is classified as StochasticSB1. 
The median RV is considered as constant, but with an extra dispersion
compared to the adopted RV uncertainties. This extra dispersion is estimated.
Properly speaking, this simple model is not an orbital one.
Some badly sampled orbits and high-order multiple systems could fall into this
category, perhaps momentarily. Some of them will certainly benefit from an increase of the
number of transits in future releases. Short-period intrinsic variables 
(e.g. $\delta$ Scuti stars) may also receive such solutions.
\subsection{The downstream combination of solutions}\label{ssec:spectroSB1_combination}
The outputs of the NSS spectroscopic pipeline are 
delivered to the combiner that operates downstream. 
The task of the combiner is to inspect the output of the astrometric, photometric, and
spectroscopic NSS channels in order to detect possible solutions
corresponding to the same object and that could be combined.
This will be described in Sect.\,\ref{ssec:spectroSB1_combiner}. 
\section{Determination of the periodicity}\label{sec:spectroSB1_periodsearch}
The first step in the analysis is the search for a periodic behaviour of the
RVs. The analysis is based on a periodogram (actually a frequencygram), 
which has the nature of the Fourier power spectrum. Since the time sampling
of the \gaia\ satellite is particularly complex, the classical Fourier 
formalism is not applicable. To aptly understand the limitation of the
method, it is necessary to have some knowledge of the impact of the
sampling and of a few characteristic timescales.
\begin{figure*}[ht]
\centerline{
\includegraphics[width=0.9\textwidth]{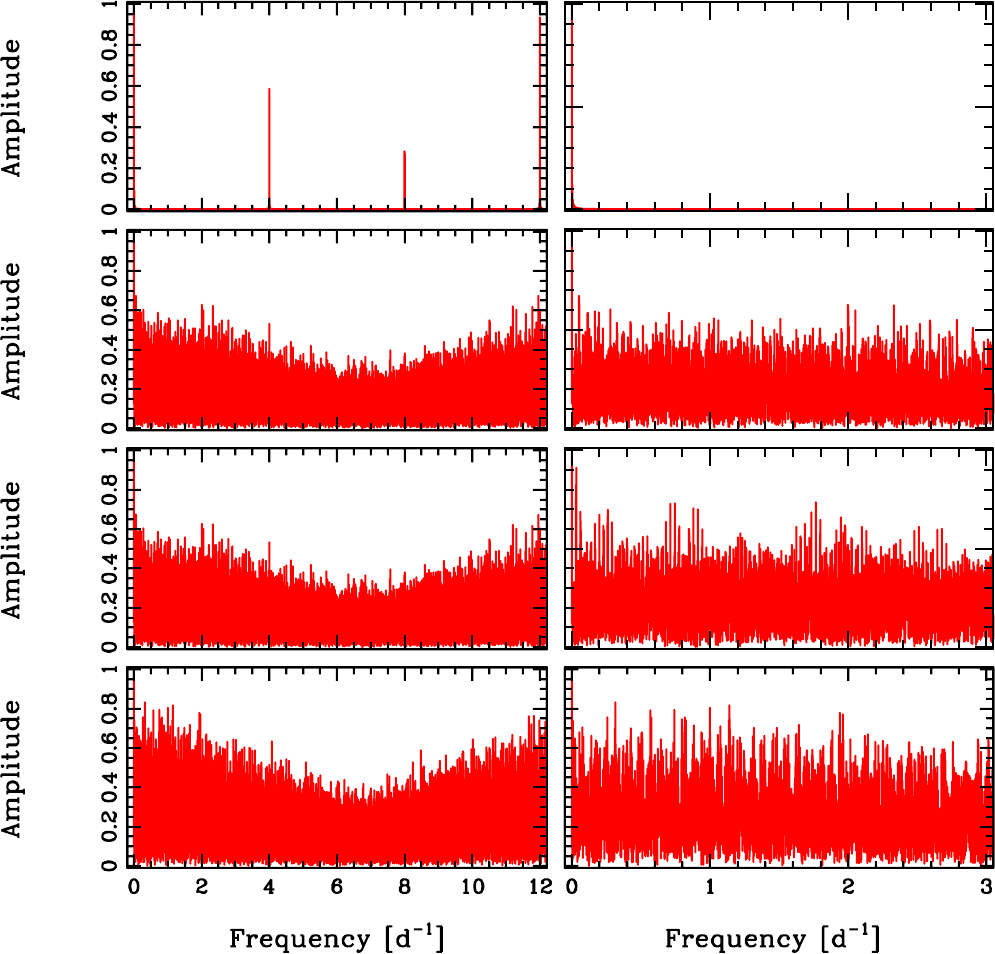}
}
\caption{Amplitude spectral window corresponding to various sampling. 
By definition, the peak at $\nu \, = \, 0$\,d$^{-1}$ 
is always going up to the value 1.
The upper panel corresponds to some 8000 data points and 
illustrates the sampling that \gaia\ would generate
if the sole effect would correspond to the spinning 
of the satellite on itself. The three other
panels represent typical spectral windows corresponding 
to objects at various positions in the sky.
The number of data points are, from the second panel to the bottom one, respectively, 
26, 19 and 14. The right panels are zooming on the low frequency domain.
All the individual peaks have by definition the 
same shape and the same width. They are not resolved in this figure.}
\label{EGfignyq}
\end{figure*}
\begin{figure}[ht]
\centerline{
\includegraphics[width=0.45\textwidth]{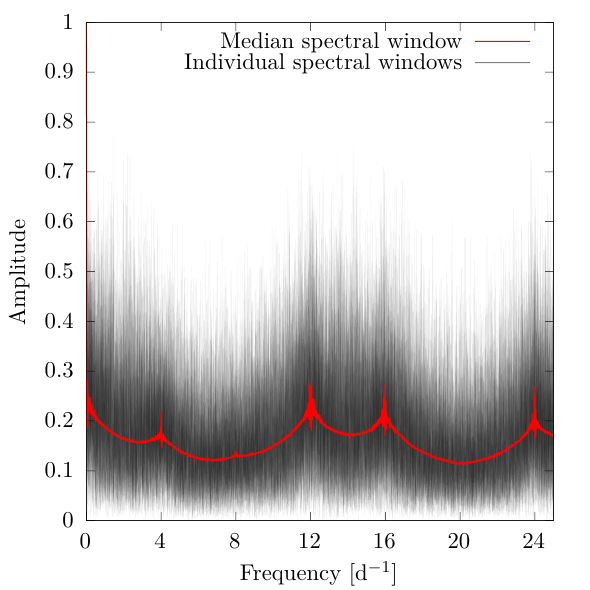}
}\caption{Median of the simulated amplitude spectral windows (in red). In black are 
shown some 32 of the individual realisations in order to 
illustrate the marked dispersion 
among the population of simulated spectral windows.
By definition, all the peaks at $\nu \, = \, 0$\,d$^{-1}$ 
are here also going up to 1.}
\label{fig:EGfigrandommedian}
\end{figure}
\subsection{Effect of the \gaia\ sampling}\label{ssec:spectroSB1_gsampling}
To explain the effect of the \gaia\ sampling, we have to start
from a single, idealistic, case and progressively add certain degrees of complexity.
We start by considering the spinning of the satellite on itself 
that makes that a star is measured every 0.25\,d 
(this is our first basic idealistic case since this simple view is never
occurring in practice). 
The existence  of the sample periodicity concerns both fields of view
individually.
Therefore, 0.25\,d is the typical, smallest regular and periodic
timescale in the sampling. 
The 0.25\,d step should generate in the time series some aliasing
with a typical 2\,d$^{-1}$ Nyquist frequency. 
This means that a progenitor frequency $\nu$
will be duplicated at $4 - \nu$, $4 + \nu$, $8 - \nu$, $8 + \nu$ and so on (in d$^{-1}$ units).
The duplication is also extended to the negative frequencies.
This duplication represents a pure aliasing, since the sampling is
in this hypothetical case regular. Therefore, the aliasing is pure
and the Nyquist frequency exists.
In such a case, intermediate and large periods could have
aliases towards periods of 0.25 to 0.5\,d (and vice versa).   

However, the existence of two fields of view
separated by 106\fdg 5 on the sky renders the situation more complex by adding
a regularity that is not properly speaking periodic. 
Thus, the aliasing is no more pure, we will call it pseudo-aliasing.
The semi-amplitude spectral window is a tool that facilitates the analysis
of such phenomenon. It has been introduced by
\citet{1975Ap&SS..36..137D, 1976Ap&SS..42..257D}. In the case of an
equidistant sampling, the spectral window is the well-known Dirac comb
\citep[or Shah function; see][]{1976fats.book.....B}.
The general spectral window represents an extension of this Dirac comb to the
general case of odd sampling. 
The property that the generalised Fourier Spectrum of the sampled data is the result
of the convolution of the underlying Fourier Spectrum 
of the signal by the spectral window
remains true for almost any kind of sampling
\citep{1975Ap&SS..36..137D, 1976Ap&SS..42..257D}.
If the satellite was limited to a rotation
on itself, the semi-amplitude spectral window corresponding to the combined
two fields of view would resemble   the plot in the upper panel of 
Fig.\,\ref{EGfignyq}. 
Contrary to the classical case of pure aliasing, the aliasing is here not pure
since all the peaks have different heights. Besides the main peak at
$\nu = 0$, the next first peak is at $\nu = 4$\,d$^{-1}$, but it is mildly
high. The peak having a height comparable to the $\nu = 0$ peak is the one at
$\nu = 12$\,d$^{-1}$ which could represent some almost pure aliasing.
This figure represents an idealistic case.
First, it is assumed that the data points in each field of view
are perfectly regularly distributed. 
This is not the case because several phenomena
induce a small perturbation of the time of the observations. 
The variations of the Basic Angle as well as
micro-clanks and meteoritic-hits are perturbing the 
regularity \citep{2016A&A...595A...1G}. 
We do not further consider these effects in our explanation.
The satellite is not only spinning on itself, but also follows
a more complicated motion. First, the rotation axis is precessing along an
axis directed towards a fictitious nominal position of the Sun
(compensating for the orbit of the satellite around the Sun-Earth 
Lagrangian point L2). In addition, the satellite
located around the Lagrangian point L2 is orbiting around the Sun with 
the Earth annual motion
\citep[change in ecliptic longitude of the Sun, ][]{2016A&A...595A...1G}.
The major consequence is that,
for a given object, the majority of the turns of the spinning satellite will not
correspond to a transit of the object and thus no observation will be acquired
for this object.
Most of the observations are not done and
medium and large gaps occur during which the field of view is not
directed towards the star. This effect is dependent on the position of the
object on the sky \citep{2016A&A...595A...1G}.
Therefore, the sky sampling is markedly non-uniform
\citep[see Sect. 5.2 of][]{2016A&A...595A...1G}.

Practically, this generates in the spectral
window a power leakage from the main peak at $\nu = 0$\,d$^{-1}$ towards other
frequencies. This effect is illustrated in the other panels of 
Fig.\,\ref{EGfignyq}. We produce typical time series of 
\gaia\ RVS with the GOST tool\footnote{\url{https://gaia.esac.esa.int/gost/}}.
Fig.\,\ref{EGfignyq} second panel represents a 
typical spectral window of a time series
containing 26 observations, the third panel 19 and the bottom one
14 observations (against the $\sim 8000$ observations for the first panel).
This small number of observations induces an aliasing but the aliasing typical
of the \gaia\ sampling is not leaking power towards a very few other frequencies
but towards numerous frequencies. Besides the $\nu = 0$ main peak,
there is a leakage towards a forest of frequencies. The contrast between the
parasitic peaks and the main one is decreasing with the number of data points
(following a $1 / \sqrt{N}$ for the semi-amplitude).
Typically, pseudo-aliasing can occur at any $\delta \nu$. 
If existing, the value of $\delta \nu$ is most probably
located below let say 3\,d$^{-1}$.   
The structure due to
the motion considered in the top panel of Fig.\,\ref{EGfignyq} almost
totally disappeared. In the power spectrum of the signal, the combination of noise 
and of aliasing is going to generate
parasitic peaks at frequencies that mainly depend on the position of the
object on the sky.  
Consequently, the time sampling is complex, and the aliasing is dependent
on the star being observed.

Looking at the third panel of Fig.\,\ref{EGfignyq}, we notice a marked aliasing
at $\delta \nu \, = \, 0.0317 \, \mathrm{d}^{-1}$ and at
$\delta \nu \, = \, 0.0634 \, \mathrm{d}^{-1}$. Therefore, in a few cases, noise
could conspire to shift the largest peak position from the progenitor peak
by this amount, thus producing pseudo-aliases of the progenitor that could dominate. 
However, in the present case, predominant aliasing seems
to be restricted to small $\delta \nu$, i.e.\ to small shifts of the progenitor peak.
Other values of $\delta \nu$ will be associated (or not)
to other regions of the sky.
Large shift aliasing is clearly possible but rather for values 
$\delta \nu \, = \, 12 \, \mathrm{d}^{-1}$. This should have no impact in the framework
of spectroscopic binaries. 

Here above, we have shown three examples illustrating the strong leakage characterising
the \gaia\ sampling (for RVS). 
An alternative view would be to compute some kind of typical spectral window. 
We obtain such a result using simulations. We generated 65\,536 positions randomly 
and uniformly distributed over all the sky. 
For each position, we extracted from the GOST software 
the series of observational epochs 
typical of this particular position and corresponding to the nominal scanning law (NSL) for RVS. 
On purpose, we discarded the first month of scanning which was characterised by the 
ecliptic pole scanning law (EPSL). For the latter, the precession
of the satellite was stopped and the scanning law clearly becomes more relevant to the scanning 
due to the spinning of the satellite, except for a slow rotation due to 
the orbital motion around the Sun.
We thus neglected the EPSL to instead concentrate on the effect of the NSL. 
For each simulation, we derived the corresponding amplitude spectral window
(for the RVS instrument).
Figure\,\ref{fig:EGfigrandommedian} illustrates (in red) the median of the amplitude 
spectral windows. Despite the fact that we restricted our simulations to the NSL,
the aliasing due to the spinning of the satellite is clearly visible. 
However, even the peaks at $\nu \, = \, 12 \,$d$^{-1}$ 
(and at $\nu \, = \, 16 \,$d$^{-1}$) are rather low, compared to the peak 
at  $\nu \, = \, 0 \, $d$^{-1}$, at most 30\% (in amplitude). 
We also plot in Fig.\,\ref{fig:EGfigrandommedian}
(in black) 32 randomly selected amplitude spectral windows. 
Clearly, the dispersion of the various realisations of 
the amplitude spectral window is rather large. 
Also, near the possible alias frequencies, the amplitude
spectral window could present values nearing zero. 
This confirms the fact that in most cases, 
the spinning pseudo-aliasing will have little, 
if any, impact on the global aliasing.
Although a few positions on the sky could still exhibit 
the effect of the spinning aliasing. 
Of course, if, over the 34 months, the EPSL scanning is dominating the definition of the 
times in the analysed time series, the effect will 
increase exhibiting a stronger spinning aliasing.
We must insist here on the fact that no signal is analysed in the present section; 
the dispersion of the amplitude spectral windows 
is originating from the randomness of the
selection of the sky positions and from the variety of nature of the aliasing 
as a function of the latter. For any time series, the observing time could be 
used to pre-calculate the particular spectral window independently of the signal.
\subsection{Search for the periodicity}\label{ssec:spectroSB1_period_search}
When a periodicity is present in a time series, the periodogram exhibits
a peak at the corresponding frequency (the progenitor), but other peaks
are also present at the location of the pseudo-aliases of the progenitor.
Contrary to the case of even sampling, the actual sampling generates
pseudo-aliases that are in principle lower than the progenitor peak. However,
the combination of the aliasing with background noise or other
phenomena (blending) induces the fact that the choice of the true progenitor
among the aliased peaks is never fully secure.

The total span of time of the DR3 time series is at most of 34 months and 
thus the largest periodicity accessible from the point of view of the Fourier
corresponds roughly to a frequency of 1/1000 d$^{-1}$. This value could actually
be larger taking into account the gaps generated by the actual sampling law.
The constraints should thus be expressed as a function 
of the actual length of the time series
and thus as a function of $\Delta T$.
It has been decided that it would be adequate to limit the range of investigation 
for periodicities to $1.5 \, \Delta T$
since beyond the possible orbit is certainly not well determined.
However, from the practical point of view, we decided 
to limit the range for period search
to $2.0 \, \Delta T$ although no orbit are computed for periods located
in this extension (see Sect.\,\ref{ssec:spectroSB1_processing} for more details).
The adopted frequency step is actually $0.01/\Delta T$ where 
$\Delta T$ is the actual total time-span of the individual time series.
This represents an oversampling of a factor of 100, which 
gives direct access to the amplitude
of the signal as the height of the peak. 
Based on the discussion detailed in the previous section, we decided that 
the adopted largest frequency to be explored is 4 d$^{-1}$.
This choice is rather atypical since it was chosen to correspond to 
the spin period. 

If the star indeed varies in a periodic way, the largest peak in the periodogram
is an estimate of the possible period, but  does not stand as definitive proof of its existence.
We should consider that a periodicity is proven if a minimum of three identical cycles
are observed. Although the long periods are indicative, they are associated
to large errors due to the small number of cycles included in
the computation.

The method used here is that of Heck-Manfroid-Mersch
\citep[hereafter HMM:][]{1985A&AS...59...63H,2001MNRAS.327..435G}, which advantageously
replaces the classical Fourier method
\citep[see e.g.][]{1976fats.book.....B}
in the case of odd sampling. In principle,
the highest peak is assumed to be the progenitor one and to correspond to the 
orbital period. This last point is particularly true for the case of
RV orbital motion, since most of the power is gathered in the fundamental
frequency; the power of the peaks corresponding to high harmonics is gradually
decreasing in the case of the Keplerian motion.
This is due to the fact that RV curves always present one 
simple cycle during the orbital period.
On the contrary, this is usually not the case for light curves of eclipsing binaries
and ellipsoidal variables where most of the power is gathered in twice the orbital 
frequency, corresponding to the average depth of the minima, whereas the power at the
orbital frequency is dominated by the difference between the primary and secondary
minima and thus is much weaker.
In their paper, HMM introduced the $\theta_{\mathrm{F}}$ statistic which is a
power related statistic \citep[see Eq.\,A.11 in ][]{2001MNRAS.327..435G}.
At a predetermined frequency and under the null-hypothesis of Gaussian white-noise,
it has the nature of the ratio of two independent $\chi^2$ statistics and it behaves
as a Snedecor F statistic with 2 and $N-3$ degrees of freedom.
It is reminiscent of a ratio of two quadratic forms, similarly to the use of nested models
(here a pure cosine model versus a constant term).
Since the global power is proportional to the variance of the white-noise process,
it is mandatory to normalise the power by a function of this variance. 
This is precisely the purpose of the $\theta_{\mathrm{F}}$ statistic.

The periodogram is explored and the 
frequencies corresponding to the various
peaks that are above a certain threshold 
are further retained. 
This threshold is defined as the limit above 
which the highest peak can be considered as
significant at a certain probability level.
This is further described in the next section.
We keep in a list up to 100 candidate
frequencies. 
This is more than the amount of 
independent ones in the classical Fourier. 
If too many peaks are listed, only the 100
highest ones are kept. 
For each of these selected candidate frequencies,
a fit is made of an eccentric Keplerian orbital model.
At this stage, there is a possibility to add one
additional candidate frequency coming for example
from CU7 (e.g.\ on the basis of a photometric curve).
In DR3 in particular, this possibility 
is only used for the subset of solutions of the type
{\tt{EclipsingSpectro}} combined with SB1 solutions 
\citep[see Sect. 7.6 by Siopis 
\& Sadowski in][]{pourbaixetaldoc,dr3-dpacp-179}. 
\subsection{Significance level of the period}\label{ssec:spectroSB1_period_sl}
Thus, to search for the period, a Fourier method is used. In power spectra, the peaks
in the periodogram are quite numerous. At a fixed frequency and
under the null-hypothesis of white-noise
with known variance, the height
of the peaks is distributed as a decreasing exponential
($\chi^2$ with a degree of freedom of two). 
A test must be made to
decide if a candidate frequency is significant by computing 
the significance level which is the probability under the 
null-hypothesis of a white-noise process of unknown variance to observe
a peak at least as high at a predetermined frequency:
this is the above-mentioned F distribution. 
Since the algorithm is considering the highest peak over all frequencies,
the adopted significance level is the probability under the null-hypothesis
of white-noise to observe the largest peak at least as high.
The actual statistic is thus
\begin{equation}
\label{EQTHETA}
\Theta_{\mathrm{F}} \, = \, \mathop{\mathrm{sup}}\limits_{{\mathrm{all}} \, \nu}\   
\theta_{\mathrm{F}}(\nu) \, .        
\end{equation}
If the significance level is very close to zero, the peak could not
be considered as generated by the noise; thus, there is considerable suspicion that
the outstanding peak is due to a deterministic periodic signal. In the case of even
sampling of the time series, such a significance could be easily computed. 
However, in the case of irregular sampling and particularly in the case 
of the \gaia\ sampling, the situation is much more complex. 
Unfortunately, no rigorous analytical formula actually exists for 
these precise circumstances. Consequently,
to test for the significance of the period,
the procedure relies on the use of Monte-Carlo simulations performed in the conditions
of the \gaia\ RVS sampling. 
In principle, this should be performed individually for each object characterised
by their own sampling, but this is not affordable for such a large number of sources.

The first dependency of the statistical distribution is
on the number $N$ of data points in the time series. 
The secondary dependency concerns the actual distribution in time of the data points.
The dispersion of the distribution decreases with $N$ increasing.
Therefore, we first attempted to perform such simulations
for each of the possibilities on this number $N$. Keeping this number constant, 
it is easy to notice through the simulations that 
the statistical distributions of the highest peak 
at fixed $N$ remain quite similar whatever is the actual distribution in time
of the data points (i.e.\ over the various simulations).
This is particularly true for the interesting regime where the significance level (SL)
is lower than 0.2, but also when $N$ is getting 
large (see Sect.\,\ref{ssec:spectroSB1_lookup}). 
Therefore, various Monte-Carlo simulations were performed and
the value of the height of the highest peak associated to various
fixed significance levels has been computed. 
From these simulations, a look-up table is built
that is used by the algorithm.

The probabilities ({\tt{conf\_spectro\_period}}) given in the 
data model (DM) are $1-S\!L$.  Despite the given name, this is thus the probability 
of rejecting the null-hypothesis of white-noise and is by no means the probability 
to have a periodicity.
Independently, it should be made clear that the
frequently encountered notion of significance in the database has nothing to do
with the presently defined significance levels. The significance rather corresponds
to the value of the parameter over its uncertainty. Such a significance is also
used in the spectroscopic channel in the particular case of the semi-amplitude 
$K$ where it is defined as $K/\sigma_K$.
\subsection{Build-up of the look-up table}\label{ssec:spectroSB1_lookup}
The computation of the empirical $S\!L$ function in our 
software can either be performed on-line
(on an individual source basis) or read from
a look-up table. The on-line computation, being necessarily more accurate 
but requiring an important amount of computational
time (more than 10 minutes per source), was no longer 
considered by the processing centre after it 
threw time-out exceptions during pre-operation tests. 
Therefore, for DR3, we adopted the method
of the look-up table which is approximate since it is based on typical average 
properties of the actual samplings (with $N$ data points) and not on the actual
individual sampling.

To build the look-up tables off-line, we gathered the observation dates from 
final RV curves of RVS sources used as standards, 
kindly provided by CU6, and we grouped them on the basis of their number of valid 
RV measurements. We preferred to rely on concrete observation dates 
rather than on the predicted NSL dates provided by the GOST tool in this exercise. 
Indeed, there are significant discrepancies between the NSL prediction and 
the final actual observation dates. The reason is that many spectroscopic 
observations fail to provide a valid RV measurement at the end of the
CU6 spectroscopic processing. This implies a sparser sampling.

For each source, we computed 10\,000 HMM periodograms 
of Gaussian white-noise realisations 
on the very same date sampling as that of the considered source, 
by adopting the same frequency step and limits as in the operation pipeline. 
For each periodogram, we retained the largest value of the $\theta_{\mathrm{F}}$, 
the so-called $\Theta_{\mathrm{F}}$ statistic. 
Then we merged all the values and ranked 
them by increasing order and build up an 
empirical cumulative distribution function (ECDF) 
with the 10\,000 values of the $\Theta_{\mathrm{F}}$ statistic.
At $N$ fixed, each realisation of the sampling provides a particular ECDF function.
The choice of 10\,000 in the framework of the ECDF allows us to assign false alarm probabilities with 
a precision better than 0.5\,\% at 95\,\% confidence.

The look-up table contains the set of median ECDFs for every 
given amount $N$ of valid transits. 
The real distribution of the number of observations per 
source is reflected in the number of ECDFs 
accounted for in each median with this scheme (i.e. sources with 17 valid observations, 
which are the most common in DR3, have a median ECDF 
covering the greatest number of individual ECDFs).
\begin{figure}[ht]
\centerline{
\includegraphics[width=0.4\textwidth]{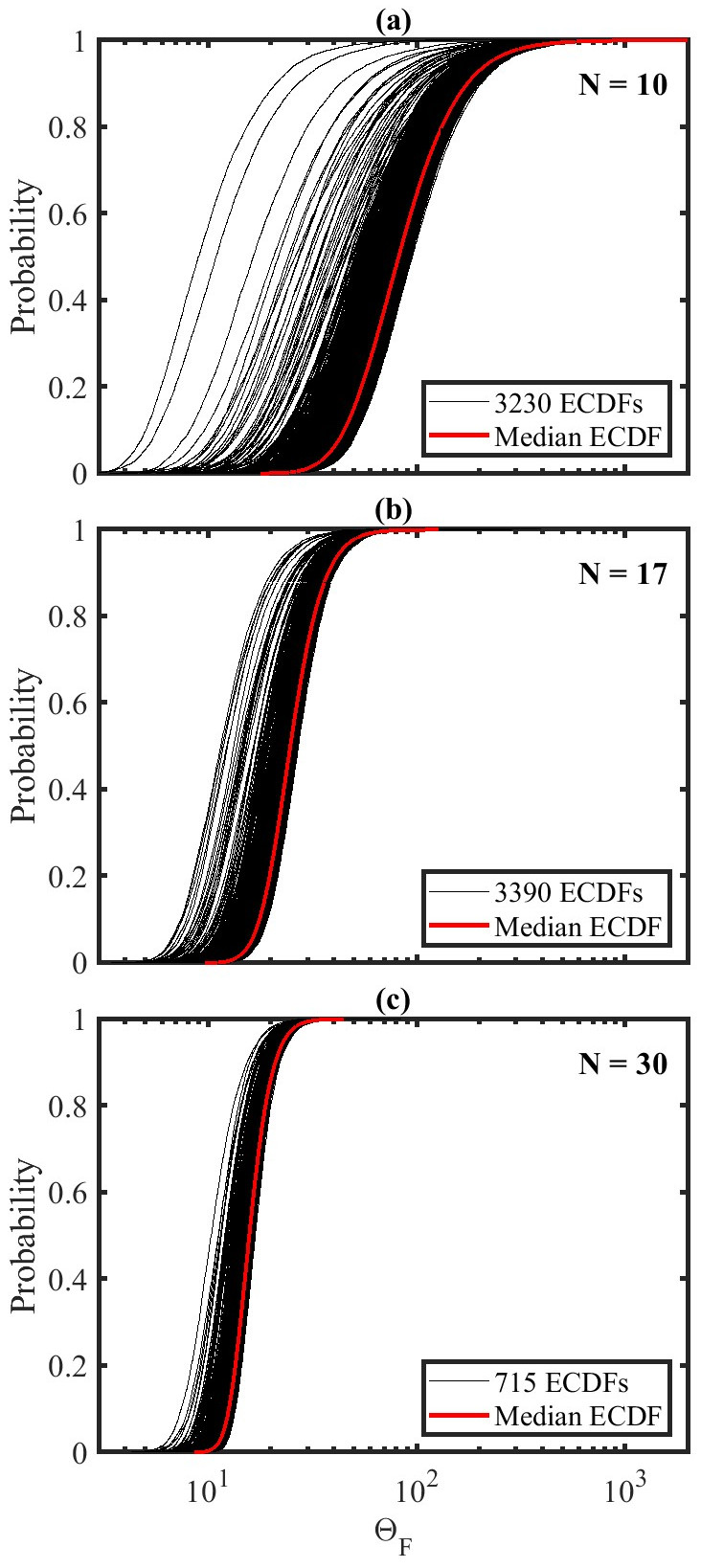}
}\caption{Examples of simulated ECDFs (in black) for different 
valid numbers of transits $N$ 
and their median (in red). 
The ordinate gives the probability to reject the null-hypothesis
of white-noise as a function of the height of the peak
($\Theta_{\mathrm{F}}$ statistic of Eq.\,\ref{EQTHETA}).
Panels (a), (b) and (c) correspond to $N$ = 10, 17 and 30  respectively. 
The number of simulated ECDFs in each panel mirrors the proportion 
of valid transits in the actual data. 
The sampling time statistically tends to be similar for large $N$ values, 
leading to a decrease in ECDF dispersion when $N$ increases.}
\label{fig:YDECDFs}
\end{figure}

Figure \ref{fig:YDECDFs} shows the simulated ECDFs for 
$N$ = 10, 17 and 30 and their corresponding median. 
As time sampling approaches the similarity of the samplings for large $N$ values, 
the ECDF dispersion decreases and their median 
approaches the regular time sampling CDF.
The values in ECDFs also decrease when $N$ increases. 
Indeed, fitting a cosine function to a large 
sample of pure noise will not significantly 
improve its variance, unlike in small samples where perfect fits may occur, 
especially when high frequencies are explored.
In addition, the signal-to-noise contrast increases with $N$.

When calculating the periodogram of an input time series with the length $N$, 
its corresponding adopted ECDF (on the basis of 
the $N$ value) is read from the look-up table.
The table has a comprehensive coverage of the number of transits, 
without gaps from $N$ = 10 to 72, but there are some entries for large $N$ values 
that are not present. In this case, the algorithm opts for 
the closest lowest $N$ input.

\section{Calculating the orbital solutions}\label{sec:spectroSB1_method}
Our main task is to derive simple orbital motion based on 
Keplerian models. The general model (see Eq.\ref{Equa_RV_Eccentric})
is highly non-linear and usually necessitates an a priori 
approximate knowledge of the true values of the parameters to fit.
This is not possible in the framework of a pipeline treating
so many objects. To define the proper method to use, 
we investigated numerous possibilities; 
they are evoked in Appendix\,\ref{sec:appA}.
Despite the many trials performed, it turned out that approaches
based on classical Levenberg-Marquardt iterations were much more
robust and precise, if appropriate precautions are taken.
As explained in Appendix\,\ref{sec:appA}, the alternative idea to perform
simulations were rejected for excessive processing time reasons.

The Keplerian fitting, even at fixed period, is time-consuming 
if it is to be performed over a vast set of periods: as for example
in a genuine Keplerian periodogram 
\citep[see e.g.\ ][their Sects. 2.3 and 5, respectively]
{2004MNRAS.354.1165C, 2009A&A...496..577Z}.

Therefore, we decided to strew subtleties over our procedure
in order to render it much more robust. The first step
is to search for periodicities using a periodogram (HMM method)
as described in Sect.\,\ref{sec:spectroSB1_periodsearch}.
This corresponds to a fit of sine/cosine, which is a linear
process (at fixed frequency). This is particularly convenient:
even if the RV curve is highly eccentric, and thus the power of the
signal is spread over several harmonics, the power related to 
the period of the orbit is still dominating the 
noiseless periodogram. If the number of cycles is sufficient,
the periodogram will always present increased power at the
correct period. However, the detection of a periodicity does not imply
that the motion is Keplerian. Therefore, after the 
establishment of a list of candidate periods,
the Keplerian model needs to be fit for each of them to 
deduce the orbital solution and corresponding 
parameters that are a better fit to the data. 
We first performed a Keplerian fit that is linearised at the
maximum (but not entirely) and represents a preliminary fit,
not definitive but robust.
The actual least-square fit was performed with the true 
anomaly as the independent variable.

The Fourier periodograms are known to be inefficient
at addressing the case of eccentric RV curves
\citep[for $e \ga 0.5$, see e.g.][]{2002A&A...392..671E, 2004MNRAS.354.1165C}.
We tend to think that this is due to the fact that
for eccentric orbits by opposition to the circular case, not all the
data points contain the same amount of information. In fact, 
for very eccentric orbits, the data points related to
observations around the maximum separation are more determining,
in particular for the semi-amplitude. If such data points are missing,
have not been acquired, this will have major impact at least
on the determination of the $K$ value. Consequently, this will have
a huge importance on the detectability of the signal: more abruptly,
we cannot detect something that is not observed.
Actually, this is not a problem of ability of the Fourier periodogram
to address eccentric orbits, this is a problem of fair sampling
of the RV curve.
The correct way to interpret things is that if a period is detected,
we are authorised to have some confidence in its existence.
Therefore, the nature of the Fourier periodogram improves the 
reliability and robustness of the algorithm at the expense
of a difficulty to have a proper estimation of the completeness.
This is a choice we made.

We finally decided to adopt the method fully described here below,
which is based on fully analytical and classical numerical
developments. After several tests, the adopted method appeared to be
faster, more stable, trustworthy and robust.
\section{Main NSS-SB1 processing chain}\label{sec:spectroSB1_procchain}
\subsection{Global structure of the chain}
\label{ssec:global_spectroSB1_procchain}
Five Java packages operate in sequential order in our pipeline. 
These are {\emph{InputChecker}}, {\emph{PeriodFinder}}, {\emph{SB1Refiner}},
{\emph{SB1CRefiner}} and {\emph{TrendSolution}}.
The package {\emph{InputChecker}} reads the input time series in the NSS data model and 
copies them into arrays of primitives that are shared with the entire pipeline.
This reduces the number of modifications required after every data model
update. It also performs sanity checks, data cleaning and some other modifications.
Finally, it passes the rearranged data to the four remaining processing packages.

The second package ({\emph{PeriodFinder}}) is working in two steps.
Its first step is to establish a list of candidate periods that are significant.
Its second task is to test each of the significant candidate periods 
by fitting the corresponding Keplerian model to the data, 
and to select the best period on the basis
of the comparison of the various models. 
The best of these models represents a first approximation of the orbit.
The third package ({\emph{SB1Refiner}}) has the charge to start from the selected
approximate solution and to refine it. As part of this computation, it derives the
variance-covariance matrix of the fit.
The fourth package ({\emph{SB1CRefiner}}) addresses the 
case of circular orbits when the
eccentric solution generates a badly configured variance-covariance 
matrix, and tries to adopt the best compromise
between a circular solution and a singular eccentric one. 
After these computations, a decision is to be taken to adopt the most adapted
periodic orbital solution.
The fifth package ({\emph{TrendSolution}}) deals with the 
RV time series exhibiting
a trend as a function of time (see Sect.\,\ref{ssec:spectroSB1_modelTrendSB1}).
This means time series that can be fitted by polynomials. 
These are acceleration models.
After the execution of all these packages, a choice is made to select the solution
that presents the best fit: SB1, SB1C, TrendSB1, or StochasticSB1.

\begin{figure}[ht]
\centerline{
\includegraphics[width=0.45\textwidth]{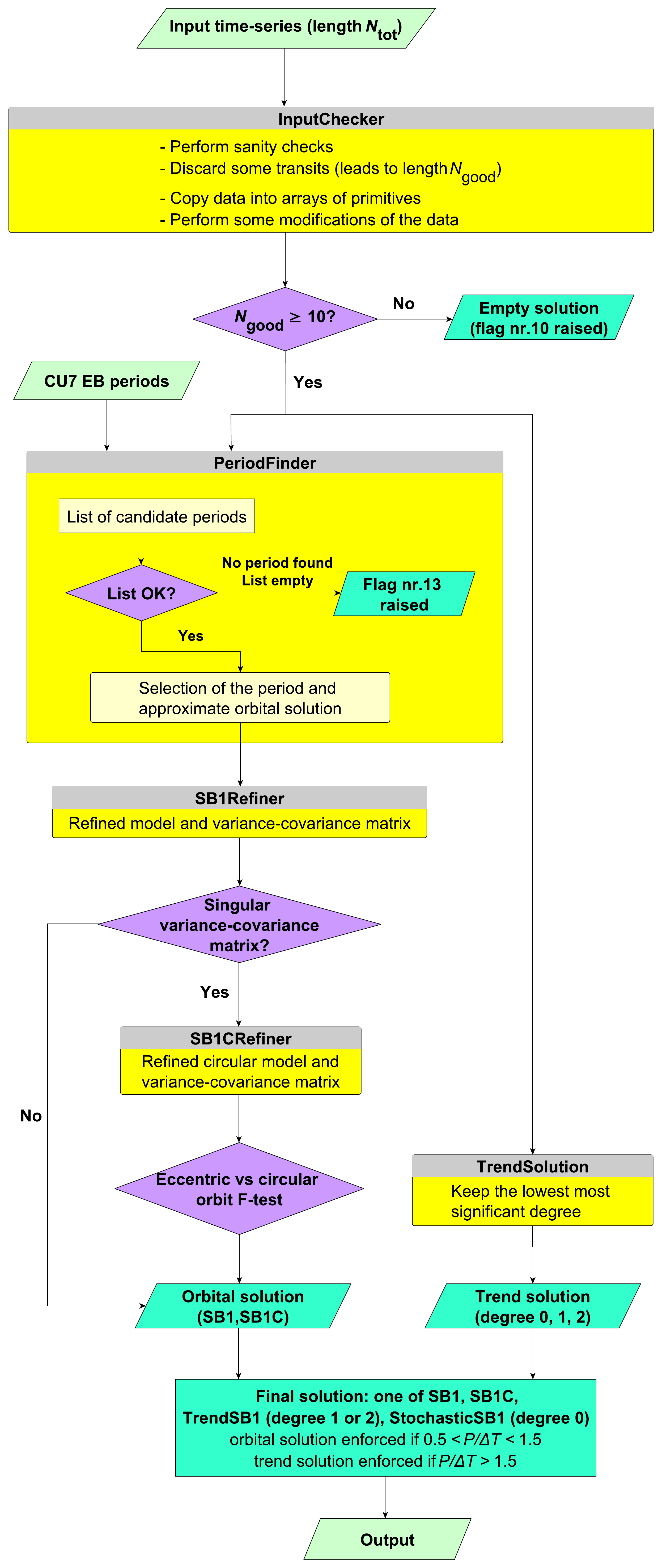}
}\caption{Flowchart of the single-line channel, as described 
in Sect.\,\ref{ssec:global_spectroSB1_procchain}. 
The details are fully explicited
in the description of the individual processing steps 
(see Sect.\,\ref{ssec:spectroSB1_processing}). 
$\Delta T$ is the actual time span of the observations.}
\label{fig:flowchart}
\end{figure}

The five packages are fully described in the respective five 
Sects.\,\ref{sssec:spectroSB1_processing_ingest}, 
\ref{sssec:spectroSB1_processing_candiper},
\ref{sssec:spectroSB1_processing_orbitalecc}, 
\ref{sssec:spectroSB1_processing_orbitalcir}, 
and \ref{sssec:spectroSB1_processing_trend}.
The flow chart of the full NSS-SB1 processing is described 
in Fig.\,\ref{fig:flowchart}, along with the
relations between the five packages mentioned above. 
It is interesting to notice that
after the upper violet lozenge that contains a test on $N_{\mathrm{good}}$
(defined in Sect.\,\ref{sssec:spectroSB1_processing_ingest}), 
there is a bifurcation towards two sub-chains that
receive the data.
One sub-chain is treating the objects exhibiting a periodicity 
that could conduct to an
orbital solution. The other sub-chain treats the cases 
where no periodicity is involved (only trends).
From the purely logical point of view, they are working 
in parallel although from the practical
point of view, they are executed in sequence.
During the processing of the full NSS-SB1 chain, some flags 
are raised to retain more information about
the path followed by the time series, as well as to 
point out a few characteristics that could be useful
for further treatment. 
In Sects.\,\ref{sssec:spectroSB1_processing_ingest}, \ref{sssec:spectroSB1_processing_candiper},
\ref{sssec:spectroSB1_processing_orbitalecc}, \ref{sssec:spectroSB1_processing_orbitalcir}, and
\ref{sssec:spectroSB1_processing_trend}, the flags are 
identified by their number and a full description
of their nature is given in Sect.\,\ref{ssec:spectroSB1_processing_flags}.
\subsection{Generalities about the fits}\label{sec:generalfits_spectroSB1_procchain}
After the {\emph{InputChecker}} intervention,
the four other NSS-SB1 processing packages are performing a set 
of linear or non-linear fits, involving linear algebra operations 
on moderate rank matrices (rank of matrices never exceeds 6).  
A home-made linear algebra toolbox, based on the Eigenvalue 
and Cholesky decompositions, has been developed for that purpose.

Although the former is known to be time-consuming compared to QR or Cholesky algorithms, 
it gives the best estimates of the condition number of involved 
algebraic systems, and thus ensures a better robustness. 
This is why it is applied all over the pipeline except for trend analysis, 
for which we adopted the Cholesky decomposition instead. 

Moreover, thanks to the eigenvectors, singular algebraic systems 
can still be solved by the Moore-Penrose pseudo-inverse 
technique \citep[Sect.\,2.6.2]{NumericalRecipes_3rdEdition}. 
This is extremely useful because it solves the discontinuity
problem between circular and eccentric models. Indeed, when 
eccentricity approaches zero, $\omega$ and $T_0$ parameters 
become progressively redundant \citep{1941PNAS...27..175S}, 
and thus the eccentric model becomes 
overstated, too powerful and thus prone to overfitting. 
The boundary between the two regimes, circular and eccentric, 
cannot be set by a fixed value of the eccentricity 
since it is governed by the signal-to-noise ratio of the input 
time series, and thus depends on many parameters 
such as the sampling, the amplitude of the signal and the 
involved noise \citep{1971AJ.....76..544L}.

Eigenvalue decompositions of matrices with rank smaller than 5 are 
performed analytically to improve robustness and reduce the processing time. 
Cardan-Tartaglia and Ferrari methods are applied to compute 
eigenvalues of rank 3 and 4 matrices, respectively.

The Cholesky decomposition is very suitable for the trend analysis 
in which we successively increment 
the polynomial degree. Since the $k^{\mathrm{th}}$ polynomial order 
is nested into the $(k+1)^{\mathrm{th}}$ one, 
we just need to update the $k^{\mathrm{th}}$ order curvature matrix 
decomposition to the new dimension instead 
of completely recomputing it.

Our home-made linear algebra toolbox defines the same 
singularity threshold for both decompositions. 
It is chosen equal to 10 $\times$ 2.22 $\times \, 10^{-16}$ $\times$ 
the square of the input matrix size. 
The value 2.22 $\times \, 10^{-16}$ is the lowest non-zero 
double precision number in current processors. 
The threshold corresponds respectively either to the ratio between the 
extreme eigenvalues (in the eponymous decomposition method), 
or to the square of the ratio of the extreme diagonal elements 
of the Cholesky matrix.
%
%
%
\subsection{Processing steps}\label{ssec:spectroSB1_processing}
\subsubsection{Ingestion of the data}
\label{sssec:spectroSB1_processing_ingest}
All stars that present RV variations
according to the following criteria
{\tt{rv\_chisq\_pvalue}}$\le$0.01
and
{\tt{rv\_renormalised\_gof}}$>$4
are processed by the pipeline starting with the {\emph{InputChecker}}
described here. Other details about 
these input data were already provided in 
Sect.\,\ref{sec:spectroSB1_input}.

In the ingestion process, the NSS input data are requested from 
the Main Data Base (MDB) and copied into {\tt{StarObject}} 
Java objects during data ingestion. 
A {\tt{StarObject}} contains the single star astrometric solution (the reference epoch, 
the mean position, the proper motions, the parallax, 
{\tt{ruwe}}, and {\tt{astrometric\_excess\_noise}}), 
the atmospheric parameters as measured by 
CU8 (Coordination Unit Astrophysical Parameters,
\citet{2023A&A...674A..27A, 2023A&A...674A..29R}), 
as well as the light curve classification 
and the orbital period if the corresponding source is suspected to be an 
eclipsing binary by CU7 (Coordination Unit Variability Analysis,
\citet{2023A&A...674A..16M}). 
In addition to these mean data, a {\tt{StarObject}} 
contains an array of $N_{\mathrm{tot}}$ 
{\tt{Transit}} Java objects, where each {\tt{Transit}} contains 
the relevant epoch data such as the source position, the integrated flux in 
the $G$, $G_{\mathrm{BP}}$ and $G_{\mathrm{RP}}$ bands, the RV 
(along with its uncertainty) as well as some related quality flags. 
Observation dates at the centre of the astrometric field are given in BJD. 
With the satellite spin, an observation at the centre of the RVS CCDs is performed
roughly 75\,s after the observation at the centre of the astrometric field:
this delay has been neglected.
The observation date, the RV and its uncertainty are extracted 
from each {\tt{Transit}} and 
copied into arrays of primitives for limiting code modifications 
required after every data model update. 
Those that are not finite numbers are filtered out. 
Observation dates and error bars must be positive numbers, 
while RVs must fall within the range of $\left[-1000,+1000\right]$\,km\,s$^{-1}$. 
The length of the time series $N_{\mathrm{tot}}$ is expected to be larger than or 
equal to 10 and if it is not the case a flag is raised (flag nr.8;
see Sect.\,\ref{ssec:spectroSB1_processing_flags}) and
the star is rejected from further processing.

After these sanity checks, we start by cleaning up non-valuable entries.
A {\tt{Transit}} is discarded when its RV is detected by CU6/MTA as an outlier, 
either on its own or together with the rotational velocity 
or the integrated flux in the $G_{\mathrm{RVS}}$ band or both. 
In addition, we also discard the {\tt{Transit}} if it exceptionally corresponds to 
a composite spectrum \citep[see][]{dr3-dpacp-161}.
The resulting time series after cleaning contains $N_{\mathrm{good}}$
data points distributed over a genuine individual time span $\Delta T$
(the difference between the two dates of observations at the extreme of the
interval).

Finally, we perform a few modifications. We correct uncertainties for sources with 
large {\tt{ruwe}}, as described in Sect \ref{sec:spectroSB1_input}.
The reference epoch J2016.0 (BJD 2\,457\,389.0) of the astrometric solutions, 
is subtracted from the observation dates for 
reducing the correlations between the derived solution parameters.
A cleaned time series of $N_{\mathrm{good}}$ RVs, still containing 
the required number of 10 observations, 
is shared with the whole software for the rest of processing, 
whilst, in the opposite case, it is rejected from the pipeline, and the corresponding 
flag is raised 
(flag nr.10; see Sect.\,\ref{ssec:spectroSB1_processing_flags} for flag descriptions).
The rejection from the pipeline is done by outputting an empty 
solution containing the flag and
no parameter values. Such an empty solution
is for internal use and statistical analysis. 
It is not further considered for the final output of the chain.
\subsubsection{Inventory of the periods and approximate model}
\label{sssec:spectroSB1_processing_candiper}
The package {\emph{PeriodFinder}} is first establishing a list of candidate frequencies.
The time series of length $N_{\mathrm{good}}$ is analysed thanks to an HMM periodogram,
as explained in Sect.\,\ref{ssec:spectroSB1_period_search}. The package reads the ECDF
corresponding to $N_{\mathrm{good}}$ (see Sect.\,\ref{ssec:spectroSB1_lookup})
and establishes a threshold above which the peak is considered significant against the 
null-hypothesis of white-noise. All the peaks in the periodogram
that are above the threshold value have their position in frequency
entering the list. This is clearly an approximation
since the test is constructed on the rigorous concept that we are 
looking at the highest peak only
(see Sect.\,\ref{ssec:spectroSB1_period_sl}). 
No fully sound counter-indication to this assumption has been identified yet.
Up to 100 frequencies/periods are accepted in the list. No additional smaller peak is 
considered above this amount. If the object under analysis is 
an Eclipsing Binary, we also add
to the list the value of the period issued from the photometry. 
We test the significance of any available external 
period (i.e.\ in DR3 from CU7) before 
adding it to the set of candidate periods.
For DR3 the periods entering the list
were selected on the basis of an adopted threshold 
probability value {\tt{conf\_spectro\_period}} larger 
than 0.95 ($S\!L$ smaller than 0.05).
If the list of significant candidate periods is empty, 
no full orbital solution is possible.
The relevant flag is raised (flag nr.13).

Since the aim of the pipeline is to compute orbital solutions, it is worth trying 
to limit the possibilities to Keplerian motion. 
For each of the frequencies $\nu_k$ in the list, a Keplerian model is fitted.
The pipeline is not performing the fit in phase
(defined by the candidate frequency under test) but in true anomaly. In this domain,
the RV curve to fit is a cosine. This of course necessitates the knowledge
of $T_{\mathrm{0}}$ and $e$. Since the eccentricity is a strongly non-linear
parameter, the pipeline proceeds as follows. 
Starting from a two-parameter grid of $e = (0, 0.5)$ 
and $T_{\mathrm{0}} = (0, 0.25, 0.5, 0.75) / \nu_k$, 
we minimise, using a Levenberg-Marquardt procedure, 
the quadratic sum of the differences between the
observed RVs and the model ones ($\chi^2$).
The vector of model RVs (coming from an expansion of Eq.\,\ref{Equa_RV_Eccentric})
is given by
\begin{equation}
RV(t_i) = \left[ \begin{array}{ccc}
1, & \cos v(t_i) + e, & \sin v(t_i)
\end{array} \right] \left[ \begin{array}{c} \gamma \\ \alpha \\ \beta \end{array} \right]
\label{RVARRAY1}
,\end{equation}
where $\alpha \, = \, K \, \cos{\omega}$ and
$\beta \, = \, K \, \sin{\omega}$.
From these $\alpha$ and $\beta$, the Campbell parameters 
$K$ and $\omega$ are easily extracted.
In Eq. (\ref{RVARRAY1}), $\gamma$, $\alpha$, and $\beta$ appear in a linear way. 
They can be optimised through an ordinary linear least-square regression
and this for every set of trial $v(t_i)$. 
This system of three linear and three non-linear parameters, 
can be efficiently solved with the 
separable variable technique \citep{Gene_Golub_2003}. 
Indeed, we optimise three non-linear parameters only: 
$e$, $T_0$ and $\nu_k$ ($\equiv$1/$P_k$). 
Consequently, the dimensional space of variables is reduced, 
leading to more robust and rapid convergences.
The pseudo-inverse implemented in the linear algebra toolbox 
smooths singularities near zero eccentricities. 
The rank of the involved linear systems is smaller than or equal to three, 
allowing them to be solved using more precise analytical computations. 
We give in Appendix\,\ref{sec:appC} a summary of the involved computations. 
The minimisation is limited to 1000 iterations for 
each of the 8 considered starting points, 
and this for every candidate frequency $\nu_k$. 
The solution is taken into account for further comparison, even when the maximum 
number of iterations is reached.
The relative convergence of the 8 different descending 
tracks are analysed to estimate the
robustness of the convergence (the arrival point should be the same in a perfect world).
Ultimately, the best solution is selected for the considered candidate
frequency. This algorithm
is quite robust and provides adequate fits for orbital curves
up to eccentricities of 0.995, well beyond what is necessary for the majority 
of the objects. The procedure is executed for every candidate frequency
and gives a $\chi^2$ of the fit that is translated into an $F_2$
statistic. 
The smallest value identifies the best fit and the best set
of parameters, including the period. The algorithm is efficient at finding the
best minimum, but the fit is performed in two steps, which is clearly an
approximation concerning the cross-covariance terms between families
of parameters (linear versus non-linear). 
Therefore, we need to refine the approximate model.

The use of a fit in the true anomaly space ensures that no analytical
solution could present a bad behaviour in the phase space.
Zones in the phase diagram where no data are present, the phase gaps,
could exist in some pathological cases. This problem is tackled in
Sect.\,\ref{ssec:spectroSB1_intfilt}.
\subsubsection{Final fit of the eccentric orbital solution}
\label{sssec:spectroSB1_processing_orbitalecc}
After having identified the best approximate orbital solution, 
the global solution refiner 
comes into play ({\emph{SB1Refiner}}). It is a Levenberg-Marquardt 
minimisation based on 
first order derivatives as given by \cite{1910PAllO...1...33S}. 
It returns the eccentric model 
as well as its variance-covariance matrix.
The algorithm thus performs, at the best minimum (the approximate model),
a classical least-square fit over all the parameters with all the
non-diagonal terms being taken into account. This has the advantage
to provide sound (but classical) estimations of the uncertainties on the parameters.
The variance-covariance matrix may be singular due to an 
over-determination of the model. 
In that case, a flag (flag nr.15) is written before activating the following package. 
If the matrix is not singular, the SB1 solution is forwarded
to the final decision module.
If the iterative process does not converge after 1000 iterations, a flag is raised
(flag nr.14).
If the numerical variance-covariance matrix is not semi-definite positive,
a flag is raised (flag nr.18).

The eccentricity is a parameter that suffers from a bias related to the noise
propagation. The RV noise could enforce a fitted value for $e$ that
is much larger than the true one.
This effect is sometimes compensated by enforcing $e=0$ for the solution 
when the eccentricity is not significant.
This is common practice in the literature to apply a test
and to decide to go circular (i.e.\ to decrease the number of free parameters)
for low-amplitudes and noisy cases. This has the pitfall to enforce
the circular solution for objects whose eccentric solution is only
marginally good (non-significant eccentricity). This produces a deficit of 
non-zero but small
eccentricity solutions.
The present algorithm does not do that. It never decides that
an orbit could be forced circular. It is allowed to possibly
(but not necessarily) go circular 
when the eccentric model cannot lead to a 
well-behaved non-singular variance-covariance
matrix.
\subsubsection{Fit of the circular orbital solution}
\label{sssec:spectroSB1_processing_orbitalcir}
The module described here is the {\emph{SB1CRefiner}}.
The best approached period obtained in the {\emph{PeriodFinder}} package is refined 
and fitted along with the linear parameters $\gamma$, $K$ and $T_0$ 
to fit a circular model to data 
(see the equations in Sect.\,\ref{ssec:spectroSB1_modelSB1C}). 
When a circular orbit is selected, $T_0$ corresponds to the maximum of velocity
by definition.
A flag is written when the derived variance-covariance matrix of the circular solution 
is singular; this is flag nr.16.
If the iterative process does not converge after 1000 iterations, a flag is raised
(flag nr.21). If the numerical variance-covariance matrix is not semi-definite positive,
a flag is raised (flag nr.19).

In some sense, the less complex circular model is nested into 
the eccentric model through the eccentricity. 
Indeed, if we put $e\,=\,0$ into the eccentric model, we obtain 
the equations of the circular model
(except for the trivial change of the definition of $T_0$).
It is thus tempting to opt for a classical statistical test for 
nested models to choose between the two. 
This is actually what is done.
However, this is not perfectly rigorous since, although the 
circular model is linear, the eccentric one is not.
Therefore, the number of degrees of freedom of the $\chi^2$ is not rigorously defined.
In any case, as a first approach, we decided to use the classical 
nested model statistic F
\citep[also described in][i.e.\ Lucy's test]{1971AJ.....76..544L}.
We thus test here the eccentric orbit model (as 
computed in {\emph{SB1Refiner}} but corresponding to a possibly unreliable
solution having necessarily at least one of the three 
flags nr.14, nr.15 or nr.18 raised) 
against the circular orbit model taken as null-hypothesis.
The eccentric model is thus retained when the Snedecor-F 
test significance level is lower 
than (or equal to) 0.05. This decision is taken regardless 
of the quality of the variance-covariance matrix of the circular model.
The {\emph{SB1CRefiner}} thus produces SB1 and SB1C type candidate solutions.
When Lucy's test has been applied, flag nr.22 is raised.
\subsubsection{Fit of trends}
\label{sssec:spectroSB1_processing_trend}
Not all the observed RV curves correspond to a true periodicity nor to
a pseudo-one. One widespread alternative possibility is the existence of a
mere trend: the data as a function of time can be represented by a polynomial
function of $t$ (see Sect.\,\ref{ssec:spectroSB1_modelTrendSB1}). 
The last computing package {\emph{TrendSolution}} applies polynomials to data 
to search for acceleration models (trends) rather than periodic orbits. 
We start by applying a zero degree polynomial (corresponding to a noisy constancy), 
then we successively increase the order at each step. 
The lower degree polynomial is thus nested in the new one. 
The choice between the lower and the higher degree polynomial 
is done through a classical
Snedecor-F test on the ratio of the improvement of the $\chi^2$ to the value of the
$\chi^2$ for the higher level model, since the polynomials given above
are indeed nested models.
A Snedecor-F test for nested models is thus applied.
The higher degree polynomial is accepted by the Snedecor-F test for a confidence larger 
than 0.99865 (3-$\sigma$ probability). 
We stop iterating at the 4$^{\mathrm{th}}$ order, or when the higher-degree model 
exhibits more than one extremum over the time baseline.
This last criterion has been adopted to avoid any possible overlaps with the orbital models
with periods between $\Delta T$ and $\mathrm{1.5} \, \Delta T$ (see below).
Most of the objects entering the trend category are probably long-term periodic
curves but with a timescale much longer than the present time span of the
mission ($\Delta T$). 
The code is built to accept polynomials of degrees
1, 2, 3, and 4 but in DR3, only degrees 1 and 2 were encountered
(at the selected probability threshold).
A flag is written when the derived variance-covariance matrix of the trend solution 
is singular; this is flag nr.17.
If the numerical variance-covariance matrix is not semi-definite positive,
a flag is raised (flag nr.20).
A regrettable mistake occurred in the code and, for the last operational run,
the flags raised were not the correct nr.17 and nr.20 but instead
flags nr.15 and nr.18, respectively. This problem has been corrected for DR4.

The coefficients of the polynomial are included in the database
but the work is performed in an Hipparcos-like formalism
\citep[see Sect. 2.3.3,][]{1997yCat.1239....0E}.
The coefficients saved in the database are 
related to the derivatives of the 
velocity curve and
are expressed with respect
to the DR3 reference epoch (J2016.0, equivalent to BJD\,2\,457\,389.0). 

We thus decided to express the trends as a function of the various derivatives.
The simple mathematical model is
\begin{equation}
V(t) \, = \, V_0 \ + \, \frac{\partial V}{\partial t} \, t \, + \,
\frac{1}{2} \, \frac{\partial^2 V}{\partial t^2} t^2
,\end{equation}
where $t$ is the time (in days) relative to the reference epoch.
We thus fit to the time series the polynomial,
\begin{equation}
a \, + \, b t \, + c t^2
\end{equation}
with the $c$ value potentially fixed to zero.
In the case of the first degree polynomial ($c$\,=\,0), we adopt the following
rigorous formulae,
\begin{equation}
V_0 \, = \, a,    \\
\frac{\partial V}{\partial t} \, = \, b  \, \, \, .
\end{equation}
The $a$ and $b$ parameters are not correlated if
$\sum t_i \, = \, 0$. This is approximated by subtracting to the
BJD dates the reference epoch (J2016.0) as a proxy
of the middle of the current mission. Practically, not all actual time series
are centred on this value since some transits might be missing
(e.g.\ more numerous before the mid-epoch or alternatively after it). 
This is in any case still an
approximation since it is implicitly assumed that the distribution
of the sample times are homogeneously apportioned. This is certainly
not true, particularly if we include the EPSL data.

In the case of the second degree polynomial ($c$ fitted), we note that the
fitted polynomial can be expressed as
\begin{equation}
a^\prime \, + \, b t \, + 
\, c \, (t^2 - \frac{(\Delta T_{\mathrm{m}})^2}{12}) \, \, \, ,
\end{equation}
where $\Delta T_{\mathrm{m}}$ is the total duration of the mission treated
in the framework of DR3.
We thus adopt the approximate formulae,
\begin{equation}
V_0 \, = \, a^\prime \, - \, c \, \frac{(\Delta T_{\mathrm{m}})^2}{12}   \,\, \, \\
\frac{\partial V}{\partial t} \, = \, b   \, \, \,  \\
\frac{\partial^2 V}{\partial t^2} \, = \, 2c   \, \, \, .
\end{equation}
The term shifting $t^2$ by $\frac{(\Delta T_{\mathrm{m}})^2}{12}$ has been introduced
such that the contribution by the second derivative is tending to be neutral
over the interval $\Delta T_{\mathrm{m}}$. This of course induces a shift on $V_0$.
The first derivative is considered as averaged over the 
full interval. This property is here also true for an homogeneous distribution
of the observing times. 
A strong disparity of the observing times could have introduced some biases.

Initially, the threshold was chosen at a probability level of 0.95
(following the Snedecor-F test of nested models) and polynomials of first, 
second, and third order (with only one extremum) were found. 
Following validation, we decided to be more conservative and we 
instead adopted a new threshold (0.99865); this
induced a downwards cascade among the different degree categories and 
restricted the palette of solutions to first and second order polynomials.
But the decision occurred too late and no more operational run took place. 
Therefore, we kept formally adopting the threshold of 0.99865 
and decided to redistribute the various
solutions and to blacklist the objects that, in case of a 
rerun, would have been discarded as trend solutions. 
This certainly induced the loss of a small number of objects in other categories.

The acceleration identification as performed in DR3 can be affected by false detection,
revealing the subtle effect of an unfortunate sampling on the polynomial fit.
In the upcoming data releases, changes to the algorithm will be made to ensure that 
trend models are first detected in the lowest frequency region of the periodogram, 
as introduced in \cite{2022MNRAS.517.1849B}.
\subsubsection{Selection of the best solution}\label{sssec:spectroSB1_processing_stoc}
Therefore, the {\emph{SB1Refiner}} and the {\emph{SB1CRefiner}} 
packages provide an SB1 or SB1C candidate
solution. They could in some cases provide no solution if no period is found.
The {\emph{TrendSolution}} package delivers in parallel polynomial 
fits and thus candidate trend solutions
(of degree 0, 1 and 2).
It is now necessary to decide which model is performing the 
best interpretation of the data.
The adopted orbital, circular or eccentric, candidate model is 
ultimately compared to the less complex 
best trend candidate  
(i.e. the null hypothesis is that the model is a trend). 
Acceptance of the orbital model is granted when the 
Snedecor-F test confidence is higher than 0.95.
If this is the case, the final delivered solution is the selected SB1 or SB1C.
If this is not the case, the trend solution is adopted. 
If the degree of the fit is 1 or 2, the object
is entering the delivered output as a TrendSB1 solution. 
If the degree is zero, we have to deal with
a constant RV with excess noise compared to the expected uncertainty.
In this last case, the solution is output as a StochasticSB1.
Thus, when no significant periodic variability nor trend are 
possible models, the algorithm
associates the star to a stochastic solution. 
These are objects that have very little chance
of being associated to a mere orbital or trend motion. 
As explained elsewhere, some badly sampled
orbits and high degree multiple systems could fall into this category,
perhaps momentarily. Some of them will certainly benefit from an increase in the
number of transits. These StochasticSB1 solutions are not provided in the DR3 catalogue.
The correct choice between an orbital solution and the trend one is
particularly difficult because the independent variable is time for the trends
and let say phase for the well covered orbits. The time entering in a non-linear way
in the Keplerian equations and the first selection of the period 
being based on a HMM periodogram, 
the statistical Snedecor-F test is not fully correct. 

Actually, the procedure described above has been applied when orbital solutions
presented a period lower than $0.5 \, \Delta T$; 
it concerns the regime where the orbital
motion is expected, when selected, to be convincingly robust. 
For longer periods, the folding with
the period does not permit us to test the perfect overlap 
homogeneously at all phases. We found it preferable to
adopt an alternative procedure. If the orbital period is comprised between
$0.5 \, \Delta T$ and $1.5 \, \Delta T$, we decided 
to enforce the choice of the orbital
solution. For periods larger than $1.5 \, \Delta T$, 
i.e.\ outside the zone of possible convincing
orbital motion, we decided to enforce the choice of 
the TrendSB1 solution. This last point explains why 
we decided to explore the periodogram
in the range between $1.5 \, \Delta T$ and $2.0 \, \Delta T$.
\subsection{Quality of the solution}\label{ssec:spectroSB1_quality_indic}
For each object processed, along with the various resulting model-parameters, 
the code also computes some statistics and various entities that might
help in estimating the quality of the solution.
The first statistic is the $F_2$ one given by
\begin{equation}
F_2 \, \, = \, \, \sqrt{\frac{9 n}{2}} \, \left[ \left( \frac{\chi^2}{n} \right)^{1/3} \, 
+ \, \frac{2}{9 n} \, - 1 \right]
,\end{equation}
where $\chi^2$ is the classical statistic of the least-square fit and where $n$ is
the related number of degrees of freedom. 
This statistic was first introduced by
\citet{1931PNAS...17..684W}.
Since the eccentric fit is non-linear in nature,
we performed various simulations to test the estimation of the number of degrees
of freedom to use. Quite surprisingly, it turned out that, for the Keplerian motion,
the number of degrees of freedom $n$ is very close 
to the number of data points minus the 
number of parameters (similarly to the linear case).

Besides the $F_2$, the code also
computes the $\delta \phi$ which is the largest gap in the phase diagram
where no data points are present. It also computes the solution efficiency
whose definition can be found in 
\citet{1990ApJ...358..575E},
and is corresponding to a measure of 
the correlation between the computed
parameters. A total absence of correlation,
and thus a diagonal variance-covariance
matrix, will give an efficiency of 1.
Values of the efficiency tending to 0
indicate strong correlation. Practically,
an efficiency identically equal to 0 is
not possible since this is associated to 
a perfectly singular matrix to invert.

In addition, we also compute the solution
significance defined for spectroscopic 
data as $K/\sigma_K$.
Finally, we should add the significance level of the periodicity as described
in Sect.\,\ref{ssec:spectroSB1_period_sl} and expressed as a confidence ($1-S\!L$).
\subsection{Flagging}\label{ssec:spectroSB1_processing_flags}
Beyond the determination of the various parameters, the code also produces
various flags that allow the user to have a good 
understanding of the results and to select some solutions that could be
better or more secure than others. 
The flags generated by our NSS-SB1 processing are incorporated 
into a more global set of flags
issued from the global NSS processing. 
We will restrict the discussion here to the set
of flags directly related to the SB1 processing chain, 
with a few information linked to flags
issued from the SB2 processing that is fully described 
in \citet{dr3-dpacp-161}.
All the NSS flags are included in a unique long variable, whose every bit
acts as an individual logical flag. All the flags involved in the
NSS spectroscopic processing are enumerated in
Appendix\,\ref{sec:appI}.
As an example, the {\emph{SB1Refiner}} processes the time series of an object
and produces a singular matrix. The flag nr.15 is thus raised, giving to the
global flag an additive value of 32\,768.
The object is then processed by the {\emph{SB1CRefiner}} and, let say,
it concluded to a singular matrix associated to the circular orbit.
The flag nr.16 is thus raised (associated to a value of 65\,536).
Consequently, the object is associated to both flags nr.15 and nr.16 raised,
giving together a value of 32\,768\,+\,65\,536\,=\,98\,304.
The object could appear in the catalogue as a {\tt{TrendSB1}}
with the two above-mentioned flags raised.
Although the majority of the flags are important for the decision tree of the
algorithm, some of them are purely indicative. This is the case of the flag nr.24
that draws the attention to the fact that the attributed eccentric orbit
appears, in the eccentricity versus log$P$ diagram, outside the zone
where the circularisation of the orbit could have partly taken place.
The borderline used is inspired from the one defined by
\citet{2005A&A...431.1129H} but modified by us\footnote{Using the formalism of
\citet{2005A&A...431.1129H}, we estimated a value of $P_{\mathrm{cut-off}}$
on the basis of a fairly crude analysis of the full SB9 catalogue 
\citep{2004A&A...424..727P}. This estimate has been deduced without proper treatment
of the dispersion of $e$ due to noise and is considered as a conservative
border. The purpose is to flag extremely pathological cases to trigger the
cautiousness of the general user.}.
Flag nr.25 just indicates that the selected adopted period was compatible with
one period suggested by the photometry. Flag nr.22 informs that the Lucy test
has been applied. Flags nr.11 and 12 concern SB2 objects that turned out
to be impossible to treat as SB2 either because the amount of 
RVs was too small for the secondary (but not for the primary) or
because the period found by the SB1 chain was not coherent with the one
of the SB2 chain \citep[see][]{dr3-dpacp-161}.
Flag nr.26 is raised by the SB2 processing to indicate that the pair of RVs
of the SB2 measurements were not anticorrelated 
\citep[more details are given in][]{dr3-dpacp-161}.
\section{Post-processing}
\label{sec:spectroSB1_postproc}
\begin{figure}[ht]
\centerline{
\includegraphics[width=0.45\textwidth]{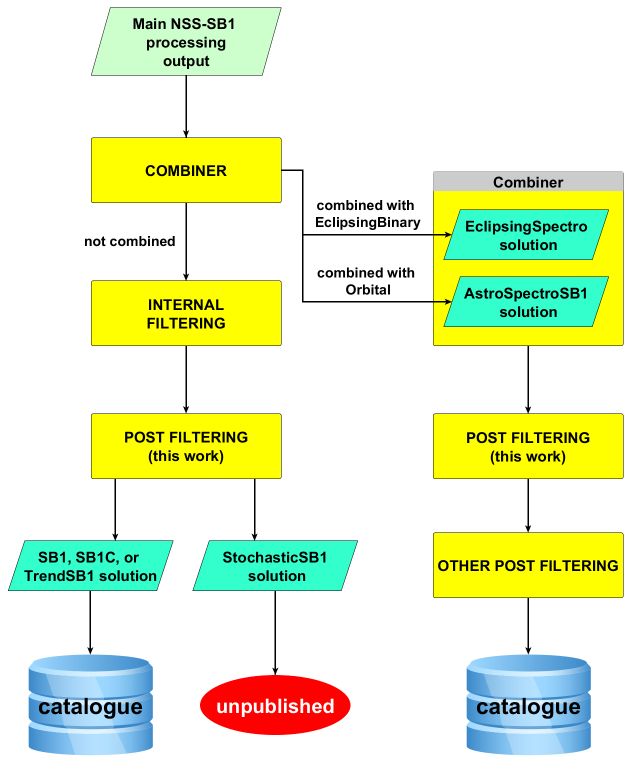}
}\caption{Flowchart of the post-processing that is applied
after the output of the flowchart of Fig.\,\ref{fig:flowchart}.
The channel on the left side solely concerns the present work.
The channel on the right side also concerns the combiner. The post-filtering
(this work) concerns the post-filtering announced in 
Sect.\,\ref{ssec:spectroSB1_postfilt} and detailed in
Sect.\,\ref{ssec:spectroSB1_add_cons_postfilt}. The other post-filtering
only concerns the astrometric and/or the photometric channels.}
\label{fig:flowchartpopo}
\end{figure}

The post-processing described here further cleans and distributes the results of the
NSS-SB1 main processing chain. The detailed flowchart is provided in
Fig.\,\ref{fig:flowchartpopo} and additional explanations are given in this section.
\subsection{Internal filtering}\label{ssec:spectroSB1_intfilt}
From summer 2020 to spring 2021, the NSS pipeline processed
the \gaia\ data in the framework of several validation runs.
At least seven of them were concerning the software 
linked to the present paper. 
This processing helped in finding remaining failures of the algorithms and
in correcting them. This also helped in fine-tuning the decision tree
of the pipeline by adjusting some statistical thresholds.
The operational runs took place in summer 2021.

In order to maintain the quality level of the various solutions,
we defined an internal filtering of these solutions that would
allow us to retain the very best and most secure ones. The aim is also to spot out 
and reject solutions that are problematic. 
Most of the filtering is done via the definition of thresholds. Very often,
the selection is made in a statistical way and by using probability thresholds
based on the distribution of some statistical variables. As always, the
selection near the borders is never perfect. We detail here the internal
filtering for SB1 solutions.

Since the individual (per transit) RVs are searched for
in the range between --1000 km\,s$^{-1}$ and +1000 km\,s$^{-1}$,
we found wise to test that all solutions have 
$ \lvert \gamma \rvert \le \, $1000 km\,s$^{-1}$. This is the case 
but this sane test is not highly restrictive.

On the opposite, one of these important statistics is the $F_2$.
The $F_2$ statistic represents a classical goodness of fit
criterion that is directly related to the objective function
of the fit and to the classical $\chi^2$ statistic.
The $F_2$ statistic is expected to have under the null-hypothesis
a Gaussian distribution. Typically, above $F_2$\,=\,5, the solution 
is not satisfactory and presents extra-dispersion of the residuals,
be it due to the data or to the model. In order to be conservative,
we decided to put the threshold such that  $F_2 \, \le \, 3$.

In the course of the NSS-SB1 processing, various flags are raised, tracking 
the details of the decision tree and of the problems encountered. 
As explained in 
Sect.\,\ref{ssec:spectroSB1_processing_flags}, some combinations of 
flags imply that the solution is not acceptable or is presumably inaccurate
or too hazardous. The corresponding selection was of course considered.
Since the internal filtering is supposed to protect the results from
spurious solutions and that, in the present purely spectroscopic
SB subcatalogue, no confirmation comes from other channels, we decided
to be highly cautious for this step.
For the SB1 solutions, we discarded objects having at least
one of the three flags nr.14, nr.15, nr.18 raised. As a consequence, this
suppresses all the SB1 solutions issued by the {\emph{SB1CRefiner}}.

We further found that a conservative attitude is to restrict
the palette of the solutions to those yielding a
$K \le \, \, $250 km\,s$^{-1}$ which eliminates a few cases
of surprisingly large semi-amplitudes that must be further validated
(a work beyond the scope of the present paper).

We also used a threshold on the solution efficiency. 
Any solution efficiency below 0.1
is rejected because the corresponding fit is considered as spurious
\citep{1990ApJ...358..575E}.
The solution efficiency is part of the catalogue, allowing the user
to adopt a more restrictive attitude.

When the supposedly periodic RV data are folded in phase, it happens
that a zone in the phase diagram is not populated by any
data points. If this region is too wide, the corresponding solution could be
pathological with respect to the sampling. Therefore, we choose
to discard the solution with a too large $\delta \phi$ where
this value represents the largest interval in phase with no data points.
We reject solutions leading to $\delta \phi > 0.3$. The value of the gap 
for remaining solutions is not propagated to the SB subcatalogue but will be 
accessible in DR4 with the release of the epoch RVs. The threshold value itself
has been optimised on the basis of the various validation runs
(see Sect.\,\ref{sec:spectroSB1_validation}
and in particular 
Sect.\,\ref{ssec:spectroSB1_validation_intern}). Due to a small bug in our software,
a small part of the objects presenting $\delta \phi > 0.3$ have not been
properly rejected in DR3.

The mere existence of the periodicity in the data is a strict test
of the validity of the orbital solution. It is reasonable to consider
values for the probability {\tt{conf\_spectro\_period}} from 0.95 to 0.995
or even stricter. During the validation, it appeared a strong population 
with short-period orbital solutions sometimes associated
to relatively large eccentricities. This is somewhat in contradiction
with the expected circularisation phenomenon (or these are not 
properly speaking orbital solutions). Therefore, we decided again to be 
conservative and to put a threshold that is period dependent.
We hope to improve this procedure for DR4.
For periods above 10 days, we selected a lower threshold of 0.95
on {\tt{conf\_spectro\_period}} (significance level $<$ 0.05).
For periods less than one day, we adopted 0.995 as the lower limit. 
In-between, the lower threshold on {\tt{conf\_spectro\_period}} is 
given by $0.995 - 0.045 \log P$.
The solutions associated to a value less than 0.95
are filtered out. The solutions associated to a value above
0.995 are kept. Solutions in-between are kept but a flag 
(nr.13, i.e.\ 8192; see Appendix\,\ref{sec:appI})
is raised if they are below the defined threshold.  

Another natural statistic is the spectroscopic significance
defined as $K/\sigma_K$. This ratio should not be confused with the
probabilistic quantity canonically named significance level.
A high significance is necessary
to adopt a solution. However, we should not forget that
$\sigma_K$ is itself an estimator and that statistical fluctuations
could give a value of $\sigma_K$ that is exceedingly small, 
boosting the significance to very large values.
The opposite case is also true. We decided to keep the significance
as a possible sorting criterion for the future data applications.
Therefore, we only rejected all the solutions with a significance
lower than 5. This threshold has been fixed for the internal
filtering, but it could be advisable to the general user to select a value
of the order of 20 to 50 depending on the details of the scientific application.
An example of this way of fine-tuning the selection is illustrated in
\citet{2023A&A...674A..34G}.

Finally, some solutions corresponding to values of the argument of the periastron
$\omega$ are accompanied by a very large value of their estimated uncertainty.
Conservatively, solutions characterised by
a 1-$\sigma$ uncertainty larger than 2$\pi$ (in radians) are considered 
as problematic, since this actually means that the $\omega$ is 
properly undefined. We preferred to discard these objects.
The filtering could have been more stringent but we preferred 
to rely on user's discernment.
As expected, the majority of large uncertainties on $\omega$ correspond to very small
eccentricities ($e \, < \, 0.05$) characterised by rather large uncertainties.
However, some have eccentricities that can be considered as significant.

An internal filtering of the kind performed here above was also applied
to the SB1C type of solutions (except of course for the last one on
$\sigma_\omega$). For the SB1C solutions, we discarded objects with either flags nr.16 or nr.19
raised. 

An internal filtering was similarly defined for the
TrendSB1 solutions. This last filtering was on $F_2$, on the bad flags and
on the equivalent of the above-mentioned gap in phase. 
Concerning bad flags for TrendSB1 solutions, we should have discarded objects with 
either flag nr.17 or flag nr.20 raised. However, these flags were never activated.
We decided to filter out on the flags nr.15 and nr.18, although there could be some
unfortunate interactions with the SB1 channel. We rejected
solutions presenting a gap (in time) in the time series larger than
0.3$\Delta T$. 

For reasons that will become evident in 
Sect.\,\ref{sssec:spectroSB1_results_illust_probl_fake},
we also decided to discard objects with flags nr.11 or nr.12 raised.
The details on the various proportions of rejected objects
are given in Sect.\,\ref{ssec:spectroSB1_results_numbers}.
\subsection{Combined solutions}
\label{ssec:spectroSB1_combiner}
After the spectroscopic processing described here above,
the solutions are delivered to the combiner.
The combiner and combinations are described elsewhere 
\citep[see Sect. 7.7 by Gavras \& Arenou in][]{pourbaixetaldoc}
and are not extensively addressed here.
This combiner explores opportunities
to unveil solutions common to other channels.
In the framework of DR3, two possibilities were considered.

Either the NSS combines objects from the spectroscopic channel
with those issued from the photometric channel 
and present in the class {\tt{EclipsingBinary}}, 
i.e. presenting eclipses in photometry
\citep[see Sect. 7.6 by Siopis \& Sadowski in][]
{pourbaixetaldoc, dr3-dpacp-179, 2023A&A...674A..34G}. 
The main criterion for the combination is the existence of a common value 
for the period. If the tentative combination is successful, a global fit
over the photometric and the spectroscopic data is performed.
In such a case, the new (improved) parameters supersede the old one
and are introduced in the catalogue in the new class 
{\tt{EclipsingSpectro}}
\citep[see Sect. 7.7 by Gavras \& Arenou in][]{pourbaixetaldoc}. 
Most of the contributors to this class
are issued from our set of SB1 solutions, although some other types
of solutions could also be considered (SB1C, SB2, SB2C).

Alternatively, NSS can try to find objects for which the results of the
astrometric channel \citep{2023A&A...674A...9H, 2023A&A...674A..34G} can be combined
with objects having a solution issued from the spectroscopic pipeline.
The relevant combined solutions and the related objects
are then redirected towards the new class labelled
{\tt{AstroSpectroSB1}}
\citep[see Sect. 7.7 by Gavras \& Arenou in][]{pourbaixetaldoc}. 
New fits are here also performed and the
old values of the parameters are also superseded.
We recall that in this case, the fit in DR3 is done in the framework
of the formalism of Thieles-Innes \citep{2023A&A...674A...9H, 2023A&A...674A..34G}.
This formalism implies that in DR3 {\tt{AstroSpectroSB1}}, the combined solutions
include two different values (one astrometric and one spectroscopic) for the argument
of periastron.
The combination with astrometry accepts several types of possible spectroscopic input
(SB1, SB1C, TrendSB1, StochasticSB1).

The objects corresponding to a successfully combined spectroscopic
solution are removed from the spectroscopic output.
Even after the combination process, it might be that an object appears
in two classes.
In the framework of DR3, these are the sole allowed and/or activated possibilities. 
The criteria to consider a combination opportunity are described in details 
in Sect. 7.7 of \gaia\ archive 
documentation\footnote{{{\url{https://gea.esac.esa.int/archive/documentation/GDR3/}}}}
\citep{pourbaixetaldoc}.

If, for an object, the combination with the spectroscopic output 
turned out to be impossible, the sole spectroscopic solution
(and the object) enters the newly created class that is
eponymous of the spectroscopic  solution type, either
{\tt{SB1}}, {\tt{SB1C}}, {\tt{TrendSB1}} or
{\tt{StochasticSB1}}.
Contrary to the first three, the {\tt{StochasticSB1}}
class is not published in DR3. The classes {\tt{EclipsingSpectro}}
and {\tt{AstroSpectroSB1}} are not further considered in the present paper.

To avoid losing any interesting objects (and in a purely 
prototypical context), the combination was performed
before the internal filtering. It could have been more secure to perform
it after the internal filtering. Also, the combination of the
astrometric solution with the StochasticSB1 one should certainly be a matter
of concern. The StochasticSB1 solutions
by essence present no actual periodicity and are not expected to be further
combined (in principle). On the opposite, solutions that are purely spectroscopic
and are ultimately populating the classes {\tt{SB1}}, {\tt{SB1C}} and {\tt{TrendSB1}} 
were necessitating increased caution and were submitted to the
internal filtering.
\subsection{Post-filtering}
\label{ssec:spectroSB1_postfilt}
On the basis of the use of results from
other origins, it is possible to further
refine the set of solutions that
is coming out of the chain. This is made
by blacklisting spurious solutions.
For a good understanding of the reader,
this post-filtering is explained
in details below in
Sect.\ref{ssec:spectroSB1_add_cons_postfilt}
after the validation sections.
\section{Results}\label{sec:spectroSB1_results}
DR3 is the first release providing orbital solutions for the RV time series. 
As such, the present results must be considered preliminary even if the 
performances are already good.
\subsection{Tallies}\label{ssec:spectroSB1_results_numbers}
Some 33\,812\,183 objects have validated (non NULL) RV values in DR3.
Among them, some 995\,315 objects obey the input criteria. We recall them:
RVs always between \mbox{--1000} and +1000 km$\, \mathrm{s}^{-1}$, 
$G_{\mathrm{RVS}}^{\mathrm{int}}$ 
brighter than 12,
effective temperature between 3875 and 8125 K, and {\tt{rv\_renormalised\_gof}}
above 4. Only the objects suspected of variability enter the orbital solution chain.
Thus, an additional cut-off of {\tt{rv\_chisq\_pvalue}} at 0.01 
(kept below 0.01) further decreases the sample.
A little more than 866 kilo-objects (866\,403)
entered the present main NSS-SB1 processing. If 
we consider the SB1+SB2 processing, the total amount is 876\,644.
The repartition of the results among the different types of solutions
introduced in Sect.\,\ref{sec:spectroSB1_model} is given in 
Table\,\ref{tab:tableresult1}. The missing objects
correspond to a lack of possible solutions.
An important contribution comes from the objects
having $N_{\mathrm{good}}$ smaller than 10.

After this preliminary main NSS-SB1 processing, the internal filtering is applied,
as explained in Sect.\,\ref{ssec:spectroSB1_intfilt}.
The corresponding numbers are given in
Table\,\ref{tab:tableresult2} for the SB1-type solutions,
in Table\,\ref{tab:tableresult3} for the SB1C-type solutions and finally,
in Table\,\ref{tab:tableresult4} for the TrendSB1-type solutions.
We give the effect of each criterion on the tallies.
The internal filtering has been actually performed in the same
sequence as given in the table.
These tallies are however only indicative because they can be dependent on the
order of the adopted filtering. Only the number of entering objects and
the number of remaining objects are not sensitive to the adopted order.
As explained in Sect.\,\ref{sec:spectroSB1_input}, some objects
can enter the SB1 processing coming from the SB2 processing
\citep[see also the details in][]{dr3-dpacp-161}.
Table\,\ref{tab:tableresult4p} deals with the very few objects coming from the
SB2 chain. Due to the filtering on flags nr.11 and nr.12, these objects
do not pervade the final SB subcatalogue. No class is associated to these objects.

We have now to estimate the number of solutions from our pipeline
that were removed because they were combined and are now appearing
in the {\tt{EclipsingSpectro}} and {\tt{AstroSpectroSB1}} classes.
Since we are only interested here in the disappearance of some solutions
from our output due to the combination process, we can formally consider 
the tallies as if the combination were made after the internal filtering.
The results of the combination mechanism for the 198\,885 
SB1-type solutions are detailed in Table\,\ref{tab:tableresult5}.
The column marked 'Single' corresponds to the solutions appearing in one
class only, whereas the column marked 'Duplicated' refers to the objects
appearing in two classes. All the 198\,885 objects are classified,
except 2201 that will not be published due to a rejection
during the process. Among these 2201 objects, 2107 correspond
to a rejection at the level of the post-filtering which is detailed 
below. Some 94 other objects were rejected for other reasons.
The majority of these 94 unpublished objects are corresponding 
to tentative combinations that
finally failed and the objects were not re-injected in our SB subcatalogue.
The {\tt{Orbital}} and Others classes are given for the sake of
completeness but they are not further considered either in the present
work nor, for some, in DR3. Only the first three classes of the 
Table\,\ref{tab:tableresult5} are of importance for the present work.
We will in addition restrict the present analysis of the SB1-type 
solutions to the {\tt{SB1}} class.
To illustrate the approach, we will further consider the case of
{\tt{EclipsingSpectro}}. According to \citet{2023A&A...674A..34G},
some 155 objects are in this class.
Over these 155 objects, 1 was originally attributed a solution of type
SB2C, 1 of type SB1C, 2 of type SB2, and 151 of type SB1. 
Among these 151 SB1 solutions coming out of our
main processing, 79 were internally filtered out. Therefore, some
72 solutions were dispatched to {\tt{EclipsingSpectro}} 
and did not reach the {\tt{SB1}} class.

The solutions of type SB1C were cross-correlated along the same
lines as the SB1 ones. 
No object was classified in another category by the NSS combiner.
One object was rejected at the level of the post-filtering.
Therefore, all the 202 (i.e.\ 203--1) SB1C solutions
entered the eponymous class {\tt{SB1C}}. 

Finally, the TrendSB1-type solutions were compared with the
astrometric channel results and 891 turned out to be redirected
into the {\tt{AstroSpectroSB1}} class. They thus are not present
in the {\tt{TrendSB1}} class, leading to an amount of 56\,808 objects. 
The details are explicited in 
Table\,\ref{tab:tableresult6}.
The majority of the
21 unpublished objects are corresponding to tentative combinations that
finally failed and the objects were not re-injected in our SB subcatalogue.
In this table, we also counted the {\tt{TrendSB1}} objects that
presented or not a significant period at the level of the main processing
(flag nr.13 raised means no period found). 
\begin{table}[t]
\caption{Preliminary results of the spectroscopic orbital pipeline.}
\centering
\begin{tabular}{|l|r|r|}
\hline
\hline
Solution types & \multicolumn{1}{|c|}{From} & Redirected from \\
& SB1 chain & SB2 chain \\
\hline
SB1 & 367328 & 107 \\
SB1C & 3533 & 2 \\
TrendSB1 & 168930 & - \\
StochasticSB1 & 322420 & - \\
 & & \\
Total & 862211 & 109 \\ 
\hline
\end{tabular}
\label{tab:tableresult1}
\end{table}
\begin{table}[t]
\caption{Tally of the results from the progressive application
of the internal filtering for the objects leading to 
an SB1-type of solutions from the SB1 chain.}
\centering
\begin{tabular}{|l|r|}
\hline
\hline
Internal filtering step & Number of objects \\
\hline
Entering objects & 367328 \, \, \, \, \, \, \, \, \\
$F_2 \, >  \, 3$ & 71227 rejected \\
Bad flags & 2840 rejected \\
$K \, >$ 250 km\,s$^{-1}$ & 89 rejected \\
Bad efficiency & 27749 rejected \\
Phase gap $>$ 0.3 & 25656 rejected \\
Non-significant period & 24251 rejected \\
Significance $<$ 5 & 15829 rejected \\
Too large $\sigma_\omega$ & 802 rejected \\
Remaining objects & 198885 \, \, \, \, \, \, \, \, \\
\hline
\end{tabular}
\label{tab:tableresult2}
\end{table}
\begin{table}[t]
\caption{Tally of the results from the progressive application
of the internal filtering for the objects leading to an 
SB1C-type of solutions from the SB1 chain.}
\centering
\begin{tabular}{|l|r|}
\hline
\hline
Internal filtering step & Number of objects \\
\hline
Entering objects & 3533 \, \, \, \, \, \, \, \, \\
$F_2 \, >  \, 3$ & 1920 rejected \\
Bad flags & 61 rejected \\
$K \, >$ 250 km\,s$^{-1}$ & 23 rejected \\
Bad efficiency & 1 rejected \\
Phase gap $>$ 0.3 & 588 rejected \\
Non-significant period & 735 rejected \\
Significance $<$ 5 & 2 rejected \\
Remaining objects & 203 \, \, \, \, \, \, \, \, \\
\hline
\end{tabular}
\label{tab:tableresult3}
\end{table}
\begin{table}[t]
\caption{Tally of the results from the progressive application
of the internal filtering for the objects leading to a TrendSB1-type 
of solutions from the SB1 chain.}
\centering
\begin{tabular}{|l|r|}
\hline
\hline
Internal filtering step & Number of objects \\
\hline
Entering objects & 168930 \, \, \, \, \, \, \, \, \\
$F_2 \, >  \, 3$ & 87223 rejected \\
Bad flags & 1381 rejected \\
Time gap $>$ 0.3$\Delta T$ & 12103 rejected \\
Passage to 0.99865 & 10503 rejected \\
Remaining objects & 57720  \, \, \, \, \, \, \, \, \\
\hline
\end{tabular}
\label{tab:tableresult4}
\end{table}
\begin{table}[t]
\caption{Tally of the results from the progressive application
of the internal filtering for the objects leading to an SB1- or SB1C-type 
of solutions and previously redirected from the SB2 chain.}
\centering
\begin{tabular}{|l|r|r|}
\hline
\hline
Internal filtering step & \multicolumn{2}{|c|}{Number of objects}  \\
\cline{2-3}
& SB1FromSB2 & SB1CFromSB2 \\
\hline
Entering objects & 107 \, \, \, \, \, \, \, \,& 
\! \! \! \! 2 \, \, \, \, \, \, \, \, \\
$F_2 \, >  \, 3$ & 64 rejected & 1 rejected \\
Bad flags & 3 rejected & 0 rejected \\
$K \, >$ 250 km\,s$^{-1}$ & 0 rejected & 0 rejected \\
Bad efficiency & 3 rejected & 0 rejected \\
Non-significant period & 3 rejected & 0 rejected  \\
Significance $<$ 5 & 4 rejected & 0 rejected \\
Too large $\sigma_\omega$ & 0 rejected & --  \, \, \, \, \, \, \, \, \\
Remaining objects & 30 \, \, \, \, \, \, \, \,& 1 \, \, \, \, \, \, \, \, \\
\hline
\end{tabular}
\label{tab:tableresult4p}
\end{table}
\begin{table} 
\caption{Repartition of the 198\,885 SB1-type solutions among 
the various classes of the catalogue.}
\centering
\begin{tabular}{|l|r|r|r|}
\hline
\hline
Class & Single & Duplicated & All \\
\hline
{\tt{SB1}} & 175725 & 5602 & 181327 \\
{\tt{AstroSpectroSB1}} & 15227 & 58 & 15285 \\
{\tt{EclipsingSpectro}} & 71 & 1 & 72 \\
{\tt{Orbital}} & 55 & 5291 & 5346 \\
Others & 17 & 371 & 388 \\
Total (all classes) & 191095 & 11323 & 202418 \\
\hline 
& & & \\
Total (first three classes) & 191023 &5661 & 196684 \\
Not published & - & - & 2201 \\
\hline 
& & & \\
Total (distributed) & - & - & 198885 \\
\hline
\end{tabular}
\tablefoot{Single: the solution appears in one class only; 
Duplicated: the solution appears in two classes and should thus be counted
only once to avoid redundancy.}
\label{tab:tableresult5}
\end{table}
\begin{table}[t]
\caption{Repartition of the 57\,720 TrendSB1-type solutions among the 
{\tt{TrendSB1}} and the {\tt{AstroSpectroSB1}} classes of the catalogue.}
\centering
\begin{tabular}{|l|r|r|r|}
\hline
\hline
Class & Count & Period & \multicolumn{1}{|c|}{No}\\
& & & period\\
\hline
{\tt{FirstDegreeTrendSB1}} & 24083 & 8391 & 15692 \\
{\tt{SecondDegreeTrendSB1}} & 32725 & 16989 & 15736 \\
Total TrendSB1 & 56808 & 25380 & 31428 \\
& & &\\
{\tt{AstroSpectroSB1}} & 891 & - & - \\
Not published & 21 & - & -\\ 
&  & &\\
Total & 57720 & & \\
\hline
\end{tabular}
\label{tab:tableresult6}
\end{table}

After the combination process,  a further step of filtering took place.
This one was called post-filtering, and the motivation and practical details
are described below in Sect.\,\ref{ssec:spectroSB1_add_cons_postfilt}.
The post-filtering was done on the basis of data coming from outside
the RVS channel and is basically a process of blacklisting aiming
at improving the purity of the catalogue. However, it should be made 
clear that it is partly constrained by the fact that the spectroscopic 
solutions are further used by the combiner, and
thus we should be careful not to eliminate good solutions
that could be useful downstream in the process. 
In the present section we restrict ourselves
to the effect of the applied post-filtering on the countings of the objects
among the various classes. 
Table\,\ref{tab:tableresult7} details the effect on the number of 
objects considered in the important classes. The amount of objects
before the post-filtering and after it are mentioned.
The column after post-filtering gives the countings that are 
corresponding to the final DR3 catalogue.

The spectroscopic channel presented here had a strong importance
in the building of the
{\tt{AstroSpectroSB1}} class and a last information of interest
must be given here. Namely, the origin of the various
spectroscopic solutions that were combined with other
types of data in order to build up the
{\tt{AstroSpectroSB1}} class.
The detailed origins are explicited in
Table\,\ref{tab:tableresult8} for the four kinds of
solutions. We dwell again on the fact that none SB1C-type
solutions were combined. We also notice that some combination has been
performed on the basis of the StochasticSB1 solutions.

A few pie charts illustrating these tallies are available
in Appendix\,\ref{sec:appD}. 
\begin{table}[t]
\caption{Results of the application of the post-filtering.}
\centering
\begin{tabular}{|l|r|r|r|}
\hline
\hline
 & Countings & Discarded & Catalogue \\
Class & \multicolumn{1}{|c|}{before} & \multicolumn{1}{|c|}{at} & \multicolumn{1}{|c|}{after} \\
 & post-filter. & post-filter. & post-filter.  \\
\hline
{\tt{SB1}} & 183434 & 2107 & 181327 \\
{\tt{SB1C}} & 203 & 1 & 202 \\
{\tt{EclipsingSpectro}} & 72 & 0 & 72 \\
{\tt{AstroSpectroSB1}} & 36082 & 2615 & 33467 \\
{\tt{TrendSB1}} & 56808 & 0 & 56808 \\
\hline
\end{tabular}
\raggedleft \tablefoot{The first four classes can be found
in table {\tt{nss\_two\_body\_orbit}} whereas the last class can be found 
in table {\tt{nss\_non\_linear\_spectro}}.} 
\label{tab:tableresult7}
\end{table}
\begin{table}[h!]
\caption{Various contributions to the class {\tt{AstroSpectroSB1}} from the point 
of view of the spectroscopic solutions.}
\centering
\begin{tabular}{|l|r|r|}
\hline
\hline
Original solution-types & \multicolumn{1}{|c|}{Before} & \multicolumn{1}{|c|}{After} \\
& filtering & filtering \\
\hline
SB1 & 16980 & 15285 \\
SB1C & 0 & 0 \\
TrendSB1 & 6104 & 891 \\
StochasticSB1 & 10383 & 10383 \\
& & \\
Total & 33467 & 26559 \\
\hline
\end{tabular}
\label{tab:tableresult8}
\end{table}

\subsection{Some illustrative results}\label{ssec:spectroSB1_results_illust}
In the present section, we want to illustrate various solutions
pointing out the efficiency of the code. Here below, we adopt the convention to express
the uncertainties as error bars of $\pm 1 \sigma$.
\subsubsection{Good results}\label{sssec:spectroSB1_results_illust_good}
\begin{figure*}[!ht]
\centerline{
\includegraphics[width=0.31\textwidth]{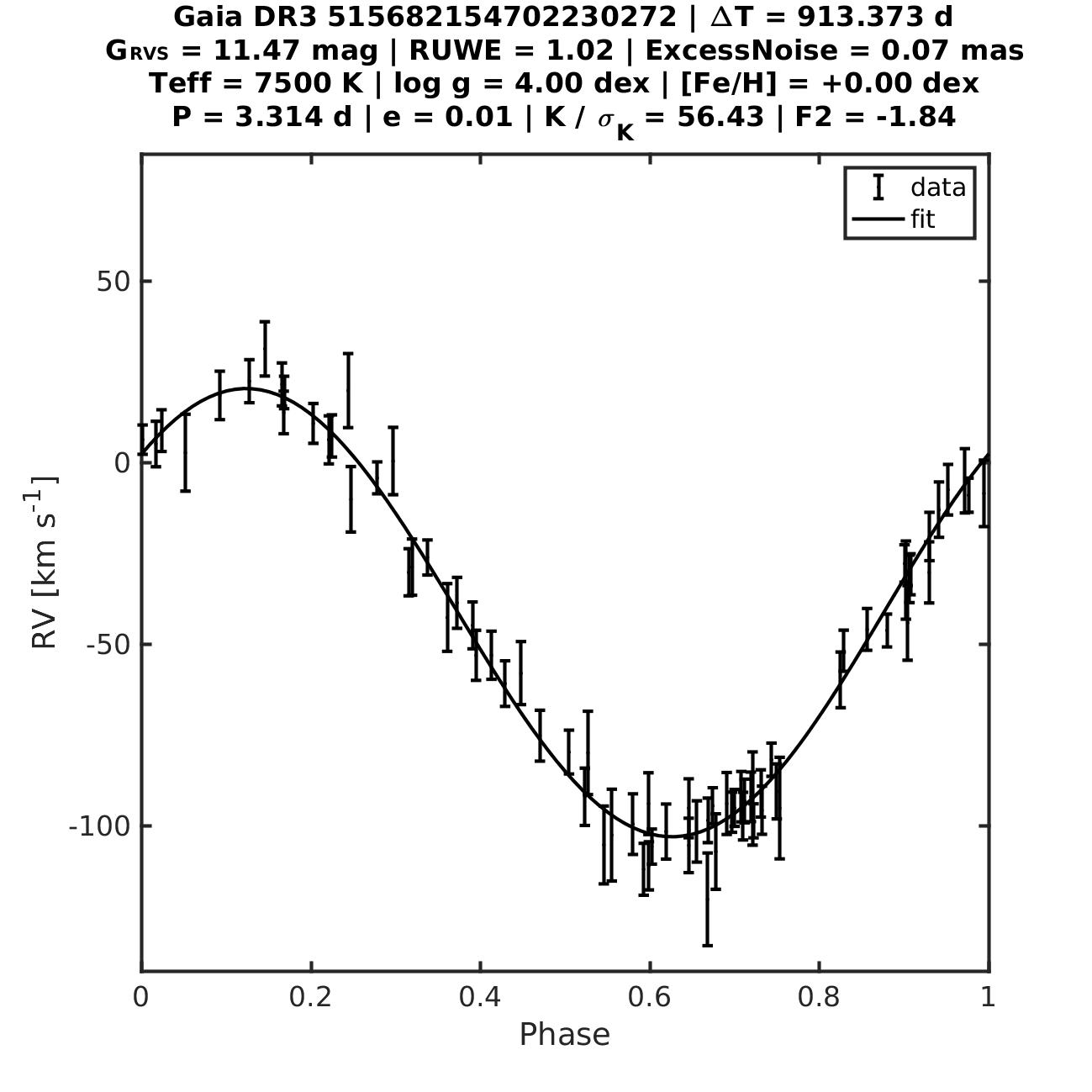}
\includegraphics[width=0.31\textwidth]{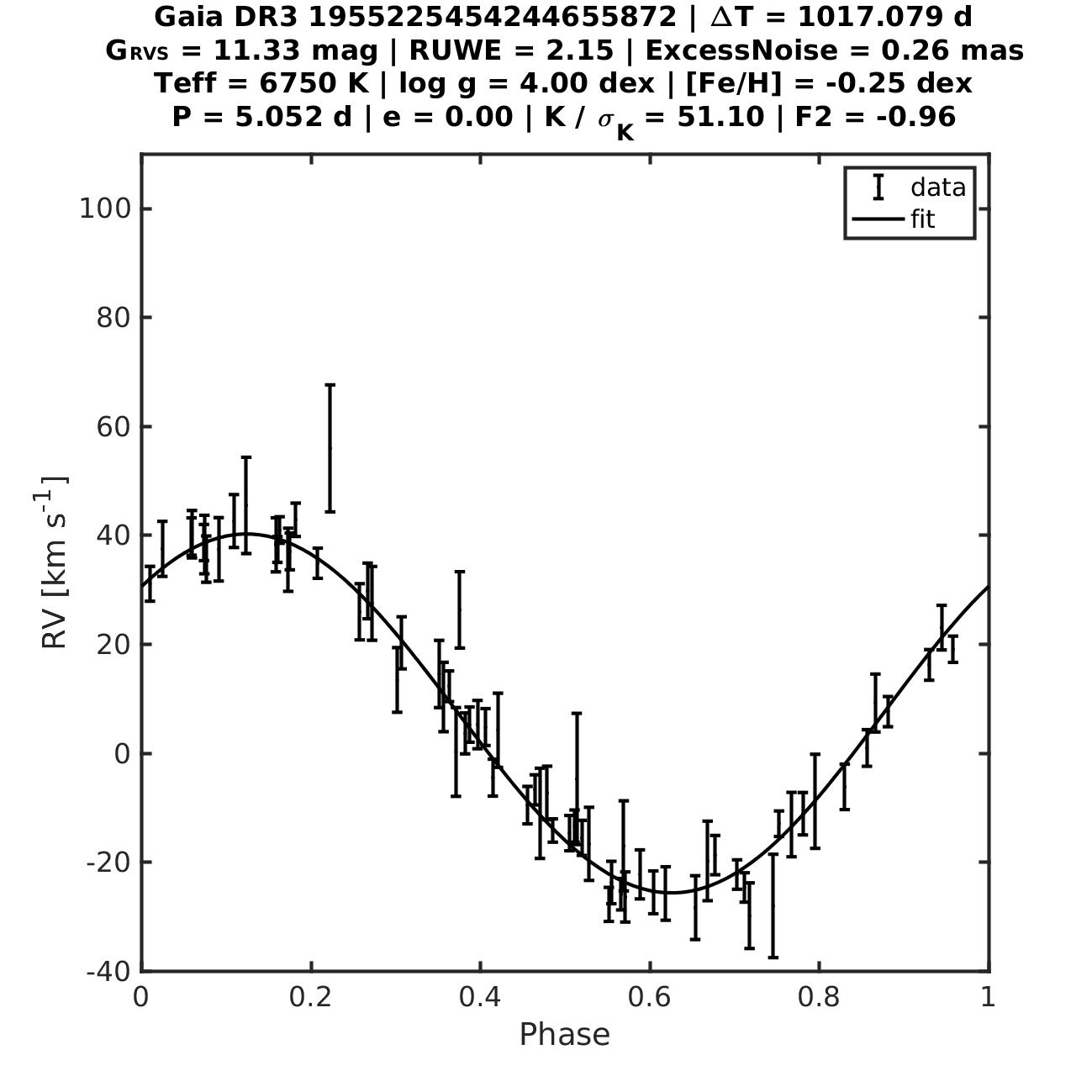}
\includegraphics[width=0.31\textwidth]{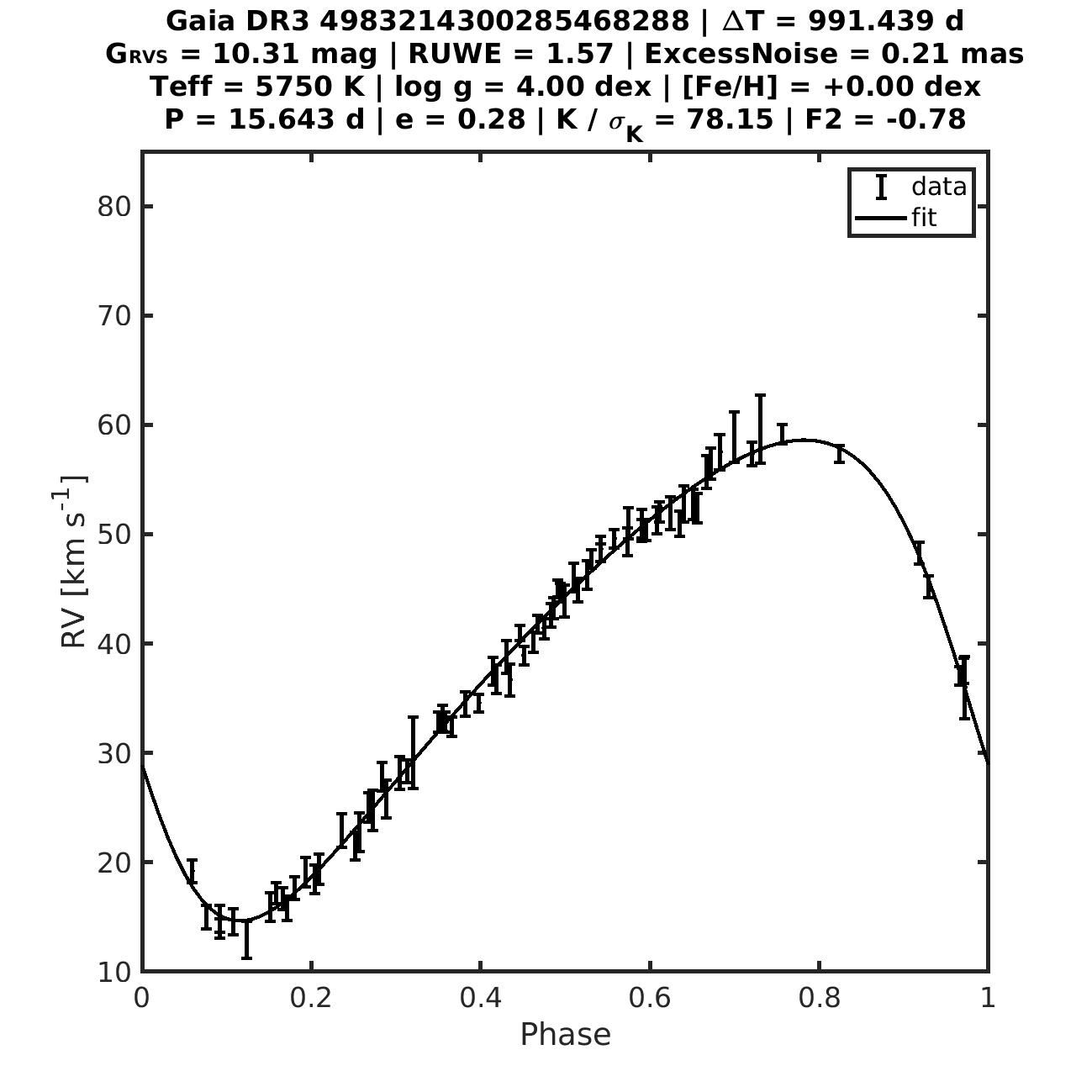}
}
\centerline{
\includegraphics[width=0.31\textwidth]{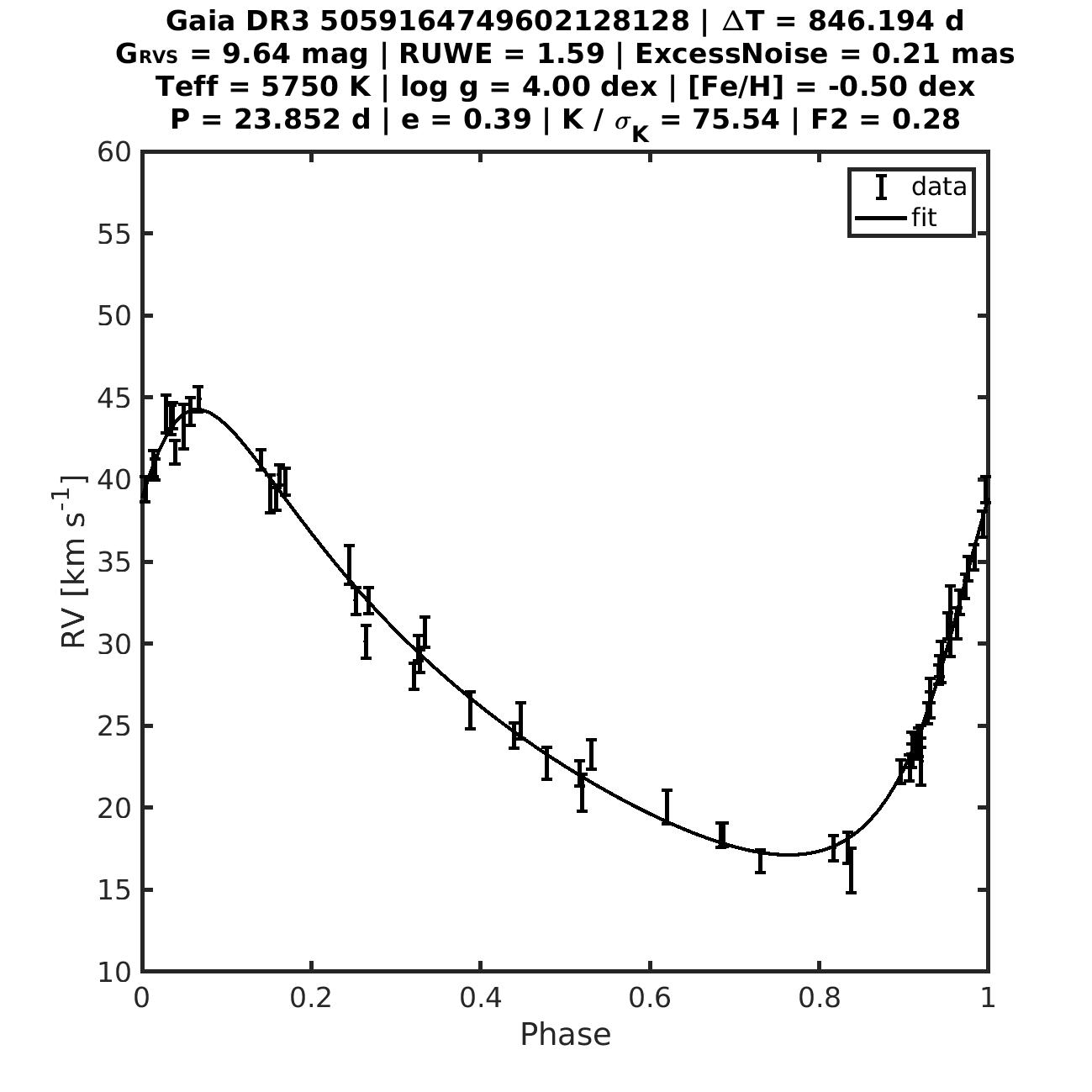}
\includegraphics[width=0.31\textwidth]{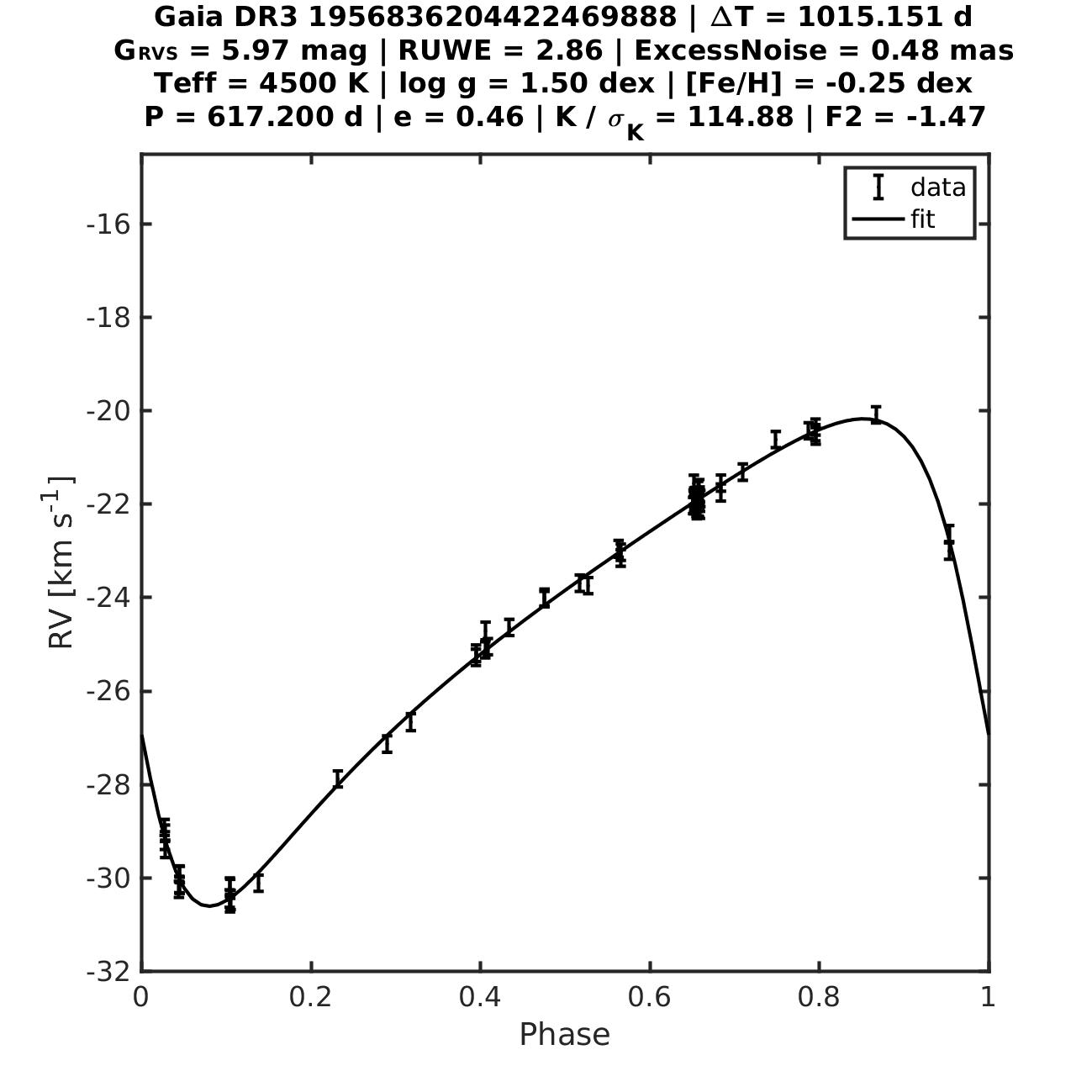}
\includegraphics[width=0.31\textwidth]{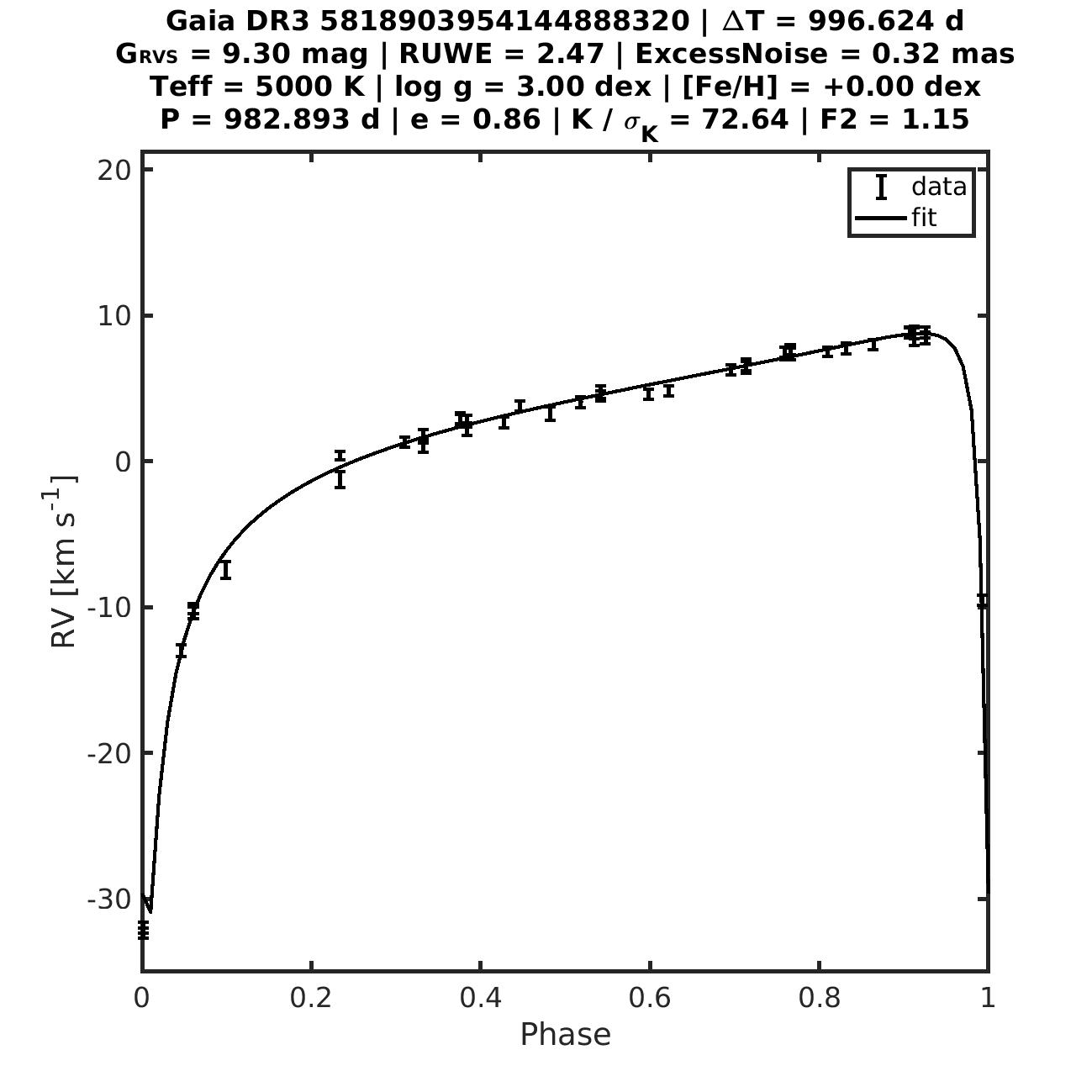}
}
\caption{Several examples illustrative of the output 
of the pipeline for SB1-type solutions. Each panel
concerns an object whose \gaia\ name is given in 
the header. Each of the panels shows the phase 
diagram containing the folded RVs (data points at 
the centre of the $\pm 1 \sigma$ error bars) 
along with the fitted orbital solution. The header
gives also the time span covered by the time series
($\Delta T$). The second line gives the 
$G_{\mathrm{RVS}}^{\mathrm{int}}$ magnitude of 
the object, the astrometric {\tt{ruwe}} and the
astrometric excess noise, whereas the third line concerns
the physical parameters identifying the template used. 
The last line indicates the period, the eccentricity, 
the significance, and the statistic
$F_2$ related to the adopted solution. The objects are
ordered by increasing period.}
\label{fig:goodressb1part1}
\end{figure*}

Here, we introduce a few examples
of the solution fitted to selected RV time series.
The selection is somewhat arbitrary and is essentially
motivated by the illustration of the efficiency
of the chain.
All these illustrating good solutions are part of the
DR3 spectroscopic orbit catalogue
(SB subcatalogue).
Figure\,\ref{fig:appgoodressb1part1}
illustrates various orbital solutions
that are part of the {\tt{SB1}} class. A small subset
of these {\tt{SB1}} solutions is also provided in 
Fig.\,\ref{fig:goodressb1part1}.
The displayed phase diagrams cover a palette
of values for the period associated to the SB1 solution.
It is a first positive point that the pipeline is able
to treat such a large domain of periodicity.
At mid to large period, we also present a few examples
of highly eccentric orbits, up to $e \sim 0.9$. 
It is a matter of fact that a good fit to a 
very eccentric orbit necessitates a large number 
of data points in the time series, 
particularly near the periastron. 
An homogeneous distribution in phase is
another necessity. 

Since the vast majority of the analysed time series
covers a span of time around 800-1000\,d, the solutions
with short periods are extremely convincing since 
the large number of cycles folded in the phase diagram
do not mitigate the goodness of fit and this can be 
associated to a high coherency of the periodicity.
This conclusion holds when more than three cycles are 
covered (i.e.\ $P \sim \Delta T / 3$, or less).
For larger periods, the coherency of the variations
is not fully guaranteed by the quality of the goodness
of fit. At larger values 
($P \sim \Delta T$), we are no more able to guarantee
the coherency of the variation, since only one cycle
is observed and this could correspond to a transitory
event, not to a periodic variation. In any case, a
tentative fit is still more indicative than no fit 
at all. However, the SB1-type solutions with such large
$P$ up to $1.5 \, \Delta T$ (the adopted upper 
value allowed) should be considered as not secure
as periodic solutions and the reported period
is at the very best an indicative timescale.
Indeed, for given periods between $\Delta T$ 
and 1.5\,$\Delta T$,
the periods are not well determined and could even be
potentially biased, due to the non-linear nature of the
fit interacting with a possible bad phase coverage.
This topic is further addressed in 
Sect.\,\ref{sssec:spectroSB1_quality_val_compGriffin}
and in Sect.\,\ref{sssec:spectroSB1_quality_val_compSB9}.
As illustrated below, the latter bias could induce biases
on other parameters (see e.g.\ the case of the eccentricity
in Fig.\,\ref{fig:cu4nss_spectro_comparison_SB9_SB1_e}). 
The objects
\gaia\ DR3 6422103351757549056,
\gaia\ DR3 2509001872917567360,
and \gaia\ DR3 5818903954144888320
are certainly concerned by this possible pitfall.

It is interesting to notice that several objects in the
sample correspond to a quasi-circular solution: 
the eccentricity is very small, even perhaps not
significant. 
A perfect illustration (see Fig.\,\ref{fig:goodressb1part1})
is the case of
\gaia\ DR3 1955225454244655872, whose eccentricity
is $e\,=\, 0.004$ with a 1-$\sigma$ error
of 0.021. This object most probably corresponds to 
a circular orbit, but the adopted solution is not
an SB1C but an SB1.
This SB1-type solution is associated to a solution
efficiency of 0.16 and is thus on the verge to be
processed with an SB1C model.
This is due to the fact that we adopted the 
philosophy of not enforcing a circular solution
even if the eccentricity turned out not to be 
significantly different from zero.
We only adopt a circular solution when the
structure matrix is such that it could 
not be properly inverted in the classical way 
and thus a full SB1-type solution is not
possible. We are consequently constrained to decrease 
the number of free parameters. In fact,
\gaia\ DR3 1955225454244655872
benefited from a large number of observations
(good transits) and this large number of data points
allowed us to fit an SB1 solution even for a very small
apparent value of the eccentricity.
Another example of truly circular orbit could be
e.g.\ \gaia\ DR3 515682154702230272.
Starting from these examples, we encourage the user 
of the catalogue to understand that the classes
{\tt{SB1}} and {\tt{SB1C}} differ by the fitted model
and NOT by the true nature of the object. For example,
someone interested in objects in circular orbits,
i.e.\ to study only genuine circular orbit, should
certainly consider both 
{\tt{SB1}} and {\tt{SB1C}} classes. 
If a user wants to transform an SB1-type solutions
into an SB1C one, they can certainly use the period,
the centre-of-mass velocity and the semi-amplitude
as given by the SB1-type solution. However, it should 
not be overlooked that the argument of periastron 
is thus modified, since the $T_0$ for SB1C solutions
corresponds by definition to the time of maximum
velocity. Therefore, the $T_0$ must be accordingly
corrected for this change of $\omega$. Otherwise,
the SB1-type solution is applicable to objects
in true circular orbits.

Another interesting and enlightening case is the one
of  \gaia\ DR3 4983214300285468288. As can be seen from
Fig.\,\ref{fig:goodressb1part1}, the fit of the 
SB1-type solution is very good and no individual 
data point deviates significantly from the model.
However, an expert eye (not the pipeline) will detect
that around phase 0.65, the observed RVs are 
systematically grouped below the model and then above
the modelled curve. Both groups of data points
contain some 10 points and the offset could thus
be significant (at least marginally) although 
the grouping is observed a posteriori.
In any case, if this anomaly at phase 0.65 is real
(and this will become clear in future releases), it is
hardly explained  in the framework of the Keplerian
motion. The time corresponds to a phase  where the 
primary is approaching the maximum velocity 
and thus the
possible lines of the secondary are close to the maximum
separation and should not influence the determination
of the primary RVs (primary meaning here the 
star that dominates the combined spectrum).
We are also (most probably) far from a possible eclipse, and this 
could not be due to a Rossiter-McLaughlin effect.
The reported anomaly, if confirmed, will necessitate
exotic effects to be explained in the context
of a Keplerian model. It is worrying that several
intrinsic variables (pulsating stars at least) could
exhibit such a phenomenon at approximately 
this location in their RV curve. It is beyond 
the scope of the present paper to make a decision for
this particular object but we would like to take this
opportunity to highlight the following statement.
It is not because a beautiful fit can be performed
with a Keplerian model by the pipeline that the
true nature of the object is the one of a binary.
The membership to the {\tt{SB1}} or {\tt{SB1C}} classes
of solutions does not automatically imply the binarity
of the object. Additional clues are necessary
to converge to a sound classification.
The binary nature of {\tt{SB2}} objects is much more
convincing
\citep[although problems could also 
exist; see][]{dr3-dpacp-161}.
Concerning {\tt{SB1}} or {\tt{SB1C}} objects, 
a conclusion could be drawn if, for example, the
orbit is confirmed by the photometric 
or the astrometric
channels (classes {\tt{EclipsingSpectro}} or
{\tt{AstroSpectroSB1}}).
The combined solutions are not part of the classes
treated in the present paper.
For the isolated spectroscopic orbital solutions
reported here, we recommend at the very least,
to build up a subcatalogue constituting some kind 
of gold sample, or to combine with additional data
(e.g.\ positions in the HR diagram) in order to
increase the veracity of the proposed classification.

The object \gaia\ DR3 1956836204422469888
is also found on the list of
{\sc{Hipparcos}}-\gaia\ proper motion anomaly of
\citet{2019A&A...623A..72K} and the list of 
\citet{2021ApJS..254...42B}.
All the objects included in
Fig.\,\ref{fig:goodressb1part1} and in
Fig.\,\ref{fig:appgoodressb1part1}
were not previously known to have periodic 
RV variations.
\begin{figure*}[!h]
\centerline{
\includegraphics[width=0.31\textwidth]{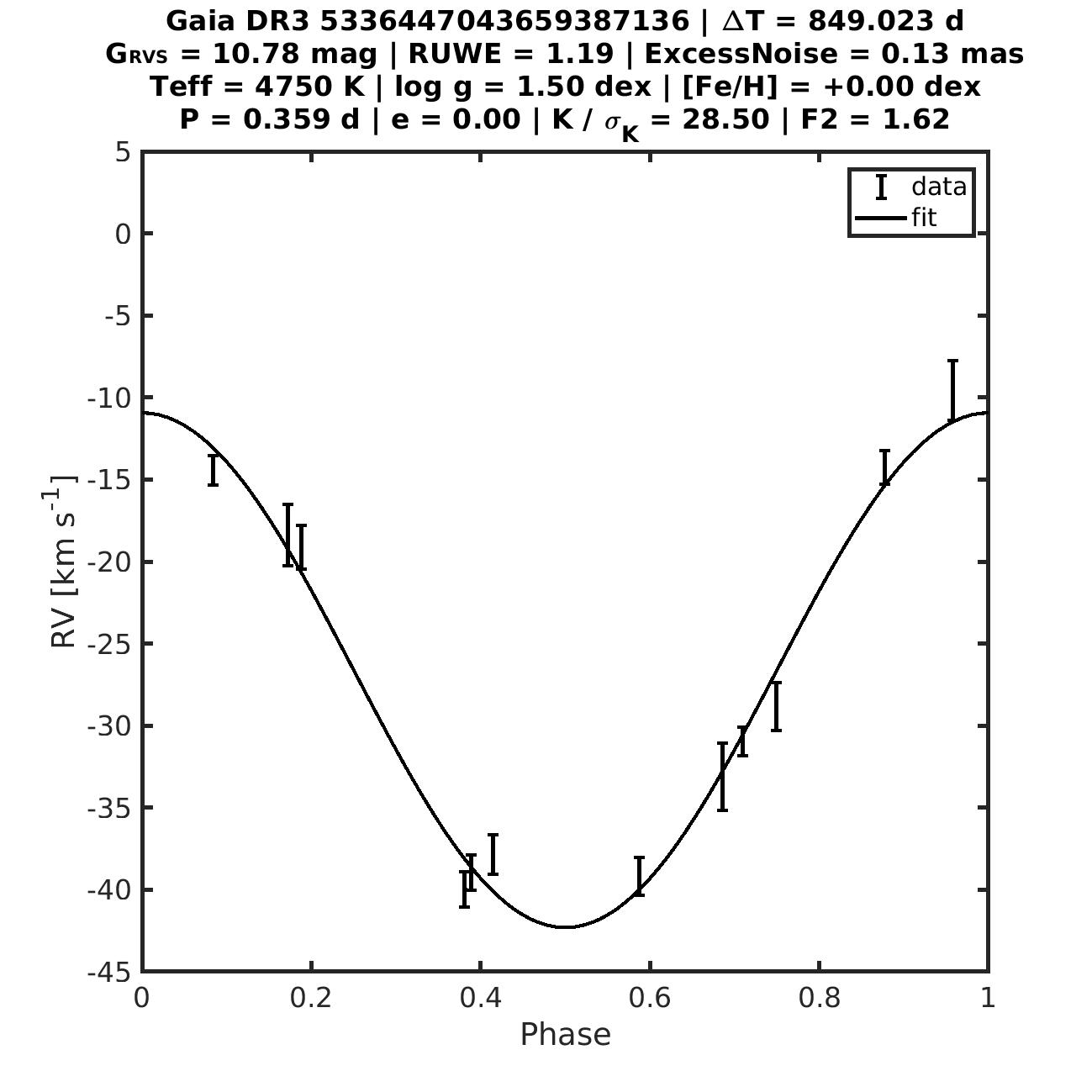}
\includegraphics[width=0.31\textwidth]{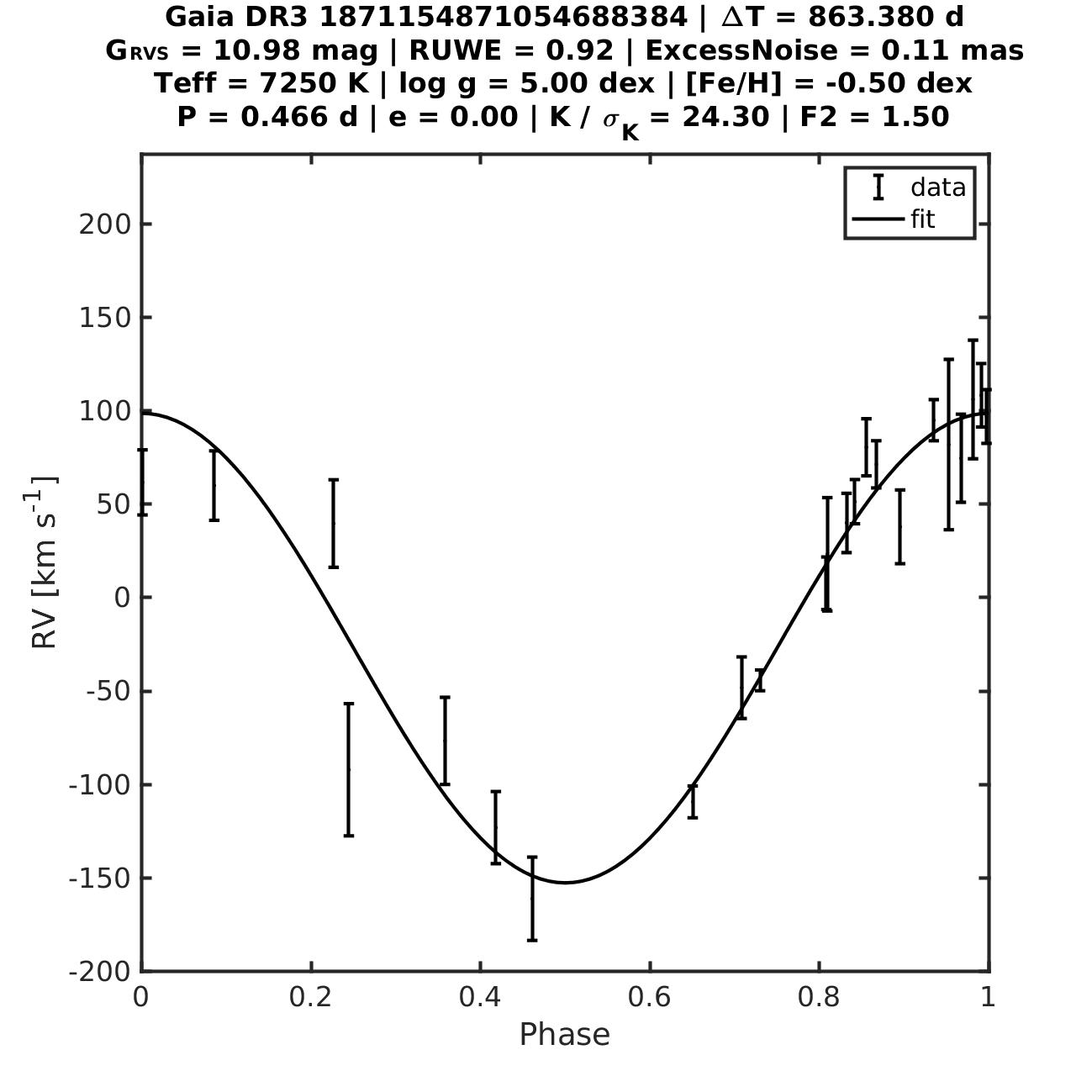}
\includegraphics[width=0.31\textwidth]{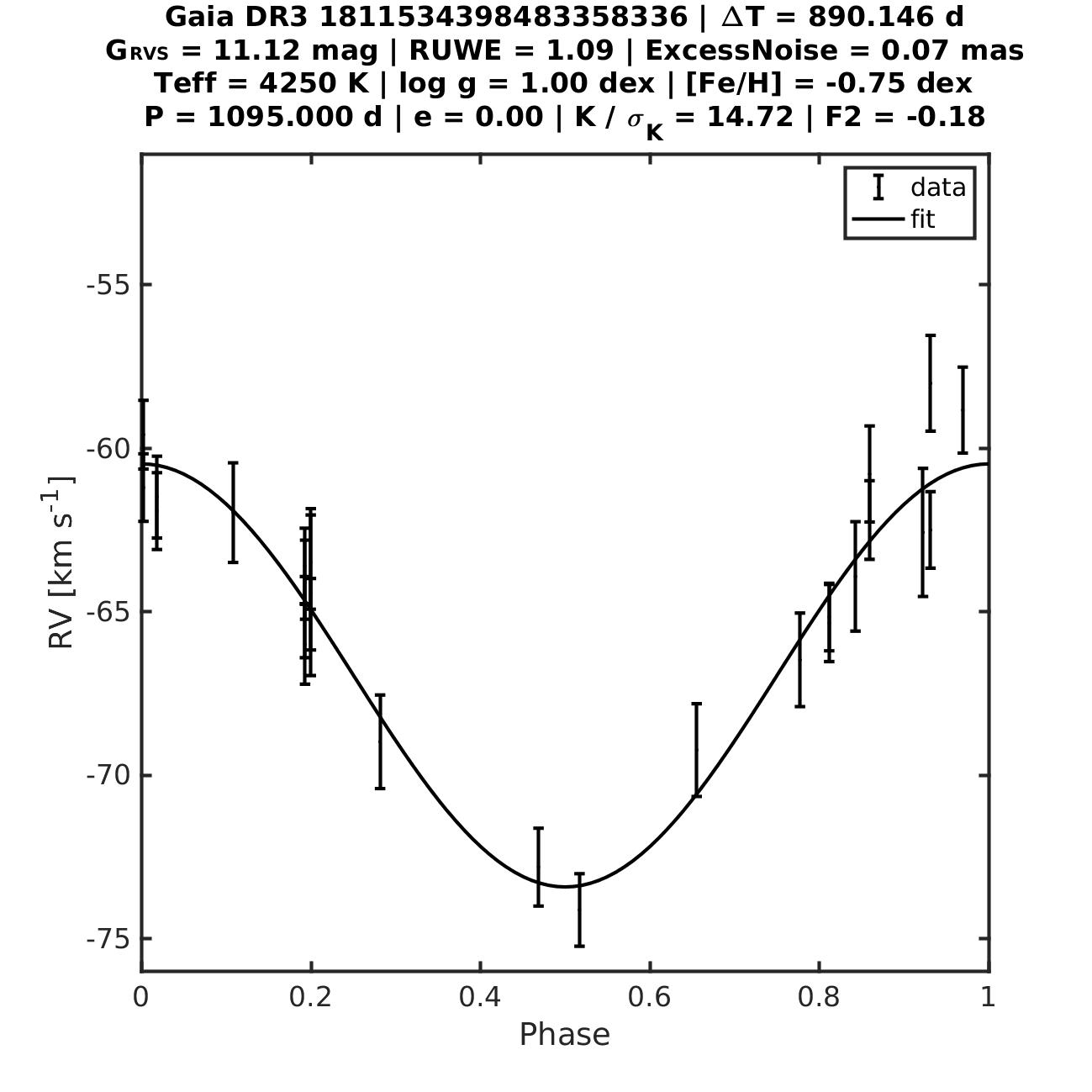}
}
\caption{Same as Fig.\,\ref{fig:goodressb1part1} but concerning 
the SB1C-type solutions.}
\label{fig:goodressb1c}
\end{figure*}
\begin{figure*}[!h]
\centerline{
\includegraphics[width=0.31\textwidth]{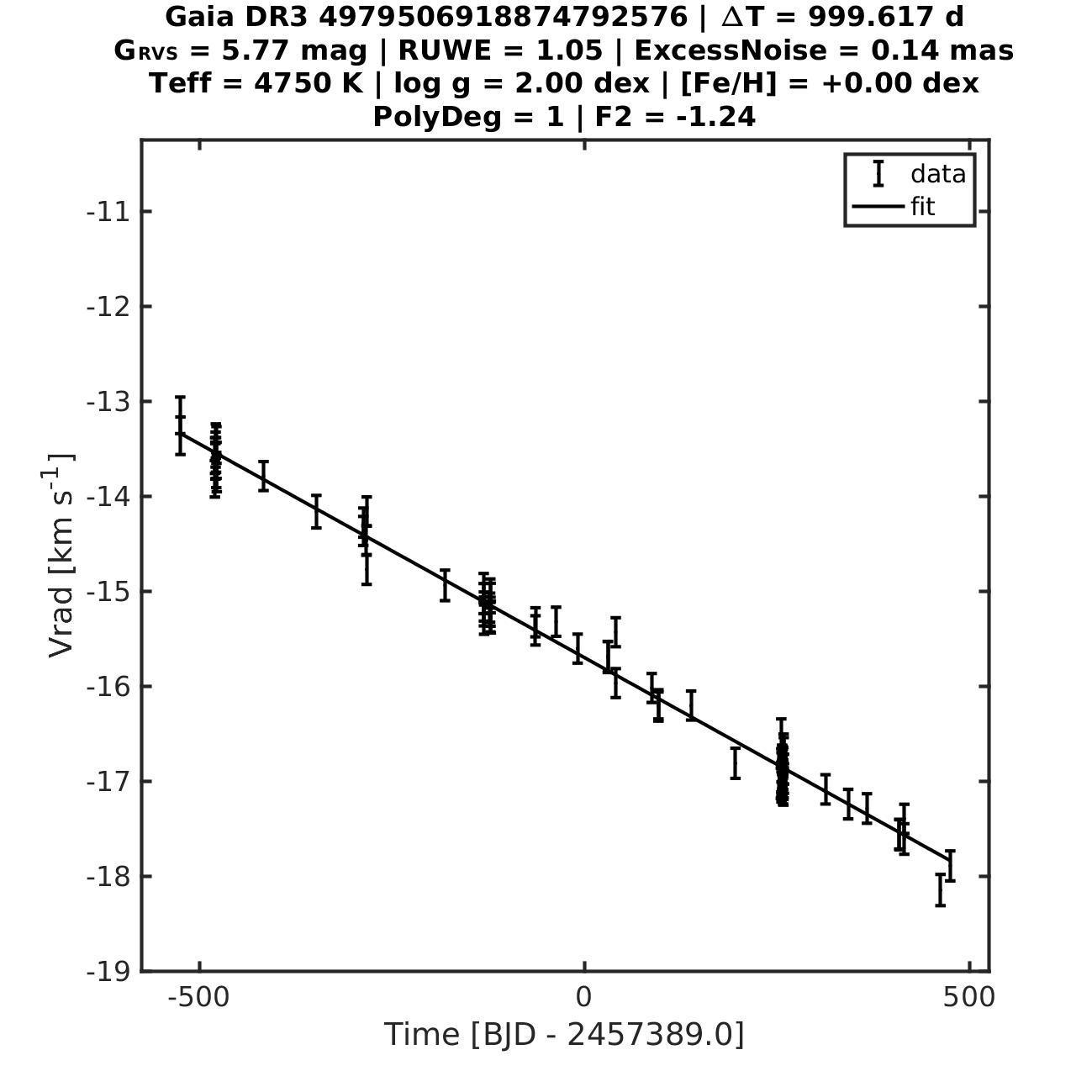}
\includegraphics[width=0.31\textwidth]{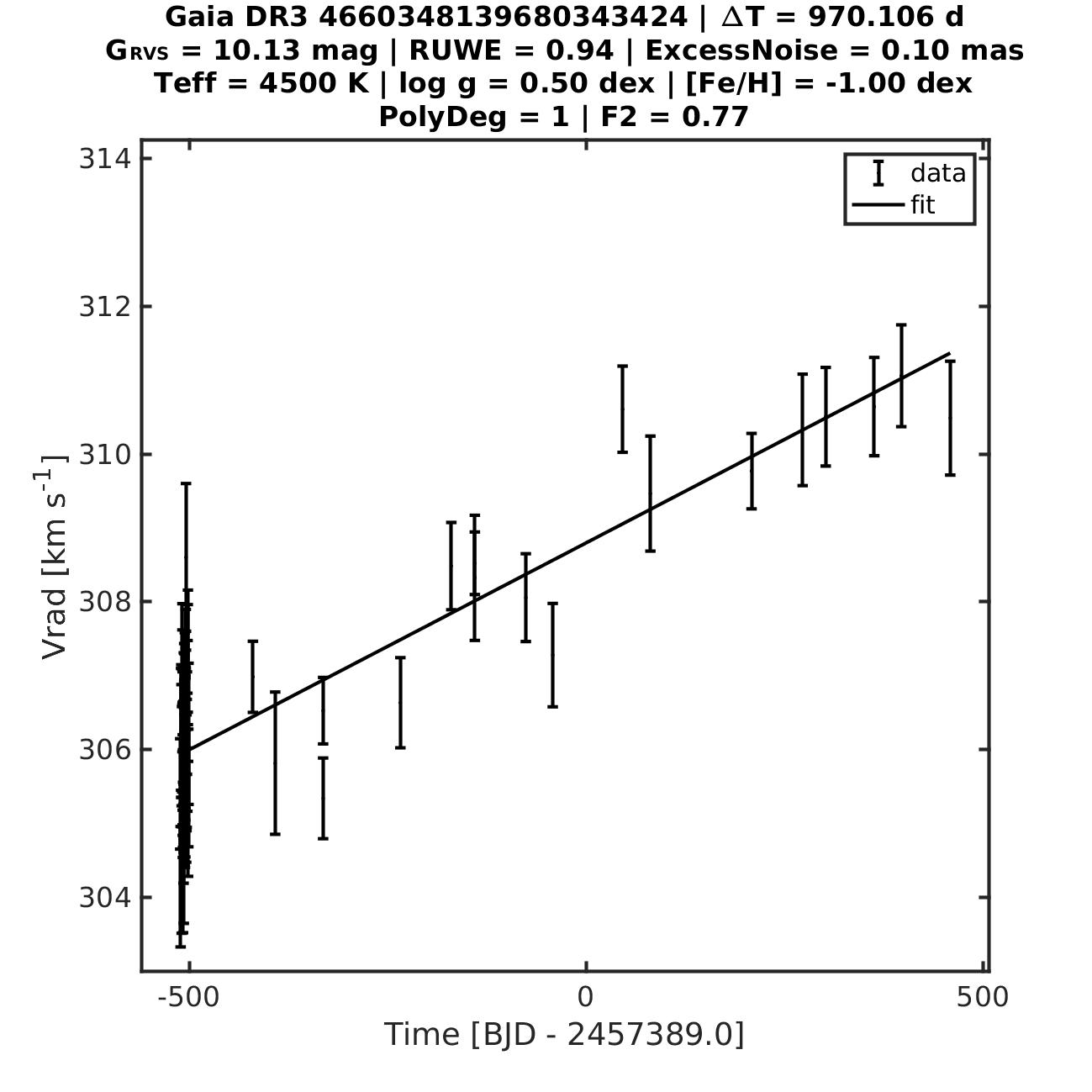}
\includegraphics[width=0.31\textwidth]{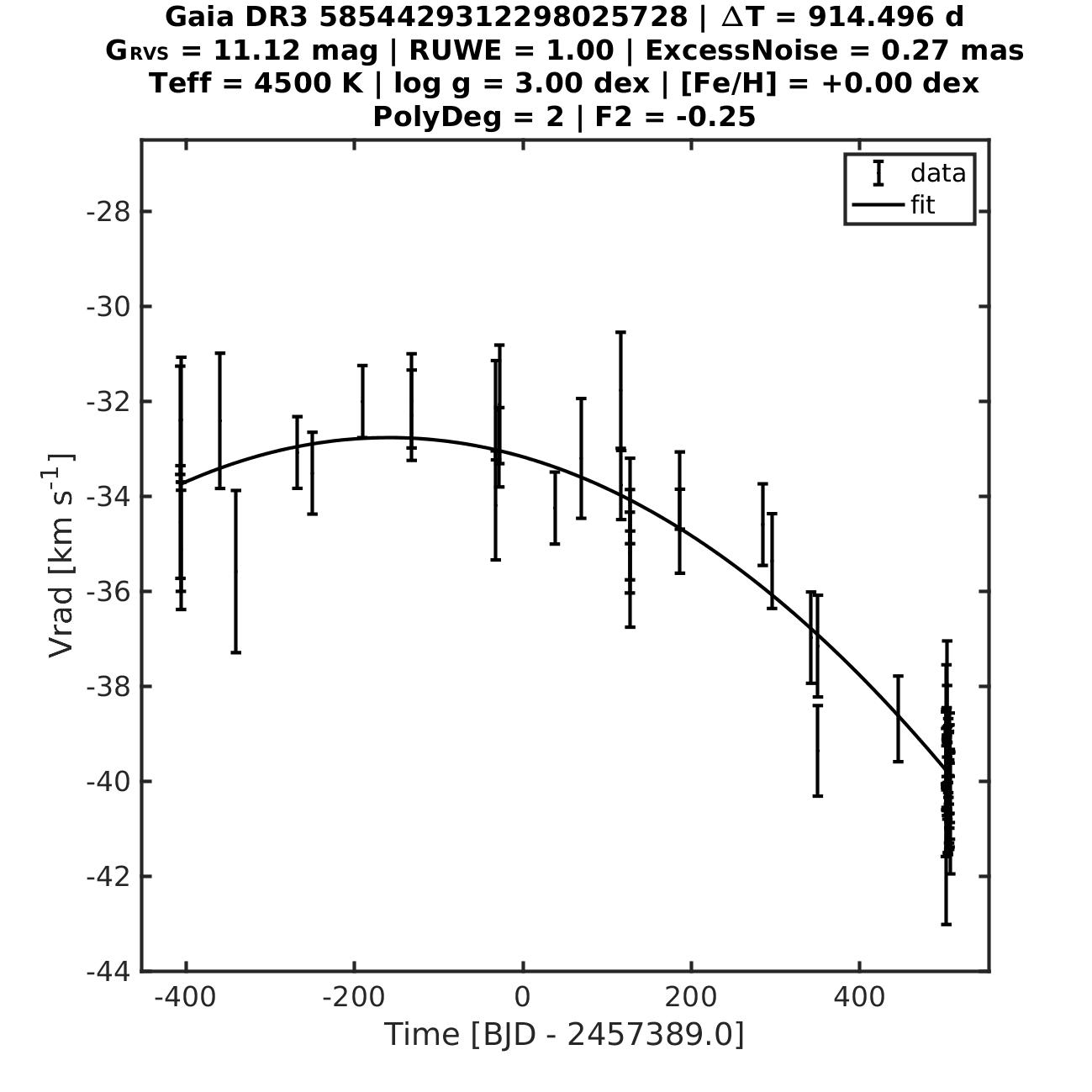}
}
\caption{Same as Fig.\,\ref{fig:goodressb1part1} but concerning 
the TrendSB1-type solutions. 
The order of the objects is arbitrary, except that the second 
degree trend has been placed at the end.}
\label{fig:goodrestrend}
\end{figure*}

In Fig.\,\ref{fig:appgoodressb1c}, we exhibit a few
examples of orbital solutions of the class {\tt{SB1C}}.
A still smaller subset is provided in Fig.\,\ref{fig:goodressb1c}.
A general finding is the fact that most of the
concerned time series are characterised by a rather
limited number of data points. They also correspond
to objects characterised by 
$G_{\mathrm{RVS}}^{\mathrm{int}}\,>\, 9$.
This is related to 
the decision we took to fit an SB1C-type solution
only when the SB1 failed: we must be constrained to limit
the amount of free parameters. 
We recall here that the convention has been
adopted that, for SB1C-type solutions, the $T_0$
represents the time of passage at maximum velocity.
We note from Fig.\,\ref{fig:appgoodressb1c} that we have
circular solutions for a few large period objects.
Here also, the majority of the objects were not 
previously known to present a periodic behaviour
in their RV.

The object \gaia\ DR3 3384500857079232512
has been observed with {\sc{Kepler}} and a period
$P \, = \, 3.74 \,$d (or twice that value) has been
detected 
\citep{2015A&A...579A..19A}.
This does not seem to be coherent with our results
($P \, = \, 0.297 \,$d).
The solution of this discrepancy will certainly
be automatically found in future releases thanks
to the important increase of the number of data points
in the time series under analysis. 
This conclusion is supported by the simulations reported in
Sect.\,\ref{ssec:spectroSB1_validation_simul}.

Another very interesting case is the one of
\gaia\ DR3 1871154871054688384, also known as
V2197\,Cyg (NSVS\,5761314). From ground-based 
observations, this star is a well-known
SB2 star. From photometric data, the system seems to be
an eclipsing system on a circular orbit with
an inclination of about 80\,\degr. It is
in near contact and has a period of
0.4657489\,d ($\sigma$=0.0000001\,d). 
This is in good agreement with the \gaia\ 
spectroscopic period
$P \, = \, 0.4657410 \,$d ($\sigma$=0.0000034\,d).
The difference amounts to 2.3\,$\sigma$.
It is the first time ever that a spectroscopic
period has been determined for this object.
From the work of \citet{2017IBVS.6203....1N},
the primary star is of type \mbox{F2-4\,V} and emits
about 80 to 90\,\% of the visible light of the system.
With 7 spectra taken at mid-resolution, these authors
derive a RV curve leading to
$\gamma \, = \, -25.2$ ($\sigma \, = \, 0.6) \,$km\,s$^{-1}$
and 
$K \, = \, 123.0$ ($\sigma \, = \, 0.8) \,$km\,s$^{-1}$.
These values compare very well with the 
$\gamma \, = \, -27.0$ ($\sigma \, = \, 4.9) \,$km\,s$^{-1}$
and 
$K \, = \, 125.7$ ($\sigma \, = \, 5.0) \,$km\,s$^{-1}$
as derived from our pipeline.
This illustrates the fact that the orbit of a
relatively bright primary in an SB2 
system observed as an SB1 one
(here within the {\tt{SB1C}} class)
can be trusted.
It is particularly interesting to look at the orbital
solution in the phase diagram for this object
(Fig.\,\ref{fig:goodressb1c}).
From phase 0.5 to 1.0, the measured RVs are 
rather accurate: this corresponds to the primary 
being in front. On the opposite, from phase 0.0 to 0.5,
the dispersion of the data points is larger; the unseen
secondary is in front. The dispersion around the model
is particularly important at phase 0.2-0.25.
However, \citet{2017IBVS.6203....1N} predict at 
these phases the presence of a Rossiter-McLaughlin
effect and at phase 0.22, their model predicts
a measured velocity of 40-50\,km\,s$^{-1}$, again in
good agreement with the \gaia\ observations. 
However, the agreement is not so good for the point 
at a slightly later phase. In any case, this could 
be the first time that this effect is detected by \gaia .
This will be further tested in future releases.
We would like to recall here that the measurement
of RVs by cross-correlation with templates is 
not the best method to address this effect.

Finally, some illustrative examples of single-line
trend solutions are proposed in
Fig.\,\ref{fig:appgoodrestrend} with the small subset
provided in Fig.\,\ref{fig:goodrestrend}.
The various objects are ordered randomly except that
the second-degree solutions occupy the last 
panels. None of these objects were previously known
to exhibit RV variations.
\gaia\ DR3 4979506918874792576
($\equiv$ HD\,4130) is also present in the
lists of \citet{2019A&A...623A..72K} and of 
\citet{2021ApJS..254...42B}.

\gaia\ DR3 6070331129187853824 and
\gaia\ DR3 5059000922370217728
have been observed a few times 
in the framework of the RAVE survey
\citep{2017AJ....153...75K}; 
the first object of the two was also
observed in the framework of the GALAH survey
\citep{2018MNRAS.478.4513B}.
No strong RV discrepancy with our orbital
solutions is observed.

The velocities of the RV variable object
\gaia\ DR3 4660348139680343424
(SV* HV5893) could be surprisingly large
(see Fig.\,\ref{fig:goodrestrend}).
Actually, this star is a known M supergiant
in the Large Magellanic Cloud. It has been observed
in photometry in the framework of the MACHO survey
\citep{2008AJ....136.1242F}
exhibiting a period of about 12.8\,d. This is not 
directly detectable in \gaia\ RVs, but this is not 
either incompatible with the trend claimed here.
\citet{1985A&AS...62...23P} reported for 
this star a velocity of
310.4\,km\,s$^{-1}$ on HJD\,2\,445\,662.77.
The large difference in time between the two
observed passages at this velocity is about
11975\,d and suggests that the variation could be
periodic with a scale of several thousands of days.

The solutions shown in 
Figs.\,\ref{fig:appgoodressb1part1}, \ref{fig:appgoodressb1c}, 
and \ref{fig:appgoodrestrend}
are just illustrative and much more objects with solutions of similar 
quality are present in the SB subcatalogue.
This subcatalogue also contains solutions of slightly less good quality 
as well as solutions of intermediate quality.
In Appendix\,\ref{sec:appE},
we provide a few examples of SB1-type solutions selected from the
SB subcatalogue at random but with $F_2$ around 0
(Fig.\,\ref{fig:randomsample}). This is to have
fits of similar quality but for which the number of data points
is lower (10 to 45) than in Fig.\,\ref{fig:goodressb1part1}. 
Although the quality of these solutions is sometimes less evident, 
only very few could be easily
questionable. 
Since we decided to push the software to its limit, solutions of 
more limited quality are equally present.
For that reason, we encourage the general user to further filter the data. 
This is possible thanks to the quality factors 
that we deliver for each object 
(see Sect.\,\ref{ssec:spectroSB1_quality_indic}).
For a general statistical use, this should generate no problem, provided 
the user adopts the filtering to its
particular application. We encourage the user to be very careful about the 
results on individual objects
that could not benefit from the statistics. 
Some further information and advice can be found below in
Sect.\,\ref{ssec:spectroSB1_catalog_weak}.
\subsubsection{Problematic cases: multi-variability}
\label{sssec:spectroSB1_results_illust_probl_multi}
\begin{figure}[ht]
\centerline{
\includegraphics[width=0.45\textwidth]{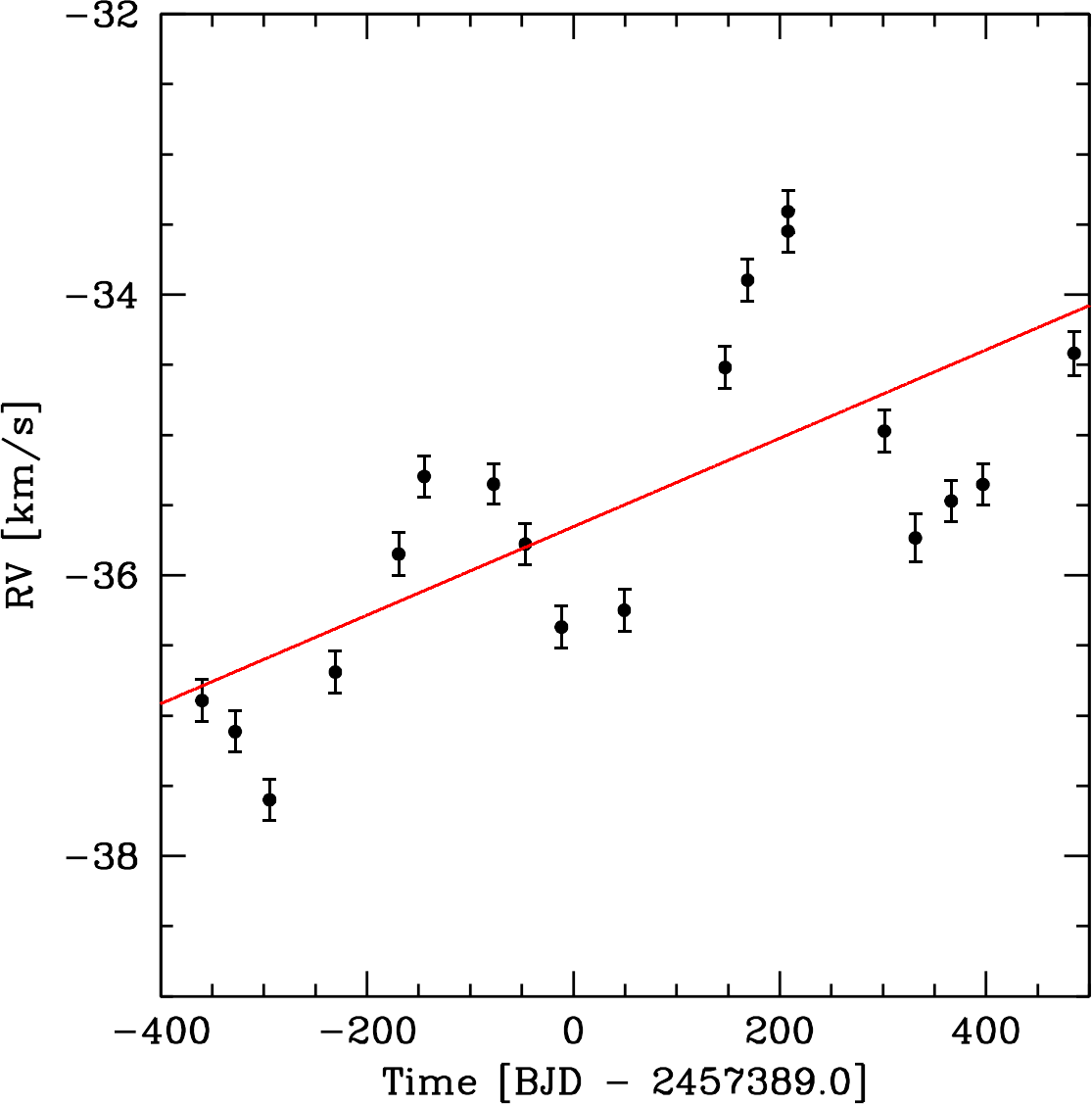}
}\caption{Illustration of a time series of RVs for an 
anonymous object (unpublished).
The dots represent the observed RVs with their error bars. The red 
straight line shows the trend fitted by the pipeline. It corresponds 
to $F_2 \, = \, 20.42$ and thus the corresponding solution is rejected.}
\label{fig:EGmultiple}
\end{figure}
In the framework of DR3, the SB1 pipeline is intended to search for
orbital solutions. It is however designed to treat the cases which 
present only one type of variability (either SB-type with a single period
or pure trends). If one object presents two types of variability, 
most of the time the algorithm could fail to find an adequate solution.
We present here a typical case of badly handled data.
Figure \,\ref{fig:EGmultiple} shows the time series  of a star observed by
\gaia\ RVS ($\Delta T \, = \, 845.177$\,d). The data are unpublished and 
the solution is not part of the DR3 catalogue. The star is thus kept
anonymous. The pipeline detected for this object a linear trend of about
1.15 km\,s$^{-1}$\,y$^{-1}$. The solution corresponds to
$F_2 \, = \, 20.42$. This is easily understood because the RV curve
exhibits extra-variations with a kind of oscillation with a period
a little over 300\,d. The variation is larger than the error bars 
and is thus significant. It necessarily induces a large value
for $F_2$ coming from the fit of the trend, and the corresponding solution
has thus been rejected from the DR3 catalogue. This object actually
presents two superposed, different, types of variability.
A trend that could be associated to a long period binary and, in addition,
a roughly sinusoidal variation that could be due to an SB1 system in
a triple system, or to an intrinsic RV variation or to something else.
If, by chance, the pipeline would have first detected and reported 
an SB1 solution, the existing trend would have also implied a large
$F_2$ associated to the SB1 fit and the solution would also have been
rejected. This is clearly a weak point of the DR3 chain.

For the coming releases, the problem will be corrected and the
pipeline will be able to process properly this kind of multiple 
variability by simultaneously fitting two types
of solutions. This will open the door to the treatment of objects like
triple or quadruple SB1 stars, as well as of intrinsic variables
in a binary system.
\subsubsection{Problematic cases: fake SB1 and spectrocentre}
\label{sssec:spectroSB1_results_illust_probl_fake}
\begin{figure}[ht]
\centerline{
\includegraphics[width=0.45\textwidth]{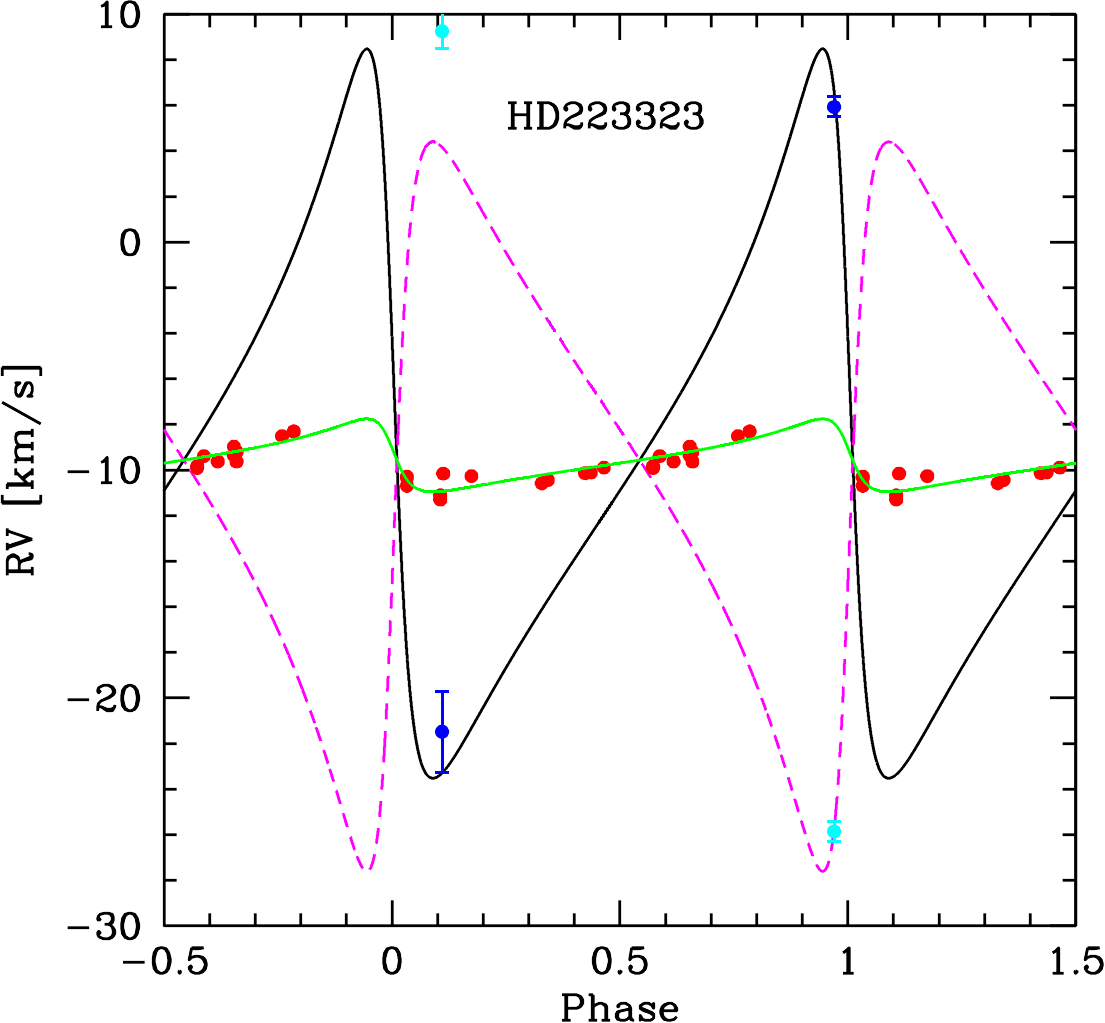}
}\caption{Known ground-based orbital solution for the binary star
HD\,223323 (\gaia\ DR3 2852594583674129152). The primary orbit is illustrated
with a black continuous line and the secondary with a magenta dashed line.
The SB1 RV measurements are indicated by red dots, whereas the SB2
ones are marked in blue (for the primary) and in cyan (for the secondary).
The expected, typical RV of the blend is plotted in green
(see text for explanations).}
\label{fig:EGhd223323}
\end{figure}
We present here the illustrative case of HD\,223323. 
This star ($G$\,=\,6.968)
is a well-known binary object of type SB2 that has been well studied
from the ground; it has a period of 1175.1\,d
\citep{2007Obs...127..113G}.
The two components are rather similar, and only a merged spectral type
in the range F2-4 IV-V is reported
\citep[see references in][]{2007Obs...127..113G}.
An accurate orbital solution has been published
by the latter author, and the elements of interest for the present
discussion are given in Table\,\ref{tab:hdversusgaia}.
Figure\,\ref{fig:EGhd223323} illustrates the fitted orbital model
published by \citet{2007Obs...127..113G}.
The ground observations at phase 0.98 
\citep[almost at larger separation;
see Fig. 5 of][]{2007Obs...127..113G} suggest that the lines 
of the two stars are almost deblended.
\begin{table}[t]
\caption{A few parameters of interest from the Griffin orbital SB2 solution
for HD\,223323 as well as from the \gaia\ equivalent (SB1). 
Note the spurious values for \gaia\ $K$ and mass function.}
\centering
\begin{tabular}{|l|c|c|}
\hline
\hline
Parameter & Griffin & \gaia\ \\
\hline
$P$ (d) & 1175.1$\pm$1.3 & 977.9$\pm$104.3 \\
$\gamma$ (km\,s$^{-1}$)\tablefootmark{*} & --9.56$\pm$0.06 & --9.94$\pm$0.10 \\
$K_1$ (km\,s$^{-1}$) & 16.38$\pm$0.11 & 1.188 \\
$K_2$ (km\,s$^{-1}$) & 16.40$\pm$0.16 & ... \\
$M_1/M_2$ & 1.001 & ... \\
e & 0.604$\pm$0.003 & 0.586$\pm$0.150 \\
$\omega$ (degree) & 77.8$\pm$0.60 & 56.0$\pm$11.0 \\
$f(M_1)$ (solar mass) & 0.272$\pm$0.006 & 9.07$\times$10$^{-5}$ \\
$f(M_2)$ (solar mass) & 0.273$\pm$0.008 & ... \\
$\sigma_{\mathrm{fit}}$ (km\,s$^{-1}$) & 0.53 & 1.61 \\
\hline
\end{tabular} 
\tablefoot{
\tablefoottext{*}{The systemic velocity is expressed in the 
Griffin RV reference system 
(see also Sect.\,\ref{sssec:spectroSB1_validation_otherset_griffin}).} }
\label{tab:hdversusgaia}
\end{table}
The star has been observed with \gaia\ and is present in the DR3 catalogue
under the name \gaia\ DR3 2852594583674129152. A total of 24 measurements
($N_{\mathrm{tot}}\,=\,24$) were collected by
the RVS over a span of time of 
$\Delta T\,=\,889.57$\,d. Two of the transits exhibit a double-line
spectrum, and RVs have been derived as explained in
\citet{dr3-dpacp-161}.
For the remaining 22 transits, the star has been measured as a
single-line object. Owing to the small percentage of time when
the star appeared as double-lined, the pipeline decided to discard
the two corresponding transits
\citep[see the treatment described by][]{dr3-dpacp-161}.
Therefore, the object entered the SB1 chain with
$N_{\mathrm{good}}\,=\,22$ data points.
An SB1 fitted model was presented in the catalogue with a period
977.9\,$\pm$\,104.3\,d. This value is in principle only indicative,
since it is larger than the total span of time 
$\Delta T$ by a factor 1.11.
In any case, the \gaia\ period is in good agreement with the precise
Griffin period since it is only at 1.89\,$\sigma$.
The value of the period will automatically improve
in the coming releases due to the increase of the
$\Delta T$.
The other parameters of interest derived by \gaia\ are available
in Table\,\ref{tab:hdversusgaia}. The measured \gaia\ RVs are plotted in
Fig.\,\ref{fig:EGhd223323}. The red dots represent the RVs derived
by the single-line channel (they are put in the Griffin reference system
by adding 1\,km\,s$^{-1}$; see Sect.\,\ref{sssec:spectroSB1_quality_val_compGriffin} 
below).
The other data points represent the primary (in blue) and the secondary
(in cyan) RVs as derived in CU6 
\citep{dr3-dpacp-161}. They were also corrected to the Griffin
reference zero point.
It is remarkable that the \gaia\ solution yields a very small value
for the semi-amplitude. The interpretation is quite straightforward.
Except for the phases near the extrema in RVs, the 
resolution of the RVS and the predominance of the broad Calcium lines of the
triplet makes that the lines never deblend at the other phases. 
After the rejection of the two double-line transits, 
the star becomes a fake SB1 which is 
actually a never-deblending SB2. 
There is a tendency for the position of the blended
lines to moderately follow the behaviour of the primary variation.
According to \citet{2007Obs...127..113G}, the lines of the secondary are slightly
broader and the lines of the primary slightly deeper. 
This thus implies that the primary lines are moderately dominating the blend.
The secondary
line depths are roughly 90\,\% of those of the primary one
(we will call this ratio B\,=\,0.9). The effect of the width of the lines
of the secondary is attenuated since the lines
are rather strongly blended.
\citet{1968MNRAS.141...43T} presented a 
computation of the position of the minimum of the blend of two Gaussian
profiles (which is unique in the case of a strong blend) as a function
of the widths and of the depths of the two profiles.
Following their work, using the same notation as in their Eq.\,5, and 
deciding to neglect the difference in broadness
of the lines, we derived a position of the minimum which is given
by
\begin{equation}
x \, = \, \frac{B}{1+B} \, \, \, .
\end{equation}
According to the latter author, $x$ adopts a value of 0 if the minimum
is at the velocity of the primary and 1 at the secondary position.
In the present case, we derive $x$\,=\,0.45 which means that the minimum
is indeed attracted by the primary. As an approximation,
we consider that the position of the minimum and the one of the blend
are similar. In the same approach, we consider that the position of the 
blend can be approximated by a weighted sum of the velocities of the
primary and of the secondary. It follows
\begin{equation}
RV_\mathrm{blend} \, = \,  0.55 \, RV_{\mathrm{pr}} \, + \, 0.45 \, RV_{\mathrm{sc}}
\end{equation}
consistently giving more weight to the primary velocity.
Starting from the modelled RVs by Griffin, we computed the evolution of the
position of the blend as a function of phase. This is illustrated
in Fig.\,\ref{fig:EGhd223323} by the green line.
Despite the fact that this line is in no circumstances a fit, the agreement
with the red dots is very good. This demonstrates the interpretation of the
\gaia\ results as a fit of the blend.
A more rigorous approach would be much  heavier and is not in the
scope of the present analysis which is intended to remain purely
illustrative.
By analogy with the notion of photocentre, we could introduce here the
concept of {\em{spectrocentre}} which would correspond to the measurement
of the RV of the blend (in the spectrum) constituted by the presence
of two sets of undeblended lines. This could thus be termed a
{\em{spectrocentre}} effect\footnote{It should however be pointed out that 
this effect is mainly dependent on the characteristics of both sets of lines
(with a different contribution from line to line) and thus
on the Astrophysical Parameters characterising the two stars.}.

It is interesting to notice that in the case of HD\,223323, the most
affected parameter is essentially the semi-amplitude. On the
contrary, it is clear from Table\,\ref{tab:hdversusgaia} that the other
parameters are not strongly perturbed.
The eccentricity in particular is well determined, as well as the
argument of periastron. This is a consequence of the slight domination
of the primary on the secondary effect.

The case of the never-deblending SB2 stars is crucial. If the two
object spectra are similar (except for the Doppler shift), 
the semi-amplitude associated to the blend  will near zero.
Therefore, the corresponding object will be characterised
by a very-small $K$ and also be associated to an abnormally small
mass function. An extreme case with two identical stars will
lead to a fake RV constant star.
On the opposite, if the similarity of the two systems
of lines for the two objects does not hold, the resulting semi-amplitude
could of course be much larger and it is to be expected that such a case
will be harder to detect. At the other extreme, if the secondary spectrum
is negligible, the solution will tend towards a true SB1.
The detection of the fake SB1 never-deblending SB2 is important, and
nothing global has been implemented for DR3.
We identify two avenues to solve the problem for future releases.
The first one is to look at the evolution with phase of the
broadness of the lines (even epoch by epoch). This could be done
directly on the spectra. For DR3, an a posteriori look at the epoch 
{\tt{vbroad}} values \citep{2023A&A...674A...8F} failed to produce results.
An alternative would be to systematically
reconstruct the mean spectrum at rest (according to the SB1 solution)
and to detect line-profile variability through inspecting the
dispersion of the reconstructed mean spectrum in the 
neighbourhood of the main spectroscopic lines. Due to the general
architecture of the \gaia\ pipeline, these possibilities were 
impossible to implement in the DR3 pipeline. Similarly,
we did not investigate in DR3 the possibility to use the
information from astrometry to alleviate the problem.
\subsubsection{Problematic cases: possible spuriously variable RVs}
\label{sssec:spectroSB1_results_illust_probl_spurious}
\begin{figure}[ht]
\centerline{
\includegraphics[width=0.45\textwidth]{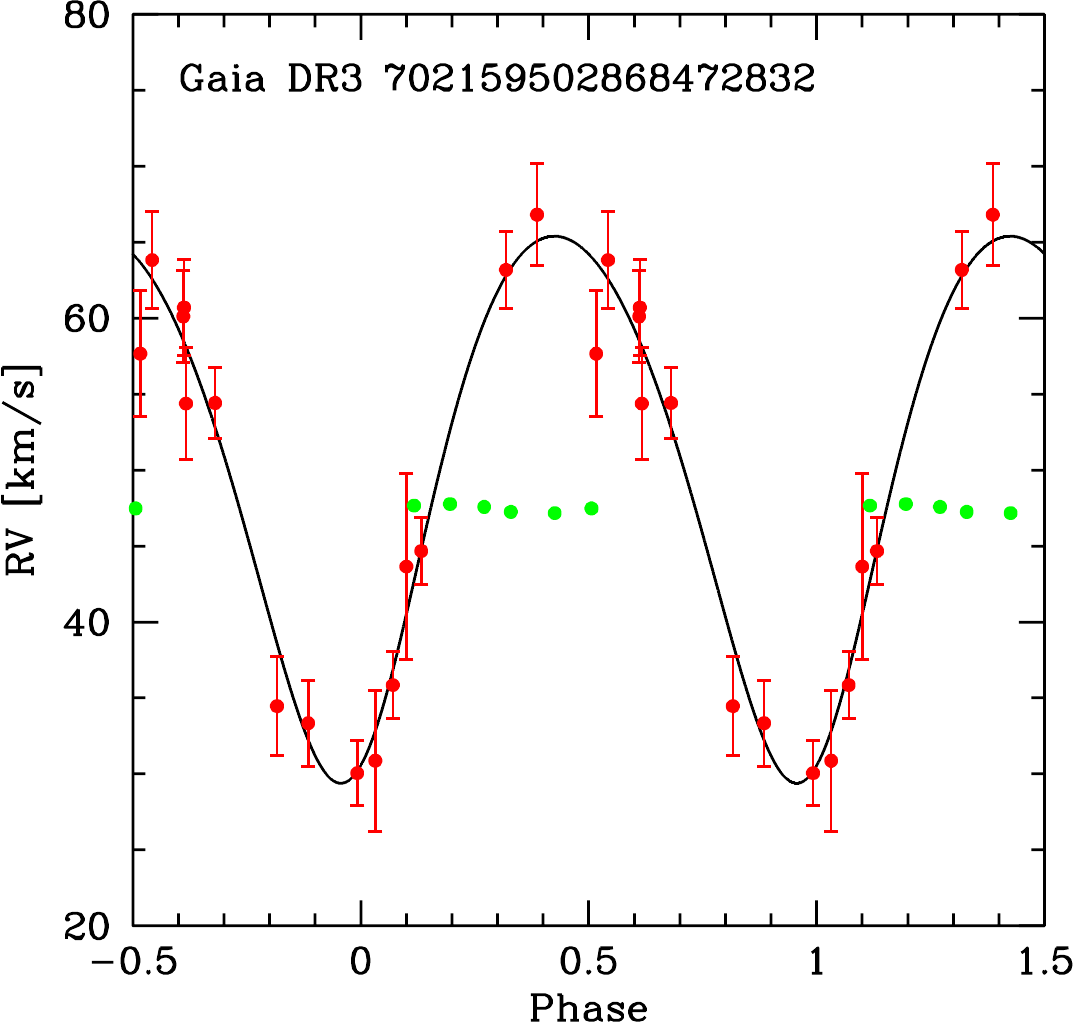}
}\caption{Folded RV curve for the star \gaia\ DR3 702159502868472832 
corresponding to the selected period of 1.084400\,d. 
The observed RVs and their errors are plotted in red, along 
with in black the fitted RV curve. The green dots represent 
the ground-based RVs of the APOGEE survey folded with the same ephemeris.}
\label{fig:EGcas2}
\end{figure}
Some stars are associated with an orbital SB1-type solution that looks very convincing,
and is identified as good according to the various quality tests. 
These stars thus enter the orbital solution catalogue. However, ground-based 
observations do not confirm these orbital variations. 
An illustrative case is the one of
\gaia\ DR3 702159502868472832. Its RV time series suggests a significant period
of 1.084400$\, \pm \,$0.000058\,d with a probability of 0.997
and a resulting folded RV curve that is given in 
Fig.\,\ref{fig:EGcas2}. With a significance of 15.41 and an $F_2 \, = \, -0.36$, 
there is no solid reason to
doubt the nature of this object. However, besides the \gaia\ RV curve and the fit, 
we plot also in the figure the few folded RVs that can be 
found in the literature from the
APOGEE survey DR17 \citep{2022ApJS..259...35A}. 
The two data sets are clearly not compatible. 
This suggests that some of the spectroscopic data from \gaia\ could not 
only present spurious RVs that 
are variable, but also that a period can be found from these measurements. This also 
suggests that the effect could be instrumental. The origin is 
currently not securely identified. A small number of objects 
enter this category, but the 
related sparsity could be due to the small percentage of objects having ground-based
observations. This is supported by the validation performed 
in Sect.\,\ref{ssec:spectroSB1_validation_otherset} below.
Therefore, it is primordial to discover the origin in order to improve
the robustness and purity of the future Data Release catalogues. 
We listed a few such doubtful cases that were detected here:
they are
\gaia\ DR3 66572718940795136,
\gaia\ DR3 286652890498864768,
\gaia\ DR3 487550398091020160,
\gaia\ DR3 164423034699251968,
\gaia\ DR3 577600125523574400,
and the illustrative case \gaia\ DR3  702159502868472832.
All of them have been compared to APOGEE DR17 data.

\begin{figure}[ht]
\centerline{
\includegraphics[width=0.45\textwidth]{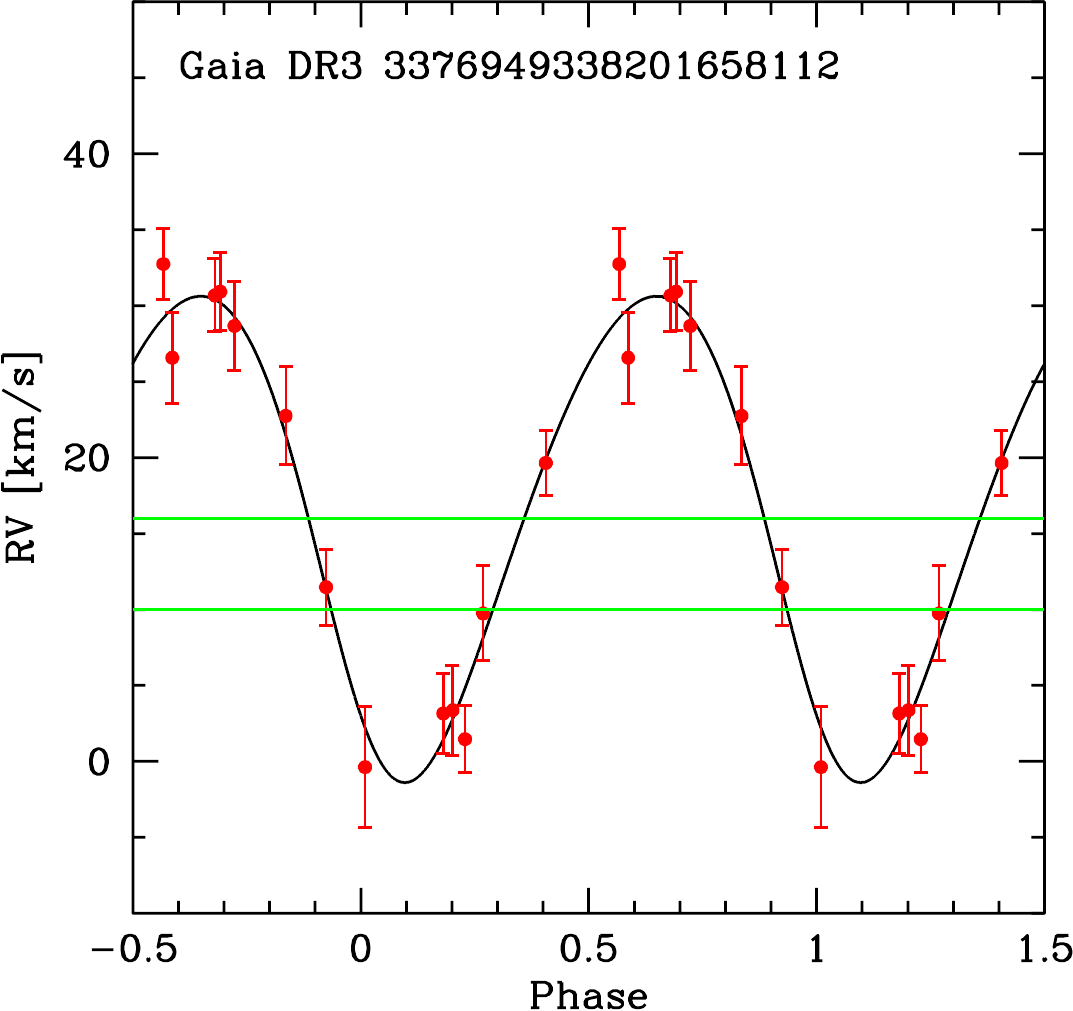}
}\caption{Folded RV curve for the star \gaia\ DR3 3376949338201658112 
corresponding to the selected period of 0.657582\,d. 
The observed RVs and their error bars are plotted in red along with 
in black the fitted RV curve. The green lines delimit the
domain where the ground-based RVs part of the LAMOST survey are distributed.
See also the figure A.3 of \citet{2022MNRAS.517.3888B}.}
\label{fig:EGcas1}
\end{figure}

Another similar case has been detected by
\citet{2022MNRAS.517.3888B} but this time issued from 
comparison with the LAMOST RV survey
\citep[][\footnote{\url{https://dr6.lamost.org}}]{2012RAA....12.1197C}.
The object is \gaia\ DR3 3376949338201658112 and a significant periodicity has here also
been identified ($P\, = \, 0.657582 \, \pm \, 0.000022$\,d) and the corresponding 
RV curve is shown in 
Fig.\,\ref{fig:EGcas1}. Again the RV variations are convincing as well as the fit.
The object is one of very few that have been observed at several epochs in the
framework of the LAMOST survey. All the corresponding RVs are contained between the
two green lines, further underlining the incompatibility of the 
ground-based data with the
\gaia\ ones. 

The fact that two ground-based surveys disagree with the results from \gaia\
underlines the fact that some instrumental effects are most probably present in the
\gaia\ data. It is particularly intriguing that the RV time series exhibit what 
could appear as a convincing periodic variability characterised by timescale not known
to be problematic. Clearly, the problem is linked to the production of the RV
time series rather than to the behaviour of the algorithm of orbital
parameter determination, except perhaps for the period.
\section{Validation of the results}\label{sec:spectroSB1_validation}
The analysis of the huge data set generated by \gaia\ and 
the establishment of the related catalogue
necessitates some consolidation mechanism.
This mechanism can be made of several approaches: they 
are described in the following subsections.
The aim is evidently to test and fine-tune the pipeline 
itself in order to address its efficiency 
but also to discern the unavoidable biases.
The understanding of the pipeline behaviour is directly related to the quality of the 
SB subcatalogue 
and looking at its characteristics
from all angles is a necessity. This validation process can be made on the basis of the 
internal consistency of the solutions but also on the basis of a comparison between 
the parameters delivered from the \gaia\ data with those obtained from the literature. 
This last approach is limited to the dramatically smaller 
amount of objects that have been 
the subject of a previous study. The validation paves the way for the determination 
of the global selection function.
\subsection{Simulations}\label{ssec:spectroSB1_validation_simul}
\begin{figure}[ht]
\centerline{
\includegraphics[width=0.25\textwidth]{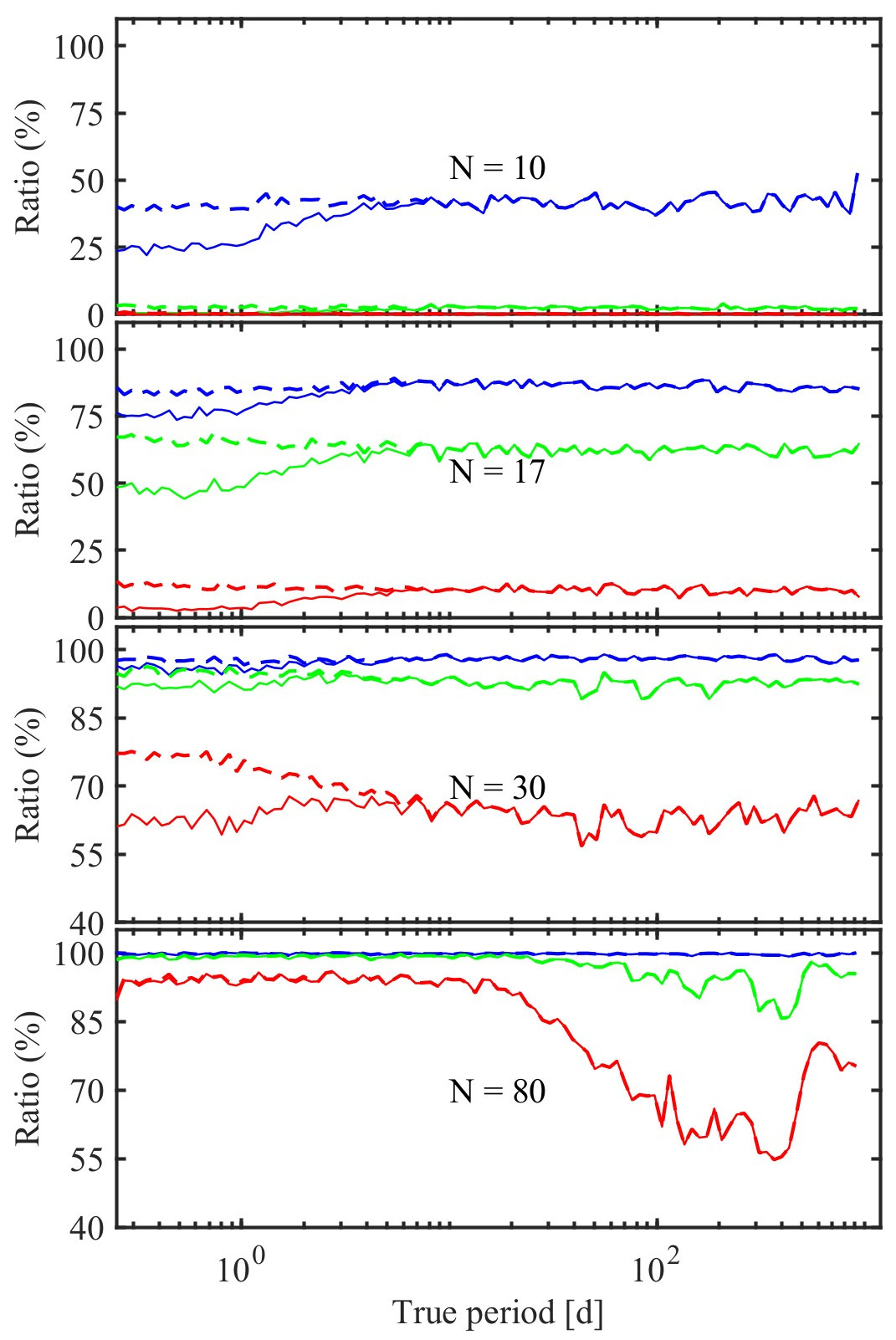}
\includegraphics[width=0.25\textwidth]{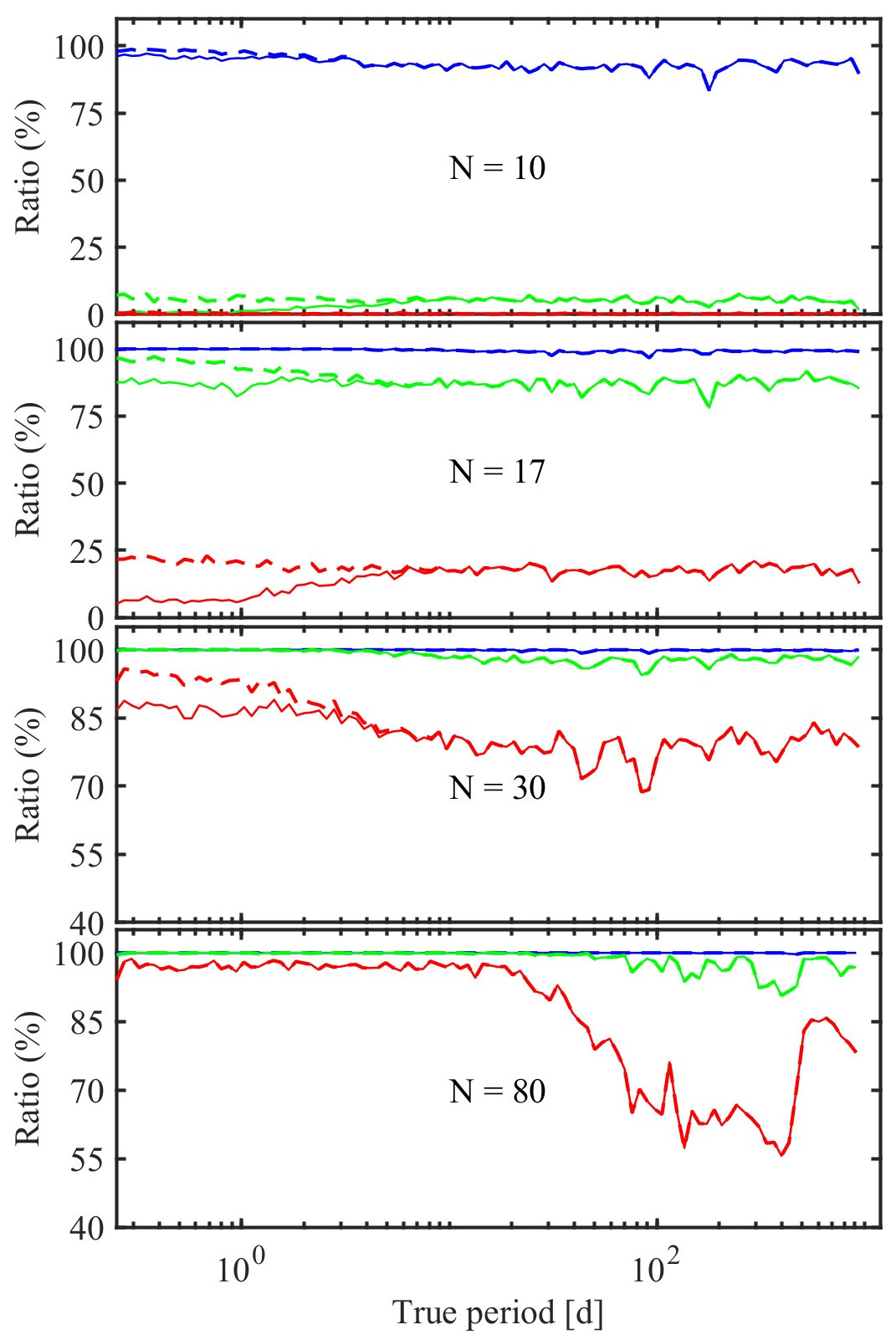}
}\caption{Performance for the period recovery of the pipeline 
applied on simple simulated data.
$N$ is the number of data points in the time series. The colour curves correspond to
different S/N ratios $s$ (blue: 50, green: 10, red: 5). 
The dashed curves correspond to the
recovery before the application of the statistical test on the period 
significance level;
the continuous curve refers to the recovery rate after having applied the selection.
Left: eccentric orbits; Right: circular orbits.}
\label{fig:EGsimulation}
\end{figure}

As a very first step, it is necessary to study the ability of the pipeline to
recover the period. The study described here is based on the generation of synthetic
RV curves of eccentric systems (with a random $e$), and of circular systems.
Both sets of RV curves are generated with a random phase.
We considered a number of data points of $N \, = \,$ 10, 17, 30, and 80. 
The times of observations
are mimicking the \gaia\ time series and are drawn in the same manner as in
Sect.\,\ref{ssec:spectroSB1_lookup}. 
The adopted S/N ratio $s \, = \, 2 \, K/ \sigma_{\mathrm{RV}}$
is considered for values 5, 10, and 50.
Figure\,\ref{fig:EGsimulation} reveals the recovery rate (in percent) 
for various true periods.
The period is assumed to be recovered if its value does not differ 
from the input period by more than $5 \, \sigma$.

For $s$ equal to or larger than 10, the recovery rate is 
mostly better than 80 percent except when the number of data points
is too low (see e.g.\ $N \, = \ 10)$. 
This suggests that a too low number $N$ is a problem for the algorithm.
It is also clear that an $s$ of 5 is equally detrimental. 
In Table\,{\ref{tab:tabsignif}},
we give the corresponding value for the equivalent {\tt{significance}} given by
\begin{equation}
\sim \frac{\sqrt{N}}{2 \sqrt{2}} \, s
,\end{equation}
as derived from Eq.12 of \citet{1971AJ.....76..544L}.
\begin{table}[t]
\caption{{\tt{Significance}} as a function of the number of data points $N$
and of the S/N ratio, $s$.}
\centering
\begin{tabular}{|l|c|c|c|}
\hline
\hline
$s\, \, \, \, \rightarrow$ & 5 & 10 & 50 \\
\hline
$N \,= \, 10$ & 5.6 & 11.2 & 55.9 \\
$N \,= \, 17$ & 7.3 & 14.6 & 72.9 \\
$N \,= \, 30$ & 9.7 & 19.4 & 96.8 \\
$N \,= \, 80$ & 15.8 & 31.6 & 159.1 \\
\hline
\end{tabular} 
\label{tab:tabsignif}
\end{table}

As an alternative, a cut-off at a {\tt{significance}} of 
about 10 is certainly a good attitude.
It is interesting to note that the recovery rate is slightly 
better for circular orbits 
(see right panel of Fig.\,\ref{fig:EGsimulation}).
\subsection{Internal verification of the results}
\label{ssec:spectroSB1_validation_intern}
\subsubsection{{\tt{SB1}} and {\tt{SB1C}} classes}
\label{sssec:spectroSB1_validation_intern_SB1}
\begin{figure}[ht]
\centerline{
\includegraphics[width=0.45\textwidth]{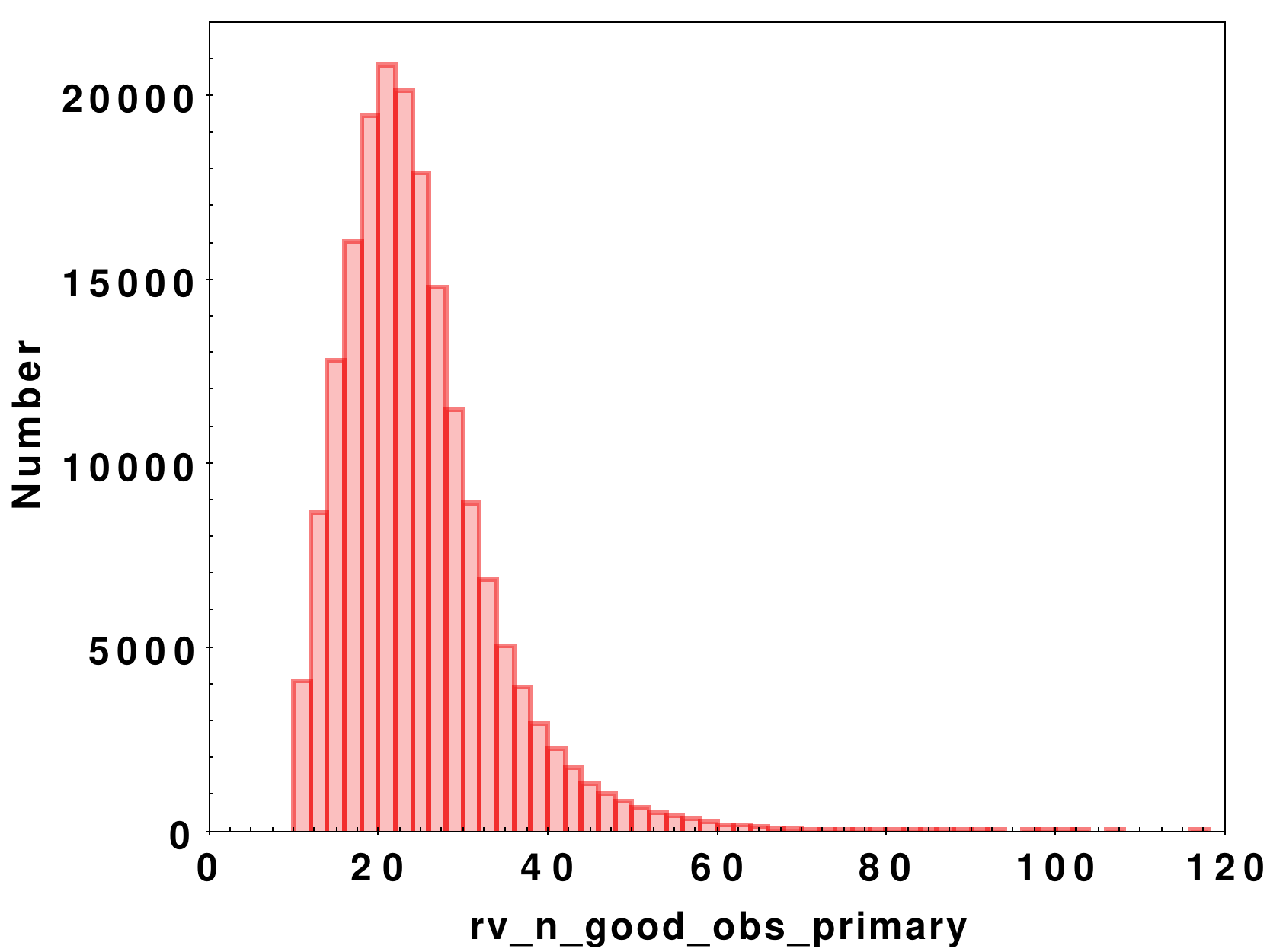}
}\caption{Histogram of the distribution of the number of good transits
({\tt{rv\_n\_good\_obs\_primary}} $\equiv N_\mathrm{good}$) associated 
to the different objects with SB1-type solutions processed
by the pipeline. The typical number of data points is slightly larger
than 20. The bin width is 2 transits.}
\label{fig:EGnumbertransits}
\end{figure}
\begin{figure}[ht]
\centerline{
\includegraphics[width=0.45\textwidth]{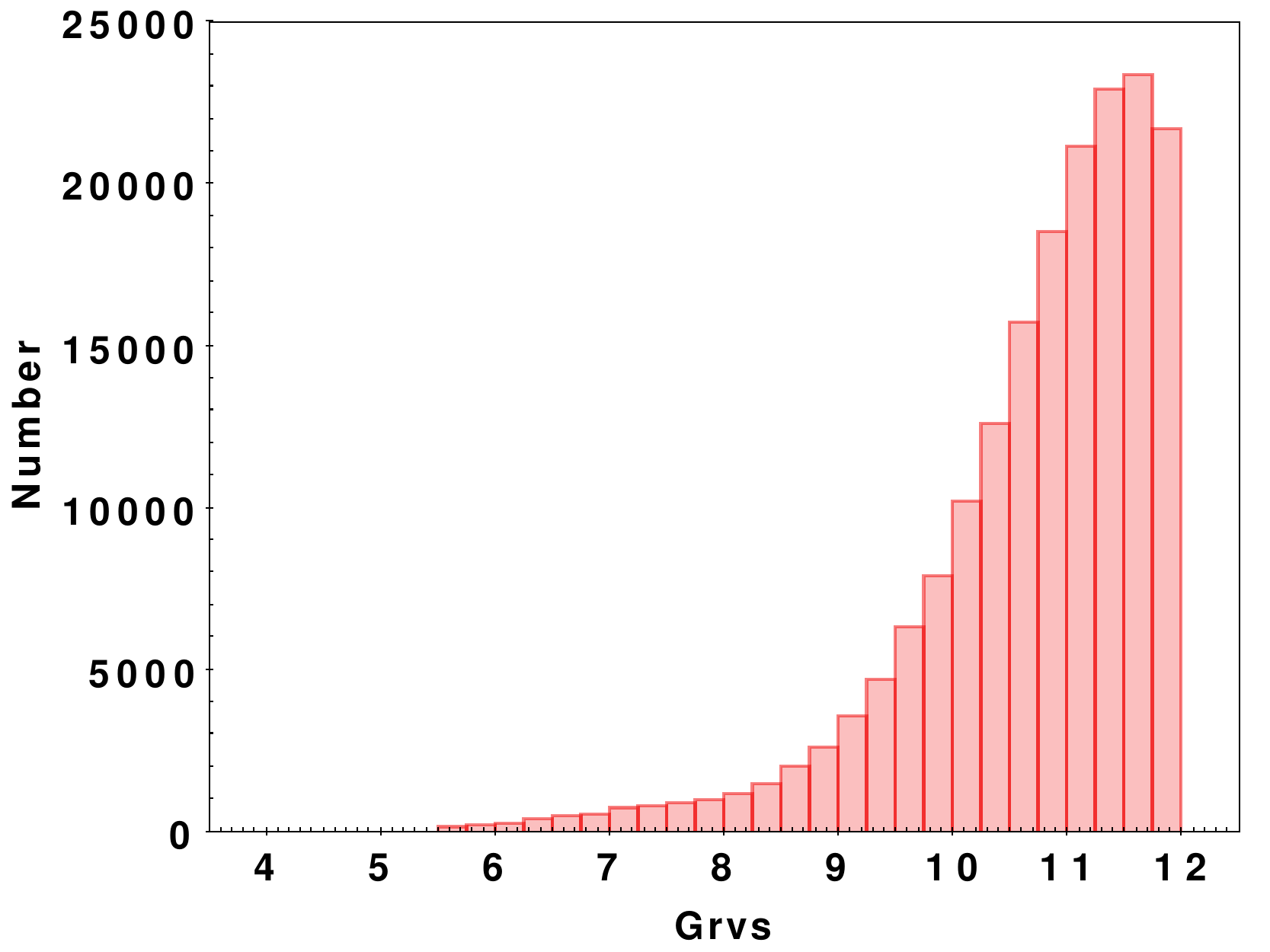}
}\caption{Histogram of the distribution of
$G_{\mathrm{RVS}}^{\mathrm{int}}$
for all the 183\,434
{\tt{SB1}} solutions. The bin width is 0.25 mag.}
\label{fig:EGhistogrvsmag}
\end{figure}%
\begin{figure}[ht]
\centerline{
\includegraphics[width=0.45\textwidth]{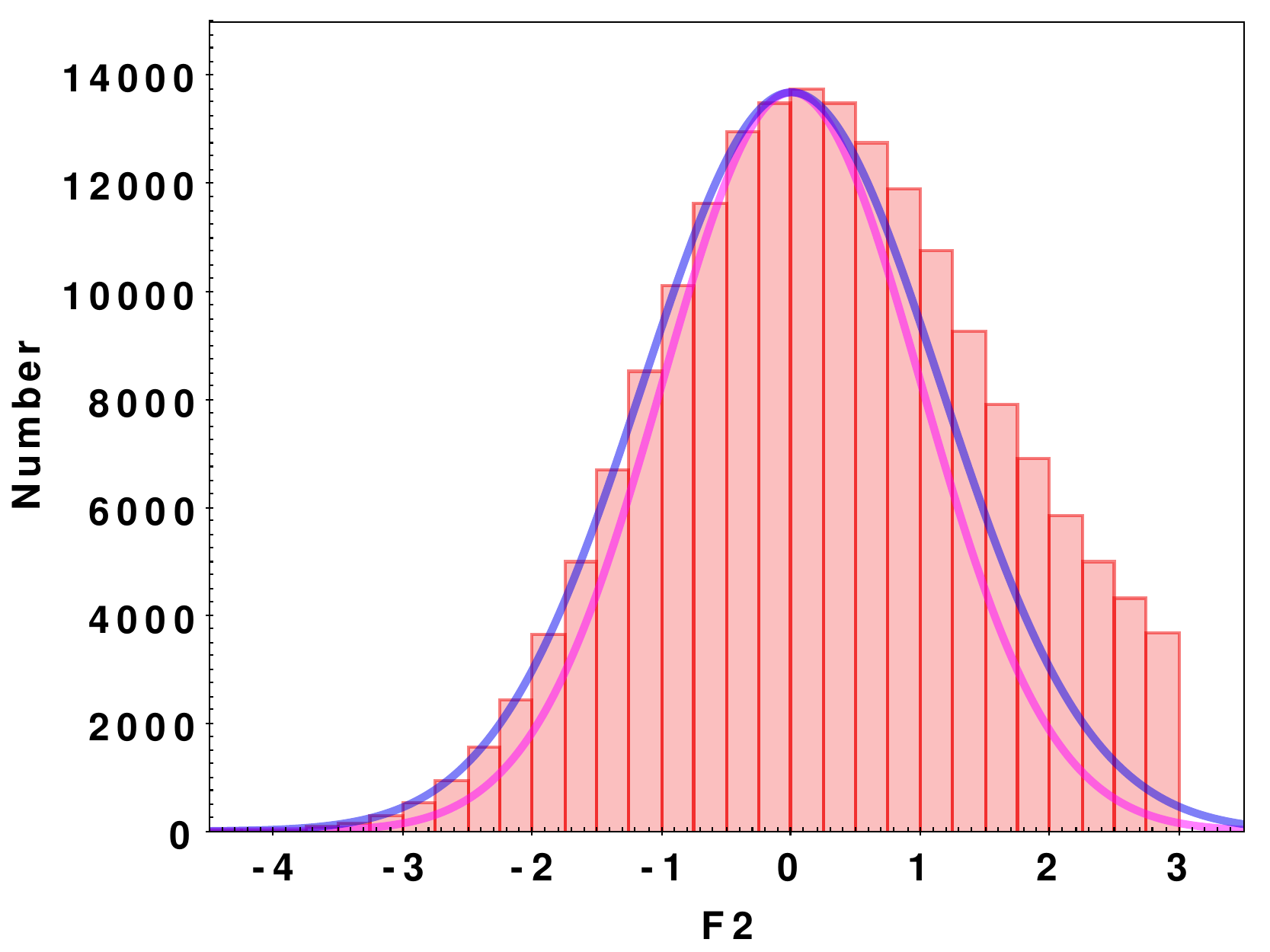}
}\caption{Histogram of the distribution of the $F_2$ statistic of the 
different fits that are adopted as good ones. The bin width is taken to be 0.25.
The cut-off at $F_2$\,=\,3 corresponds 
to the adopted threshold. The magenta curve is a Gaussian function corresponding 
to a zero mean and a dispersion $\sigma = 1$. This is the expected behaviour
of the $F_2$ statistic.
The blue curve represents
a Gaussian with $\sigma = 1.15$. This latter function follows rather well the slope
of the $F_2$ histogram on the left side.  Broadly speaking, the $F_2$ function is 
behaving as expected except for this larger dispersion. 
Both Gaussian curves were adjusted
on the maximum of the histogram and are consequently not normalised on their area.
They are drawn to illustrate the explanation in the text.}
\label{fig:EGhistoF2}
\end{figure}
\begin{figure}[ht]
\centerline{
\includegraphics[width=0.45\textwidth]{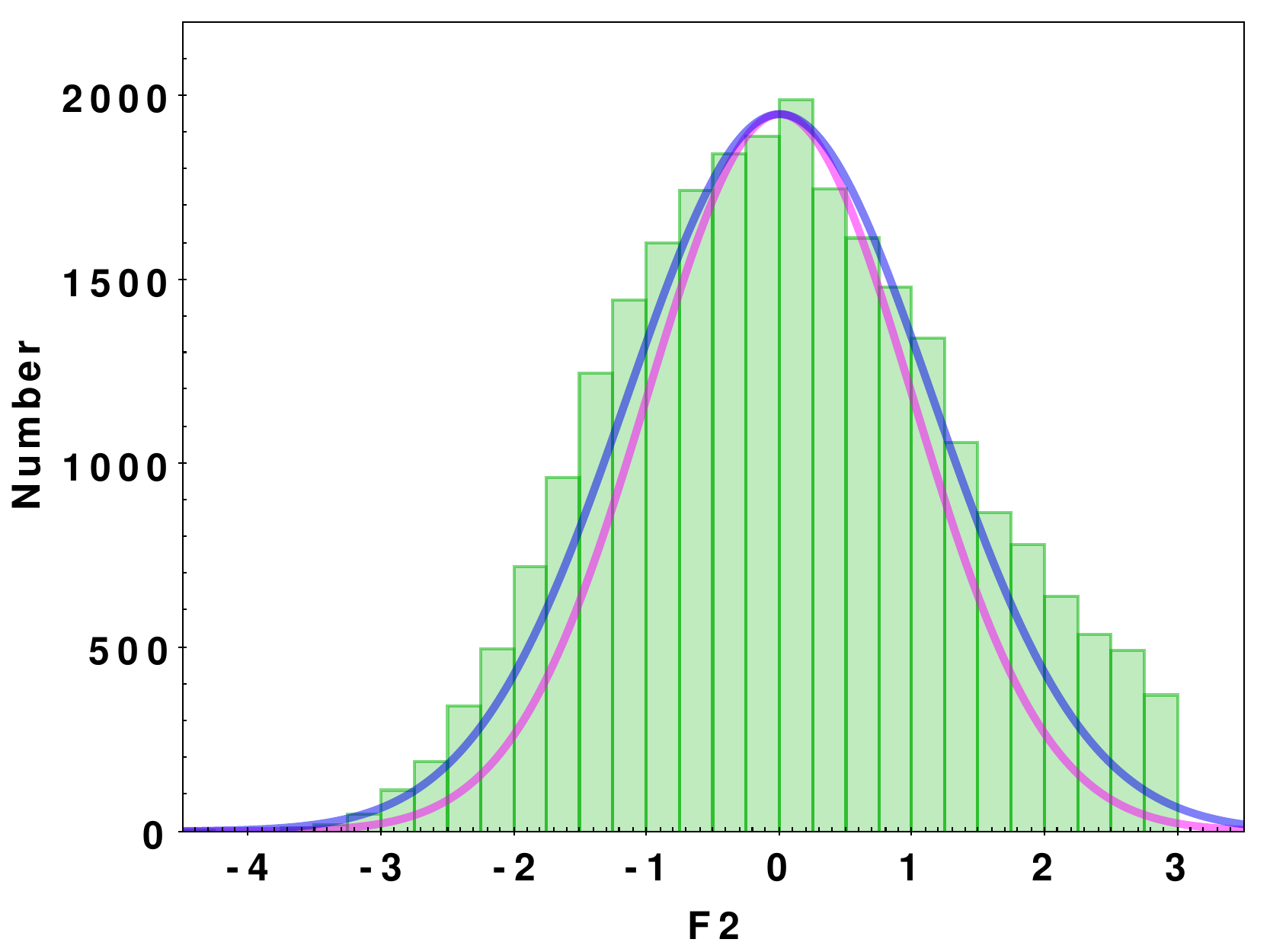}
}\caption{Same as Fig.\,\ref{fig:EGhistoF2} but restricted to 
solutions with $n$ in the range 4-9, i.e.\ 
$N_{\mathrm{good}}$ in the range 10 to 15.}
\label{fig:EGF2lowdof}
\end{figure}
\begin{figure}[ht]
\centerline{
\includegraphics[width=0.45\textwidth]{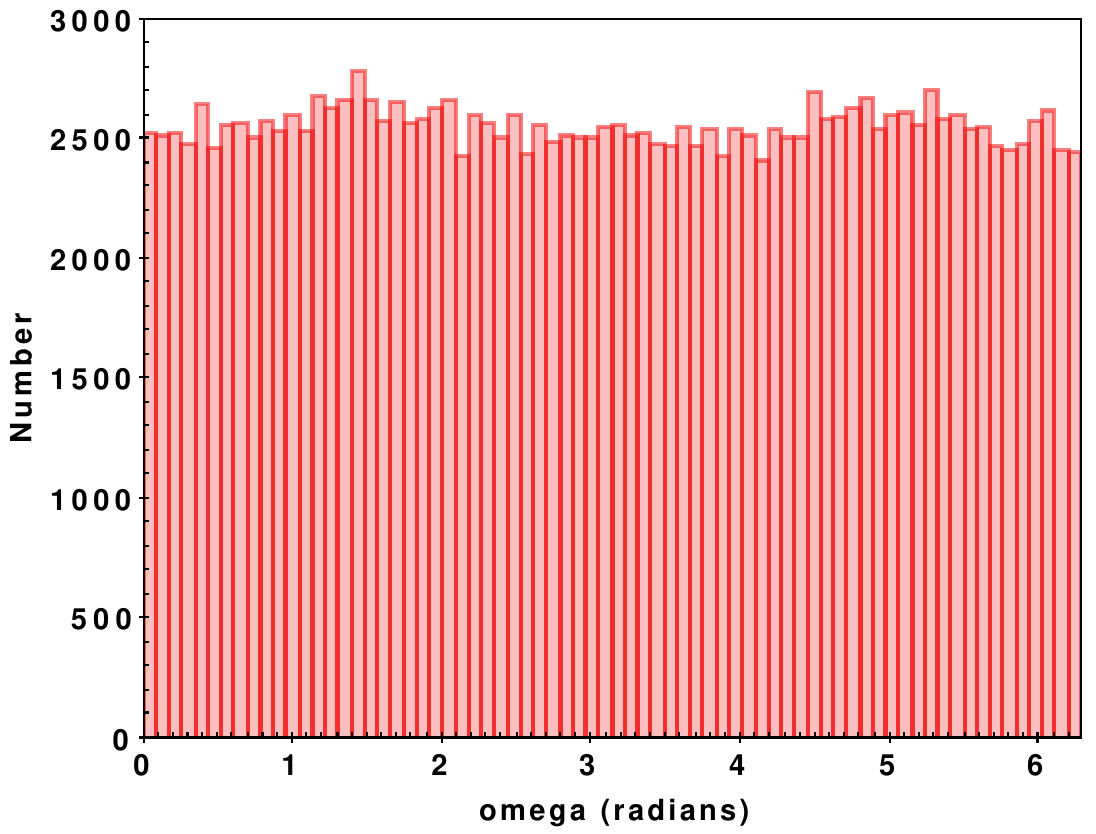}
}\caption{Distribution of the argument of periastron over the whole 
{\tt{SB1}} sample. The
distribution is pretty well homogeneous. The bin width is 0.0873 radian (5\,\degr).}
\label{fig:EGhistomega}
\end{figure}
\begin{figure}[ht]
\centerline{
\includegraphics[width=0.45\textwidth]{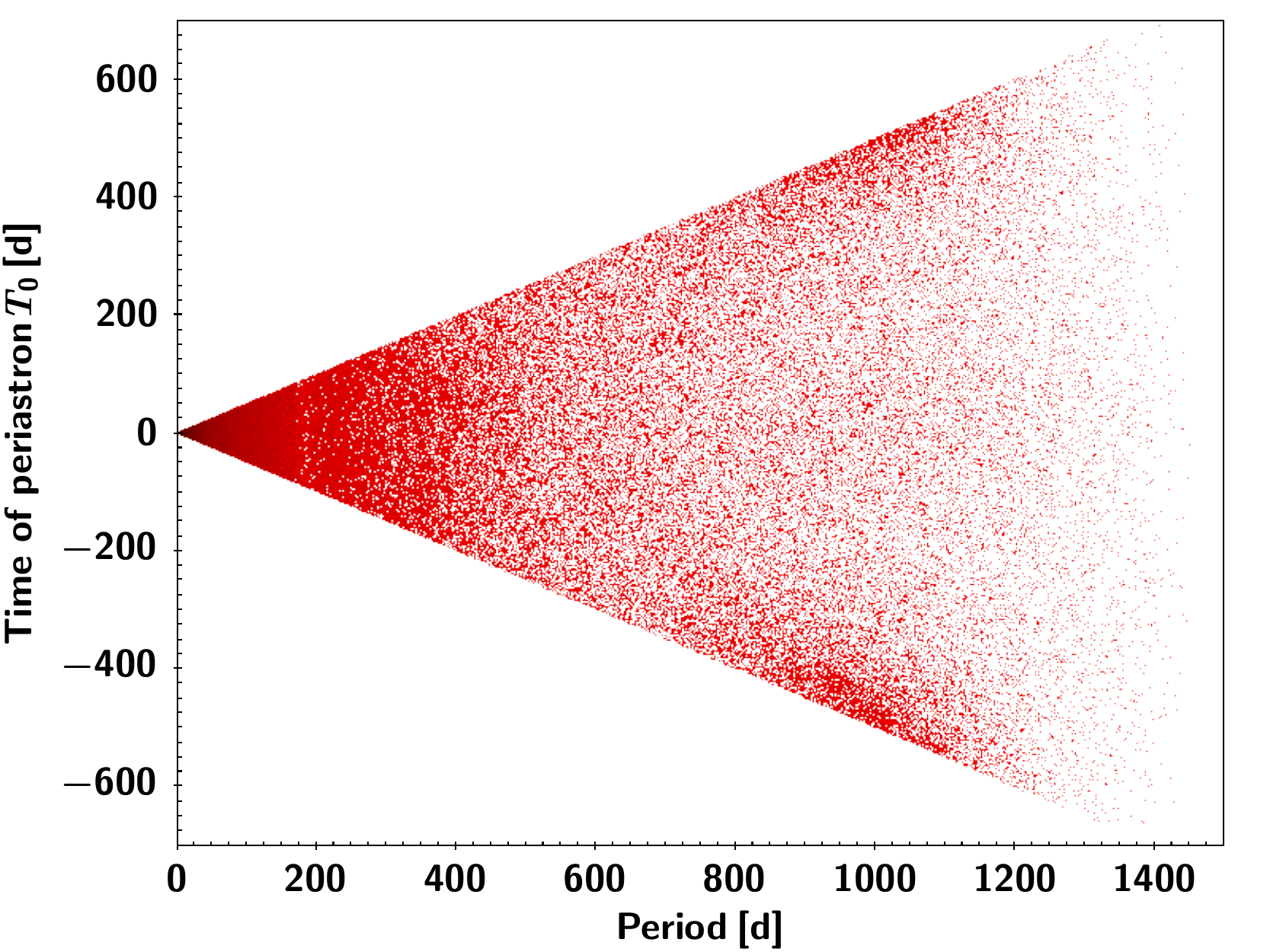}
}\caption{Plot of $T_{\mathrm{0}}$ versus the period for all the 183\,434 objects.}
\label{fig:EGtperversusperiod}
\end{figure}
\begin{figure*}[ht]
\centerline{
\includegraphics[width=1.1\textwidth]{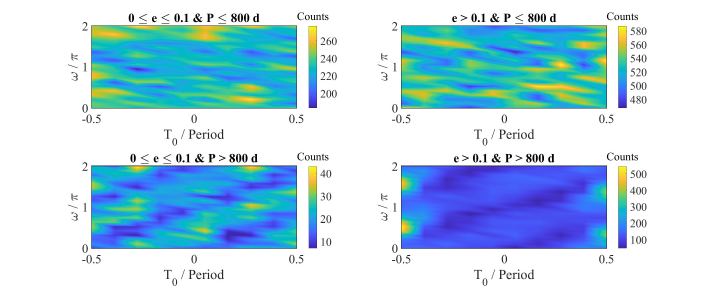}
}\caption{Graph of the distribution of the density of
solutions as a function
of two parameters for various ranges in eccentricity and period.
The abscissae denote $T_0$ normalised by the period whereas the
ordinate denotes the value of $\omega$ over $\pi$. The four panels
correspond to eccentricities below and above 0.1 (left and right, 
respectively) and periods below and above 800\,d 
(up and bottom, respectively). 
}
\label{fig:EGOmegaT0}
\end{figure*}
The analysis presented here below is essentially based on the output of the 
spectroscopic pipeline after the internal filtering.
The SB1 data set contains 183\,434 objects.
The basic reason for this choice is that the principal 
motivation is to first scrutinise the 
behaviour of the pipeline.
The post-filtering is based on information from other channels and could thus 
be more complex to interpret, although it is necessary in order
to constitute the DR3 SB subcatalogue. In any case, for DR3, the differences 
between the analysed data set and the data set after
post-filtering that contains 181\,327 objects, are almost negligible 
from a statistical point of view.
The resulting data coming out of the pipeline were systematically inspected
in order to verify the conformity to the expectation and the expected statistical 
distribution of the fitted parameters. Basically we are mainly treating the
{\tt{SB1}} class excluding the SB1 solutions that were combined downstream  
with the solutions from other channels. The {\tt{SB1C}} class contains 
too few objects to allow us to perform a fully sound 
statistical analysis. However, these 
solutions correspond to a very simple model that 
should not bear any particular problem.

The distribution of the number of good transits (i.e.\ data points) for the
SB1-type solutions processed by the pipeline is illustrated in
Fig.\,\ref{fig:EGnumbertransits}. 
The objects having less than 10 transits were not further considered
as explained above.
This cut-off below 10 transits
should not discard too many objects. Typically, we have slightly over
20 data points in our various time series. The largest value is 228 
(not shown here) but these objects are exceptional and 
are corresponding to the EPSL observations.
As a first test, we examine the distribution
of the magnitude
$G_{\mathrm{RVS}}^{\mathrm{int}}$
of all solutions in the {\tt{SB1}} class.
Figure\,\ref{fig:EGhistogrvsmag} exhibits the histogram of
$G_{\mathrm{RVS}}^{\mathrm{int}}$
for all the 183\,434 objects. We immediately notice that no object 
is present between $G_{\mathrm{RVS}}^{\mathrm{int}}\,=\,2.76$
and 5.5. 
This is an artefact of the selection 
based on the {\tt{rv\_renormalised\_gof}} 
at the level of the input data. 
This quantity is defined when
$G_{\mathrm{RVS}}^{\mathrm{int}}\,>\, 5.5$
and thus below this threshold, 
the value for these objects contains NaN
even if these bright objects should 
be kept. This is a weakness of DR3 
and it will be corrected in future releases.
On the other hand, it is also clear that signs of incompleteness are starting 
to appear above magnitude 11.5. A minimum of
$\sim 2000$ objects are lost between magnitude 11.75 and 12.0. 
This is most probably 
an effect of the increased importance that noise
could have for the fainter objects, thus leading to 
badly determined solutions.
However, an incompleteness also appears in the global RV 
output of the CU6 pipeline 
\citep[see fig.\,4 of][]{2023A&A...674A...5K} for objects fainter than 
$G \, \sim \, 12.15$ (corresponding for a solar-like star to 
$G_{\mathrm{RVS}}^{\mathrm{int}}\,>\, 11.5$). 
Therefore, the difficulties to measure RVs on noisy spectra,
the selection of variable objects in noisy time series and
the subsequent fit of orbital solutions
conspire to induce this incompleteness.

The observed distribution of the quality-of-fitting statistic deduced from 
the chi-square, the goodness of fit $F_2$, is given for the filtered data set 
of solutions (see Fig.\,\ref{fig:EGhistoF2}). 
The recommended and applied cut-off at $F_2 = 3$ is clearly visible. Otherwise, 
the original decay of the histogram is going up to values of several tens. 
Large values of the statistic are to be associated with bad, unacceptable solutions. 
This is certainly so for
$F_2$ larger than 5. The cases between 3 and 5 could be further discussed. 
Conservatively, we only kept values below 3.
This induces the rejection of some acceptable
solutions, and consequently of the related objects.
Along with the observed distribution, we also plot a Gaussian
around the mean zero and with a $\sigma = 1$ dispersion (in magenta). 
This curve represents 
the expected behaviour of $F_2$ under the null-hypothesis 
of a perfect Keplerian model
with additive Gaussian noise 
(corresponding to the uncertainties
associated to the individual RVs).
Another Gaussian curve
with a $\sigma = 1.15$ (in blue) is also shown.
This latter better follows the histogram on its left tail. 
The side of the negative values of $F_2$ corresponds to solutions for which   
the realisation of the noise conspires to produce better fitting. 
On the other hand, the right tail is associated 
to the opposite effect where the solutions 
are noisier although still acceptable. It is interesting to notice that, in this 
region, the histogram reveals a remarkably too large number of objects. 
This is true even comparing to the blue Gaussian curve; an excess is thus present.
They most probably correspond to solutions for objects following a Keplerian orbit 
(thus generating a rather good fit) but affected 
by additional variations (pulsations, activity, spots, …), not due to 
the observational noise, but at a 
sufficiently low level such that the dominating effect 
is still the orbital variation. 
Among these objects, multiple systems and intrinsic variables could be 
the main contributors to this effect. 
A more sophisticated study, beyond the scope of the present paper, is necessary.
However, we should not be misled by the normalisation
of the comparison Gaussian curves. Actually, 
if we normalised with respect to the area, there 
would appear a deficit (roughly 25\%) of objects
with solutions characterised by an $F_2$ broadly
located around zero
(essentially between --1. and +1.). 
These solutions are displaced
towards the right tail.
The effect is present both on the individual $\chi^2$
(with a selected value of $n$, the number
of degrees of freedom)
and on the  global $F_2$ statistics. The transformation
of the $\chi^2$ values into the global $F_2$ statistic
is not the cause of the deviation.

Coming back to the anomaly of the left tail,
it could be due to a general over-estimation of the
individual uncertainties on the measured RVs. This would
displace the whole distribution by 15\% to the left
but the above-mentioned effect on the right tail
would prevent any confirmation based on this 
right tail. This explanation would induce an effect
independent of the number of data points.

Around $F_2 \, \sim \, -3.$, there is a 
deficit of solutions
with large significance (roughly above 40).
Consequently, any attempt to correct the shift to
the left of the histogram by
filtering out low-significance solutions, implies 
the appearance of a shift of the maximum of the 
peak in the histogram towards positive values of $F_2$.
This is not the solution either.
It is interesting to notice that, if we limit the
population of the solutions to those with
$N_{\mathrm{good}} \, \geq \, 21$ and thus to a number 
of degrees of freedom $n \, \geq \, 15$, the tail
is only shifted by a factor 1.07 (7\% of excess). 
Therefore, the observed shift seems to be more
a matter of the amount of data points.
The distribution of solution efficiencies
is peaking at a value of 0.44.
Indeed, if
alternatively we select solutions with $n$ between
4 and 9, we obtain the $F_2$ distribution exhibited
in Fig.\,\ref{fig:EGF2lowdof}, further strengthening
our conclusion.
For low $n$, the efficiencies are peaking
at values lower than 0.3.
The behaviour of the left tail as a function of
$n$ could be due to the apparition for
small amount of data points of overfitting 
or of effects due to the
intrinsic non-linearities of the model that are
inducing correlations among the parameters due to
particularly bad distributions in phase of the
observations and/or effects due to the fact that the
individual uncertainties on RVs are estimators with a
random distribution far from Gaussian.

We further inspected (not shown here) the distributions
of the solutions as a function of two parameters, in particular
$F_2$ versus {\tt{rv\_n\_good\_obs\_primary}} or versus
$G_{\mathrm{RVS}}^{\mathrm{int}}$. Nothing particular is deduced from
these plots. The reason is that the 2D distributions of the solutions
are well interpreted as the product of the marginal 1D distributions. 

The next step we adopted was to create the
histogram of the argument of periastron distribution 
(see Fig.\,\ref{fig:EGhistomega}). 
The distribution is rather uniform, 
with a very slight tendency to have more objects at 
90\degr\ ($\pi$/2) and at
270\degr\ (3$\pi$/2). This results from a very small bias because 
saw-tooth RV curves are 
easier to detect than shark-tooth RV curves, particularly at high eccentricity. 
The effect is practically negligible.

Another interesting plot is the distribution of $T_{\mathrm{0}}$ versus the period.
It is given in Fig.\,\ref{fig:EGtperversusperiod}. 
Since $T_0$ is a cyclic variable of modulo $P$ 
(see Sect.\,\ref{sec:spectroSB1_model}), an infinity of possible values
are available. The data points being located around $t \, = \, 0$, the value
presenting the best precision and a minimum correlation
is necessarily situated between $-P/2$ and $+P/2$.
It is clear that the graph is essentially
as expected with a rather homogeneous density of 
data points slowly decreasing towards the
largest periods. 
For periods around 900 to 1100 days,
there is a slight tendency to have a larger density of solutions
near the values $T_{\mathrm{0}} \sim \pm P/2$. 
A part of the solutions located there correspond
to low significance and rather large eccentricities. 
This is dominantly an artefact
of the decision we took to enforce an SB1 solution in this 
region of periods against the possibility to consider trends.
Indeed, let us consider a time series of a noisy linear trend.
The enforced SB1 solution will deliver a period roughly about
$\Delta T/2$ (only one `cycle' but this depends on the exact sampling). 
The trend folded in phase will very often mimic an eccentric
curve with a jump between the last data point of the time series
and the first one. Therefore, this jump, after folding, will be associated
to the passage at periastron and, consequently, the returned
$T_0$ will be attributed a value around $\pm \, P/2$.
This conclusion is further supported by the 2D distributions 
of solutions (i.e.\ versus $\omega$ and $T_0$)
shown in Fig.\,\ref{fig:EGOmegaT0}. We can clearly see that the
excess of solutions is effectively due to the range of periods
enforced to an SB1 solution (prominent zone in yellow in the 
bottom right panel, not present on the upper left and right ones). 
It is also clear that the effect is absent
for quasi-circular orbits (which are fully correct) 
and is associated to larger eccentricities. 
In addition, we can clearly see that the solutions at the basis of the excess
are confined to values of $\omega$ around $\pi/2$ and $3\pi/2$
(very same zone in yellow). This indicates
that the trends enforced into an SB1 solution generate a RV curve
that is predominantly of the saw-tooth type, further confirming
our conclusions. The excess partly contributes to the excess 
at specific $\omega$ noticed in Fig.\,\ref{fig:EGhistomega}.
This problem will be corrected for DR4; it will also occur at
another period since $\Delta T$ will be larger. Some other effects
might also be present. Note that not all
the solutions in this region are problematic.  

In the next figure, Fig.\,\ref{fig:EGhistogamma}, 
we plot the histogram of the distribution
of the RVs of the centres of mass. This histogram is very similar to the one
built for the assumed constant stars for EDR3 
\citep[see figure 4 of][]{2021A&A...653A.160S} and for DR3
\citep[see figure 11 of][]{2023A&A...674A...5K}, 
which is a strong indication of the quality 
of these $\gamma$ centre of mass RVs. The latter curve from DR3
is overplotted in
Fig.\,\ref{fig:EGhistogamma} to facilitate the detailed comparison. 
The histogram for the constant stars is slightly more dispersed, 
but the agreement is actually good. 
The remaining discrepancy is a matter of sampling different stellar populations. 
Indeed, if we compare with a population restricted to stars cooler than
8125~K and with $G_{\mathrm{RVS}}^{\mathrm{int}}$ brighter than 12, 
the agreement becomes almost perfect.

The distribution of the derived semi-amplitudes is given in
Fig.\,\ref{fig:EGhistoK}.
The major part of the distribution is
decreasing for increasing $K$. The decay law seems 
to be essentially linear on this 
logarithmic scale. Taking into account 
small number statistics (as seen above
150 km\,s$^{-1}$), no break is visible 
in the distribution function and we will
retain that the distribution is rather
smooth, which indicates a very good quality
of the results.
No anomaly is detected.

In the case of noiseless data, the difference between the two extrema of the individual 
RV curves is expected to be a good estimator of twice the semi-amplitude.
We show in Fig.\,\ref{fig:EGKversusamplirobust} a plot
of the {\tt{rv\_amplitude\_robust}} \citep{CU6documentation}
versus the $K$'s. We conclude that the expected relation 
is certainly present. The distribution below the 
reference line (in blue) is essentially explained by the noise on the two values.
The dispersion over the reference line is markedly larger. This is due to the fact 
that the {\tt{rv\_amplitude\_robust}} has a larger dispersion, 
being the difference of the two extreme values (even after some cleaning).
This means that it is more sensitive to persisting outliers or even slight 
extreme values (or noise) whereas the 
semi-amplitude is the result of a fit on all the data points. 
It appears that the discrepant data are more numerous at large $K$.
This very small population of solutions (in red) at large $K$ is associated
to solutions with a large $\sigma_K$; the corresponding objects are characterised
by large values of {\tt{vbroad}} \citep{2023A&A...674A...8F}.
Based on a similar picture (not shown here) for 
the objects in the {\tt{SB1C}} class, 
the conclusions are quite identical.

Figure \,\ref{fig:EGsigmagammaversussigmaK}
presents the distribution of the uncertainty on the 
semi-amplitude $K$ as a function of the
uncertainty on the RV of the centre of mass $\gamma$. The ratio of the two is 
expected to be the square root of 2 \citep{1971AJ.....76..544L}
which is what is observed despite a rather large dispersion.  
No particular anomaly
is present on the uncertainties of the considered parameters.
\begin{figure}[ht]
\centerline{
\includegraphics[width=0.455\textwidth]{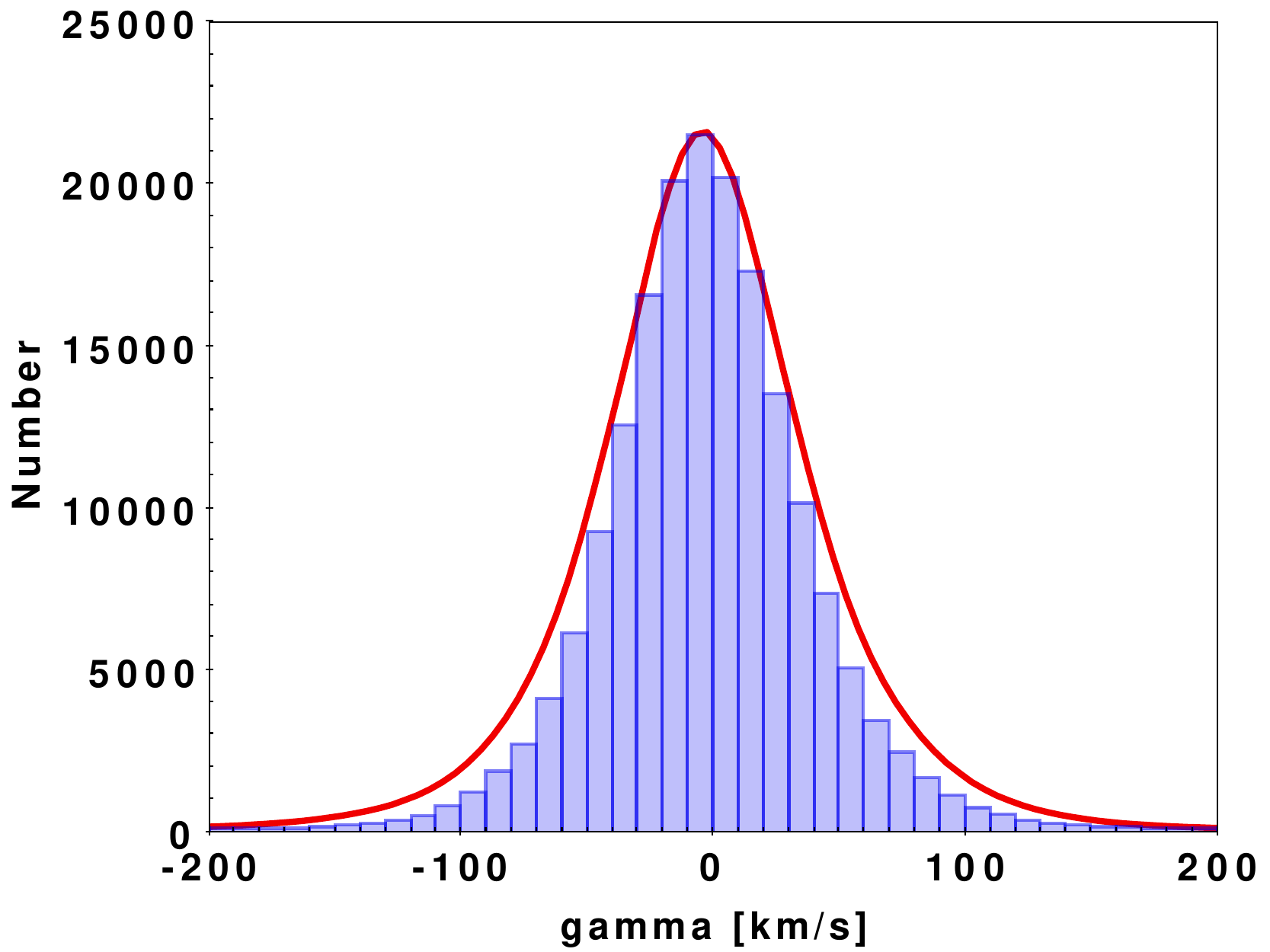}} 
\centerline{
\includegraphics[width=0.45\textwidth]{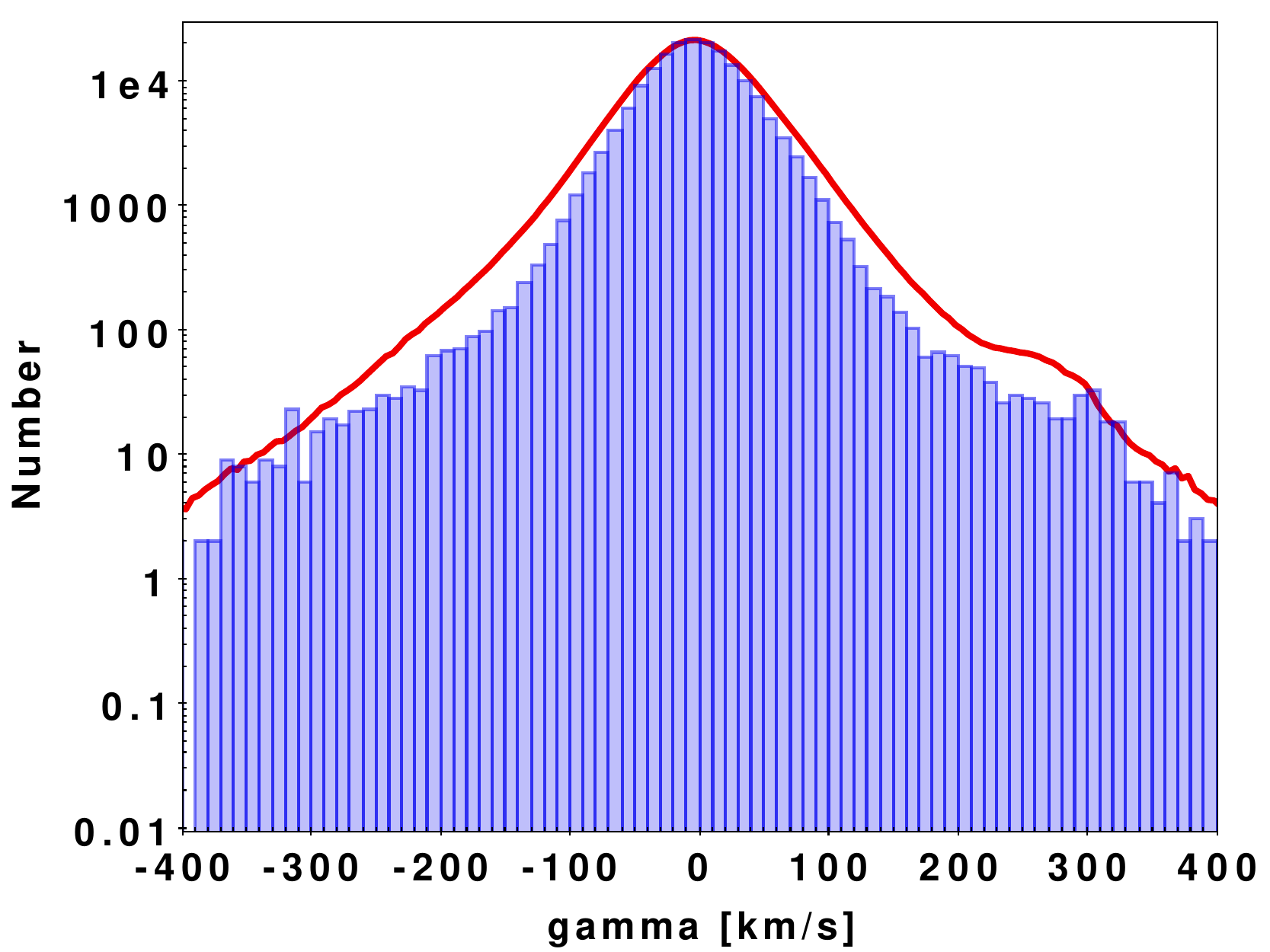}
}\caption{Histogram (above) and histogram in log (below) of the RVs of the centre of 
mass $\gamma$ for all the {\tt{SB1}} objects. The bin width is 10 km\,s$^{-1}$. 
The red curves represent the 
distribution observed for the RVs of the assumed constant stars
\citep{2023A&A...674A...5K}. 
These red curves have been normalised to the maximum of the histogram.}
\label{fig:EGhistogamma}
\end{figure}
\begin{figure}[ht]
\centerline{
\includegraphics[width=0.45\textwidth]{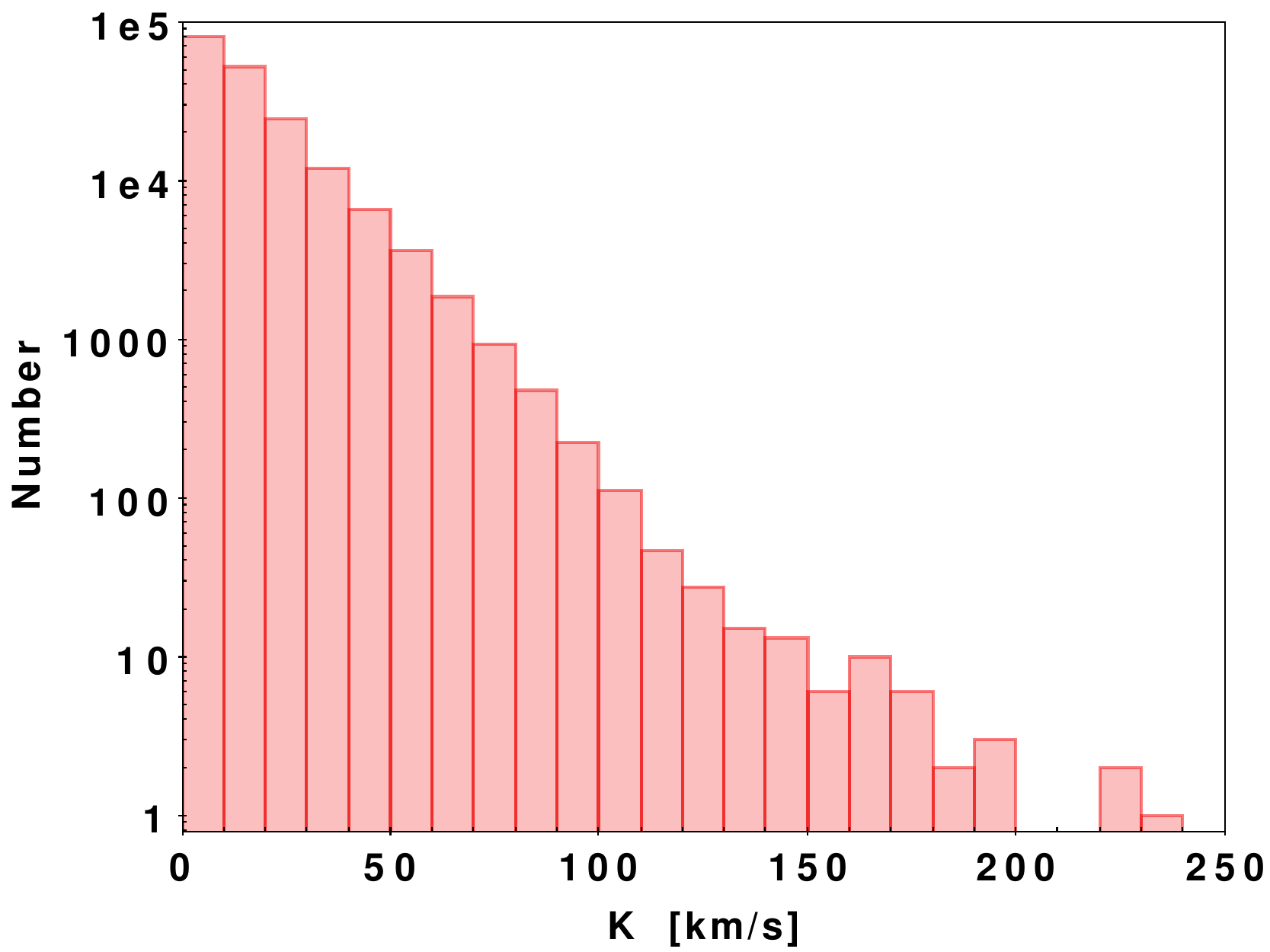}
}\caption{Histogram of the various semi-amplitudes $K$ (in km\,s$^{-1}$). 
The ordinates are given on a logarithmic scale.
The number decreasing with $K$ is to first order linear on this log scale.
The bin width is 10 km\,s$^{-1}$. 
}
\label{fig:EGhistoK}
\end{figure}
\begin{figure}[ht]
\centerline{
\includegraphics[width=0.45\textwidth]{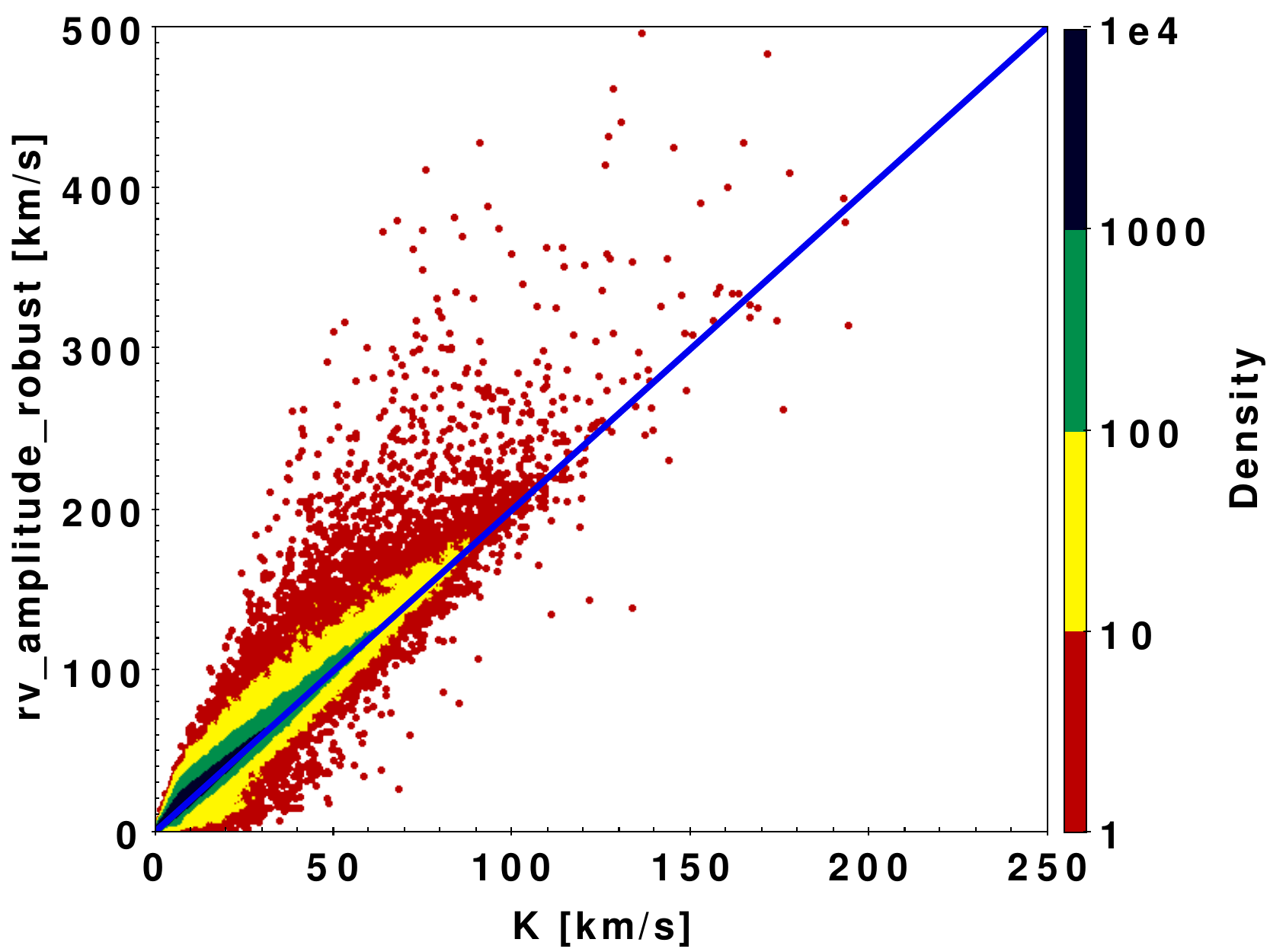}
}\caption{{\tt{rv\_amplitude\_robust}} measured by
MTA as a function of the semi-amplitude $K$ (in km\,s$^{-1}$). 
The blue line indicates the expected value (factor of 2) in absence of noise. 
Due to the large amount of objects (183\,434),
we preferred to adopt a colour-code scale that underlines the high density
of solutions close to this blue line (see on the right).}
\label{fig:EGKversusamplirobust}
\end{figure}
\begin{figure}[ht]
\centerline{
\includegraphics[width=0.45\textwidth]{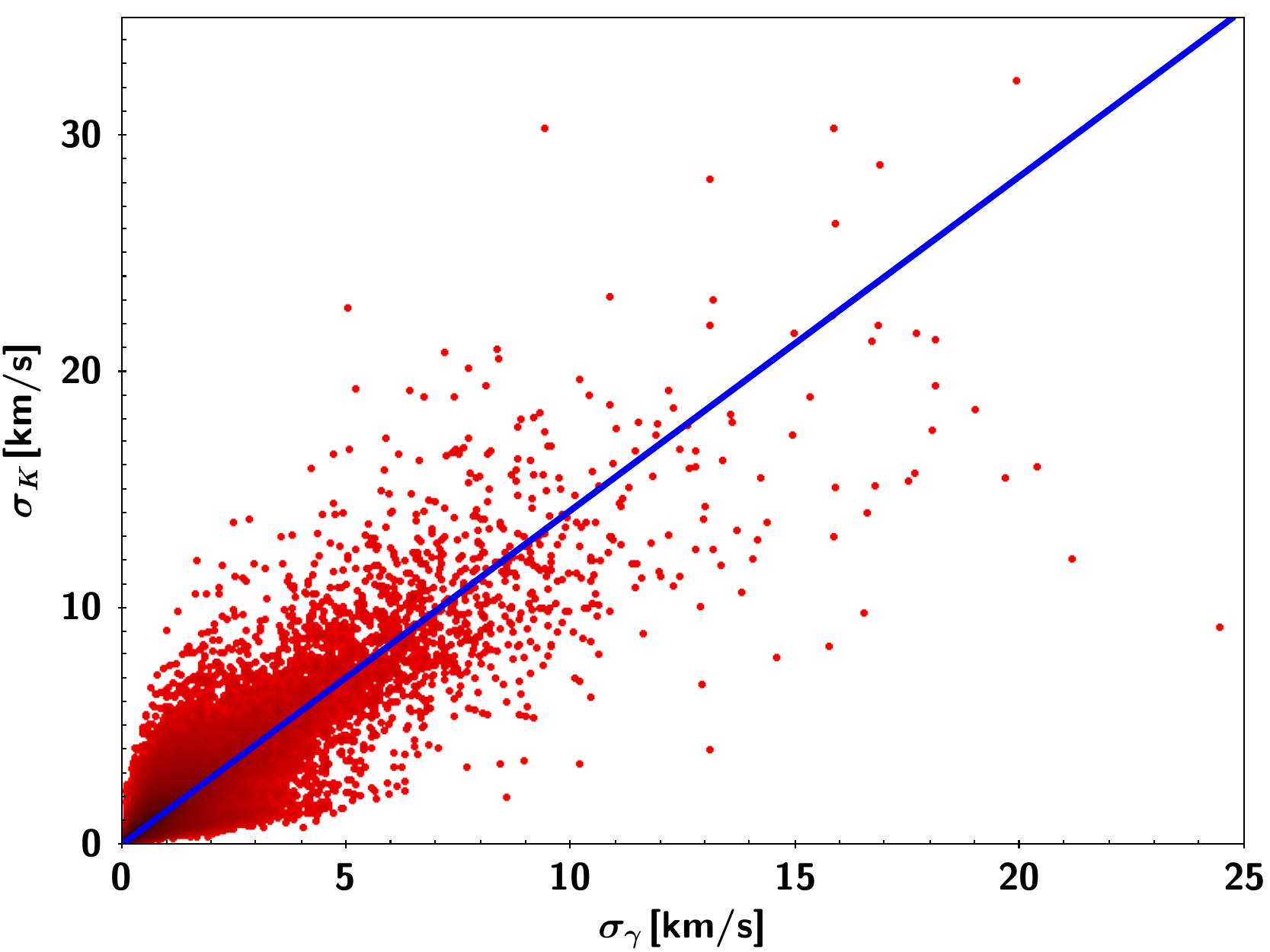}
}\caption{Uncertainty (one standard deviation) on the semi-amplitude $K$ as a function
of the uncertainty on the RV of the centre of mass $\gamma$
(systemic velocity). 
Their uncertainties should be in a ratio of
the square root of two which is marked by the blue straight line. 
Although the dispersion
is rather large, the ratio law is well respected.}
\label{fig:EGsigmagammaversussigmaK}
\end{figure}
\begin{figure}[ht]
\centerline{
\includegraphics[width=0.45\textwidth]{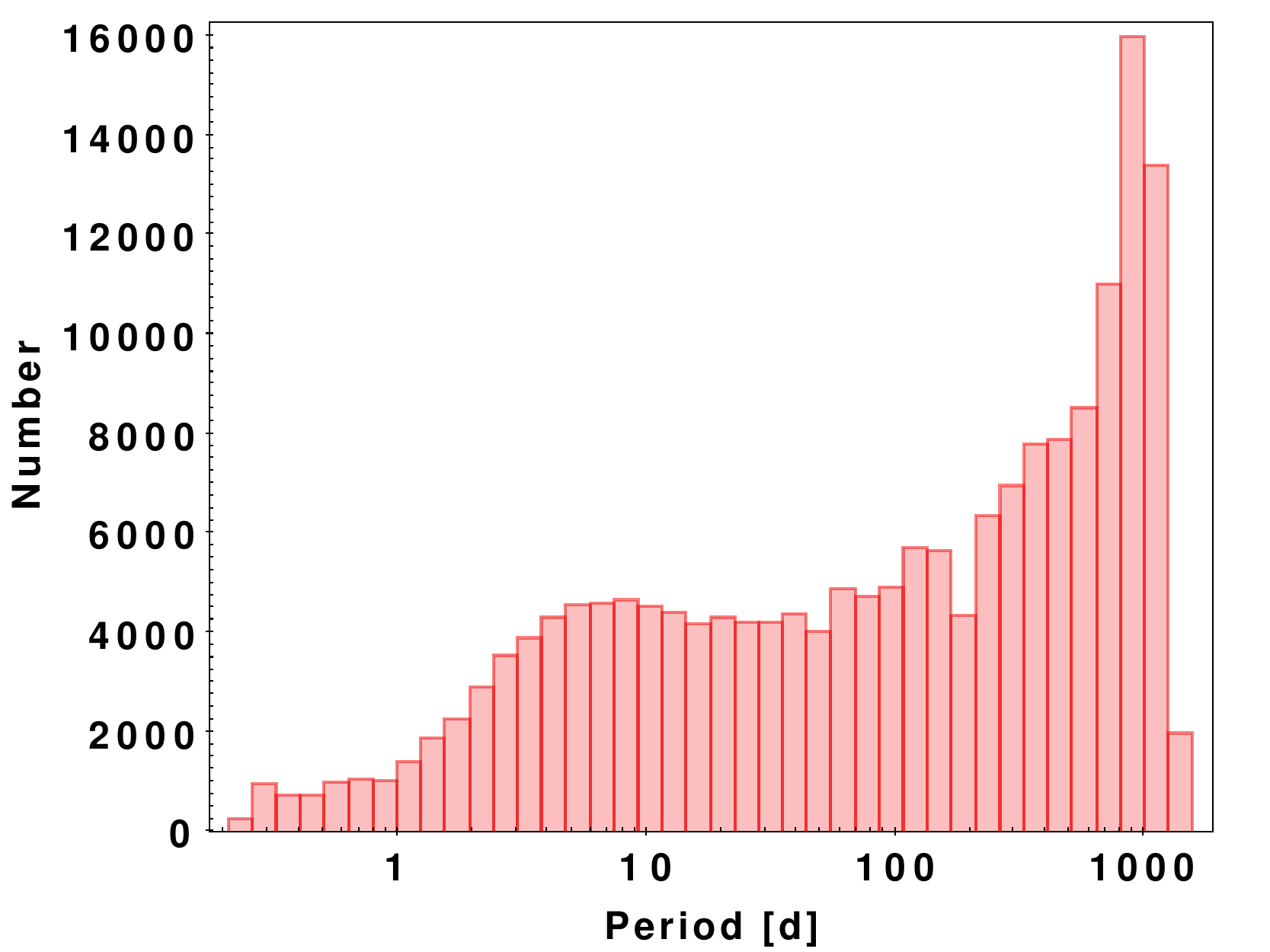}
}\caption{Histogram of the observed period distribution. We note that 
the period is on a logarithmic scale.}
\label{fig:EGhistoperiod}
\end{figure}
\begin{figure}[ht]
\centerline{
\includegraphics[width=0.45\textwidth]{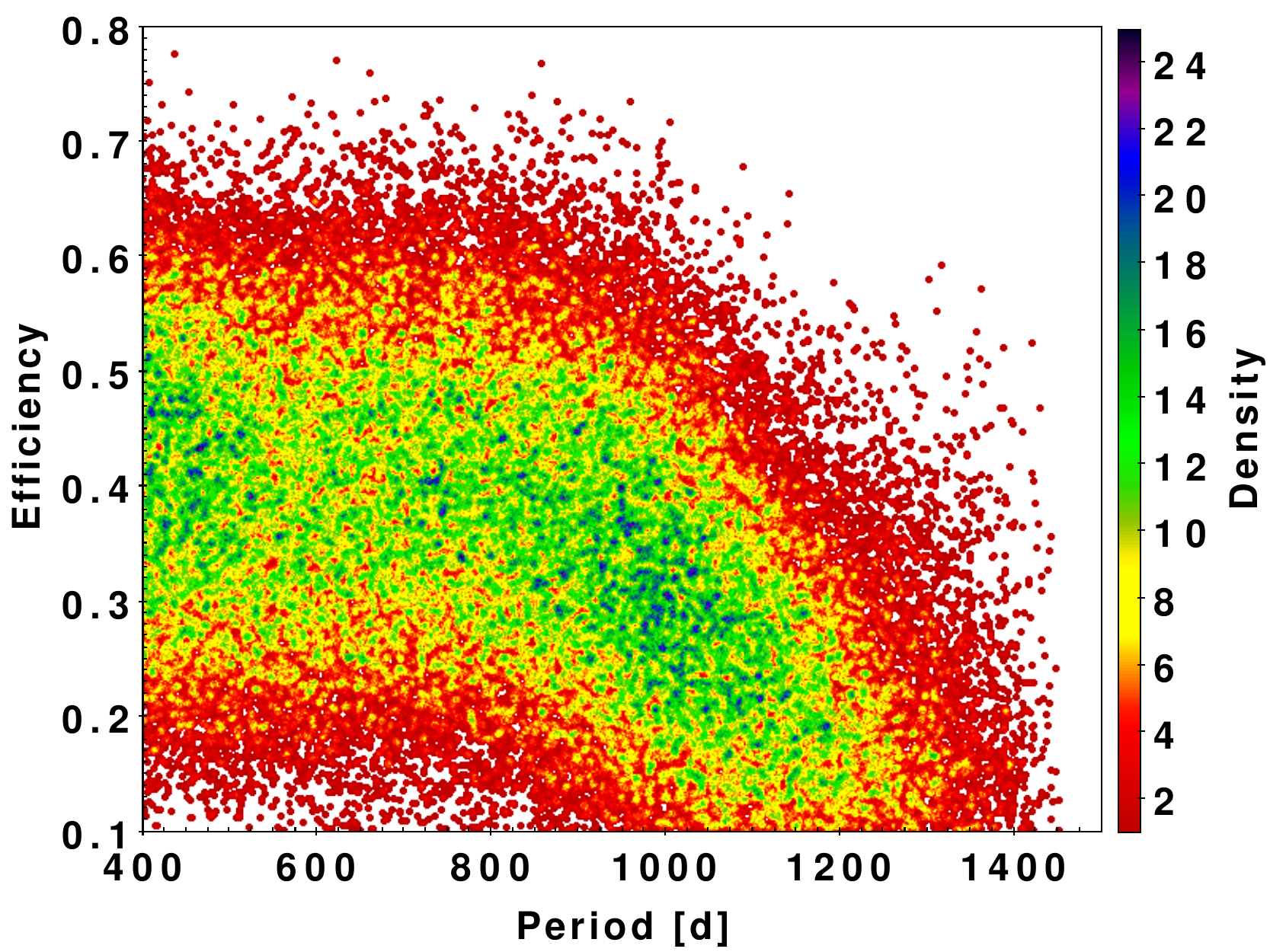}
}\caption{Solution efficiency
as a function of the period. The colour bar represents the density
of objects on a linear, arbitrary scale.}
\label{fig:EGpereffic}
\end{figure}
\begin{figure}[ht]
\centerline{
\includegraphics[width=0.45\textwidth]{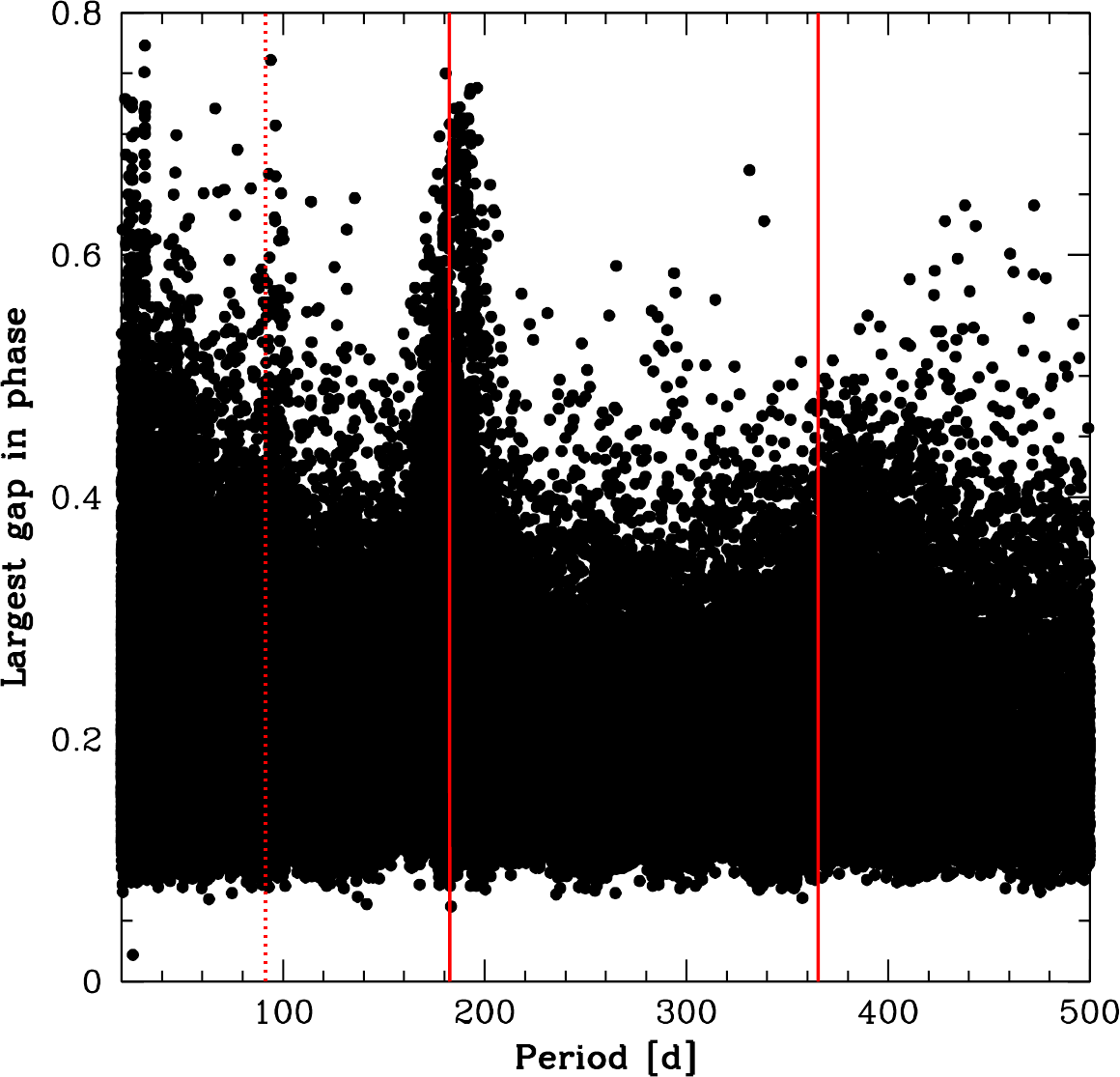}
}\caption{Largest gap observed in the phase diagram as a function 
of the period.  Periods near 365.25\,d and particularly
182.6\,d (vertical red lines) are corresponding to the presence
of some large gaps. It could also be true for the
period of 91.3 d (vertical red dotted line).}
\label{fig:EGinvestgap}
\end{figure}
\begin{figure}[ht]
\centerline{
\includegraphics[width=0.55\textwidth]{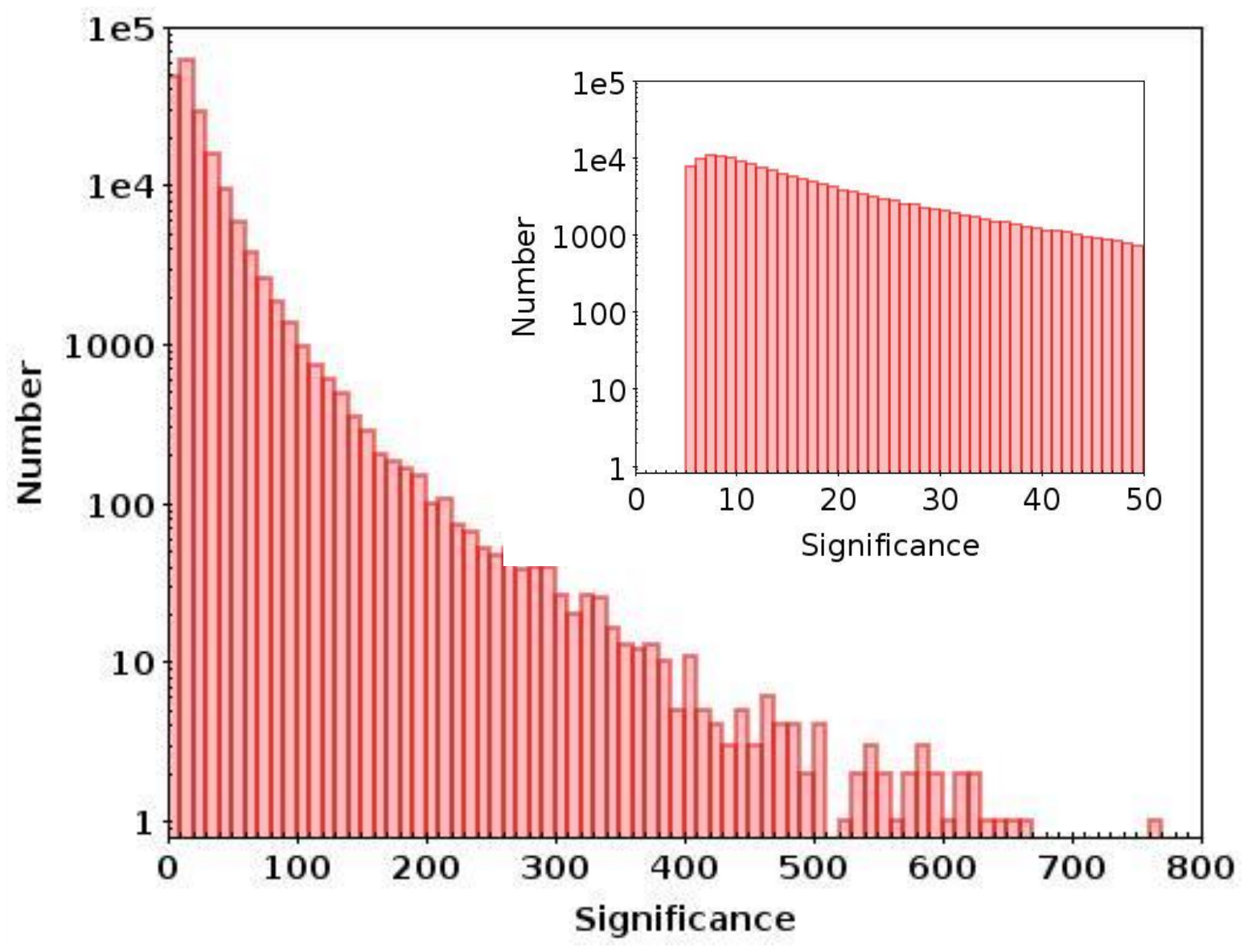}
}\caption{Histogram of the significance
$\frac{K}{\sigma_K}$. The ordinates are on a logarithmic scale; the bin width is 10. 
No break is seen in the
distribution, which prevents us from defining a possible selection criterion.
The insert presents a zoom on small significance: the related bin width is here
1.}
\label{fig:EGhistosignificance}
\end{figure}
\begin{figure}[ht]
\centerline{
\includegraphics[width=0.45\textwidth]{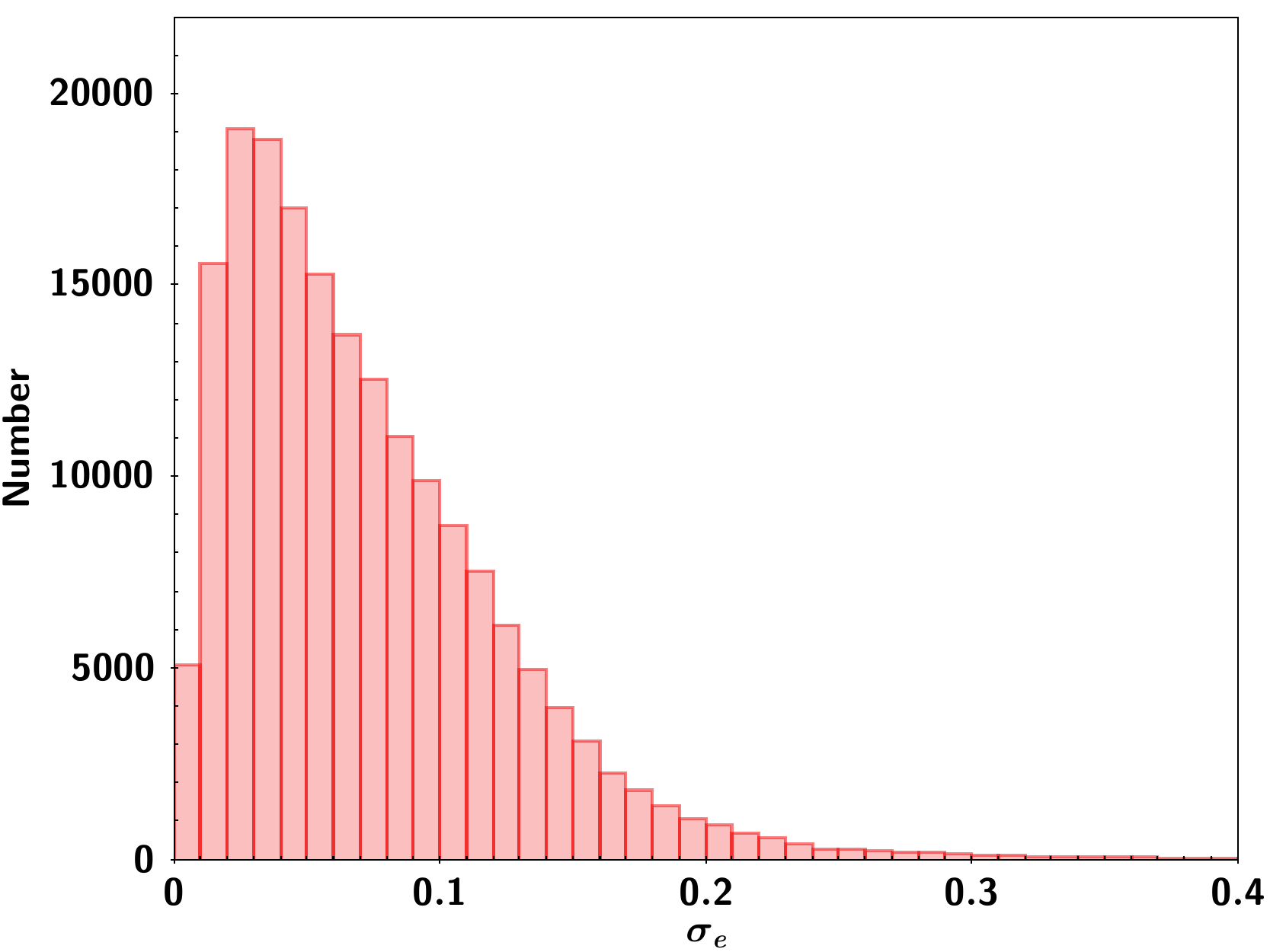}
}\caption{Histogram of the derived standard deviations associated to the eccentricity.
The bin width is 0.01.}
\label{fig:EGhistosigmaecc}
\end{figure}
\begin{figure}[ht]
\centerline{
\includegraphics[width=0.45\textwidth]{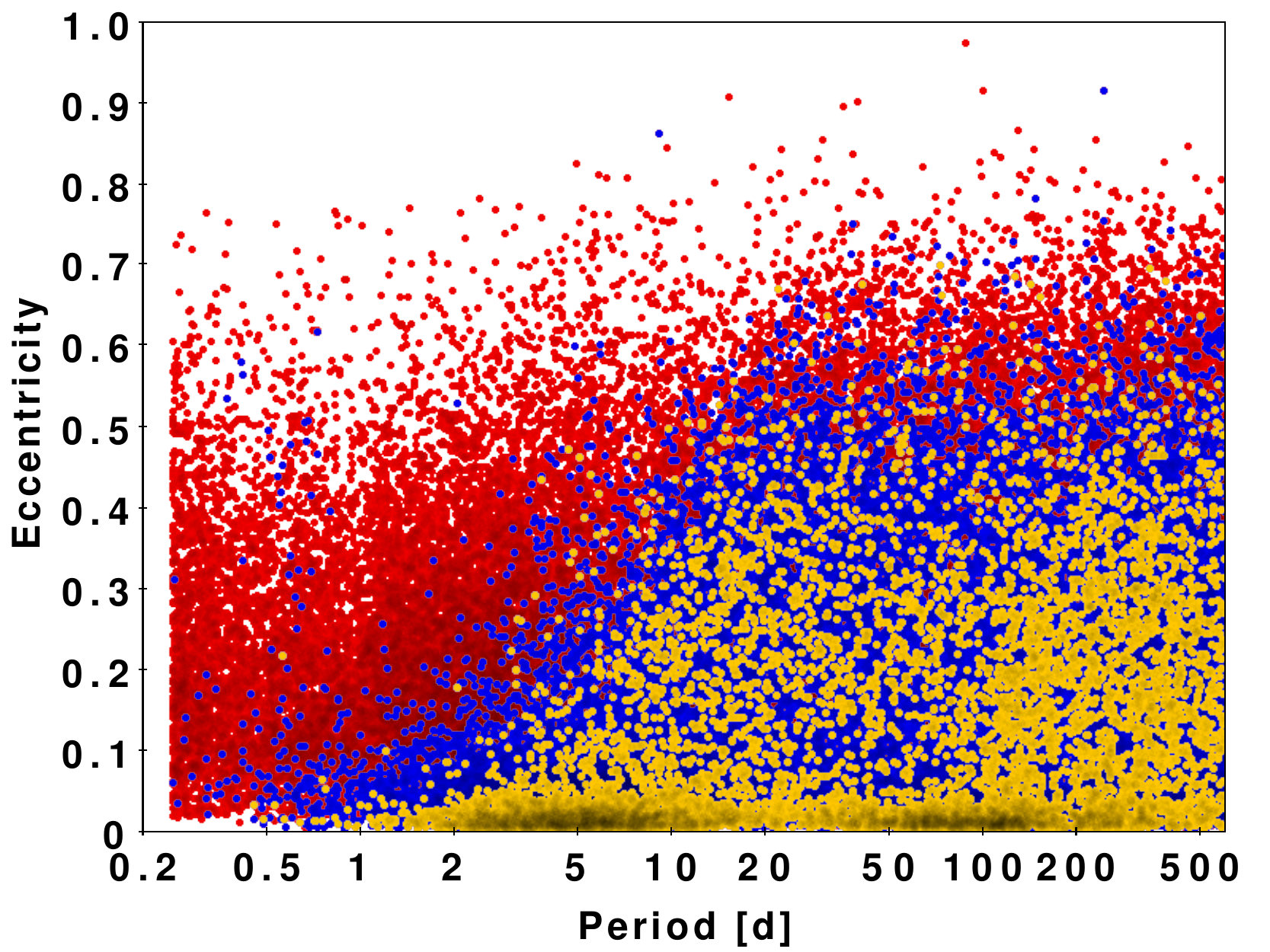}
}\caption{Eccentricity as a function of the period. 
The red points correspond to the full sample (significance
larger than 5). The blue (respectively orange) points correspond to a significance
larger than 30 (respectively 80). 
The blue points already reveal the likely circularisation
phenomenon that is expected to be present.}
\label{fig:EGperiodversuseccent}
\end{figure}
\begin{figure}[ht]
\centerline{
\includegraphics[width=0.45\textwidth]{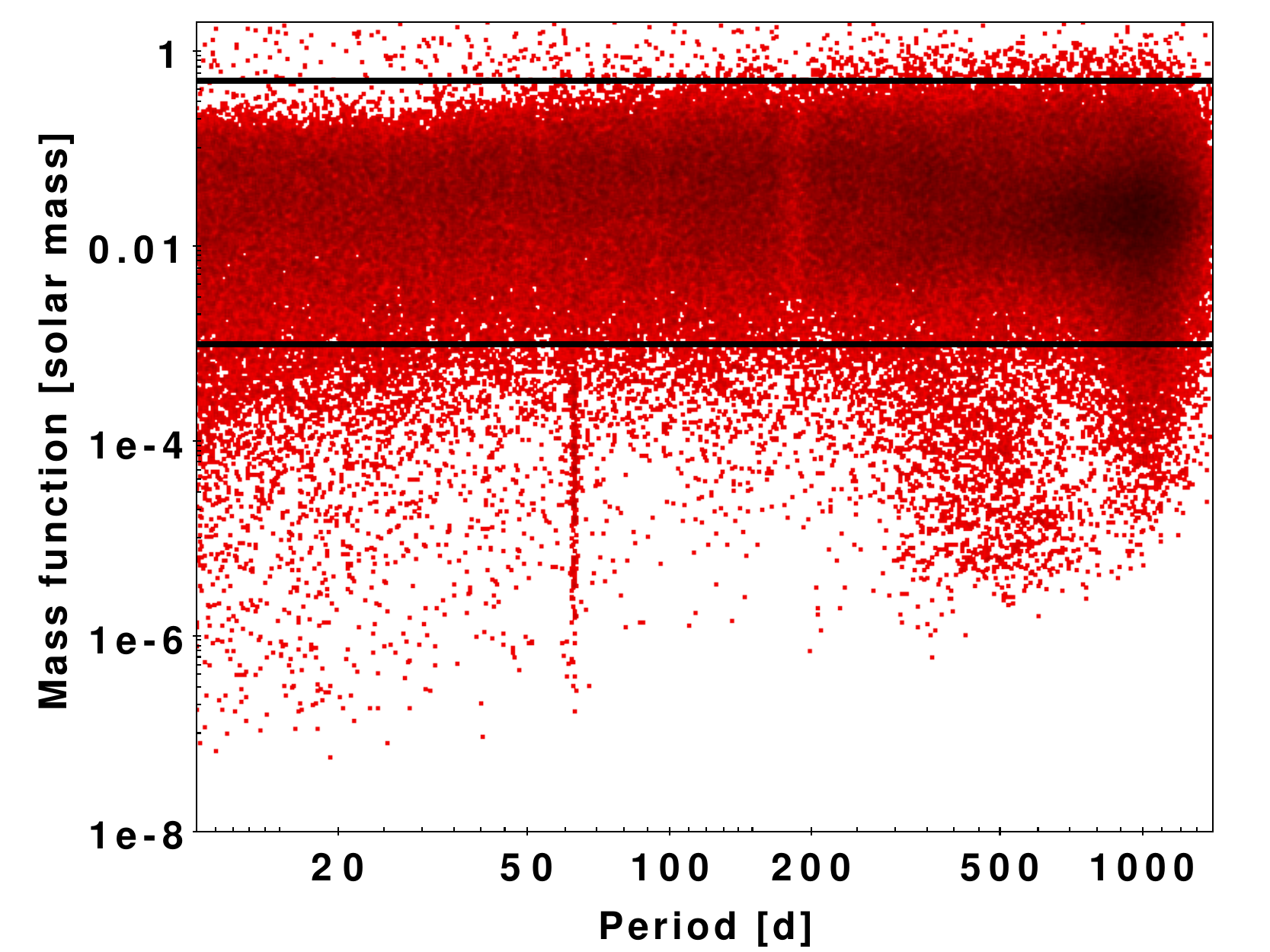}
}\caption{Distribution of the observed SB1 mass 
function (M$_{\sun}$ in log scale) 
as a function of the period (days in log scale). Guiding black lines are given and
correspond to 0.5 M$_{\sun}$ and 0.001 M$_{\sun}$.
Clearly the majority of the objects
is in-between the two lines, the maximum populated area is just half-way (in log)
between the two lines.}
\label{fig:EGperiodversusfdm}
\end{figure}
The critical point when fitting the orbital SB1 models 
is the determination of the period
(see above). Figure\,\ref{fig:EGhistoperiod} displays the 
histogram of the period (in log).
Several features are visible. The log-distribution is rather flat, with a marked excess
at very large periods. This excess can be due to the fact (see above) that the periods
of the same order of magnitude as the span of time of the mission are rather indicative
and certainly not a perfect estimator of the true period. This uncertainty implies the
concentration of several solutions near these long periods. 
This excess is very sensitive to the
threshold separating SB1 and TrendSB1 solutions. 
Another explanation is the fact that the orbital solutions 
are enforced in this region.
On the other hand, the deficit of solutions around the
period of 182.6 days is due to the yearly 
Earth orbit that the \gaia\ satellite follows
by staying at L2. The satellite's avoidance of the direction of the Sun produces
regular gaps with a timescale of half a year. These gaps presented by the sampling 
are associated to large gaps in phase leading to the removal of the solutions 
because they correspond
to badly determined periods and orbits. 
Figure\,\ref{fig:EGinvestgap} is showing a diagram of the value of the largest gap 
in the solution as a function of the period. Exceptionally, here we
use the data set of 367\,328 objects constituting the set of SB1 solutions 
before the application of the internal filtering.
It is clear from this figure that some of the solutions at 
periods near 182.6\,d generate
phase diagrams that are associated to the presence of vast gaps. 
This situation seems also to be the case, but to a lesser extent, for the period 
around one year. The phenomenon is perhaps also present for periods of 91.3\,d. 
A detailed look at this 
Fig.\,\ref{fig:EGinvestgap} suggests that 
a conservative choice for the threshold on the gap of 0.3 removed all the potentially 
spurious or at least unsafe solutions.
The decrease of density of solutions for lower periods, as seen in
Fig.\,\ref{fig:EGhistoperiod},
is an artefact of the application of an ad hoc selection criterion 
(see the paragraph on the internal filtering of the period
in Sect.\,\ref{ssec:spectroSB1_intfilt})
that prevents the 
proliferation of small periodicities. This effect should be further investigated
in the next data releases.

In Fig.\,\ref{fig:EGpereffic}, we present
a diagram showing the efficiency of the
solutions as a function of the adopted
period. We highlight here the long-period objects. 
For periods between 200 and 1000\,d,
the efficiency is essentially distributed
between 0.2 and 0.8, independently of the
period with a maximum of occurrence
around 0.44 (as already reported above).
The values of solution efficiency
above 0.8 are basically absent, indicating
that all the solutions present correlations
among the final parameters.
These values are part of the catalogue.
There are very few solutions with
efficiencies between 0.1 and 0.2,
and all these solutions could be considered
as doubtful, although we preferred to
maintain them in the catalogue and to let
the user the possibility to reject 
them or not. Above periods of 1000\,d,
the typical efficiencies are lessening and
a population below 0.2 is appearing.
This is a direct consequence of the fact
that the associated period is larger than
$\Delta T$. It is evident that these
solutions should not be blindly trusted,
but in any case the reported periods
can be considered as indicative.
Again, the low value of the efficiency
unveils existing correlations among the
final parameters due to bad phase coverage.
As mentioned above, for variabilities at
still larger timescales, the trend solution
is taking over. However, addressing
solutions slightly above $\Delta T$
would necessitate a too large degree
of the fitted trend polynomial,
introducing a lack of robustness of the 
fit. The orbital solution is thus more
informative despite the possible biases
in the derived parameters. 
For future releases, the range of 
concerned periods will change accordingly
with the increasing value of 
$\Delta T$.

Figure\,\ref{fig:EGhistosignificance} presents the histogram of the spectroscopic 
significance $\frac{K}{\sigma_K}$. 
The significance is distributed as a decreasing power law vanishing at 800.
The maximum accepted value for $K$ being 250 km\,s$^{-1}$, the maximum 
significance value corresponds to a typical uncertainty 
of about 0.3 km\,s$^{-1}$ which is coherent with expectation.
The distribution shows no inflection locus that could be useful 
in setting a selection threshold. In this
context, we decided to introduce a threshold
at 5 but not more restrictive ones. Below a value of 5, the corresponding 
solutions are clearly not trustworthy. We leave the possibility to the 
catalogue users to define a threshold that could be appropriate for their own work.

Figure\,\ref{fig:EGhistosigmaecc} exhibits 
the distribution of the standard deviations on 
the eccentricity. The typical uncertainty on the 
eccentricity is of $\sigma$\,$\sim$\,0.03-0.10\,.
It is necessary to draw the attention to the fact
that with such an uncertainty a derived eccentricity of
0.3-0.4 is still compatible with a circular orbit
(this is within three $\sigma_e$ keeping in mind that
the distribution of eccentricities is not Gaussian).
RVs are well-known to be a poor source
of determination of eccentricities in contrast
to the astrometric orbital solutions. Errors larger
than 0.35-0.4 (there are a few) could be associated to insecure 
adjustments and we encourage the catalogue user 
to be careful with these questionable solutions.

As an example, Fig.\,\ref{fig:EGperiodversuseccent} shows a plot of the
eccentricities as a function of the corresponding period. 
The plot of all the objects (in red) includes short period solutions with 
some eccentricity up to 0.4. As mentioned just above, these values of $e$ 
up to 0.3-0.4 are not particularly anomalous. 
However, the general aspect of the diagram is not the expected one in terms
of the theory of circularisation of the orbits
(close binaries are not expected to remain eccentric for long). 
If we restrict the data set to values with a 
significance larger than 30, we obtain the distribution in blue where the 
expected effects of the circularisation start
to be well visible. A restriction to the significance values above 80 generates 
the distribution in orange. Another study of this topic can be found in
\citet{2023A&A...674A..34G}. Evidently, the impact of the selection on the basis of the 
significance is a delicate topic and we advise the greatest caution for the selection
of the threshold to be applied. 
This remark can also be extended to other quality factors
than just the significance.

Finally, Fig.\,\ref{fig:EGperiodversusfdm} represents the plot of the
derived SB1 mass function as a function of the adopted period. 
We can clearly see that the majority of the data points 
are between 0.001 and 0.5 M$_{\sun}$. 
Objects outside this range are very
few, in particular on the side of the high mass function. 
From the statistical point of view,
the diagram is rather close to the expected one. 
Data points at an unexpected location
could be due to problems in the derivation of the period or of the eccentricity.
In particular, we should be very careful while searching 
for abnormal mass function values
or massive black holes. 
In addition, the overdensities tracing vertical lines at $P=62.97\,$days and at
one year (much broader) are due to objects with a varying apparent position inducing
systematic effects on the RVs associated to the variation of the scan-angle. These are
objects presenting small amplitudes at these particular periodicities (bad astrometry
inducing a wrong wavelength calibration). These objects
could easily be filtered out but they were not before the application of the combiner
with the astrometric channel in order not to miss interesting objects. The effect 
in astrometry is described in \citet{LL:LL-125}.
A more detailed analysis of the scan-angle dependent signals is given
in \citet{2023A&A...674A..25H}.
\subsubsection{{\tt{FirstDegreeTrendSB1}} and {\tt{SecondDegreeTrendSB1}} classes}
\begin{figure}[ht]
\centerline{
\includegraphics[width=0.46\textwidth]{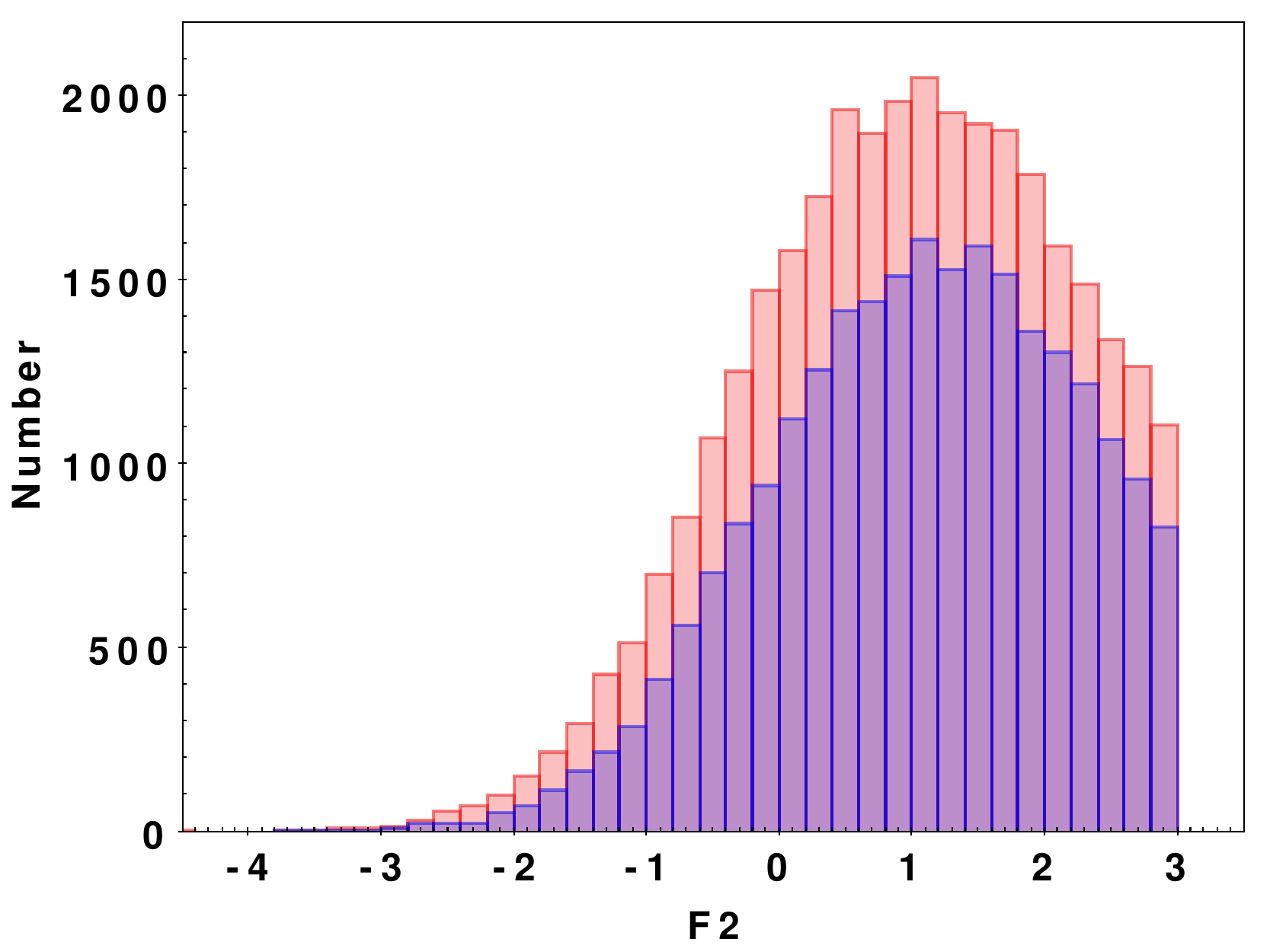}
}\caption{Histogram of the goodness of fit $F_2$ for the trend fits.
The bin width is 0.2.
In blue the solutions are associated to the 
{\tt{FirstDegreeTrendSB1}} 
class and in red with the
{\tt{SecondDegreeTrendSB1}} one. The cut-off above 3 is outstanding.
The centring of the distribution is rather located near 1 instead of 0.}
\label{fig:EGhistoF2trendD2D1}
\end{figure}

Concerning trend solutions, we show in
Fig.\,\ref{fig:EGhistoF2trendD2D1} the distribution of the goodness of fit $F_2$ 
for all the TrendSB1 solutions. It appears that the expected distribution that 
should be centred on zero, is actually rather centred on 1. 
This points out to an anomaly of this distribution, tending to suggest that the 
fits are not perfect although still good.
The reason of this shift is for the 
moment unknown; to understand it is very important 
because we should study if the cut-off at 3 is still justified
in the context of the observed shift of the distribution.
The characteristics of the distribution is similar for the
two subtypes of solutions.

The trend is certainly not a purely physical model, since it is basically induced 
by the shortness of the time span of the observations compared to the true cycle
of the star considered. Statistical decisions are adopted to select a model
(orbital, linear trend, high-order trend) but it seems unlikely that the adopted 
thresholds could induce the shift of the $F_2$ distribution in such a global manner.
Another explanation for the shift could be the presence of some intrinsic
variability additional to the trend signal, as already evoked for orbital solutions.
Again, it is very unlikely that this could affect such a 
large percentage of the objects.
Thus, the last possibility  and most likely hypothesis 
is that the individual uncertainties
associated to RVs are underestimated. The trend models are certainly a good way
to investigate this possibility. Indeed, the linear trend is a very simple model
which is linear in the parameters and the fit should be robust and unambiguous
at least compared to the non-linear orbital model whose behaviour could be more
complex (see Sect.\,\ref{sssec:spectroSB1_validation_intern_SB1}).
The overfitting is here not a concern, and the number of degrees of freedom
of the $\chi^2$ of the fit is precisely established.
We thus further investigated the global distribution (over all the objects)
of the observed minus calculated residuals normalised by the associated
individual uncertainties. To be strictly conservative, we only considered objects
with an astrometric {\tt{ruwe}} less than 1.4 for which no additional contribution
to the uncertainties was applied; only the values delivered by CU6 are involved.
The distribution of the normalised residuals follows very closely a Gaussian
probability density function, centred on zero, but with a standard deviation
larger than $\sigma \, = \, 1$: actually we deduced $\sigma \, = \, 1$.08.
Therefore, an increase by 8 percent of the uncertainties could solve the reported
anomaly of the $F_2$ distribution except for a small effect that could be
explained by some additional variability. This statement needs further
confirmation.

\begin{figure}[ht]
\centerline{
\includegraphics[width=0.45\textwidth]{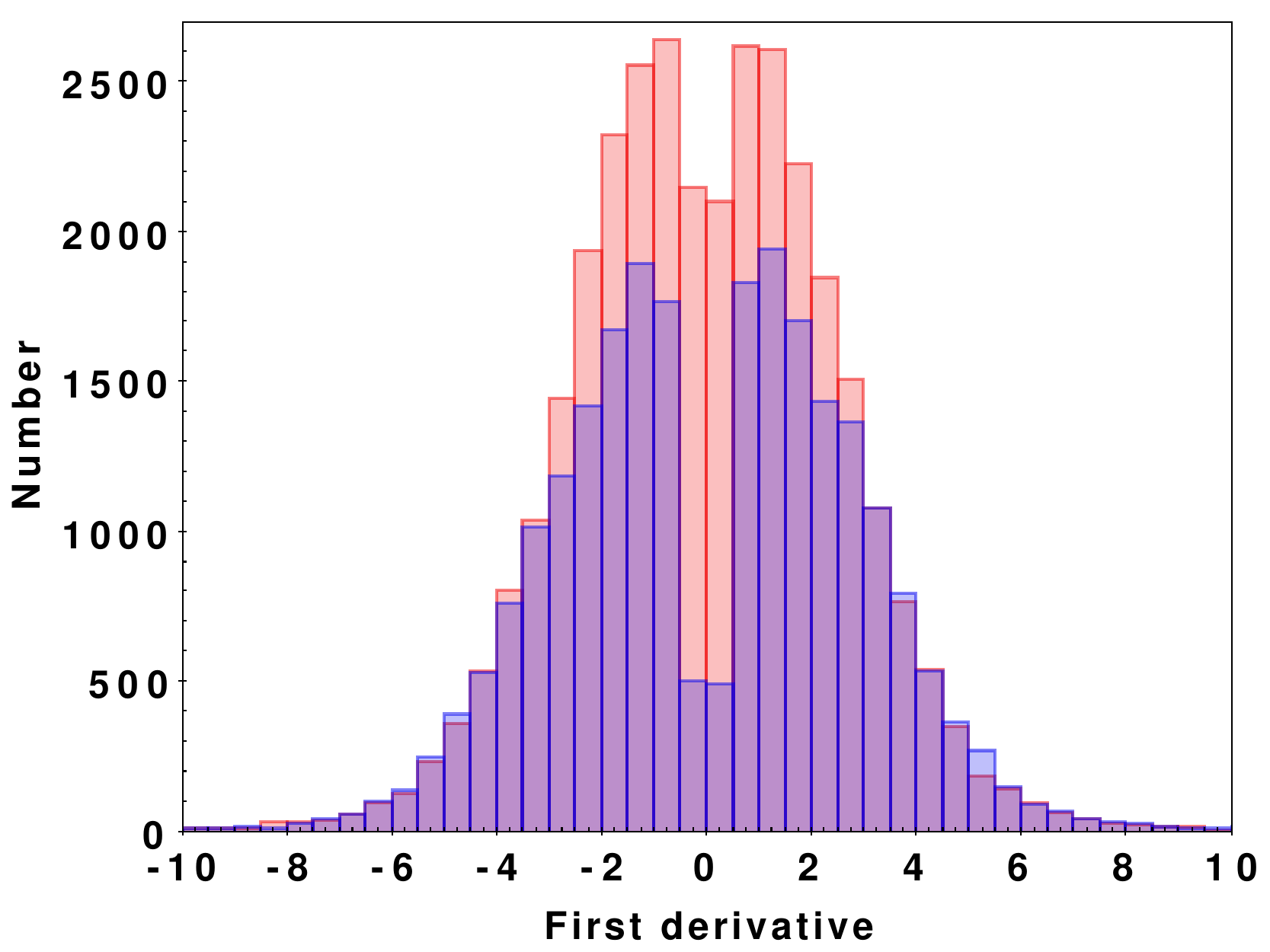}
}\caption{Histogram of the first derivative normalised by its derived
uncertainty for the trend fits.
In blue the solutions are associated to the 
{\tt{FirstDegreeTrendSB1}} 
class and in red with the
{\tt{SecondDegreeTrendSB1}} one. The bin width is 0.5.
The central dip is due to fits where the first derivative is too small 
and thus to objects considered as constant
in RVs.}
\label{fig:EGtrendfirstderivnorm}
\end{figure}

Figure\,\ref{fig:EGtrendfirstderivnorm} illustrates the distribution
of the first derivative normalised by its uncertainty. The distribution is 
very similar for the two classes and presents
an almost perfect symmetry, as expected. 
The deficit of solutions around the
zero derivative is due to the fact that these solutions are not
included because they correspond to constant objects. 
A lot of the solutions are well above $\pm \, 3 \, \sigma$.
A few well-defined derivatives are further away up to
$10 \, \sigma$.

\begin{figure}[ht]
\centerline{
\includegraphics[width=0.45\textwidth]{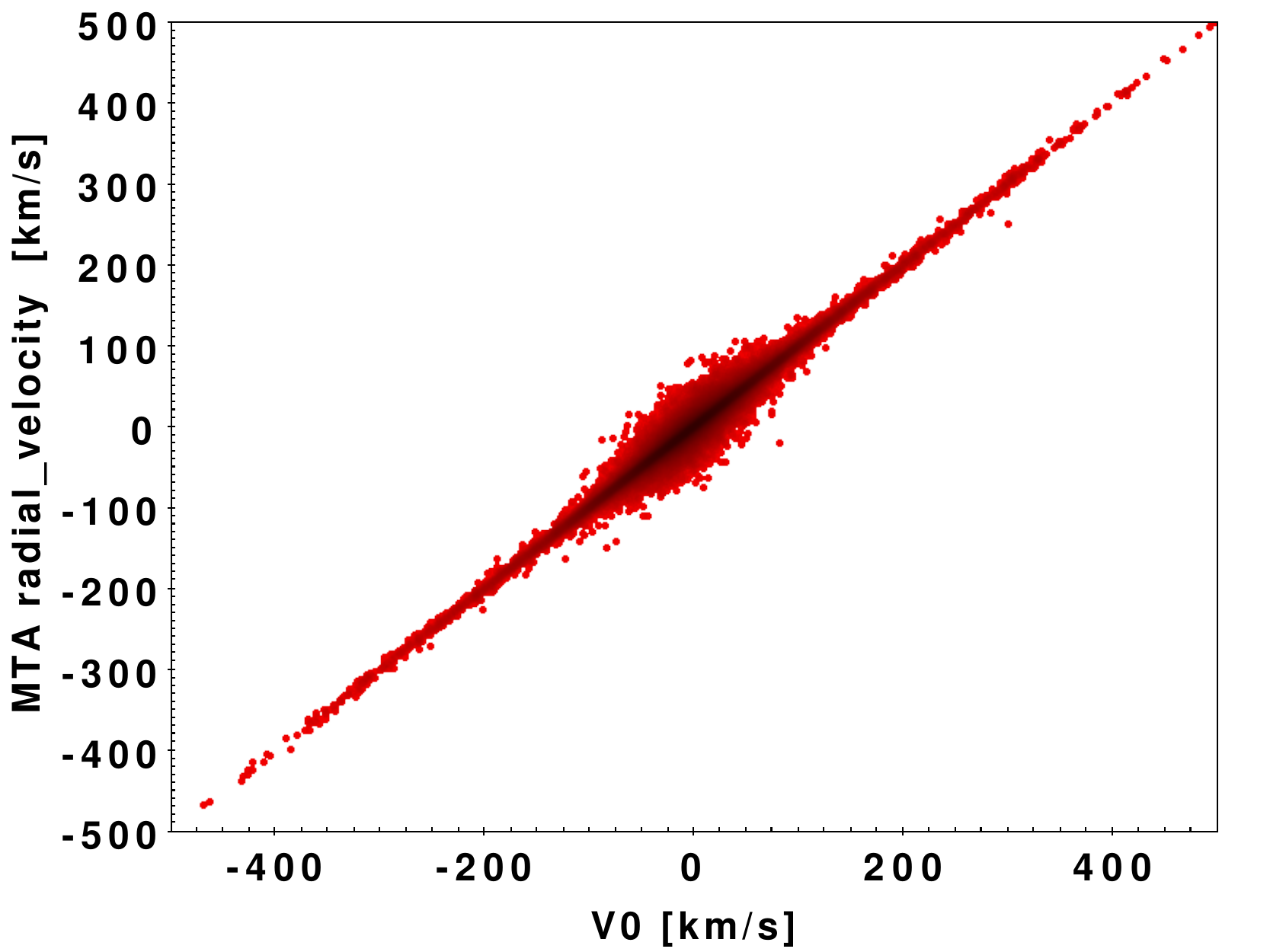}
}\caption{Correlation between the $V_0$ derived
in our pipeline and the mean RV as computed
by MTA assuming the objects are constant.}
\label{fig:EGmeanversusradial}
\end{figure}

Finally, we try to validate the values of the derived
$V_0$. They are plotted in 
Fig.\,\ref{fig:EGmeanversusradial} with in ordinates the mean velocity 
computed by MTA. We observe that the two parameters
are very well correlated, providing further support to the
$V_0$ values we are delivering.
\subsection{Validation against CU7 photometric periods}\label{ssec:spectroSB1_validation_cu7}
\begin{figure}[!htp]
\centerline{
\includegraphics[width=0.5\textwidth, trim = 40 160 40 130, clip]
{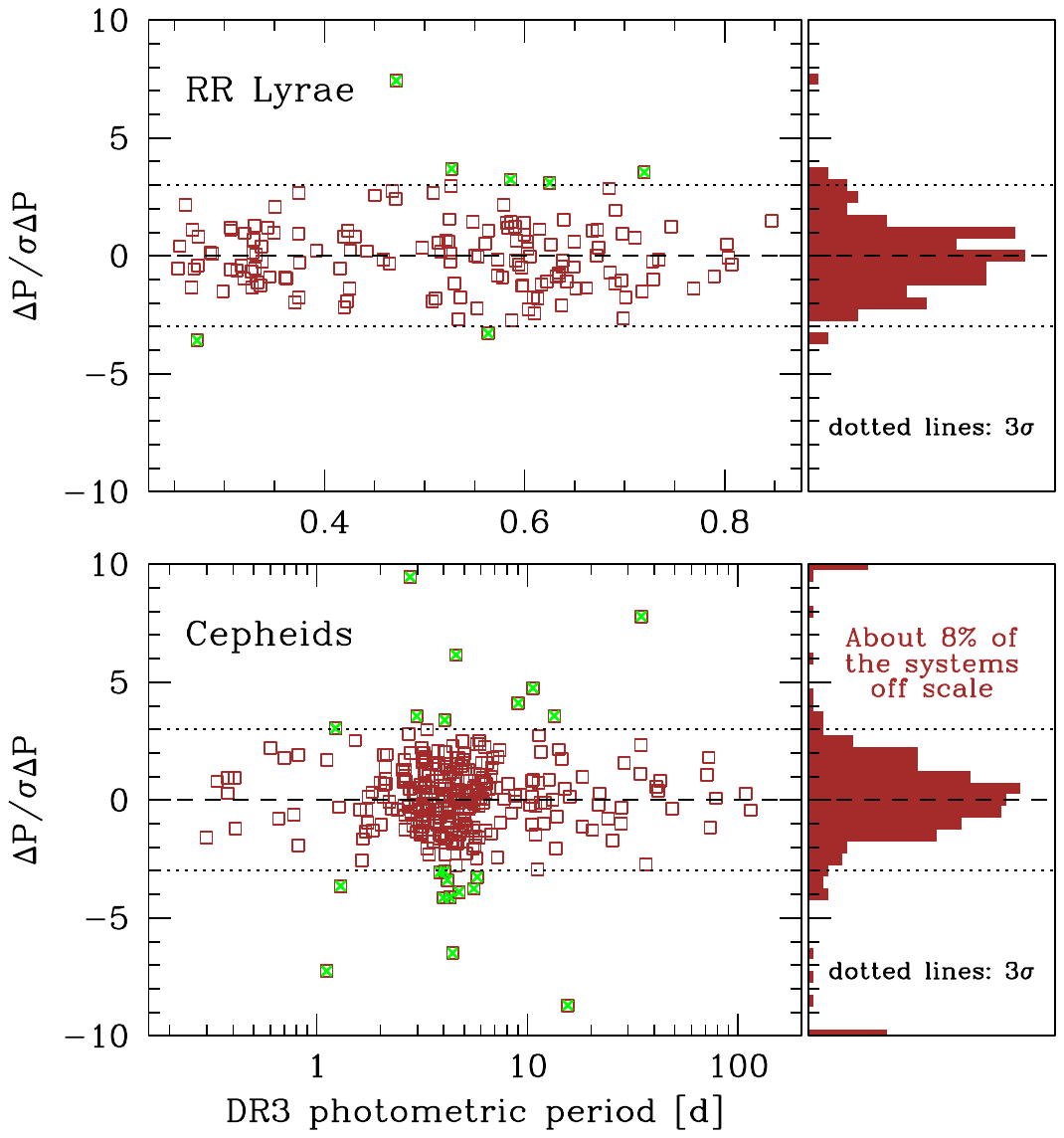}
}\caption{Deviations in $P$ expressed in $\sigma$ units for the RR Lyrae 
(top panels) and Cepheid (bottom panels) variables excluded 
from the SB subcatalogue 
(see post-filtering in Sect.\,\ref{sssec:spectroSB1_add_cons_postfilt_intvar}), 
as a function of the photometric period. Objects for which the reference 
period is not recovered to within 3$\sigma$ are flagged with green crosses. 
We note that a small fraction of the Cepheids are off scale.}
\label{fig:cu4nss_spectro_DeltaP_over_sigmaDeltaP_vs_P_pulsators_SB1}
\end{figure}

Some intrinsic variables can exhibit RV variations
that could be mistaken as the effect of an 
orbital motion. As explained in
Sect.\,\ref{sssec:spectroSB1_add_cons_postfilt_intvar},
a sample of securely identified Cepheids
\citep[333,][]{2023A&A...674A..17R} and RR Lyrae
\citep[147,][]{2023A&A...674A..18C} have also
an SB1 solution. They should be eventually excluded
from our SB subcatalogue but they also offer the unique
opportunity to have an independent test of the
spectroscopic periods by comparing them to the
photometric ones. Indeed,
we compared our spectroscopic periods to those
determined from DR3 photometric data for the 147
RR Lyrae and the 333 Cepheids. 
The difference is expressed as 
[P(spectroscopy)--P(photometry)]/$\sigma$, where $\sigma$ is 
the square root of the sum in quadrature of both uncertainties.  
We find a recovery success rate to within 3$\sigma$ of $\sim$95\% 
for the RR Lyrae's. It is not only lower for the Cepheids ($\sim$85\%), 
but the outliers are also much more severely discrepant 
(Fig.\,\ref{fig:cu4nss_spectro_DeltaP_over_sigmaDeltaP_vs_P_pulsators_SB1}). 
We find that the solution efficiency is better for the 
RR Lyrae's than for the Cepheids.

\subsection{Validation against other data sets}\label{ssec:spectroSB1_validation_otherset}
\begin{figure}[!ht]
\centerline{
\includegraphics[width=0.7\textwidth, trim= -20 490 170 165, clip]
{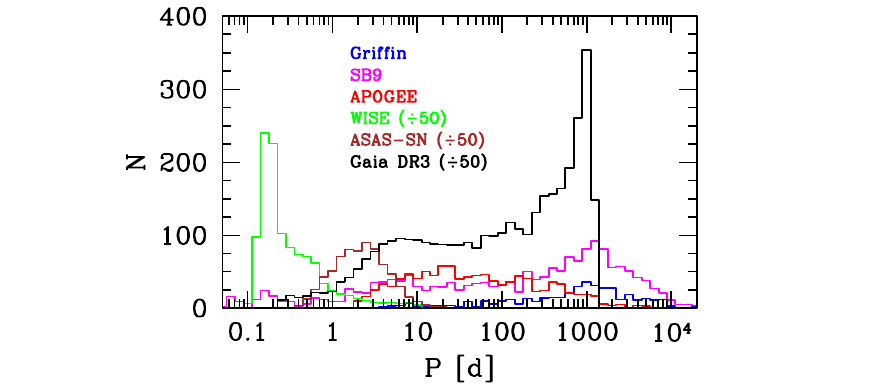}
}\caption{Breakdown of the orbital periods for the DR3 (black), Griffin (blue), 
SB9 (magenta), APOGEE (red), WISE (green), and ASAS-SN (brown) binary samples. 
Only SB1s are shown for the first four catalogues that 
are based on spectroscopic data. 
The counts for DR3, WISE, and ASAS-SN are divided by 50.}
\label{fig:cu4nss_spectro_validations_catalogues_P_SB1}
\end{figure}

We present below a comparison between the DR3 orbital 
parameters and those from external catalogues, 
either relying on spectroscopic (Griffin, SB9, and APOGEE) or
photometric (WISE and ASAS-SN) data. 
The most distinctive feature between the various validation 
samples is the widely different 
range of orbital periods that span about five orders of magnitude 
(see Fig.\,\ref{fig:cu4nss_spectro_validations_catalogues_P_SB1}). 
The WISE and ASAS-SN catalogues are made up of candidate short-period 
binaries that are useful 
for assessing the reliability of our orbital solutions in this regime.  
We consider exactly the same DR3 data set as discussed in 
Sect.\,\ref{ssec:spectroSB1_validation_intern}. 
In the following, except for ASAS-SN, literature results without 
an uncertainty given are ignored. 
The difference for a given orbital parameter is expressed as 
[value(NSS)--value(reference)]/$\sigma$, 
where $\sigma$ is the square root of the sum in quadrature of both uncertainties. 
\subsubsection{Comparison against Griffin's results}\label{sssec:spectroSB1_validation_otherset_griffin}
\label{sssec:spectroSB1_quality_val_compGriffin}
We first considered the orbits determined in a series of papers 
(not only in The Observatory) 
by Roger Griffin and collaborators. The orbital elements were taken 
from the 2 March 2021 version of the SB9 
catalogue\footnote{\url{https://sb9.astro.ulb.ac.be/}}
\citep[original paper:][]{2004A&A...424..727P}, 
which is made up of 4021 systems. 
The association to the relevant DR3 ID
was simply tracked from the DR2 alias available in 
SIMBAD\footnote{\url{https://simbad.u-strasbg.fr/simbad/}} at that time. 
Since for validation we needed to be confident in our reference catalogue,
we remained conservative at this level, avoiding adverse and insecure
cross-matching.
This procedure can now certainly be much improved. 
To avoid any ambiguities in this cross-match, 
or a possible bias in centre-of-mass velocity, $\gamma$, 
visual/multiple systems were excluded. 
They were identified as having a \texttt{component} 
field in SB9 not empty or to have a SB9 ID entry 
in the 23 June 2020 version of the Multiple Star Catalog 
(MSC)\footnote{\url{http://www.ctio.noirlab.edu/~atokovin/stars/}}. 
This validation catalogue contains the orbits of 414 SB1's and 101 SB2's. 

The breakdown of DR3 spectroscopic solution types is 
as follows for Griffin's SB1 catalogue:
\begin{itemize}
\item
\tt{SB1}: 83 systems,
\item
\tt{SB1C}: 0 system,
\item
TrendSB1: 10 systems,
\item
\tt{SB2}: 1 system,
\item
\tt{SB2C}: 0 system,
\item
Without any DR3 spectroscopic deterministic solutions: 320 systems.
\end{itemize}

The period of the ten systems with a trend solution is
much longer than the time span of the 
RVS observations, which is as expected. 
Five and one systems among the SB2 validation catalogue have received 
a {\tt{SB1}} and {\tt{TrendSB1}} solution, respectively. 

\begin{figure}[!htp]
\centerline{
\includegraphics[width=0.5\textwidth, trim= 40 300 40 250, clip]
{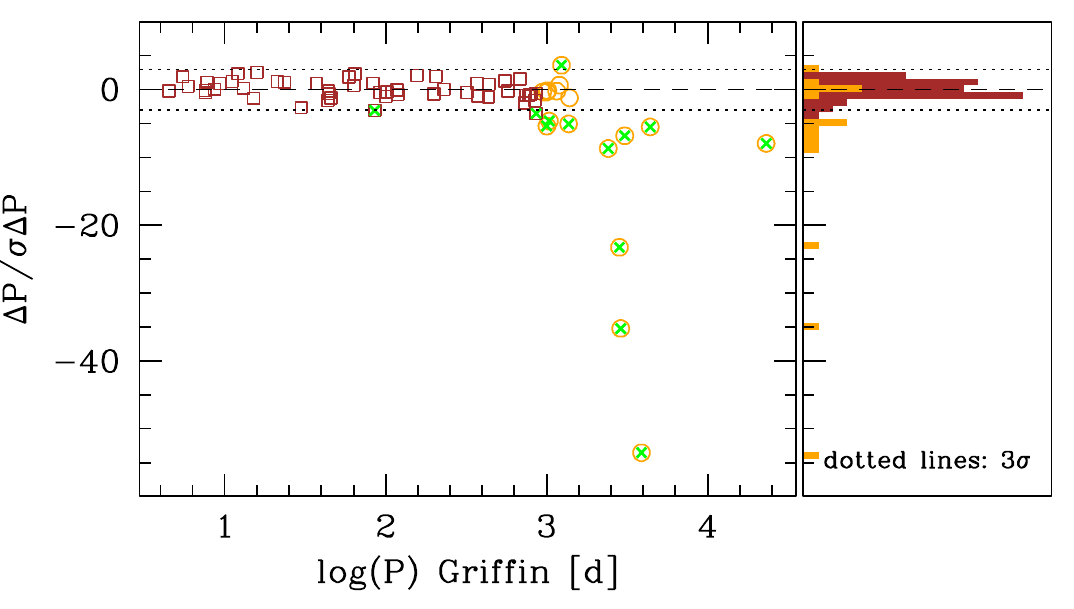}
}\caption[Deviation in period with Griffin]{Deviations in $P$ 
expressed in $\sigma$ units with respect 
to Griffin, as a function of the reference period. Systems 
for which the period, $P$, 
in the literature is unlikely to be found (i.e.\ $\Delta T$ $<$ $P$) 
are indicated with orange circles. 
Systems for which $\Delta T$ $\geqslant$ $P$ are shown as brown squares. 
Binaries for which the reference 
period is not recovered to within 3$\sigma$ are flagged with green crosses.}
\label{fig:cu4nss_spectro_DeltaP_over_sigmaDeltaP_vs_P_griffin_SB1}
\end{figure}

Figures\,\ref{fig:cu4nss_spectro_comparison_griffin_SB1_e}, 
\ref{fig:cu4nss_spectro_comparison_griffin_SB1_gamma0}, 
\ref{fig:cu4nss_spectro_comparison_griffin_SB1_K}, 
and \ref{fig:cu4nss_spectro_comparison_griffin_SB1_P} 
show the comparison between the DR3 and Griffin's SB1 orbital parameters, 
as a function of various DR3 parameters. 
The deviations in $P$ expressed in $\sigma$ 
units are shown in 
Fig.\,\ref{fig:cu4nss_spectro_DeltaP_over_sigmaDeltaP_vs_P_griffin_SB1}, 
as a function of the reference period. Systems for which the latter is 
difficult to find because of the limited length of the 
RVS observations (i.e. $\Delta T$ = time span 
observations $<$ $P$) are indicated with a different colour. 
Systems for which this  
period is not recovered to within 3$\sigma$ are also flagged. 

\begin{figure}[!htp]
\centerline{
\includegraphics[width=0.5\textwidth, trim= 40 230 0 260, clip]
{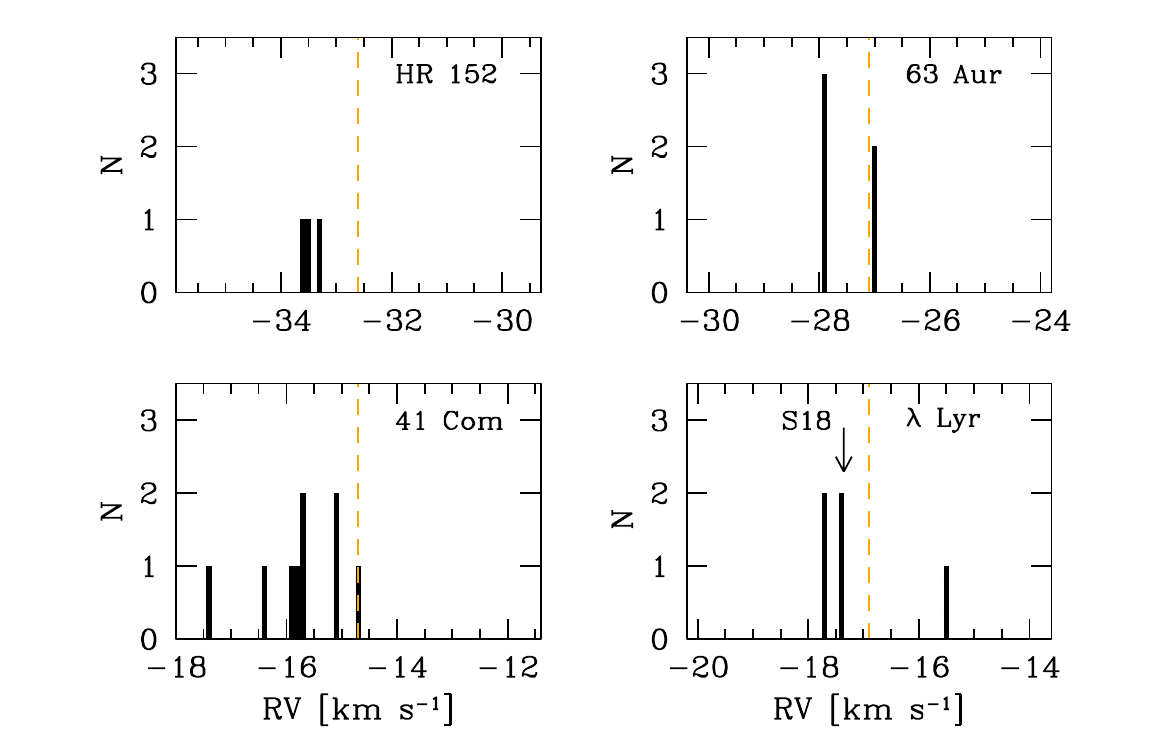}
}\caption[Comparison between Griffin and more recent RVs]{Comparison 
for the reference stars adopted 
by \citet{griffin69} between his RVs (dashed lines) 
and those in the more recent literature (histograms). 
The value of \citet{2018A&A...616A...7S} for $\lambda$ Lyr is indicated.}
\label{fig:cu4nss_spectro_histogram_vr_reference_stars_griffin}
\end{figure}

Some deviating points can be noticed in 
Figs.\,\ref{fig:cu4nss_spectro_comparison_griffin_SB1_e}, 
\ref{fig:cu4nss_spectro_comparison_griffin_SB1_gamma0}, 
\ref{fig:cu4nss_spectro_comparison_griffin_SB1_K}, 
and \ref{fig:cu4nss_spectro_comparison_griffin_SB1_P}, 
especially when the period is not recovered. 
The incidence of outliers is quantified below. 
In some cases, the agreement is satisfactory in absolute terms 
and the deviation is only 
due to (very) small uncertainties. 
There is clear evidence for a systematic $\gamma$ offset, 
although the discrepancy only amounts to $\sim 1$\,{km\,s$^{-1}$} on average. 
It might be due to the fact that the RV values adopted by Griffin for his four 
reference stars back in the late sixties \citep{griffin69} 
appear to be larger by roughly the same 
amount than those in the recent literature, 
including \citet{2018A&A...616A...7S} for $\lambda$ Lyr 
(Fig.\,\ref{fig:cu4nss_spectro_histogram_vr_reference_stars_griffin}). 

Table\,\ref{tab:cu4nss_spectro_statistics_griffin_SB1} 
summarises the recovery rate of the SB1 parameters 
to within 3$\sigma$.
 Cases where the time span of the RVS observations is longer/shorter 
than the reference period are listed separately. 
As may be expected, the performance of the pipeline greatly improves 
when the observations cover more than one orbital cycle. 

\begin{table}[!htp]
\caption[Recovery rate to within 3$\sigma$ of each parameter wrt Griffin]
{Recovery rate to within 
3$\sigma$ of each parameter with respect to Griffin expressed in per cent. 
\label{tab:cu4nss_spectro_statistics_griffin_SB1}}
\centering  
\begin{tabular}{lcccc}
\hline\hline
                   & $e$ & $\gamma$ & $K$ & $P$ \\
\hline
$\Delta T$ $\geqslant$ $P$ & 96.4 (55) & 12.0 (50) & 94.1 (51) & 96.1 (51) \\
$\Delta T$ $<$ $P$ & 88.2 (17) & 17.6 (17) & 58.8 (17) & 35.3 (17) \\
All                & 94.4 (72) & 13.4 (67) & 85.3 (68) & 80.9 (68) \\
\hline 
\end{tabular} 
\tablefoot{The number in brackets is the number 
of systems the estimate is based on. 
Cases where the time span of the observations is longer/shorter 
than the reference period are listed separately.}
\end{table}

\subsubsection{Comparison against SB9
catalogue}\label{sssec:spectroSB1_validation_otherset_SB9}
\label{sssec:spectroSB1_quality_val_compSB9}

As a next step, the analysis above was repeated for the whole SB9. 
Although the sample is now considerably larger, it is much less 
homogeneous and contains solutions of generally lower quality. 
The most recent orbit was selected when a given system had 
multiple entries in the catalogue. 
As described above, visual and/or multiple systems were also excluded. 
Six SB2 orbits were replaced 
by those determined by \citet{lester20}. 
This validation catalogue contains the orbits of 1750 SB1's and 647 SB2's. 
The breakdown of DR3 spectroscopic solution types is as 
follows for SB1s in the SB9 catalogue:
\begin{itemize}
\item
\tt{SB1}: 226 systems,
\item
\tt{SB1C}: 0 system,
\item
TrendSB1: 45 systems,
\item
\tt{SB2}: 3 systems,
\item
\tt{SB2C}: 0 system,
\item
Without any DR3 spectroscopic deterministic solutions: 1476 systems.
\end{itemize}

All (but one) of the 45 systems with a trend solution have very long periods. 
Furthermore, 18 and 2 systems in the SB2 validation catalogue have received 
a SB1 and a TrendSB1 solution, respectively.

\begin{figure}[!htp]
\centerline{
\includegraphics[width=0.5\textwidth, trim= 40 300 40 250, clip]{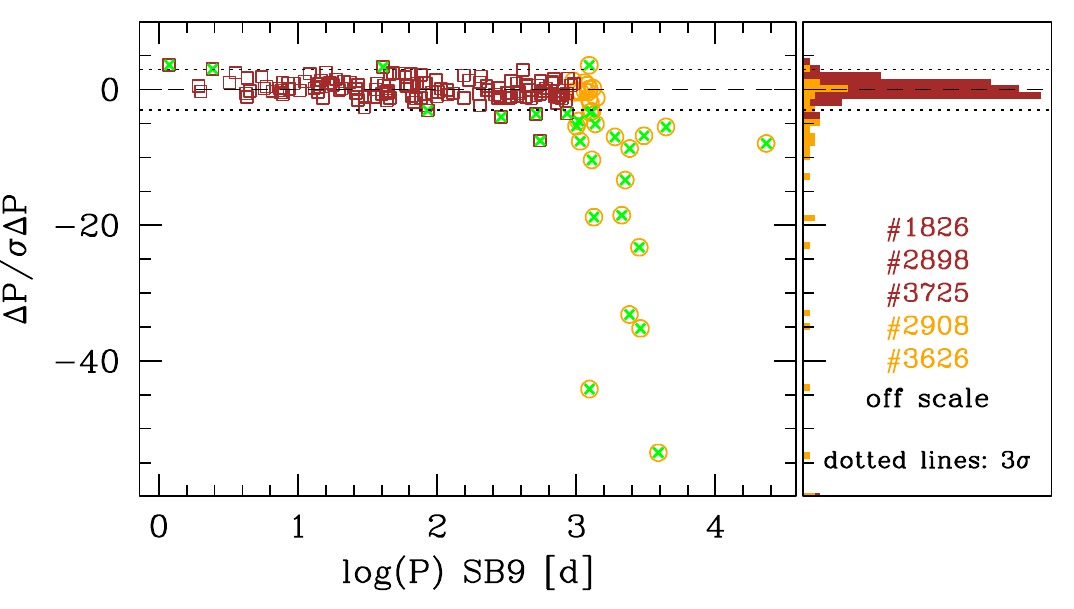}
}\caption[Deviation in period with SB9]{Same as 
Fig.\,\ref{fig:cu4nss_spectro_DeltaP_over_sigmaDeltaP_vs_P_griffin_SB1}, 
but for the SB9. As indicated, a few systems have very 
negative $\Delta P/\sigma\Delta P$ values and are off scale. 
The SB9 IDs are given.}
\label{fig:cu4nss_spectro_DeltaP_over_sigmaDeltaP_vs_P_SB9_SB1}
\end{figure}

Figures\,\ref{fig:cu4nss_spectro_comparison_SB9_SB1_e},
\ref{fig:cu4nss_spectro_comparison_SB9_SB1_gamma0}, 
\ref{fig:cu4nss_spectro_comparison_SB9_SB1_K}, 
and \ref{fig:cu4nss_spectro_comparison_SB9_SB1_P} show the 
comparison between the orbital parameters. The deviations in $P$ 
expressed in $\sigma$ units are shown 
in Fig.\,\ref{fig:cu4nss_spectro_DeltaP_over_sigmaDeltaP_vs_P_SB9_SB1}, 
as a function of the reference period. 

The results are consistent with those obtained for Griffin's sample. 
Some features are expected. 
For instance, the difficulty to find the correct period 
for (very) wide, low-amplitude systems: 
the orbital periods are generally systematically underestimated 
(Fig.\,\ref{fig:cu4nss_spectro_comparison_SB9_SB1_P}). 
Once again, the large discrepancies may be misleading in some cases. 
For instance, some short-period 
systems are outliers in $P$ because the uncertainties are very 
small ($\sim$10$^{-4}$ d). 
The same conclusion holds for SB2's \citep[see][]{dr3-dpacp-161}. 
The slight systematic discrepancy and large dispersion for $\gamma$ 
is likely due to the fact 
that a significant fraction of the orbits are actually taken from 
Griffin (see discussion 
about offset in Sect.\,\ref{sssec:spectroSB1_quality_val_compGriffin}). 
The $\gamma$ deviations show some evidence for a bimodal 
distribution (Fig.\,\ref{fig:cu4nss_spectro_comparison_SB9_SB1_gamma0}), 
and it indeed appears that 
the smaller peak is largely dominated by results from Griffin. 
The RV bias of up to 75 m s$^{-1 }$ 
affecting the faint end (Sect.\,\ref{sec:spectroSB1_input}) is 
negligible in this respect. 
Furthermore, there are very few objects with 
$G_\mathrm{RVS}^\mathrm{int}$ $>$ 11 mag 
(Fig.\,\ref{fig:cu4nss_spectro_comparison_SB9_SB1_gamma0}).

Table\,\ref{tab:cu4nss_spectro_statistics_SB9_SB1} summarises 
the recovery rate of the SB1 parameters  
to within 3$\sigma$. Once again, not surprisingly, 
the incidence of outliers decreases 
when the time span of the observations exceeds one orbital cycle. 

\begin{table}[!htp]
\caption[Comparison of recovery with SB9]{Same as Table~\ref{tab:cu4nss_spectro_statistics_griffin_SB1}, but for the whole SB9.
\label{tab:cu4nss_spectro_statistics_SB9_SB1}}
\centering  
\begin{tabular}{lcccc}
\hline\hline
                   & $e$ & $\gamma$ & $K$ & $P$ \\
\hline
$\Delta T$ $\geqslant$ $P$ & 89.2 (167) & 53.9 (152) & 90.4 (156) & 92.7 (150) \\
$\Delta T$ $<$ $P$ & 81.1 (37) & 43.2 (37) & 64.9 (37) & 37.8 (37) \\
All                & 87.7 (204) & 51.9 (189) & 85.4 (193) & 81.8 (187) \\
\hline 
\end{tabular} 
\end{table}

\subsubsection{Comparison against APOGEE}
\label{sssec:spectroSB1_quality_val_APOGEE}

Orbital parameters for $\sim$20\,000 SB1's have been released as part of 
APO Galactic Evolution Experiment 
(APOGEE)\footnote{\url{http://www.sdss3.org/surveys/apogee.php}} 
DR16 \citep[][hereafter PW20]{price_whelan20}. 
Our validation sample is their 'Gold Sample' that contains 1032 systems 
with high-quality, unimodal posterior samplings.

The breakdown of DR3 spectroscopic solution types is as follows 
for the SB1s in PW20:
\begin{itemize}
\item
\tt{SB1}: 126 systems,
\item
\tt{SB1C}: 0 system,
\item
TrendSB1: 4 systems,
\item
\tt{SB2}: 3 systems,
\item
\tt{SB2C}: 0 system,
\item
Without any DR3 spectroscopic deterministic solutions: 899 systems.
\end{itemize}

All the systems with a trend solution have a period much longer 
than $\Delta T$.

\begin{figure}[!htp]
\centerline{
\includegraphics[width=0.5\textwidth, trim= 40 300 40 250, clip]
{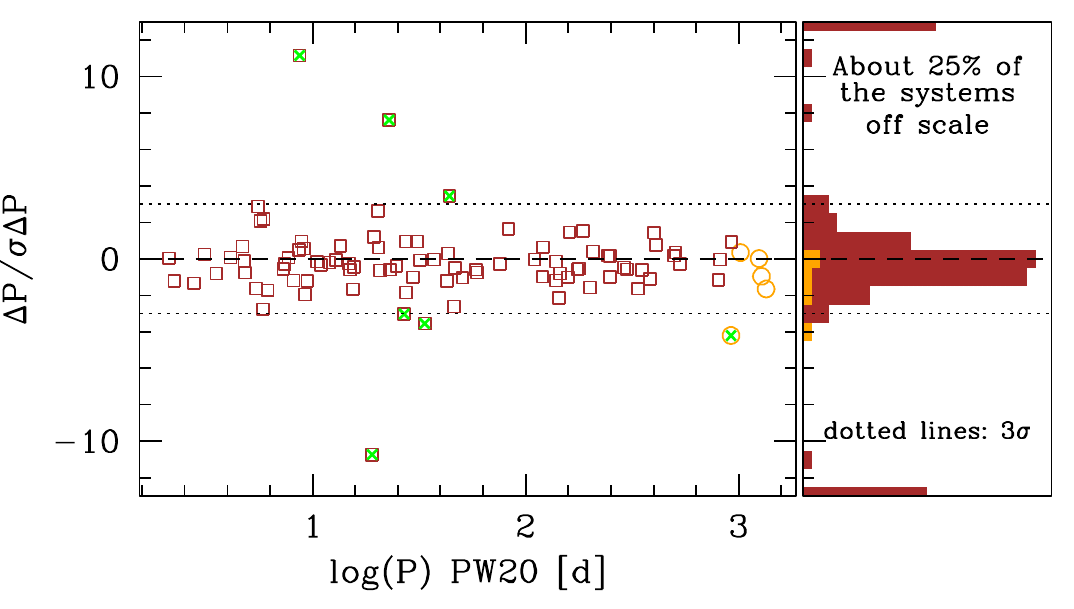}
}\caption[Deviation in period with \citep{price_whelan20}]
{Same as Fig.\,\ref{fig:cu4nss_spectro_DeltaP_over_sigmaDeltaP_vs_P_griffin_SB1}, 
but for the APOGEE sample. We note that a significant number of 
systems are off scale.}
\label{fig:cu4nss_spectro_DeltaP_over_sigmaDeltaP_vs_P_PW20_SB1}
\end{figure}

Figures\,\ref{fig:cu4nss_spectro_comparison_PW20_SB1_e}, 
\ref{fig:cu4nss_spectro_comparison_PW20_SB1_gamma0}, 
\ref{fig:cu4nss_spectro_comparison_PW20_SB1_K}, and 
\ref{fig:cu4nss_spectro_comparison_PW20_SB1_P} 
show the comparison between the orbital parameters. 
The deviations in $P$ expressed in $\sigma$ units are shown in 
Fig.\,\ref{fig:cu4nss_spectro_DeltaP_over_sigmaDeltaP_vs_P_PW20_SB1}, 
as a function of the reference period. 
For the period, two distinct situations are encountered: 
either there is a reasonable 
correspondence or, in about $\sim$25\% of the cases, 
the values are dramatically discrepant. 
For those, systematic differences are not observed. 
The solutions in PW20 for the majority of these binaries rely on 
much fewer RVs (less than ten). 
It is likely that the paucity of measurements prevents a 
robust determination of the orbital period. 
We caution that these systems are off scale in 
Figs.~\ref{fig:cu4nss_spectro_DeltaP_over_sigmaDeltaP_vs_P_PW20_SB1} 
and \ref{fig:cu4nss_spectro_comparison_PW20_SB1_P}.

Table\,\ref{tab:cu4nss_spectro_statistics_PW20_SB1} summarises the 
recovery rate of the SB1 parameters  
to within 3$\sigma$. The lower incidence of outliers when a full 
orbital cycle is not covered 
may be attributed to small number statistics. 

\begin{table}[!htp]
  \caption[Statistics compared with \citep{price_whelan20}]{Same as 
  Table~\ref{tab:cu4nss_spectro_statistics_griffin_SB1}, 
  but for the APOGEE gold sample of PW20. 
  \label{tab:cu4nss_spectro_statistics_PW20_SB1}}
  \centering  
  \ \\
  \begin{tabular}{lcccc}
 \hline\hline
                   & $e$ & $\gamma$ & $K$ & $P$ \\
\hline
$\Delta T$ $\geqslant$ $P$ & 87.6 (121) & 73.6 (121) & 85.1 (121) & 69.4 (121) \\
$\Delta T$ $<$ $P$ & 100.0 (5) & 100.0 (5) & 100.0 (5) & 80.0 (5) \\
All                & 88.1 (126) & 74.6 (126) & 85.7 (126) & 69.8 (126) \\
\hline 
\end{tabular} 
\end{table}

\begin{figure}[!htp]
\centerline{
\includegraphics[width=0.5\textwidth, trim= 40 340 0 220, clip]{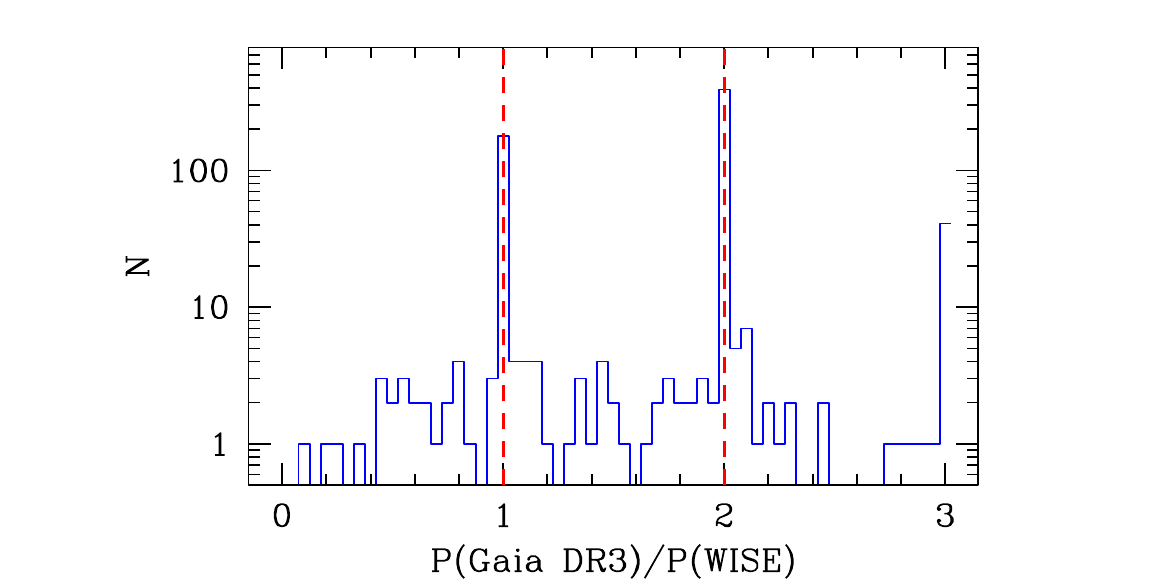}
}\caption[Ratio of periods with WISE]{Ratio between spectroscopic DR3 and 
photometric WISE periods. 
DR3 values that can be identified with the first and second WISE harmonics 
are indicated as vertical, dashed lines.}
\label{fig:cu4nss_spectro_histogram_period_ratios_WISE_SB1}
\end{figure}
\subsubsection{Comparison against WISE}
\label{sssec:spectroSB1_quality_val_WISE}

A total of $\sim$56\,000 short-period ($<$\,10\,d) binary systems observed by the 
Wide-field Infrared Survey Explorer 
(WISE)\footnote{\url{https://www.jpl.nasa.gov/missions/wide-field-infrared-survey-explorer-wise}} 
have been discussed by \citet{petrosky21}. Most of them are 
\mbox{(near-)contact} EBs. 
However, as will be shown below, a non-negligible amount are 
classical pulsators, especially RR Lyrae stars. 
The infrared photometric measurements are made through the W1 filter 
centred at about 3.4\,$\mu$m. 
A subset of 49\,465 systems can be unambiguously associated to a DR3 
source\footnote{\url{http://cdn.gea.esac.esa.int/Gaia/gedr3/cross_match/}}. 
This sample allows us to test the reliability of our orbital 
periods for very close binaries, although other orbital parameters 
are not available.

The breakdown of DR3 spectroscopic solution types is as follows 
for the SB1s in \citet{petrosky21}:
\begin{itemize}
\item
\tt{SB1}: 704 systems,
\item
\tt{SB1C}: 3 systems,
\item
TrendSB1: 12 systems,
\item
\tt{SB2}: 69 systems,
\item
\tt{SB2C}: 23 systems,
\item
Without any DR3 spectroscopic deterministic solutions: 48\,654 systems.
\end{itemize}

\begin{figure}[!htp]
\centerline{
\includegraphics[width=0.5\textwidth, trim= 40 170 0 400, clip]{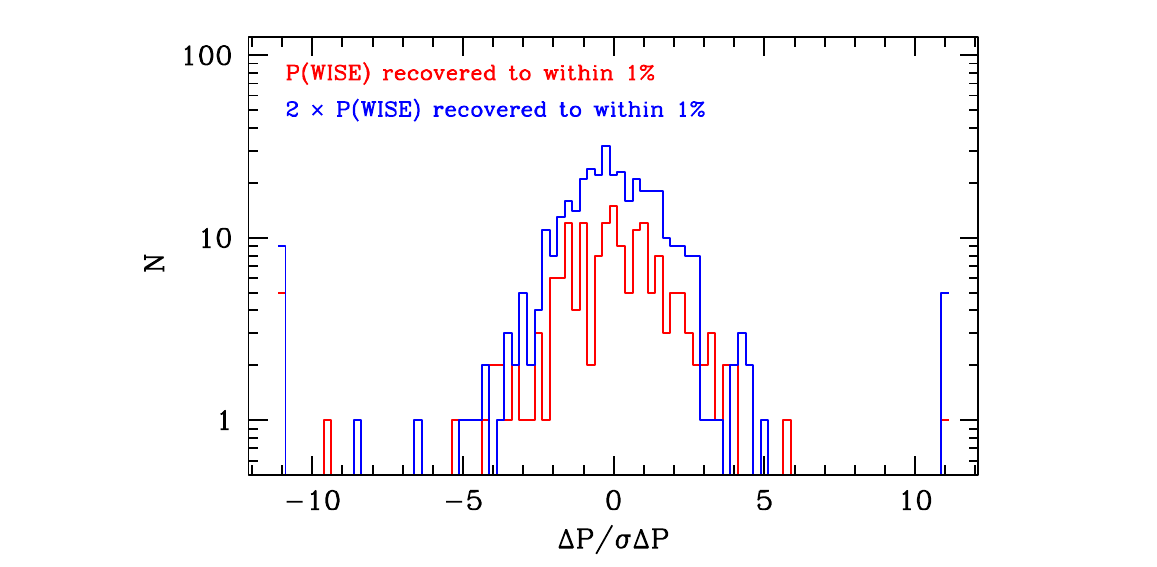}
}\caption[Comparison of periods with WISE]{Deviations in $\sigma$ units 
between the DR3 and WISE periods when the latter (or its harmonic) 
are recovered to within 1\%.
}
\label{fig:cu4nss_spectro_histogram_deltaP_over_sigmaDeltaP_WISE_SB1}
\end{figure}

We find a close correspondence in most cases with either 
the main WISE period or its second 
harmonic (Fig.\,\ref{fig:cu4nss_spectro_histogram_period_ratios_WISE_SB1}). 
As discussed by \citet{petrosky21}, the orbital period they 
found for the binaries is usually half the true 
value because of the nearly symmetrical nature of the WISE light curves. 
The WISE period is 
recovered to within 1\% for 25.3\% and 55.4\% of the cases 
for the first and second harmonic, 
respectively. The overall recovery rate at this level is thus about 81\%. 
It remains quite high for lower tolerances (e.g. $\sim$78\% for 0.1\%). 
The deviations lie in the vast majority of cases within 
five times the mutual uncertainties 
(Fig.\,\ref{fig:cu4nss_spectro_histogram_deltaP_over_sigmaDeltaP_WISE_SB1}).

\begin{figure}[!htp]
\centerline{
\includegraphics[width=0.45\textwidth, trim= 0 180 180 130, clip]{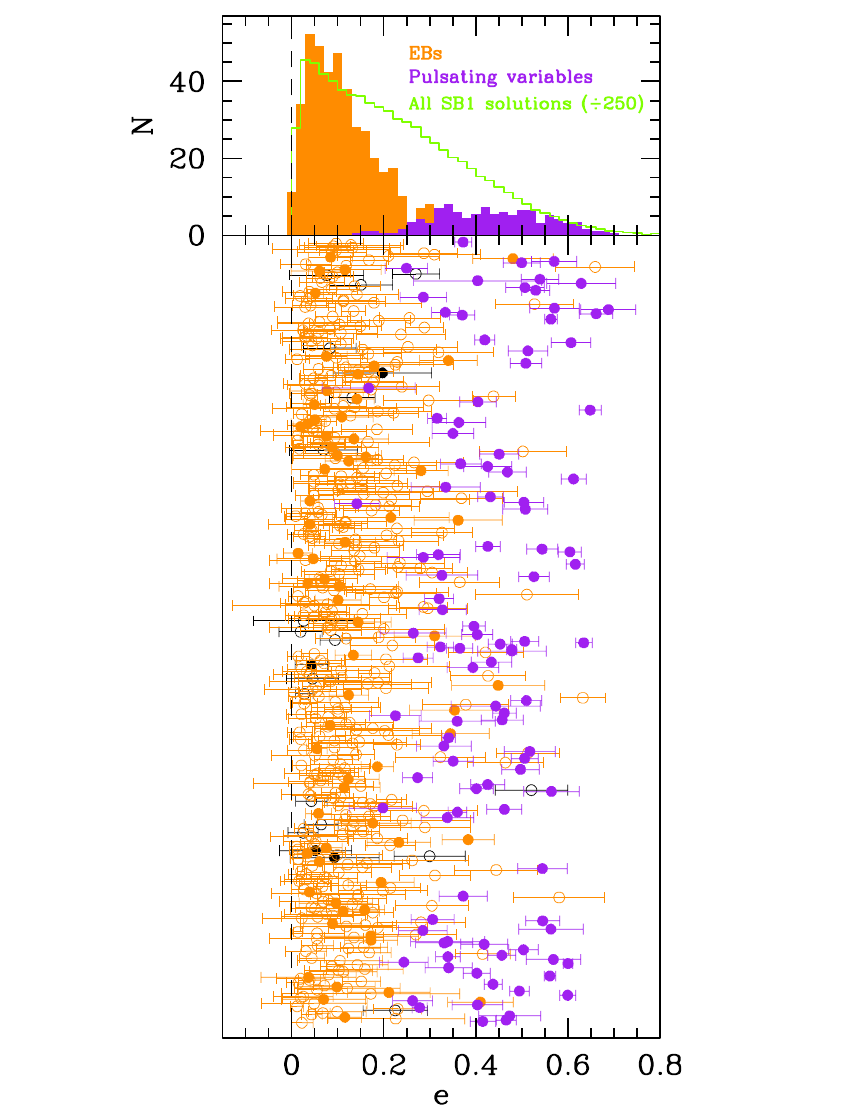}
}\caption[Distribution of the eccentricities for the WISE sources 
with the period recovered]
{Distribution of eccentricities for the WISE sources 
for which the period (filled symbols) 
or its harmonic (open symbols) is recovered to within 1\%. 
Lower panel: eccentricities 
for the systems classified as EBs \citep[orange;][]{2023A&A...674A..16M}, 
Cepheids or RR Lyrae \citep[purple;][]{2023A&A...674A..17R,2023A&A...674A..18C}. 
Those without a classification are shown in black. 
The systems are displayed in no particular order along the $y$-axis. 
Upper panel: histogram of 
the distributions for the EBs, pulsating variables, and all SB1 
solutions (the last one is divided by 250).}
\label{fig:cu4nss_spectro_distribution_e_WISE}
\end{figure}

Most of the WISE sources were assigned a classification in DR3 
based on their light variations 
\citep{2023A&A...674A..16M,2023A&A...674A..17R,2023A&A...674A..18C}. 
As can be seen in Fig.\,\ref{fig:cu4nss_spectro_distribution_e_WISE}, 
a (nearly) circular solution is found for 
the majority of the EBs, as may be expected for such close binaries. 
In contrast, the systems with high eccentricities are mostly classical 
pulsators (Cepheids or RR Lyrae) mistaken as binaries. 
As described in Sect.\,\ref{sssec:spectroSB1_add_cons_postfilt_intvar}, 
those pulsating variables were rejected from the final DR3 catalogue 
during the post-filtering process.
\subsubsection{Comparison against ASAS-SN}
\label{sssec:spectroSB1_quality_val_ASAS-SN}

The physical parameters of 35\,464 detached eclipsing binaries (DEBs) 
observed in the framework 
of the All-Sky Automated Survey for Supernovae 
\citep[ASAS-SN;][]{kochanek17}\footnote{\url{https://www.astronomy.ohio-state.edu/asassn/}} 
have been determined by \citet{rowan22}. A total of 35\,307 systems 
have a \gaia\ DR3 ID. 
The data are based on the analysis of $g$- and $V$-band light curves. 
The orbital periods nicely bridge the gap between the WISE sample of 
close binaries and other 
more general catalogues, such as APOGEE or the SB9 
(Fig.\,\ref{fig:cu4nss_spectro_validations_catalogues_P_SB1}). 
Contrary to WISE, other orbital elements (e.g. eccentricity) 
are provided besides the period, 
although without any uncertainties. 
For $P$, we assume a typical relative uncertainty 
of 10$^{-6}$ \citep[e.g.][]{holanda18}.

The breakdown of DR3 spectroscopic solution types is as 
follows for the binaries in \citet{rowan22}:
\begin{itemize}
\item
\tt{SB1}: 626 systems,
\item
\tt{SB1C}: 3 systems,
\item
\tt{TrendSB1}: 6 systems,
\item
\tt{SB2}: 92 systems,
\item
\tt{SB2C}: 35 systems,
\item
Without any DR3 spectroscopic deterministic solutions: 34\,545 systems.
\end{itemize}
Most objects have $G$ $\gtrsim$ 13 mag and are fainter than our magnitude cut-off.

\begin{figure}[!htp]
\centerline{
\includegraphics[width=0.5\textwidth, trim= 28 200 40 360, clip]{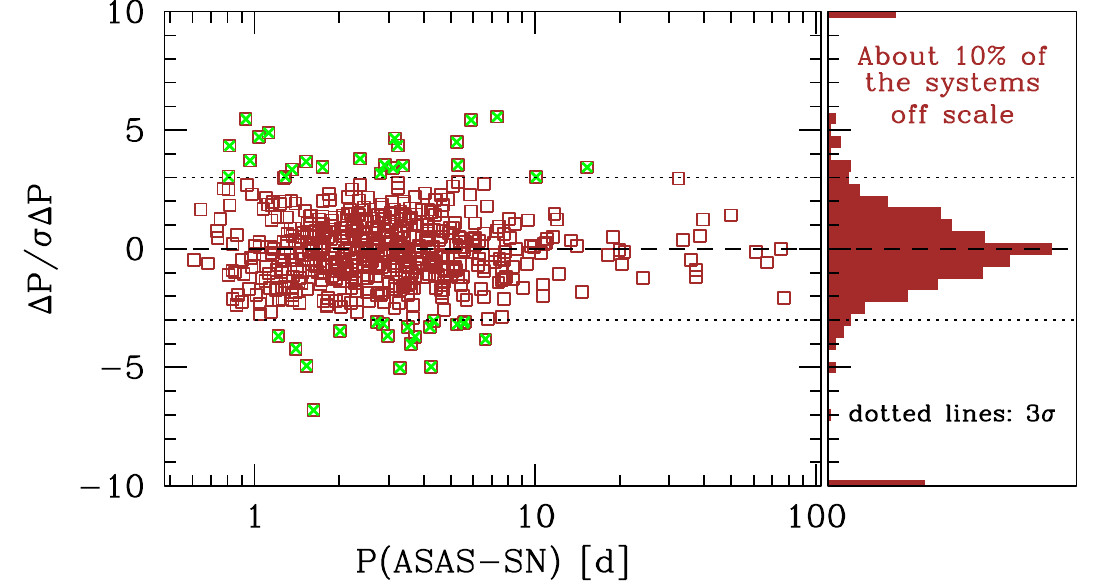}
}\caption[]{Same as 
Fig.\,\ref{fig:cu4nss_spectro_DeltaP_over_sigmaDeltaP_vs_P_griffin_SB1}, 
but for the ASAS-SN sample. We note that a significant 
number of systems are off scale.}
\label{fig:cu4nss_spectro_DeltaP_vs_P_ASASSSN}
\end{figure}

The majority of the 626 objects are classified in DR3 
as EBs based on their photometric 
behaviour \citep{2023A&A...674A..16M}. 
Conversely, none are identified as classical 
pulsators \citep{2023A&A...674A..18C,2023A&A...674A..17R}. 
For this sample with SB1-type solutions, 
the overall recovery rate of the photometric period to within 
3$\sigma$ amounts to $\sim$83\% 
(Fig.\,\ref{fig:cu4nss_spectro_DeltaP_vs_P_ASASSSN}). 
This estimate includes one system where the ASAS-SN period 
is actually the harmonic (as discussed for WISE above) or for nine 
objects twice the true value 
because one of the two eclipses is too shallow to be detected 
(see figure\,6 of \citealt{rowan22} for an example). 
Visual inspection of the ASAS-SN light curves for the nine 
DEBs supports the latter explanation. 

\begin{figure}[!htp]
\centerline{
\includegraphics[width=0.6\textwidth, trim = 0 175 150 135, clip]{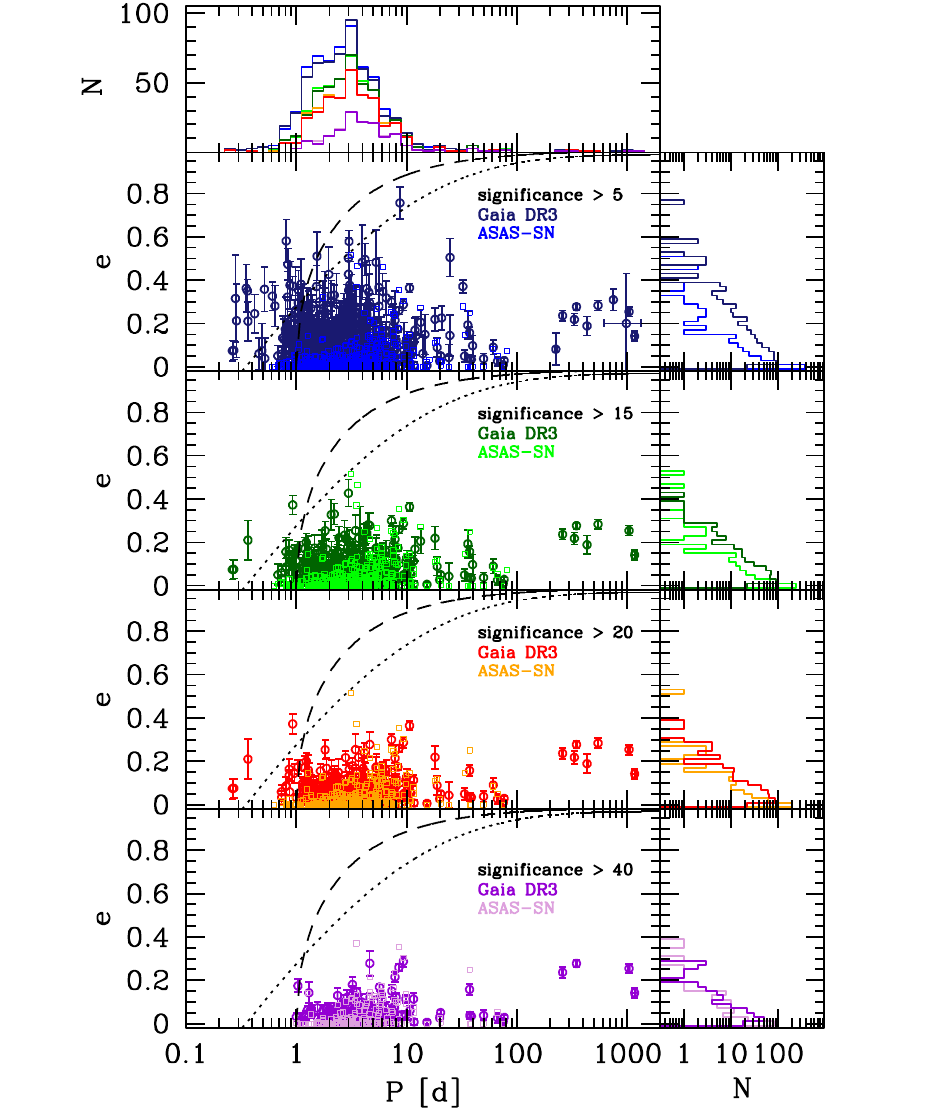}
}\caption{$e$-$\log P$ diagrams for the 626 systems in common 
between DR3 and ASAS-SN. 
Each panel corresponds to a different threshold on the solution significance. 
The dotted and dashed lines show the upper envelopes defined by 
\citet{mazeh08} and \citet{2005A&A...431.1129H}, respectively. 
A cut-off period of one day was adopted for the latter. 
The position of the curve is only 
illustrative as this choice is somewhat arbitrary.}
\label{fig:cu4nss_spectro_validations_catalogues_logP_e_ASASSN_SB1}
\end{figure}

The relevance of filtering out short-period systems 
with a low solution significance 
was discussed in Sect.\,\ref{ssec:spectroSB1_validation_intern}. 
We show in Fig.\,\ref{fig:cu4nss_spectro_validations_catalogues_logP_e_ASASSN_SB1} 
the DR3 and ASAS-SN data in a $e$-$\log P$ diagram for various cuts 
in solution significance (see also Fig.\,\ref{fig:EGperiodversuseccent}). 
The upper envelopes defined by \citet{mazeh08} and 
\citet[][modified to $P_{\mathrm{cut-off}}$ at 1\,d]{2005A&A...431.1129H} are overplotted. 
The former is based on an empirical fit of the SB9 data,
but the latter rests on theoretical arguments. 
While $P$ is recovered in $\sim$67\% of the cases 
for $K$/$\sigma_ K$ values up to 15, 
the proportion sharply increases for higher 
significances (success rate of $\sim$90\% on average). 
The differences are generally small in absolute terms, 
which explains the close match 
in the period distributions shown in 
Fig.\,\ref{fig:cu4nss_spectro_validations_catalogues_logP_e_ASASSN_SB1}. 
However, an excess of seemingly eccentric systems with respect
to \citet{rowan22} is readily visible. 
The eccentricity distributions become more and more similar 
when the significance cut-off 
increases until they become statistically indistinguishable above about 50. 
Rather than a poor determination of the period for close binaries, 
the issue thus seems to rather 
lie in solutions that are too eccentric when they are ill-constrained 
\citep[see also][]{2023MNRAS.521.5927J}.
\subsubsection{Summary}
\label{sssec:spectroSB1_quality_val_summary}
\begin{figure*}[!htp]
\centerline{
\includegraphics[width=1.\textwidth, trim= 80 230 90 230, clip]{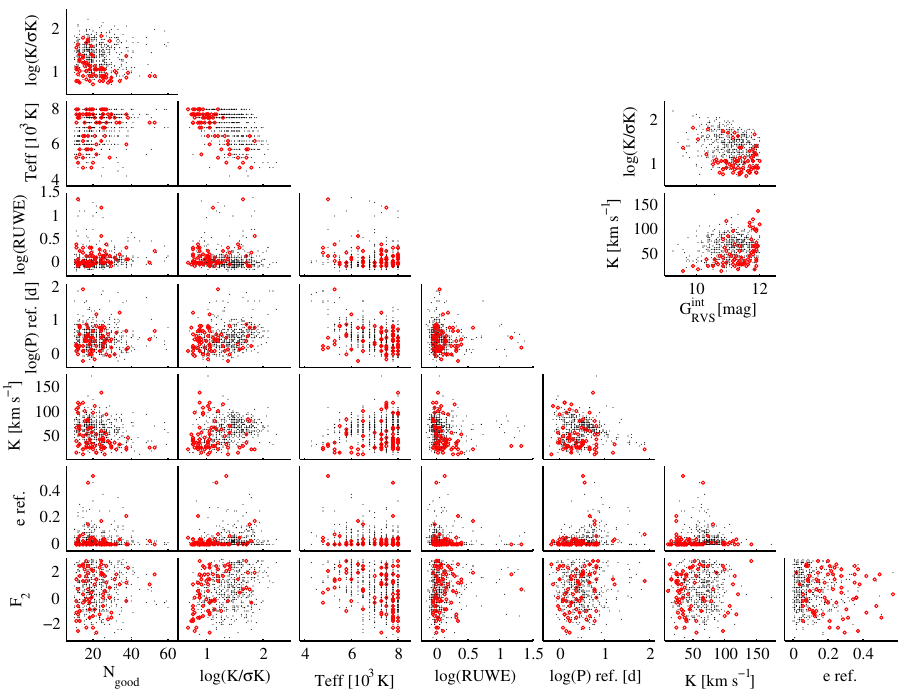}
}\caption[Corner plot for ASAS-SN]{Corner plot for the ASAS-SN sample, 
with the strong outliers in $P$ 
deviating by more than 5$\sigma$ highlighted in red. 
They constitute about 11\% of the whole sample.}
\label{fig:cu4nss_spectro_comparison_ASASSN_SB1_corner_plot}
\end{figure*}

As a preamble to summarising the main conclusions of our analysis, 
it is worth emphasising that some 
of the validation data sets discussed above are of limited 
usefulness in the sense that they do not 
offer stringent tests of our results. In particular, the APOGEE
orbits from PW20 
are based on fewer epochs (about 14 on average concerning the 126 objects) 
than most NSS sources (see Fig.\,\ref{fig:EGnumbertransits}). 
As a consequence, their orbital solutions are not necessarily 
more robust and precise. 
We indeed find a poorer match for the periods with respect to 
the other validation catalogues, 
even at high significance. It might be no coincidence that, 
except for $\gamma$, the highest 
recovery rates for systems monitored over a sufficiently long baseline are found 
for the high-quality Griffin solutions
(Table~\ref{tab:cu4nss_spectro_statistics_griffin_SB1}).

As discussed above, the success rate for the main orbital 
parameters ($P$, $K$, $e$, and $\gamma$) is 
usually well above 80\% for the whole set of validation catalogues considered here 
(Griffin, SB9, APOGEE, WISE, and ASAS-SN). 
As expected, the performance 
increases for relatively short-period systems for which at least one 
orbital period is covered 
by the observations (about 2.5 years for the 
RVS spectroscopic processing data set considered here). 
In this case, the recovery rate 
commonly exceeds 85-90\%. The WISE sample of (near-)contact 
binaries allows us to investigate 
the performance of our period search at high frequencies. 
We recover the WISE photometric 
periods at a level of about 80-85\%. It increases to about 90\% for the 
DEBs with wider orbits discussed by \citet{rowan22}. 
We can also note the excellent success rate for the pulsation periods 
of the RR Lyrae stars that are below one day 
(Sect.\,\ref{ssec:spectroSB1_validation_cu7}). 
The good agreement on the period with both EB data sets advocates 
for a deeper study using these data. 
Their combination will certainly result in an extensive 
analysis that could reveal various 
kinds of objects such as EW binaries and/or ellipsoidal variables. 
Such a task is beyond the scope of the present paper.
Our analysis globally suggests that the orbital periods below about 
10 days are generally robust 
provided the solution significance is not too low. We find a roughly 
constant success rate at the 
level of about 90-95\% once the significance exceeds 15-20. 
However, although they can often be regarded as compatible with 
zero within the uncertainties, 
it appears that the eccentricities may be overestimated 
even if the significance is larger.

We ascribe the poor correspondence between the DR3 
and Griffin/SB9 $\gamma$'s to a slight zero-point offset 
at the $\sim 1$\,{km\,s$^{-1}$}  level. 
There is indeed no evidence for a significant bias with respect to APOGEE 
(Fig.\,\ref{fig:cu4nss_spectro_comparison_PW20_SB1_gamma0}). 
For the other parameters, no obvious systematic discrepancies are observed. 

In order to obtain an order-of-magnitude estimate of the period 
whenever possible, a choice is made to systematically 
favour an {\tt{SB1}} solution over a {\tt{TrendSB1}} one 
when the period is estimated to be in the range 0.5 to 1.5\,$\Delta T$. 
As a result, an {\tt SB1} solution could have been assigned 
despite the fact that the length of the observations 
is insufficient. However, although the proportion strongly depends on 
the validation data set, 
it concerns less than 15\% of the cases. Conversely, overall 
about 79\% of the {\tt TrendSB1} solutions 
are associated to binaries with periods significantly 
longer than the observations.

Our ability to recover the orbital parameters quite significantly 
hinges on the retrieval of the true period. 
When only counting the binaries for which this condition is satisfied 
in the reference catalogues discussed in 
Sects.~\ref{sssec:spectroSB1_validation_otherset_griffin} 
to \ref{sssec:spectroSB1_quality_val_APOGEE}, we find an increase 
of $\sim$5-10\% in the recovery rate of $e$, 
$\gamma$, and $K$. For $e$ and $K$, it typically reaches $\sim$95\%.
It is evident that other limitations, either related to the characteristics of 
the observations (e.g. number of 
RVS transits) or the very nature of the binary system 
(e.g.\ velocity semi-amplitude) can also 
potentially play a significant role in the quality of the solutions. 
Some clues about the sensitivity (or lack thereof) to the combination of some 
parameters can be gauged from plots such as 
Fig.\,\ref{fig:cu4nss_spectro_comparison_ASASSN_SB1_corner_plot}. 
Besides suggesting a lack of dependency between the recovery 
rate of the period and 
some quantities (e.g.\ {\tt{ruwe}}), 
it illustrates, as discussed above, the difficulty in successfully 
finding the period at low significance values. 
However, there may be more complex dependencies lurking in the data 
that cannot be easily and 
reliably quantified given the quite limited number of
cross-matches with the validation 
catalogues (a few hundreds systems at most). 

In this respect, the fact that 
only a relatively small fraction of the SB1s in the validation 
catalogues have a spectroscopic deterministic DR3 solution
is the consequence of the various filtering steps, 
both at CU4 and CU6 levels, described previously. 
In addition, we note that for the SB9 and Griffin's sample, only a 
subset of the reference catalogue was used. The case of the SB9, 
and the steps that led to the relatively small overlap 
with respect to DR3 (see Sect.\,\ref{sssec:spectroSB1_quality_val_compSB9}) 
are discussed in detail in Appendix\,\ref{sec:appG}. 
Similar arguments apply to the other external catalogues and explain 
the limited number of benchmark binaries with a spectroscopic  
deterministic solution in DR3, although the proportion of systems 
lost at every filtering step may markedly vary.
As an illustration, only the very brightest ASAS-SN eclipsing binaries 
get a DR3 solution. While the $T_\mathrm{eff}$ range is not an issue, 
we find that $\sim$90\% have a (combined) $G$ magnitude fainter than 12 mag. 
Similar figures apply to the WISE sample. Regardless of these intrinsic 
limitations, it is naturally expected that the completeness level will 
substantially improve in DR4 thanks to the larger amount of RVs 
collected and improvements in data analysis. 

We finally note that a few systems securely reported 
as double-lined in the literature 
have received an SB1-like solution (either eccentric or trend). 
The causes for this are multiple. For instance, the
spectral signature of the secondary could be too weak
to be detected by \gaia\ or the lines in the composite spectrum
cannot be fully deblended at the moderate resolution
of the RVS, as discussed in 
Sect.\,\ref{sssec:spectroSB1_results_illust_probl_fake}. 
The various cases which can be encountered are
discussed in \citet{dr3-dpacp-161}.
\section{Additional considerations}\label{sec:spectroSB1_add_cons}
\subsection{Post-filtering}\label{ssec:spectroSB1_add_cons_postfilt}
The results of the sole NSS-SB1 processing
(see Sect.\,\ref{sec:spectroSB1_results})
are providing a plethora of orbital solutions
covering a large domain of the values for the orbital parameters. 
The pipeline nature of the processing necessarily implies that 
part of the solutions could be wrong or spurious 
(whatever could go wrong will go wrong at least once). 
Therefore, we scrutinised a subset of the results in order 
to detect surviving problems. Since the DPAC policy with 
respect to the final production of the DR3 catalogue is 
intended to favour purity of the list rather than 
completeness, it is necessary to blacklist solutions 
that are most certainly spurious. We call this procedure 
post-filtering and it intervenes between the internally filtered 
output of the pipeline and the publication of the catalogue. 
This post-filtering can be based on information coming 
from outside our internal validation. A clear flaw
of this procedure is that this analysis will never 
identify all the pitfalls. Conversely, some objects could be 
blacklisted whereas the solution was correct. 
In the following, we address the identified problematic cases
in the context of the classes associated to the sole spectroscopic processing.
\subsubsection{Intrinsic variables}
\label{sssec:spectroSB1_add_cons_postfilt_intvar}
Since the NSS-SB1 processing is applied to time series of variable RVs, 
it is expected to also treat some intrinsic variables that are 
exhibiting well-defined RV variations. This could be the case 
for example for Cepheids (actually the bright ones). 
The only firewall preventing the processing from delivering 
spurious orbital solutions for intrinsic variables is 
the fact that the processing only considers an harmonic content of the 
RV curve that is Keplerian (or nearly so). 
Unfortunately, Cepheids RV curves, for example, are renowned for their 
harmonic content being quite similar to the Keplerian one. 
This has been pointed out by
\citet{1985A&AS...61..259I, 1987A&A...175...30I, 1989A&AS...81..339I}.
The problem generated by intrinsic variables has been pointed out
in Sect.\,\ref{sssec:spectroSB1_quality_val_WISE} 
and in Sect.\,\ref{sssec:spectroSB1_quality_val_ASAS-SN} where some of them 
were detected in the binary catalogue, mainly at short periods 
and high eccentricities.

As part of DR3, \citet{2023A&A...674A..17R} published a complete analysis 
of the \gaia\ photometric curves for a little more than 15\,000 
Cepheids (of all classes), 799 having time series of RVs.
This list was cross-correlated with our 
{\tt{SB1/SB1C}} list. 
They found 338 objects in common.
Among them, the Cepheid nature is confirmed from the literature for 276 objects.
After an eye inspection of photometric and RV curves,
five of them were found 
doubtful Cepheids and have been rejected. 
Finally, 333 spurious SB1 solutions 
due to the Cepheid nature of the objects have been 
blacklisted from the SB subcatalogue. 

Along these lines, we also considered the case of RR Lyrae stars
(of all classes too).
\citet{2023A&A...674A..18C} gathered a list of a little more than 270\,000
such objects, among which some 1096 have time series of RVs.
The cross-match with our {\tt{SB1}} list yields 152 objects
in common. Among these, 5 turned out to have a discrepancy
between the period from our SB1 pipeline and the one derived
by \citet{2023A&A...674A..18C}. Their RR Lyrae classification were 
thus considered as insecure.
Therefore, we rejected (blacklisted), from the {\tt{SB1}} list,  
147 objects considered
as bona fide RR Lyrae. In summary,
we rejected from our {\tt{SB1}} list some 480 sources as being
intrinsic variables. 
Although a few pulsating stars (Cepheids, RR Lyrae) are rejected
from our SB subcatalogue, they are nevertheless studied
as pulsators
elsewhere \citep[see e.g.][]{2023A&A...674A..13E}.

Another class of intrinsic variable objects exhibiting large
photometric amplitudes are the long-period variables (LPVs).
No similar secure list exists in DR3 for the LPVs.
A tentative preliminary list exists that is based on the
analysis of the photometry: the second \gaia\ catalogue
of long-period variable candidates \citep{2023A&A...674A..15L}.
However, the related results were not entirely accessible
at the time of the execution of our pipeline. In addition,
from the later provisional analysis, it became clear
that this preliminary list most probably also contains
a lot of other objects and in particular
ellipsoidal variables.
\begin{figure}[!ht]
\centerline{
\includegraphics[width=0.4\textwidth]{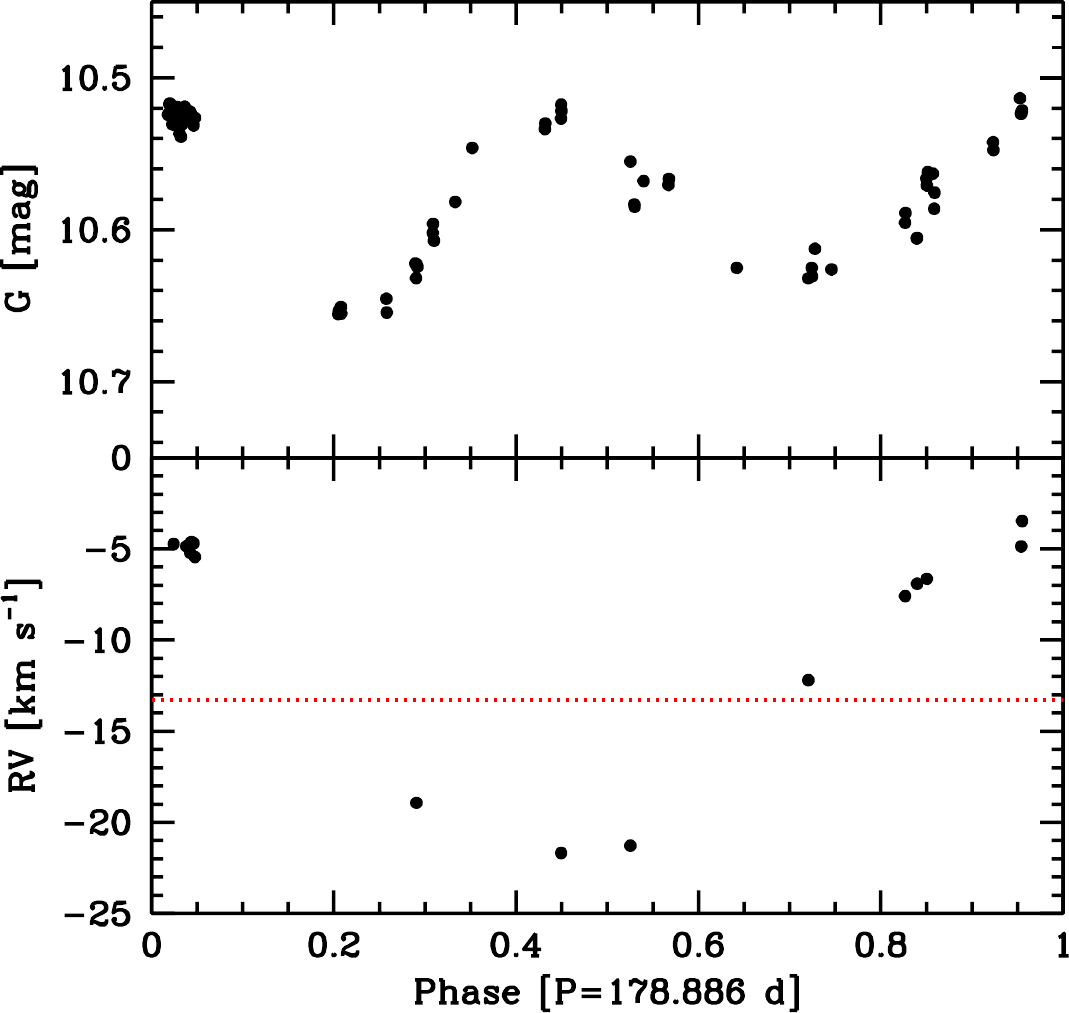}
}\caption{Variability of \gaia\ DR3 5829160851462523008.
Upper panel: $G$ lightcurve of this object
plotted as a function of the phase.
Lower panel: RV curve of the same object. The red dotted line
represents the systemic velocity. The zero of phase is arbitrary
and has been chosen to improve the readability of the graph.
Error bars are not given since they have sizes of the order of the 
symbols.
}
\label{fig:EGfigLPV}
\end{figure}
We illustrate here our concerns by further studying
the case of \gaia\ DR3 5829160851462523008. This object was not 
known to vary before \gaia .
It appears in the catalogue of LPV candidates with a derived 
photometric period of 90.5\,d. This object also appears
in the SB subcatalogue, produced in the present work,
as a binary with a period of 178.886\,d.
Figure\,\ref{fig:EGfigLPV} exhibits both the RV curve and the photometric
light curve of this object with the phase computed according to
the RV period. From this figure, it is clear that the RV curve is
unimodal, as it should be, whereas the $G$ light curve presents
two cycles in good agreement with the period attributed 
by \citet{2023A&A...674A..15L}. If the two maxima are very similar,
the two minima are significantly different.
In addition, the minima happen at phases where the RV curve
crosses the systemic velocity. All this strongly suggests
that this star is an ellipsoidal variable.
Therefore, at the time of DR3, it appeared hazardous to filter
our solutions on the basis of the existing LPV catalogue. Objects like
\gaia\ DR3 5829160851462523008 would have been unduly rejected and
we cautiously decided to not reject LPVs on this basis.
Certainly, a sound analysis necessitates the concomitant study
of RV and photometry, which was not foreseen for DR3.
This should be done in the subsequent FPR
\citep[see][]{2023A&A...680A..36G}
where some 20\,\% of the
LPV candidates are expected to be ellipsoidal variables.
We encourage the user of the SB subcatalogue
to filter out true intrinsic variables thanks to these subsequent results.

Yet another source of contaminants are main-sequence pulsators. 
The main periods of $\delta$ Scuti stars are typically below 0.2 d 
\citep{2023A&A...674A..36G} and therefore lie outside the frequency 
range explored by our pipeline (even considering aliasing). 
However, $\gamma$ Doradus stars are a more serious concern because 
they pulsate with periods in the range 0.3-1.4 d \citep{2023A&A...674A..36G} 
and exhibit RV variations of the order of a few km s$^{-1}$ 
\citep[e.g.][]{mathias04}. 

The problem of the rejection of intrinsic variables is certainly
a difficult one, and much progress is still to be made on this topic.
As mentioned above, the measurement of RVs by cross-correlating
the spectra with template prevents us from identifying
the origin of the RV variations.
\subsubsection{Double-line transits}
\label{sssec:spectro_add_cons_postfilt_double_line_transits}
Since the RVs are originally measured by the STA processing
on the basis of a per transit approach, there is 
initially no constraint from transit to transit.
In particular, the same object could exhibit a single-line
spectrum at a given time and a composite spectrum at another
moment. As part of the spectroscopic processing, a value
representing the percentage of transits leading to double-line
detection is computed. If the star is constant or assumed non-variable,
the median velocity is computed, a cut-off is then applied on this
percentage in order to avoid the propagation of the effect of suspicious
secondary RVs. If no transit exhibits a composite spectrum,
the object presenting variations is processed as a single star. On the opposite, if a 
sufficient number of transits show a double-line spectrum,
the object is processed as an SB2
\cite[see][for details]{dr3-dpacp-161}.
If the  assumed constant object presents only a minority of double-line transits,
the latter are ignored and the object is analysed on the basis
of the sole single-line data. The presence of 10\,\% or more
double-line transits is considered as a source of difficulties
for the processing of single constant objects. It has been proposed a posteriori
that the same rule (as the one applied to constant
stars) should apply to the NSS-SB1 chain. 
This led to the rejection
of 1475 {\tt{SB1}} objects and of one 
{\tt{SB1C}} object.
\subsubsection{Scan-angle dependent signals}
\label{sssec:spectroSB1_add_cons_postfilt_scan angle}
\begin{figure}[!ht]
\centerline{
\includegraphics[width=0.45\textwidth]{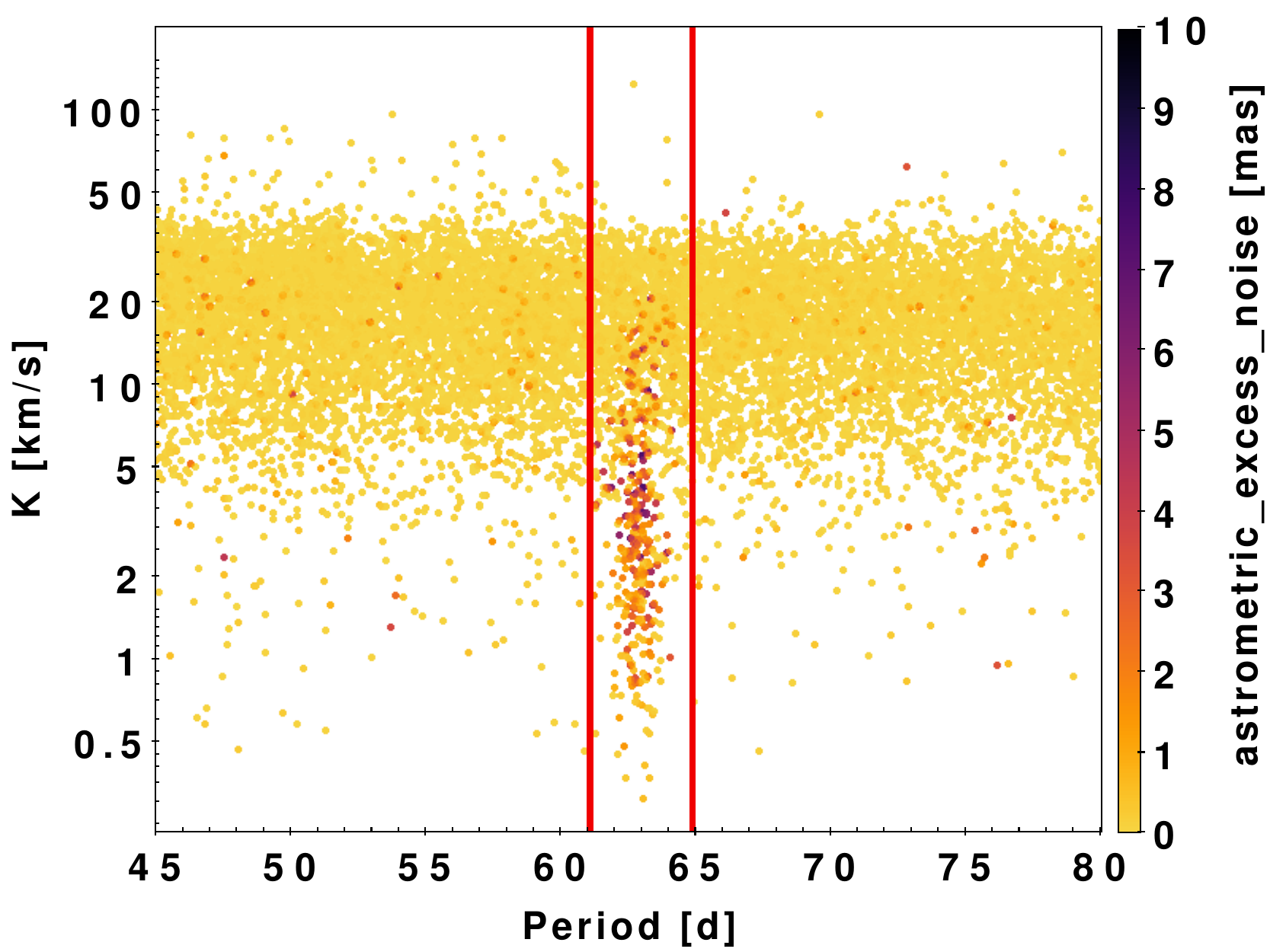}
}\caption{Semi-amplitude $K$ as a function of the period for the
SB1 solutions. The data points are coloured as a function of the
{\tt{astrometric\_excess\_noise}}. A population of solutions very close
to the 62.97\,d (within $\pm$\,3\,\% as underlined by the vertical red lines)
precession period exhibits small $K$ and large such noise.
The polluted domain spans values of $K$ between 0.8 and 16
km\,s$^{-1}$.
}
\label{fig:EGfilt63A}
\end{figure}
\begin{figure}[!ht]
\centerline{
\includegraphics[width=0.45\textwidth]{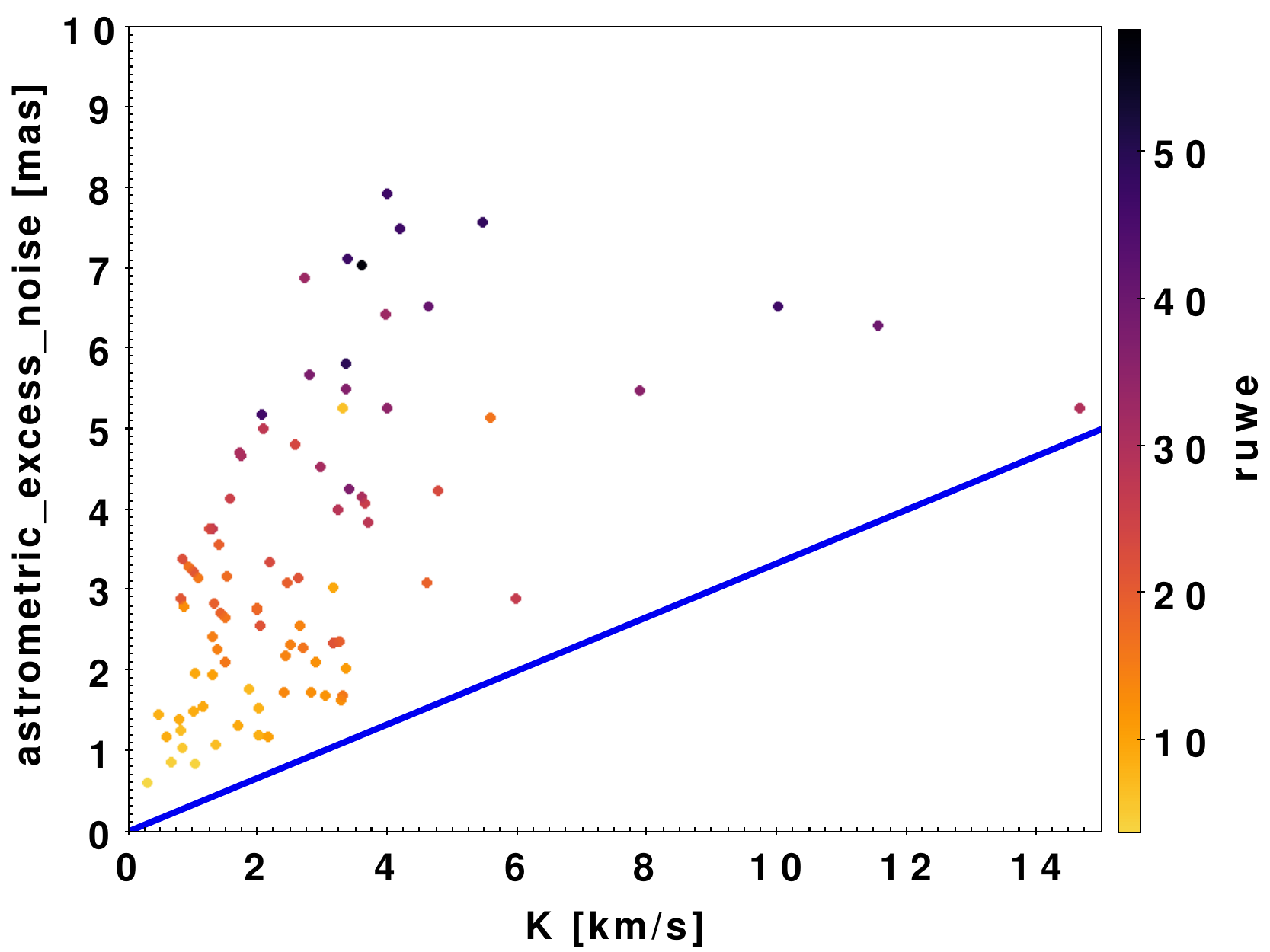}
}\caption{{\tt{astrometric\_excess\_noise}} as a function
of the semi-amplitude, $K$, for the selected subsample. 
The data points are coloured according to the
{\tt{ruwe}}. 
The blue line indicates the separation above which the $K$ is
lower than three times the excess noise.
}
\label{fig:EGfilt63B}
\end{figure}
The RVS being a slitless spectrometer, the knowledge of the exact position of
a star is necessary to have a correct calibration of the wavelength
for its observed spectrum
\citep{2018A&A...616A...6S}. 
The astrometric solution assumes that all the objects
are point-like, and the objects presenting incompatibility with this hypothesis
have potentially wrong RVs.
They are thus associated to large values of the 
{\tt{ruwe}} parameter. This point is also mentioned in
Sect.\,\ref{sec:spectroSB1_input}. 
To these deviating objects is associated an excess noise compared 
to the global astrometric solution 
(labelled {\tt{astrometric\_excess\_noise}}).
One of the main categories of the non-point-like objects are close pairs
(physical or not) unresolved or partially resolved.
A blend of objects that presents some elongated image will induce
the following effect.
The measured position of the star will have a position dependent on how
the elongated axis of the pair enters the astrometric CCD field.
Thus, the resulting RVs are dependent on the orientation of the
main axis of the elongated objects with respect to the scan.
It depends on the scan angle in a rather smooth manner. 
This potentially could generate, from transit to transit, RV variations
that are a simple function of the scan angle. In particular, solutions
for RV-constant objects could be associated to solutions
mimicking a spectroscopic orbit at a period of 62.97\,d;
these orbits are consequently spurious. During the precession cycle,
the scan angles of the various transits are performing a complete
round. The phenomenon manifests itself as an excess of solutions
with small $K$ and thus small mass function
(because the period is fixed). It is visible in
Fig.\,\ref{fig:EGperiodversusfdm} and is further illustrated in
Fig.\,\ref{fig:EGfilt63A}. Clearly, most of the solutions are spurious
and are associated to large values of the astrometric excess noise.
The suspicious $K$ span the approximate interval between 0.8 and 16\,km\,s$^{-1}$.
Another illustration can be found in
\citet{2023A&A...674A..34G}, where it is also shown that the phenomenon
could be related to a particular behaviour of
{\tt{ipd\_gof\_harmonic\_amplitude}}. A detailed investigation of this
behaviour is to be found in
\citet{2023A&A...674A..25H}.

Thus, the epoch RVs are affected by unmodelled astrometric shifts along the
scan direction. In order to alleviate the problem, 
the objects in the class {\tt{SB1}} 
were first selected on the 
basis of an astrometric {\tt{ruwe}} larger than 1.4 and 
a solution leading to a period
within 3\% of the precession period (62.97\,d).
In addition, we selected objects obeying the condition:
\begin{align}
{\tt{astrometric\_excess\_noise}}\,({\mathrm{km\,s}^{-1}}) \, = \,   \nonumber  \\
\,0.146 \, \times \, {\tt{astrometric\_excess\_noise}}\,({\mathrm{mas}})  
> 0.5 \times \epsilon,   
\end{align}
where $\epsilon$ is the mean 
of the uncertainties on the individual epoch RVs
over the full time series. 
Figure \ref{fig:EGfilt63B} illustrates the distribution of the
{\tt{astrometric\_excess\_noise}} as a function of the parameter $K$
for the selected sample.
We can clearly see that all the $K$ values are lower than three times
the {\tt{astrometric\_excess\_noise}} suggesting that these orbital solutions
are spurious and are due to a scan-angle effect. The semi-amplitude is 
too small relatively to the error induced
by the astrometric solution. It has been decided to reject and blacklist
the 87 objects presenting these characteristics.

A similar approach concerning the satellite revolution period
around the Sun (i.e.\ within 10\% of one year) led to the rejection
of another 77 sources. 
\subsubsection{Further remarks}
All in all, 2107 objects were blacklisted
for the above-mentioned three reasons.
The mere sum of the individual rejected objects
amounts to 2119, but 12 objects appear in two lists.
The post-filtering described here above does not concern classes that have been
combined. It could happen that a successful combination 
leads to a removal
of some objects from the purely spectroscopic classes. 
If later the global fit
is not successful, the object will disappear from the catalogue and will not be
re-injected in any purely spectroscopic classes. Thus, some countings could be 
difficult to perform.

Although the post-filtering has been applied with great caution, it also happens
that some problems persist. One of these cases is the following. Some objects
exhibit a good SB1 or SB1C solution with a well established 
significant orbital semi-amplitude for \gaia . 
However, these solutions could sometimes
be doubtful because the same object observed 
in the framework of the
APOGEE DR17 survey \citep{2022ApJS..259...35A} 
or LAMOST survey \citep{2012RAA....12.1197C, 2022MNRAS.517.3888B} 
appears to be constant in RV 
as detailed in Sect.\,\ref{sssec:spectroSB1_results_illust_probl_spurious}. 
This anomaly has not been
fully explained but it could be related in some cases to the presence of
a nearby neighbour (see below). These objects have not been rejected. We encourage the
catalogue user to be very cautious when using the present SB subcatalogue despite
the fact that persisting problems should only represent a minority.
All the remaining and identified problems should be corrected in the future
DR4.
\subsection{An explanation for \gaia\ DR3 3376949338201650112}\label{ssec:spectroSB1_add_cons_3376}
\begin{figure}[!ht]
\centerline{
\includegraphics[width=0.4\textwidth]{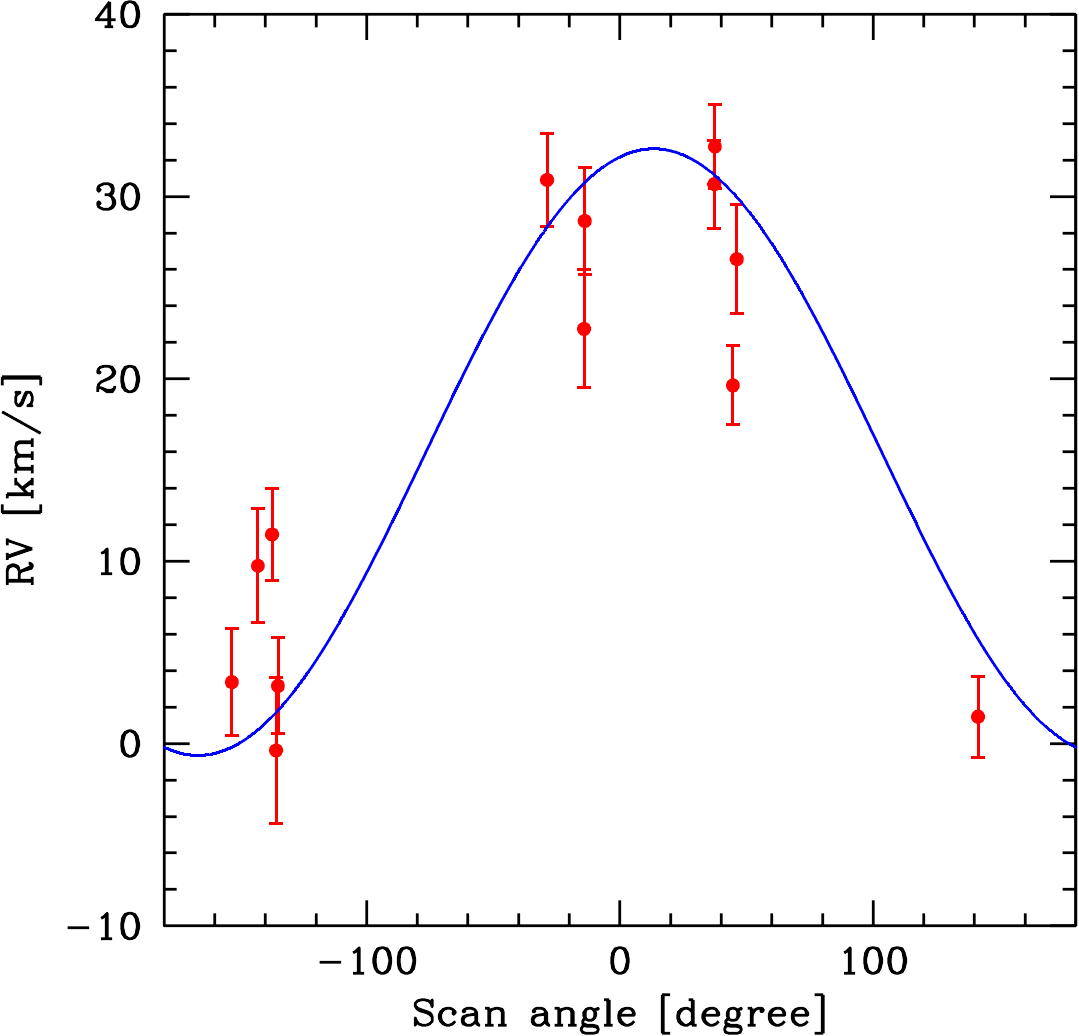}
}\caption{RVs of \gaia\ DR3 3376949338201650112 as a function of the
scan angle at the time of the observations. The blue curve represents the function
given by Eq. 2 in \citet{2023A&A...674A..25H} with the derived $K$ taken from
the catalogue.
}
\label{fig:EGscan3376}
\end{figure}
\begin{figure}[!ht]
\centerline{
\includegraphics[width=0.45\textwidth]{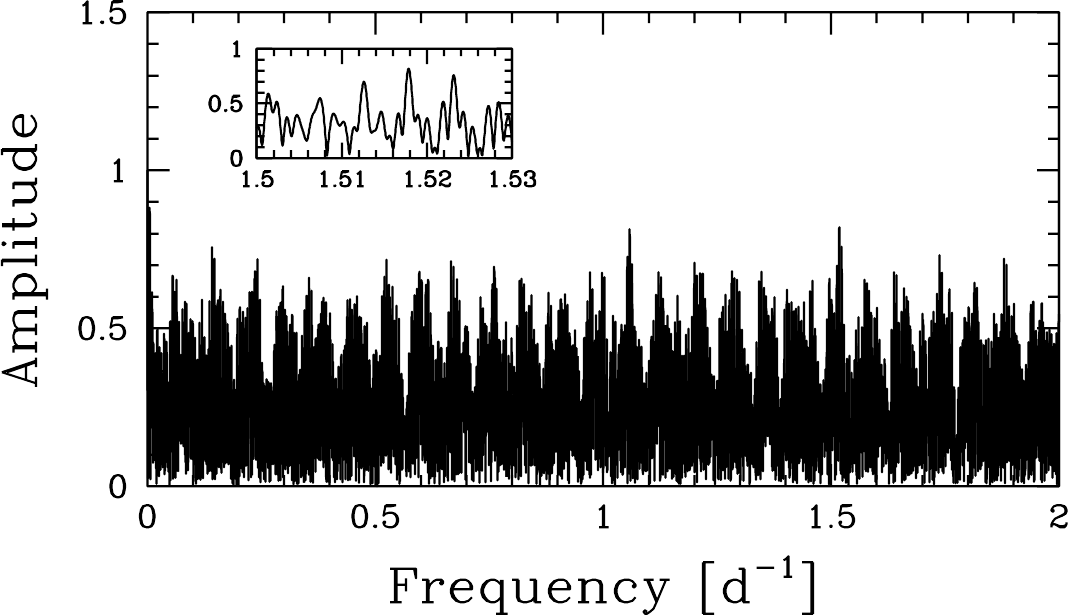}
}\caption{Amplitude spectral window as a function of
frequency as computed from the acquired time series for
\gaia\ DR3 3376949338201650112. The insert is a zoom on the important
second-highest peak, the first being at the origin as usual.
}
\label{fig:EGspwin3376}
\end{figure}
In Sect.\,\ref{sssec:spectroSB1_results_illust_probl_spurious},
we report the existence of SB1 solutions presenting a marked periodic variation,
whereas ground-based data conclude to the constancy of the related object.
Here below, we propose an explanation for this discrepancy by investigating
in details the case of \gaia\ DR3 3376949338201650112.
This example has the advantage of being enlightening and will allow us to correct 
the pipeline for future releases.

The astrometric solution for this object is characterised by a {\tt{ruwe}} of
25.35, pinpointing to a problem with the fit. Referring to the IPD parameters
\citep[image parameter determination; see e.g.][]{2023A&A...674A..25H}, we remark that
{\tt{ipd\_gof\_harmonic\_amplitude}} is equal to 0.164 which is certainly different
from zero. This is a clue that the object is badly treated as a single object.
The value of {\tt{ipd\_frac\_multi\_peak}} turns out to be 84 suggesting that
the object is expected to be double or multiple on the plane of the sky:
most probably a resolved or, more precisely, a partially resolved
pair \citep[according to the definition of][]{2023A&A...674A..25H}.
As explained by these authors, such pairs (or higher order) have a predisposition
to suffer from scan-angle effects that induce a bias on the measured
RVs that is function of the relative orientation of the scanning direction
on the sky at the time of the observations. This is confirmed for
\gaia\ DR3 3376949338201650112 by Fig.\,\ref{fig:EGscan3376} 
where it becomes clear that the measured RVs are function of the 
scan angle. The parameter {\tt{ipd\_gof\_harmonic\_phase}} adopts a value of
103.4\,\degr\ that suggests that the line joining the two stars of the pair
has a position angle on the sky of 13.4\,\degr, after the application of a
90\,\degr\ correction as explained in Sect. 3.1.2 of 
\mbox{\citet{2023A&A...674A..25H}}.
The semi-amplitude $K$ of the fake orbital solution is about 16.6 km\,s$^{-1}$
which translates according to Eq. 2 of \citet{2023A&A...674A..25H} into a 
separation of the pair of 114\,mas. The existence of such a pair is confirmed
by a preliminary run of AGIS (AGIS 4.1) linked to the future DR4 that actually detects
two objects \gaia\ DR4 3376949342501945344 and \gaia\ DR4 3376949342501945472 
that are separated by some 200\,mas and are exhibiting
an orientation of 9\,\degr, thereby supporting our interpretation.

In principle, the bias due to the scan-angle effect is intended to produce
variability with a period either of 62.97\,d or of 1\,y. This property has been
used in Sect.\,\ref{sssec:spectroSB1_add_cons_postfilt_scan angle} to apply 
post-filtering. However, in the present case, the period is 0.657582\,d which does 
not appear to be compatible and could shed some doubt on our interpretation.
The spectral window associated to the DR3 RV time series corresponding to this object
is given in Fig.\,\ref{fig:EGspwin3376}. Besides the main peak at 
$\nu \, = \, 0$\,d$^{-1}$ as usual, it exhibits an important peak at
$\nu \, = \, 1.51784$\,d$^{-1}$ that is exceeding a value of 0.8;
this is an indication that isolated aliasing could be important for this
particular time series. An aliasing with $\delta \nu \, = \, 1.51784$\,d$^{-1}$
explains the ambiguity between a period of 0.657582\,d and that of 1 year.
The combination of the aliasing phenomenon and of the scan-angle effect
explains the existence of such fake SB1 objects in our SB subcatalogue: such objects
were not post-filtered, since the post-filtering was based on the
62.97\,d or 1\,y timescales. We conclude that it would be more appropriate
to include a statistical test inspired by graphs like the one
shown in Fig.\,\ref{fig:EGscan3376} rather than on a criterion
based on the period. However, a true periodicity with these 
timescales will imply a similar behaviour of the RV dependency
on the scan angle even if no scan-angle effect is present.

The explanation given here does not apply to the other object cited
in Sect.\,\ref{sssec:spectroSB1_results_illust_probl_spurious}
(\gaia\ DR3 702159502868472832).
\subsection{The case of astrometric binaries}\label{ssec:spectroSB1_add_cons_astrom}
In Sects.\,\ref{sssec:spectroSB1_add_cons_postfilt_scan angle} and 
\ref{ssec:spectroSB1_add_cons_3376}, we discussed the effect
of astrometric close pairs on the determination of RVs.
A close pair fixed on the sky could often generate
a scan-angle effect and lead to corrupted RV values
but also could generate fake orbital motions.
An additional degree of complexity is reached if we consider
astrometric binaries. Due to their binarity, these close pairs
are not fixed on the sky and the main axis of the pair is rotating
due to the orbital motion.
Therefore, there is interaction between the above-mentioned
scan-angle effect and the astrometric orbital motion itself.
The situation still increases in complexity if the astrometric
orbital motion and the spectroscopic RV changes correspond to
two aspects of the very same orbital motion and thus have the same
period. In this latter case, the correction presents a timescale
equal to the phenomenon we are studying.
The correcting scheme is explained in Sect. 3.3.1 of
\citet{2023A&A...674A..25H} and is also illustrated there with the 
(somewhat extreme) example of \gaia\ DR3 6631710607341412096.
For details, we refer to this paper.

The effect addressed here was not anticipated and was not fully understood
at the time of the final delivery of the NSS code for DR3 processing.
As a consequence, some of our solutions are damaged by this
phenomenon. In addition, those pertaining to the
{\tt{AstroSpectroSB1}} class should be the most affected.
For DR4, a preliminary run of the NSS astrometric channel
has been included in order to compute a correction before
the CU6 processing performing the RV measurements.
\section{The SB subcatalogue}\label{sec:spectroSB1_catalog}
\subsection{Description of the output}
The present first catalogue of pure spectroscopic 
solutions is part of the \gaia\ Data Release 3.
The orbital solutions (in class {\tt{SB1}} or {\tt{SB1C}} as indicated in the
entry {\tt{nss\_solution\_id}}) are part of an auxiliary table named
{\tt{gaiadr3.nss\_two\_body\_orbit}}. We output in this table the classical Campbell
orbital parameters introduced in Sect.\,\ref{ssec:spectroSB1_modelSB1} 
for the {\tt{SB1}} class ($P$, $\gamma$, $K$, $e$, $\omega$, $T_0$) along 
with their estimated uncertainties (1$\sigma$).
The same is done for circular orbit parameters 
($P$, $\gamma$, $K$, $T_0$) introduced in 
Sect.\,\ref{ssec:spectroSB1_modelSB1C} and corresponding to the {\tt{SB1C}} class.
Beyond the orbital parameters themselves, this table also contains some indicators
of the quality of the solution (see Sect.\,\ref{ssec:spectroSB1_quality_indic}). 
The flags associated to the solution
(see Sect.\,\ref{ssec:spectroSB1_processing_flags}) 
are also provided. The boolean mask {\tt{bit\_index}} 
defines the parameters that are actually fitted,
and corresponds for the {\tt{SB1}} class (respectively 
{\tt{SB1C}}) to the value 127 (respectively 31).
In addition, is also provided the upper triangle of the 
correlation matrix of the parameters.
It is expressed as {\tt{corr\_vec}} that contains the 
cross-coefficients folded into a vector
and corresponding to the ordering of the adopted list of parameters. The first term
of the vector $C(0)$ gives the element (0,1) of the matrix, the element $C(1)$
gives the element (0,2) of the matrix and element $C(2)$ the (1,2) one. Elements
$C(3)$, $C(4)$, $C(5)$ correspond respectively to the matrix elements 
(0,3), (1,3), (2,3), and so on. This is thus a column-major storage scheme.
The details of the output are described in 
Table\,\ref{tab:appcatalogsb1} for the class {\tt{SB1}} and
in Table\,\ref{tab:appcatalogsb1c} for the {\tt{SB1C}} one.
The table contains 181327 objects (with solutions) in the {\tt{SB1}} class
and 202 in the {\tt{SB1C}} one. The marked difference in the counting
is due to our decision to consider
eccentric solutions down to as small eccentricities as possible
before going circular. This was described in Sect.\,\ref{ssec:spectroSB1_processing}.

We also released the trend solutions as discussed 
in Sect.\,\ref{ssec:spectroSB1_modelTrendSB1} and 
in Sect.\,\ref{sssec:spectroSB1_processing_trend}.
The objects belonging to the {\tt{TrendSB1}} class are part of the auxiliary table
{\tt{gaiadr3.nss\_non\_linear\_spectro}}.
The subtype of trend solutions (linear or second-order trends) are indicated in 
{\tt{nss\_solution\_id}} (respectively {\tt{FirstDegreeTrendSB1}}
and {\tt{SecondDegreeTrendSB1}}). They are characterised by a
{\tt{bit\_index}} of 7 and 15, respectively.
A few statistics about the quality of the fit
are also given. The table contains 24\,083 objects with solutions
of type {\tt{FirstDegreeTrendSB1}} and 32\,725 objects with
{\tt{SecondDegreeTrendSB1}} ones,
thus totalling 56\,808 objects with solutions
in the class {\tt{TrendSB1}}. The details of the content of the latter table are 
described in Table\,\ref{tab:appcatalogtrendsb1}.
We recall that the StochasticSB1 solutions are not delivered in
DR3.

We should also point out that the auxiliary file
{\tt{gaiadr3.nss\_two\_body\_orbit}} also contains the orbital parameters
of the classes {\tt{AstroSpectroSB1}} and {\tt{EclipsingSpectro}} that result
from the action of the combiner 
(see Sects.\,\ref{ssec:spectroSB1_combination}
and \ref{ssec:spectroSB1_combiner}).
\subsection{Weaknesses, caveats, and recommendations}\label{ssec:spectroSB1_catalog_weak}
The present section gathers the caveats 
dispersed all over the paper
and delivers a summary of the characteristics of the catalogue
that could be beneficial to the general user.
The RVs studied here have the quality and shortcomings 
of the measurements of individual (epoch)
RVs for constant stars. 
The properties and precision of these RVs 
for SB1 studies should be of the
same order. Therefore, we refer the reader to the work of
\citet{2023A&A...674A...5K} concerning the characteristics 
and condition of utilisation of these data. 
Seabroke et al. (in preparation) is also a work to be consulted.

The information extracted from variations of the RVs could
be delicate to use. Firstly, it must be clear that the way the measurements are
performed (cross-correlation with a template) makes the implicit assumption that 
the line is supposed to vary without changing its shape.
This is not necessarily true. Two textbook cases at least come to mind.
Some stars, notably the intrinsically variable ones, could exhibit
dominant variations of their line shapes. In such a case, the variations do not
represent a Doppler shift of the object as a whole and thus are not representative
of an SB1 (see Sect.\,\ref{sssec:spectroSB1_add_cons_postfilt_intvar}). 
In principle, the fact we are fitting models where the harmonic content
is typical of a Keplerian motion is certainly a protection against
variable stars appearing in the SB1 class. However, some variable stars, like
Cepheids are well-known to display RV changes that closely mimic the
Keplerian motion \citep[e.g.][]{1989A&AS...81..339I}.
As explained in Sect.\,\ref{sssec:spectroSB1_add_cons_postfilt_intvar},
efforts have been made to post-filter the Cepheid and the RR Lyrae stars,
but this post-filtering is certainly not complete for the time being.
In addition, no such step was made for other intrinsic variables in DR3
(notably Long-Period Variables).

Another tricky case is that of true SB2 stars, where the two lines never
fully deblend. They are appearing most probably in the SB1 channel.
The RV variations are then due to the corresponding change
of the full profile made by the blended lines. This has little effect on
the period determination but the impact on the eccentricity and mainly
on the semi-amplitude might be huge 
(Sect.\,\ref{sssec:spectroSB1_results_illust_probl_fake}).
The related solutions are characterised by a too small value
of $K$ and of the mass function. They could be considered as suspicious
but we cannot have any certitude that all of these types of solutions
are necessarily wrong.   

The algorithm used is intended to fit orbital solutions to the RV curves
that it analyses. It is doing so but is not protected against some anomalies. 
The RV data set suffers from all the problems not attenuated by the CU6 processing
and is not particularly protected against outliers, miss-classified 
double-line objects, intrinsic variables, emission-line objects
and other problematic situations (e.g. badly normalised continuum, etc...).
In addition, it is not (for the moment) able to properly deal with
higher degree systems. It is possible that the code detects a binary motion, but 
this one could be quantitatively corrupted by the presence of additional components
(see Sect.\,\ref{sssec:spectroSB1_results_illust_probl_multi}).
In the future (DR4), we hope to include a cleaning (whitening) of the dominant
signal followed by a new search for a second period or a trend. However, the
presence of two signals renders the statistical analysis much more complicated.

Instrumental effects are also present.  The scan-angle effect notably
induces periodic RV variations for constant stars that do not fulfil
the assumptions used in astrometry (objects are assumed to be isolated and thus single).
Close astrometric pairs could thus mimic SB1 RV variations 
\citep[see the work by][]{2023A&A...674A..25H}, 
and are polluting the present sample (even if some post-filtering took place;
see also Sect.\,\ref{sssec:spectroSB1_add_cons_postfilt_scan angle}).
If, in addition, the close pair is rotating on the plane of the sky as is the case
for astrometric binaries, the measured RV could be corrupted
(see Sect.\,\ref{ssec:spectroSB1_add_cons_astrom}).
This problem will be corrected for DR4.

In addition, it should be noted that all decisions 
are taken in a statistical way; this means
that the output is statistically correct but that some incorrect 
individual fits can survive
near the borders where the decisions are taken (e.g. between an orbital fit and
a trend solution for example). 

We recommend to the user of the catalogue to keep in 
mind the weaknesses and caveats
described here above. We also strongly recommend, when possible, the use of 
the various quality factors accompanying each solution.
If the catalogue is expected to be well-behaved from the statistical
point of view, the situation could be more intricate for individual objects
or particular set of objects.
It remains that biases in the catalogue could be present
from the point of view of stellar population studies. Such an analysis
is beyond the scope of the present paper. For the time being, the genuine
selection function of the catalogue is poorly known and any user should
be aware of this and behave accordingly.
The user is strongly encouraged to adapt the selection of solutions to their particular
objective instead of using it blindly. 
They should combine the various delivered quality factors
according to their aim. We advise a cautious approach. 
An example of interesting complex 
combinations is detailed in \citet{2022MNRAS.517.3888B}.
To some extent, other information can be considered as quality factors
(such as flags but, mainly, the number of points in the time series
$N_{\mathrm{good}}$), which could also be included in the selection process.

The global weak point of the algorithms 
could be the determination of the
period even if here above it is shown that the success rate is at 
the very least 80\% and is more often located in the range
85-95\%.
There is evidently at a very low number of transits, 
a difficulty to correctly determine 
the period that forced us to neglect data with $N_{\mathrm{good}}$ less than 10.
However, at slightly larger $N_{\mathrm{good}}$, the situation is not so simple.
From the simulations described in Sect.\,\ref{ssec:spectroSB1_validation_simul},
we conclude that a value of $N_{\mathrm{good}}$ smaller than 17 is somewhat
problematic (although it depends on the signal-to-noise ratio) for the period recovery.
We also show, from the validation (see Sect.\,\ref{ssec:spectroSB1_validation_intern}),
that an overfitting phenomenon characterises 
part of the solutions associated to an
$N_{\mathrm{good}}$ in the range 10-15. 
We should further add that, at small
$N_{\mathrm{good}}$, the periodogram has a larger propensity to produce
isolated aliasing peaks (generating ambiguity due to aliasing).
While interpreting the simulations, it appears that 
it is unclear if the determining
parameter is the $N_{\mathrm{good}}$ or the significance. 
According to these simulations,
significance lower than 10-15 can be associated with 
difficulties to have the 
proper recovery rate. This seems to be confirmed by the validation in
Sect.\,\ref{ssec:spectroSB1_validation_otherset}, 
in particular in the corner plot
of Fig.\,\ref{fig:cu4nss_spectro_comparison_ASASSN_SB1_corner_plot}.
It is also clear, from Fig.\,\ref{fig:EGperiodversuseccent}, 
that the circularisation
phenomenon is not outstandingly visible for significance 
much lower than 30. Therefore,
we consider that depending on the studies performed, the significance
should be cut between 15 and 25, or more conservatively at 30. 
If 10 is still affordable in some cases, the
solutions characterised by a significance between 5 (our cut-off) 
and 10 are most probably hazardous to accept.

The spectroscopic solutions could be merged with 
photometric or astrometric ones, and
in these cases the resulting output is certainly very secure for the period. 
If the user of the
catalogue only considers the sole results of the spectroscopy fits, it is advisable to
restrict the sample to some kind of gold sample. This can be done by performing 
additional selections in the catalogue: 
for example, a larger number of transits, a higher
significance, periods in a dedicated range (between ten days and
$\Delta T$, or allowing it to go down to
smaller periods but at the expense of a selection on the basis
of a larger significance), periods
with small uncertainties, solutions with negligible phase gaps
(although the gap values are not provided in the DR3 catalogue), and so on.
In Appendix\,\ref{sec:appJ}, we present indicative 
but simple examples of extraction of a gold sample from the catalogue.
\section{Conclusions}\label{sec:spectroSB1_conclusion}
Within the framework of the \gaia\ Data Release 3, we present the first version of 
the \gaia\ catalogue of spectroscopic orbital 
solutions (SB subcatalogue). We present 181\,327 objects
that have been found to be characterised by class {\tt{SB1}} and 202 objects by class
{\tt{SB1C}}. We also present objects exhibiting trends of variations of their
RVs over time.

The present catalogue constitutes the largest collection of objects with 
a spectroscopic orbital solution by far, based on a comparison
with the notorious SB9 compilation\footnote{\url{https://sb9.astro.ulb.ac.be/}}.
The improvement in terms of number of objects is more 
than one order (nearly two) of magnitude.
The present SB subcatalogue has also the major advantage 
of being homogeneous from the point of view
of both data acquisition (properties) and data processing, 
surpassing the characteristics
of existing compilation. The only exception is the typical number of measurements
that constitutes the individual time series, which remains limited
(at least for the present DR3 catalogue). 
The situation will undoubtedly improve for future data releases.

After a description of the data entering the NSS-SB1 processing, we describe in detail
the logic and structure of the related code. 
We reveal the general philosophy behind, 
along with the mathematical formalism used. 
We also describe, following the main processing details,
the filtering of the results performed to 
discard bad solutions and improve the catalogue.
We give a number of examples illustrating the quality of the output,  
as well as a few examples of
detected problematic cases. A full validation of the output 
is performed both on the basis
of internal consistency and on the basis of external 
existing catalogues from the literature.
We further analyse the characteristics of the catalogue and also 
describe the way the SB subcatalogue has been included 
in the \gaia\ Data Release 3.
We additionally scrutinises 
the weaknesses of the approach and provide the necessary caveats 
and recommendations for scientific applications of the catalogue.
\section*{Data availability}
The epoch RVs constituting the time series analysed here have not been
published in the context of the \gaia\ Data Release 3.
The orbital solutions and trends we derived are available since 
13 June 2022, in the ESA \gaia\ DR3 archive accessible at
{\url{https://gea.esac.esa.int/archive}}. The same results can also
be found at CDS Strasbourg via the link
{\url{https://cdsarc.cds.unistra.fr/viz-bin/Cat?I/357}}.
Appendices D to J  are only available on Zenodo
(see \url{https://zenodo.org/records/13990211}).
\begin{acknowledgements}
The authors are particularly thankful to Alain Detal who maintained over 
the years the Li\`ege computer park with great efficiency. 
EG is also thanking F.\,M\'elen for her consequent help
in the manipulation of various computer codes. 
The authors are acknowledging the numerous scientific discussions 
that took place with Christos Siopis.
The participation in this endeavour has been
made possible through an Impulsion Credit 
from the University of Li\`ege to EG.
This work presents results from the European Space Agency (ESA) space 
mission \gaia. \gaia\ data are being processed by the \gaia\ Data 
Processing and Analysis Consortium (DPAC). 
Funding for the DPAC is provided by national 
institutions, in particular the 
institutions participating in the \gaia\ MultiLateral Agreement (MLA). 
The \gaia\ mission website is \url{https://www.cosmos.esa.int/gaia}. 
The \gaia\ archive website is \url{https://archives.esac.esa.int/gaia}.
The \gaia\ mission and data processing 
have, in Li\`ege, been continuously supported 
by the Belgian Politique Scientifique F\'ed\'erale (BELSPO) 
through PRODEX contracts 
{\emph{Gaia DPAC: Binaires, \'etoiles extr\^emes et objets du syst\`eme
solaire}} (convention PEA 4200090289; years 2007-2013), 
{\emph{Gaia DPAC Early Mission Belgian consolidation (Gaia-be)}}
(convention PEA 4000110152: years 2014-1016),
{\emph{Gaia DPAC Early Mission Belgian consolidation (Gaia-be2)}}
(convention PEA 4000119840: years 2017-1019; years 2020-2022).
TM acknowledges financial further support 
from BELSPO for contract PLATO mission development.
We acknowledge support from an ARC grant 
(13/17-06; {\emph{Massive Stars: drivers of the evolution
of the Universe}}) for concerted research actions
financed by the French community of 
Belgium (Wallonia-Brussels Federation).
We are also largely indebted to the Fonds de 
la Recherche Scientifique (F.R.S.-FNRS, Belgium)
for various supports. 
YD acknowledges support to the Algerian Centre de 
Recherche en Astronomie, Astrophysique et
G\'eophysique of Bouzareah Observatory.
GS acknowledges support from BELSPO/PRODEX under various contracts
and in particular convention PEA 40000119826.
AS acknowledges support from the Italian Space Agency (ASI) 
under contract 2018-24-HH.0 
"The Italian participation to the Gaia Data Processing and 
Analysis Consortium (DPAC)" in collaboration with 
the Italian National Institute of Astrophysics.
\end{acknowledgements}

\bibliographystyle{aa}
\bibliography{papercu4}
\begin{appendix}
\onecolumn
\section{Historical developments}\label{sec:appA}
Once the period has been evaluated, the general model 
to fit (Eq.\,1) remains non-linear in some of the
parameters (mainly the eccentricity, $e$). This implies that
the $\chi^2$ of the fit could present a few or even numerous
minima in the parameter space, and to find the global one is a true challenge.
In particular, the derivation of the eccentricity is a notorious
pitfall. The search for the global minimum might be performed 
empirically by guessing its approximate position and then by descending
the $\chi^2$ valley by classical numerical methods (such as the 
Levenberg-Marquardt algorithm). However, in a pipeline as the one
developed for \gaia, identifying the location of the global minimum
is uncertain and a specific type of search is necessary.
This is particularly frustrating because a well-trained expert eye
is able to estimate a rough value for the eccentricity
with great efficiency. In this context, we explored the possibilities
to adopt alternative approaches. We acted along two main axes.
On the one hand, 
we tried to invent a method based on pattern recognition
in the phase diagram of the RVs folded with the candidate periods.
Such an approach would allow us to take a first guess for the eccentricity,
which could then be used for further classical derivation.
In order to apply pattern recognition techniques, it is mandatory to
define an interpolation function in the phase diagram to obtain
a robust description of the shape of the RV curve. Several tools have been
explored such that the nearest-neighbour interpolation, 
the linear interpolation, the interpolation by cosines, 
the local Akima periodic interpolation, the periodic interpolation by
cubic splines, and the periodic interpolation by Catmull-Rom splines.
We also explored the possibilities of performing classical adjustment by
least-square methods. In particular, we considered classical polynomials,
Fourier series, B\'ezier curves (using De Casteljan algorithm) and finally,
cubic B-splines in their open uniform version.
Once the interpolation curve is chosen and the shape is defined, we tested
various pattern recognition methods such as algorithms for regression
based on the nearest neighbourhood, algorithms for regression based
on trees and in particular on extremely randomised trees and, finally,
algorithms for regression based on neural networks.
Full details of this exploration and detailed references
are given in the work of \citet{delcha}.

On the other hand,
we developed an approach based on genetic algorithms.
These algorithms have the advantage to circumvent any preliminary guess of the
solution. This is at the expense of being obliged to explore (even cleverly) the
space of the parameters using the generation of random numbers
(Monte Carlo methods).
This work is described in details in
\citet{delcha}.
The use of genetic algorithms is a trade-off between the probability to
effectively detect the global $\chi^2$ minimum and the consumption
of computer time. 
The codes using genetic algorithms were written, delivered and tested.
The detection of the global minimum was possible but was implying
a prohibitively large amount of computer time despite 
strong efforts made to accelerate it.
The large amount of objects to treat in 
the \gaia\ processing suggests using much more efficient algorithms.
The genetic algorithm code was not activated for the operational runs.

Other methods are now available to derive orbital solutions.
They became widespread in the domain of exoplanets,
where the orbits are particularly incompletely covered and
complex. The majority of these methods are based on Monte Carlo simulations.
Very often, they use the Markov chain Monte Carlo
(MCMC) method \citep[see e.g.][]{2017ARA&A..55..213S}
often utilised in conjunction with Bayesian approaches.
These algorithms necessitate a dense coverage of the full space of priors,
at the expense of the computer time. 
Comparing the works of \citet{2017ApJ...837...20P} 
and \citet{price_whelan20} 
pinpoints the strong impact that could have the choice of 
priors on the resulting posteriors.
The Bayesian methods have admittedly the big advantage
to potentially provide marginal posterior probabilities for a series of models
and thus estimate how much the observational data could favour a particular model.
According to \citet{2007ASPC..371..189F}, the challenge 
of estimating marginal posterior
probabilities is still a crucial problem and obvious estimators are 
suffering from poor convergence properties.
In addition, for the preliminary DR3 catalogue, the directives were that we should
output one and only one model per treated object, thus mitigating the interest
of using Bayesian methods.
In addition, these Monte Carlo methods are computer time-consuming
and were not acceptable for the DR3 processing. Finally, we adopted 
a method based on fully analytical and classical numerical
developments.
\section{Practical details on the processing}
\label{sec:appB}
The Data Processing Centre at CNES 
(DPCC\footnote{\url{https://www.cosmos.esa.int/web/gaia/dpac}}), located at the Centre 
National d'Etudes Spatiales in Toulouse (France), ran the 
spectroscopic processing chain (CU6) and the object processing chain (CU4).
Globally, it ran the whole set of NSS codes and, in particular, the spectroscopic
orbital solution chain described in the present paper. Our NSS-SB1 
processing chain delivers the results to the combiner.
The Main Data Base (MDB) is fed with solutions from this combiner. 
If the combiner finds no other
solutions to combine with, the spectroscopic results are directly transmitted
to the MDB. 
If a combination is possible, only the output
of the combination is saved. 

In the framework of the \gaia\ endeavour, the DPCC is hosting a cluster
of 250 computers (6000 cores, 40 TBytes of RAM and 7 PBytes of HDFS disk storage).
The resources are managed by an Hadoop system. The codes are
integrated into a framework SAGA (System of Accomodation of \gaia\ Algorithms)
developed by Thales. The spectroscopic processing was executed over 2500 cores and
occupied 17 TBytes of RAM. Some 3$\times$10$^6$ hours of CPU were necessary, 
equivalent in real time to 120 days. Concerning the NSS chain, the full treatment
(Ingestion, Processing, and Post-Processing) was executed 
on the Hadoop cluster over 1400 cores and 
necessitated 16\,105 days of CPU corresponding to an effective duration of 17 days.
All the codes were written in Java 8.0. The version of the code tagged and executed for
the operational run was 20.10.
\clearpage
\section{Separable variable technique applied to the Keplerian model fit}
\label{sec:appC}
\subsection{A reminder of the Levenberg-Marquardt method}
Eq.\,\ref{RVARRAY1} links the $N$ dependent variables $RV(t_i)$ 
(the RVs of the model) to the independent ones $t_i$
(the observing times). Hereafter, we represent these two variables by 
the column arrays $\mathbf{RV}$ and $\mathbf{t}$ respectively. 
$f$ represents the Keplerian function, such that
\begin{equation}
\mathbf{RV} = f\left(\mathbf{t},\mathbf{p}\right),
\end{equation}
with $\mathbf{p}$ being the array of 6 parameters $\left[\gamma,\alpha,\beta,\nu_k,e,T_{\mathrm{0}}\right]^T$
where the superscript $T$ denotes the transpose operation.
The Newton-Raphson algorithm is based on the Taylor first order development 
\begin{equation}
\mathbf{RV} = f\left(\mathbf{t},\mathbf{p}^{\{n\}}\right) + \mathbf{J}^{\{n\}} ~ \mathbf{\Delta p}^{\{n\}}
\label{TAYLOR:appA}
,\end{equation}
where $\mathbf{\Delta p}^{\{n\}} = \mathbf{p}^{\{n + 1\}} - \mathbf{p}^{\{n\}}$, and $\{n\}$ denotes the iteration number. 
$\mathbf{J}$ is the Jacobian matrix which elements are 
\begin{equation}
J^{\{n\}}_{i,j} = \frac{\partial f\left(t_i, p_j^{\{n\}}\right)}{\partial p_j^{\{n\}}} \, \, \, \, 
\mathrm{for} \, \, \,   1 \leq i \leq N, \, 1 \leq j \leq 6 \,\,\, .
\end{equation}
The objective function to be minimised is the quadratic sum of residuals as 
\begin{equation}
\chi^2 = \left[\mathbf{RV_{\mathrm{obs}}} - f\left(\mathbf{t},\mathbf{p}\right)\right]^T \mathbf{W} \left[\mathbf{RV_{\mathrm{obs}}} 
- f\left(\mathbf{t},\mathbf{p}\right)\right]
\label{CHI2:appA}
,\end{equation}
with $\mathbf{W}$ being the inverse of the data variance-covariance matrix. The $\chi^2$ is given at the $n^{\mathrm{th}}$ iteration by 
\begin{equation}
\chi^{2,\{n\}} = \left[\mathbf{RV}_{\mathrm{obs}} - f\left(\mathbf{t},\mathbf{p}^{\{n\}}\right) - \mathbf{J}^{\{n\}} ~ \mathbf{\Delta p}^{\{n\}}\right]^T 
\mathbf{W} \left[\mathbf{RV_{\mathrm{obs}}} - f\left(\mathbf{t},\mathbf{p}^{\{n\}}\right) - \mathbf{J}^{\{n\}} ~ \mathbf{\Delta p}^{\{n\}}\right] \, \, \, \,, 
\label{CHI2ITERATION:appA}
\end{equation}
where $\mathbf{RV}_{\mathrm{obs}}$ contains the observed RVs.
The optimal parameter step $\mathbf{\Delta p}^{\{n\}}$ is obtained by solving 
\begin{equation}
\frac{\partial \chi^2}{\partial\mathbf{\Delta p}^{\{n\}}} = 0 \,\,\, .
\end{equation}  
In the case of Newton-Raphson, this leads to the following algebraic system:
\begin{equation}
\left(\mathbf{J}^{\{n\}T} ~ \mathbf{W} ~ \mathbf{J}^{\{n\}} \right) \mathbf{\Delta p}^{\{n\}} 
= \mathbf{J}^{\{n\}T} \mathbf{W} \left[\mathbf{RV}_{\mathrm{obs}} - f\left(\mathbf{t},\mathbf{p}^{\{n\}}\right)\right]. \, \, \, \, 
\label{NEWTON:appA}
\end{equation}
The matrix $\mathbf{H}^{\{n\}} = \mathbf{J}^{\{n\}T} ~ \mathbf{W} ~ \mathbf{J}^{\{n\}}$ is a 
positive semi-definite matrix called the hessian matrix (it is not a positive definite matrix because it might be singular). 

The Levenberg-Marquardt algorithm consists in rescaling the diagonal elements 
of $\mathbf{H}^{\{n\}}$ by the damping factor $ 1 + \lambda$. 
The system (Eq.\,\ref{NEWTON:appA}) becomes 
\begin{equation}
\left[\mathbf{H} + \lambda ~ \mathsf{diag}\left(\mathbf{H}\right)\right]^{\{n\}} \mathbf{\Delta p}^{\{n\}} 
= \mathbf{J}^{\{n\}T} \mathbf{W} \left[\mathbf{RV_{\mathrm{obs}}} - f\left(\mathbf{t},\mathbf{p}^{\{n\}}\right)\right]; \, \, \, \, 
\label{LEVENBERG:appA}
\end{equation}
where $\lambda \, \, $ is the damping factor, chosen to be large when $\mathbf{H}$ approaches the singularity.

\subsection{Separable variable technique}
The parameters $\mathbf{p_l} = \left[\gamma, \alpha, \beta\right]^T$ are linear because they satisfy the following property:
\begin{equation}
f\left(\mathbf{t},\mathbf{p}\right) = \sum_{k = 1}^{3}{\frac{\partial f\left(\mathbf{t},\mathbf{p}\right)}
{\partial \mathbf{p_l}(k)} ~ \mathbf{p_l}(k)} = \mathbf{J_l} \, \mathbf{p_l} \, \, \, \,.
\label{LINEAR:appA}
\end{equation}
Hereafter, subscripts $\mathbf{l}$ and $\mathbf{nl}$ refer to linear and non-linear respectively. Then, 
$\mathbf{J_l}$ is the Jacobian matrix whose columns are the derivatives of $f$ with respect to $\mathbf{p_l}$.

While in both Newton-Raphson and the (commonly used) Levenberg-Marquardt (LM) algorithms, the linear and non-linear 
parameters are treated the same way, the separable variable technique has the advantage 
that $\mathbf{p_l}$ can be obtained for every non-linear parameter 
$\mathbf{p_{nl}}^{\{n\}} = \left[\nu_k, e, T_{\mathrm{0}}\right]^{\{n\} T}$ 
by a linear least-square regression, such that 
\footnote{The inversion of the normal equations is known to be numerically unstable 
and should be avoided in favour of orthogonal decomposition approaches.}
\begin{equation}
\mathbf{H_l}^{\{n\}} \, \mathbf{p_l}^{\{n\}} = \mathbf{Z_l}^{\{n\}}
\label{LINEARPARAMS1:appA}
,\end{equation}
with 
\begin{eqnarray}
\mathbf{H_l}^{\{n\}} &=&  \mathbf{J_l}^{\{n\}T} ~ \mathbf{W} ~ \mathbf{J_l}^{\{n\}} \label{LINEARPARAMS2:appA} ,\\
\mathbf{Z_l}^{\{n\}} &=&  \mathbf{J_l}^{\{n\}T} \, \mathbf{W} \, \mathbf{RV}_{\mathrm{obs}} \, \, \, \, . \label{LINEARPARAMS3:appA}
\end{eqnarray}

Therefore, Eq.\,\ref{TAYLOR:appA} becomes a function of $\mathbf{p_{nl}}^{\{n\}}$ only, such that 
\begin{equation}
\mathbf{RV} = f\left(\mathbf{t},\mathbf{p}^{\{n\}}\right) + \mathbf{\tilde{J}_{nl}}^{\{n\}} ~ \mathbf{\Delta p_{nl}}^{\{n\}}
\label{TAYLOR2:appA}
,\end{equation} 
where the step of the non-linear parameters 
is $\mathbf{\Delta p_{nl}}^{\{n\}} = \mathbf{p_{nl}}^{\{n + 1\}} - \mathbf{p_{nl}}^{\{n\}}$. \\
The Jacobian matrix $\mathbf{\tilde{J}_{nl}}^{\{n\}}$ encompasses both the derivatives of $f$ 
with respect to $\mathbf{p_{nl}}^{\{n\}}$ and $\mathbf{p_{l}}^{\{n\}}$ 
\begin{eqnarray}
\mathbf{\tilde{J}_{nl}}^{\{n\}} &=& \frac{\partial f\left(\mathbf{t}, \mathbf{p}^{\{n\}}\right)}
{\partial \mathbf{p_{nl}}^{\{n\}}} + \frac{\partial f\left(\mathbf{t}, \mathbf{p}^{\{n\}}\right)}
{\partial \mathbf{p_{l}}^{\{n\}}} \frac{\partial \mathbf{p_{l}}^{\{n\}}}{\partial \mathbf{p_{nl}}^{\{n\}}} \\
 & = & \mathbf{J_{nl}}^{\{n\}} + \mathbf{J_{l}}^{\{n\}} \mathbf{A}^{\{n\}} \, \, \, \, .
\label{JACOBIAN:appA}
\end{eqnarray} 
$\mathbf{A}^{\{n\}} = \frac{\partial \mathbf{p_{l}}^{\{n\}}}{\partial \mathbf{p_{nl}}^{\{n\}}}$ 
is a 3$\times$3 matrix containing the derivatives of the linear parameters $\mathbf{p_{l}}^{\{n\}}$ 
with respect to the non-linear ones $\mathbf{p_{nl}}^{\{n\}}$. 
It is obtained by computing the derivative of Eq.\,\ref{LINEARPARAMS1:appA} with respect to $\mathbf{p_{nl}}^{\{n\}}$ 
\begin{equation}
\mathbf{A}^{\{n\}} = \left[\mathbf{H_l}^{\{n\}} \right]^{-1} 
\left[\frac{\partial \mathbf{Z_l}^{\{n\}}}{\partial \mathbf{p_{nl}}^{\{n\}}} 
- \frac{\partial \mathbf{H_l}^{\{n\}}}{\partial \mathbf{p_{nl}}^{\{n\}}} \mathbf{p_l}^{\{n\}}\right]  \, \, \, \, .             
\label{MATRIXPASSAGE:appA}
\end{equation}
Using Eq.\,\ref{LINEARPARAMS2:appA} and Eq.\,\ref{LINEARPARAMS3:appA}, we deduce the derivatives as 
\begin{eqnarray}
\frac{\partial \mathbf{H_l}^{\{n\}}}{\partial \mathbf{p_{nl}}^{\{n\}}} &=&  
\frac{\partial \mathbf{J_l}^{\{n\} T}}{\partial \mathbf{p_{nl}}^{\{n\}}} ~ \mathbf{W} ~ \mathbf{J_l}^{\{n\}} 
+  \mathbf{J_l}^{\{n\}T} ~ \mathbf{W} ~ \frac{\partial \mathbf{J_l}^{\{n\}}}{\partial \mathbf{p_{nl}}^{\{n\}}} 
\label{LINEARPARAMS4:appA} \\
\frac{\partial \mathbf{Z_l}^{\{n\}}}{\partial \mathbf{p_{nl}}^{\{n\}}} &=&  \frac{\partial \mathbf{J_l}^{\{n\} T}}
{\partial \mathbf{p_{nl}}^{\{n\}}} ~ \mathbf{W} ~ \mathbf{RV}_{\mathrm{obs}} \, \, \, \, . \label{LINEARPARAMS5:appA}
\end{eqnarray}
Therefore, the system (Eq.\,\ref{LEVENBERG:appA}) becomes 
\begin{equation}
\left[\mathbf{\tilde{H}_{nl}} + \lambda ~ \mathsf{diag}\left(\mathbf{\tilde{H}_{nl}}\right)\right]^{\{n\}} 
\mathbf{\Delta p_{nl}}^{\{n\}} = \mathbf{\tilde{J}_{nl}}^{\{n\}T} \mathbf{W} \left[\mathbf{RV}_{\mathrm{obs}} 
- f\left(\mathbf{t},\mathbf{p}^{\{n\}}\right)\right]
\label{NEWLEVENBERG:appA}
\end{equation}
with $\mathbf{\tilde{H}_{nl}} = \mathbf{\tilde{J}_{nl}}^{\{n\}T} \mathbf{W} \, \mathbf{\tilde{J}_{nl}}^{\{n\}}$ 
being the new hessian matrix of the system.
We have then split the 6$\times$6 system (Eq.\,\ref{LEVENBERG:appA}) into two smaller systems, 
each one of 3$\times$3, in Eq.\,\ref{LINEARPARAMS1:appA} and Eq.\,\ref{NEWLEVENBERG:appA}, respectively.

\subsection{Analytical expressions of the involved matrices}
To perform the above computations, we need the expressions of the Jacobian matrices 
$\mathbf{J_l}^{\{n\}}$ and $\mathbf{J_{nl}}^{\{n\}}$.
The former, relative to the linear system of Eq.\,\ref{LINEAR:appA} is given by 
\begin{equation}
\mathbf{J_l}^{\{n\}} = \left[1, \cos(\mathbf{v}) + e, \sin(\mathbf{v})\right]^{\{n\}}  \, \, \, \,.
\label{JL:appA}
\end{equation}
Its derivatives with respect to the non-linear parameters are
\begin{eqnarray}
\frac{\partial \mathbf{J_l}^{\{n\}}}{\partial \nu_k} &=& 
\left[0, -\sin(\mathbf{v}), \cos(\mathbf{v})\right]^{\{n\}} \, \frac{\partial \mathbf{v}^{\{n\}}}{\partial \nu_k},\\
\frac{\partial \mathbf{J_l}^{\{n\}}}{\partial e} &=& 
\left[0, 1-\sin(\mathbf{v})\,\frac{\partial \mathbf{v}}{\partial e}, \cos(\mathbf{v})\,\frac{\partial \mathbf{v}}{\partial e}\right]^{\{n\}} ,\\
\frac{\partial \mathbf{J_l}^{\{n\}}}{\partial T_{\mathrm{0}}} &=& 
\left[0, -\sin(\mathbf{v}), \cos(\mathbf{v})\right]^{\{n\}} \, \frac{\partial \mathbf{v}^{\{n\}}}{\partial T_{\mathrm{0}}}  \, \, \, \, .
\label{JLDERIVATIVE:appA}
\end{eqnarray} 
Very similarly, the Jacobian matrix $\mathbf{J_{nl}}^{\{n\}}$ is given by 
\begin{eqnarray}
\frac{\partial \mathbf{J_{nl}}^{\{n\}}}{\partial \nu_k} &=& 
\left[0, -\alpha \, \sin(\mathbf{v}), \beta \, \cos(\mathbf{v})\right]^{\{n\}} \, \frac{\partial \mathbf{v}^{\{n\}}}{\partial \nu_k} 
\label{JNLDERIVATIVEJNLNU:appA},\\
\frac{\partial \mathbf{J_{nl}}^{\{n\}}}{\partial e} &=& 
\left[0, \alpha \left(1-\sin(\mathbf{v})\,\frac{\partial \mathbf{v}}
{\partial e}\right), \beta \, \cos(\mathbf{v})\,\frac{\partial \mathbf{v}}{\partial e}\right]^{\{n\}}
\label{JNLDERIVATIVEJNLE:appA} ,\\
\frac{\partial \mathbf{J_{nl}}^{\{n\}}}{\partial T_{\mathrm{0}}} &=& 
\left[0, -\alpha \, \sin(\mathbf{v}), \beta \, \cos(\mathbf{v})\right]^{\{n\}} \, 
\frac{\partial \mathbf{v}^{\{n\}}}{\partial T_{\mathrm{0}}} \, \, \, \, .
\label{JNLDERIVATIVEJNLT0:appA}
\end{eqnarray}
The derivatives of the true anomalies $\mathbf{v}$ with respect to the three non-linear parameters are 
\begin{eqnarray}
\frac{\partial \mathbf{v}}{\partial \nu_k} &=& 
2 \, \pi \, (\mathbf{t} \, - \, T_{\mathrm{0}}) \, (1 \, + e \, \cos(\mathbf{v}))^2 \,  / \, (1 \, - \,e^2)^{3/2} 
\label{VDERIVATIVENU:appA} ,\\
\frac{\partial \mathbf{v}}{\partial e} &=& 
\sin(\mathbf{v}) \, (2 \, + e \, \cos(\mathbf{v})) \, / \, (1 \, - \,e^2) \label{VDERIVATIVEE:appA} ,\\
\frac{\partial \mathbf{v}}{\partial T_{\mathrm{0}}} &=& 
-2 \, \pi \, \nu_k \, (1 \, + e \, \cos(\mathbf{v}))^2 \,  / \, (1 \, - \,e^2)^{3/2} \, \, \, \, .
\label{VDERIVATIVET0:appA}
\end{eqnarray}
\subsection{Computation steps}
The parameter step $\mathbf{\Delta p_{nl}}^{\{n\}}$ in Eq.\,\ref{NEWLEVENBERG:appA} is computed as follows:
\begin{enumerate}
    \item Given the preliminary non-linear parameters $\mathbf{p_{nl}}$, we compute the mean and eccentric 
    anomalies given by Eq.\,\ref{ECCENTRICANOMALIES}. This transcendental equation is solved 
    following the algorithm proposed by \citet{1996AJ....112.2858F}, and implemented in a common \gaia\ tool library.
    \item The true anomaly trigonometric functions are computed through Eq.\,\ref{COSV} and Eq.\,\ref{SINV}.
    \item The Jacobian matrix $\mathbf{J_l}^{\{n\}}$ is filled given Eq.\,\ref{JL:appA}. 
    The linear parameters $\mathbf{p_{l}}^{\{n\}}$ are then derived by the linear least-square regression 
    given by Eq.\,\ref{LINEARPARAMS1:appA}.
    \item The derivatives of the true anomalies are computed using Eq.\,\ref{VDERIVATIVENU:appA}, Eq.\,\ref{VDERIVATIVEE:appA}, 
    and Eq.\,\ref{VDERIVATIVET0:appA}.
    \item We fill the Jacobian matrix $\mathbf{J_{nl}}^{\{n\}}$ following Eq.\,\ref{JNLDERIVATIVEJNLNU:appA}, 
    Eq.\,\ref{JNLDERIVATIVEJNLE:appA}, 
    and Eq.\,\ref{JNLDERIVATIVEJNLT0:appA}.
    \item The matrix $\mathbf{A}$ is computed thanks to Eq.\,\ref{LINEARPARAMS4:appA}, Eq.\,\ref{LINEARPARAMS5:appA}, 
    and Eq.\,\ref{MATRIXPASSAGE:appA}.
    \item Finally, the total Jacobian matrix $\mathbf{\tilde{J}_{nl}}^{\{n\}}$ is obtained by Eq.\,\ref{JACOBIAN:appA}, 
    and the linear system (Eq.\,\ref{NEWLEVENBERG:appA}) is solved for $\mathbf{\Delta p_{nl}}^{\{n\}}$.
\end{enumerate}

\clearpage
\section{Pie charts illustrating the tallies}
\label{sec:appD}
\begin{figure*}[!htp]
\centerline{
\includegraphics[width=0.55\textwidth]{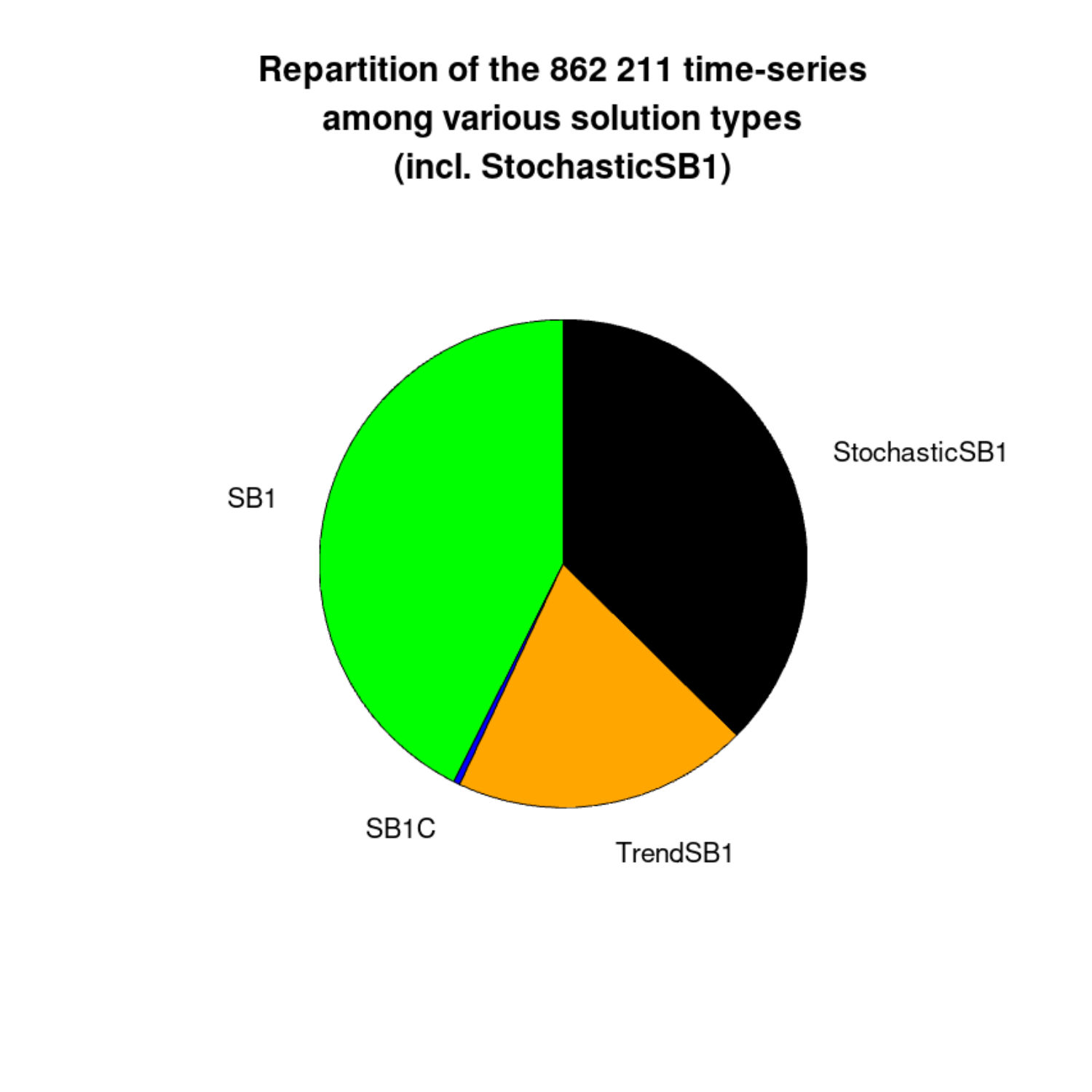}
\includegraphics[width=0.55\textwidth]{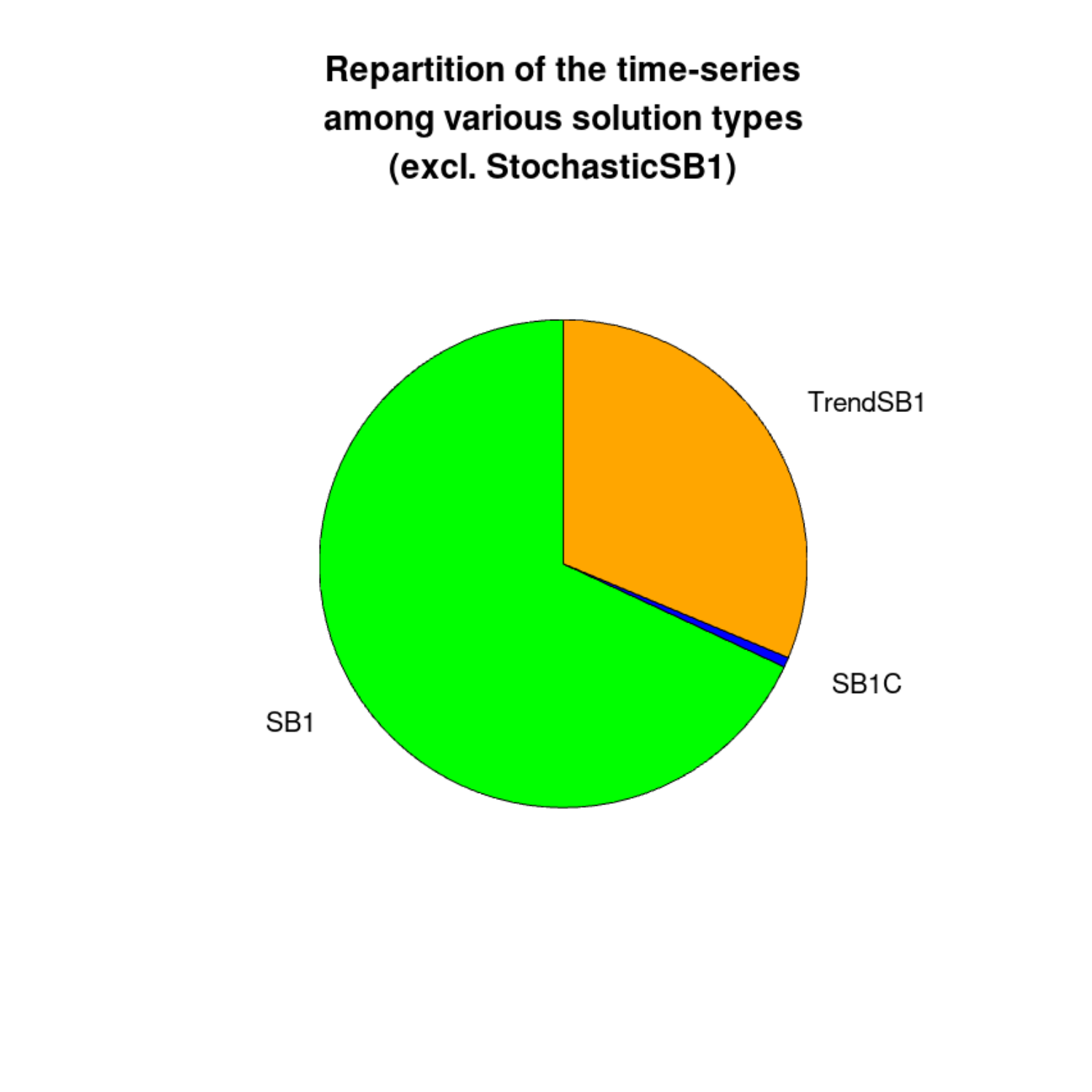}
}
\centerline{
\includegraphics[width=0.55\textwidth]{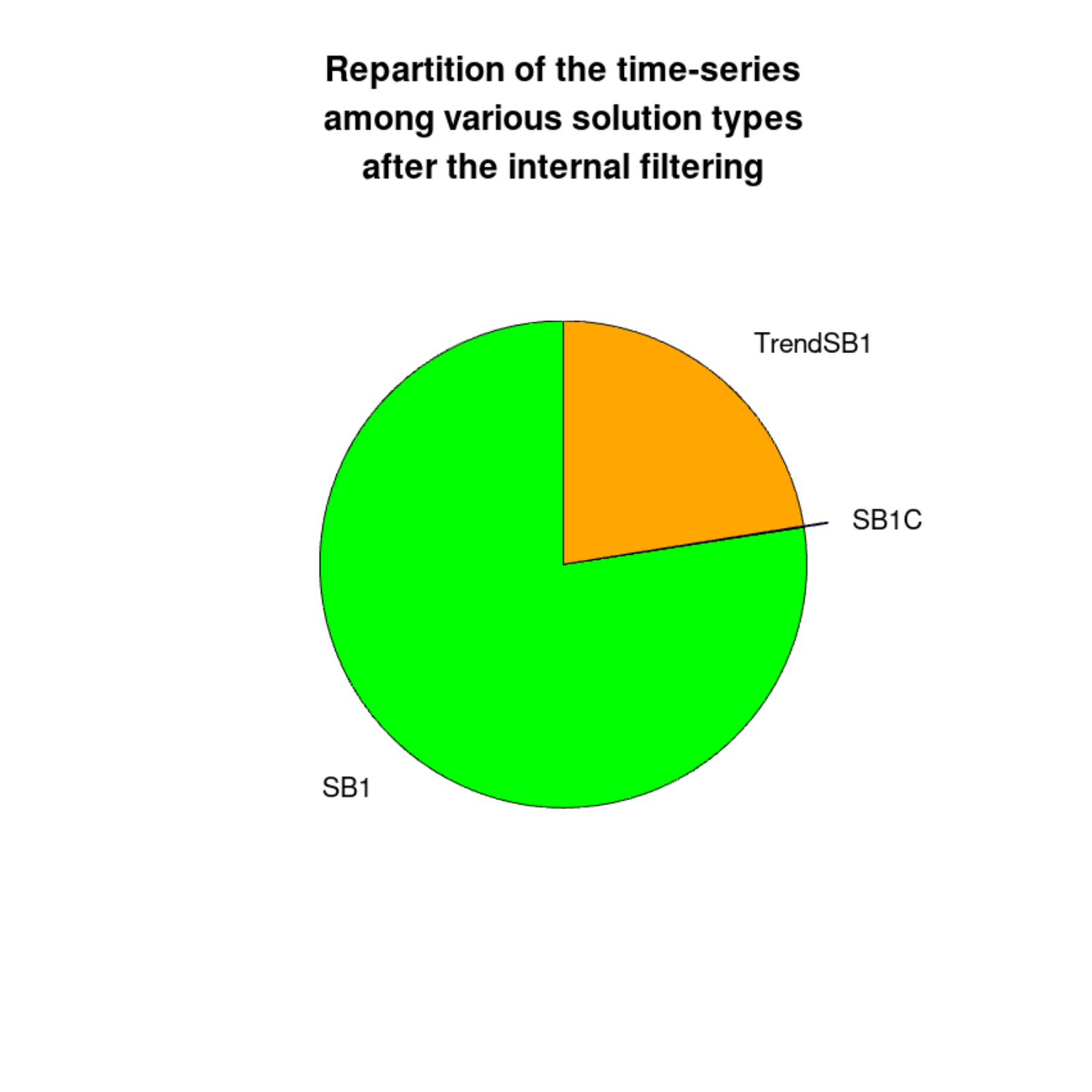}
\includegraphics[width=0.55\textwidth]{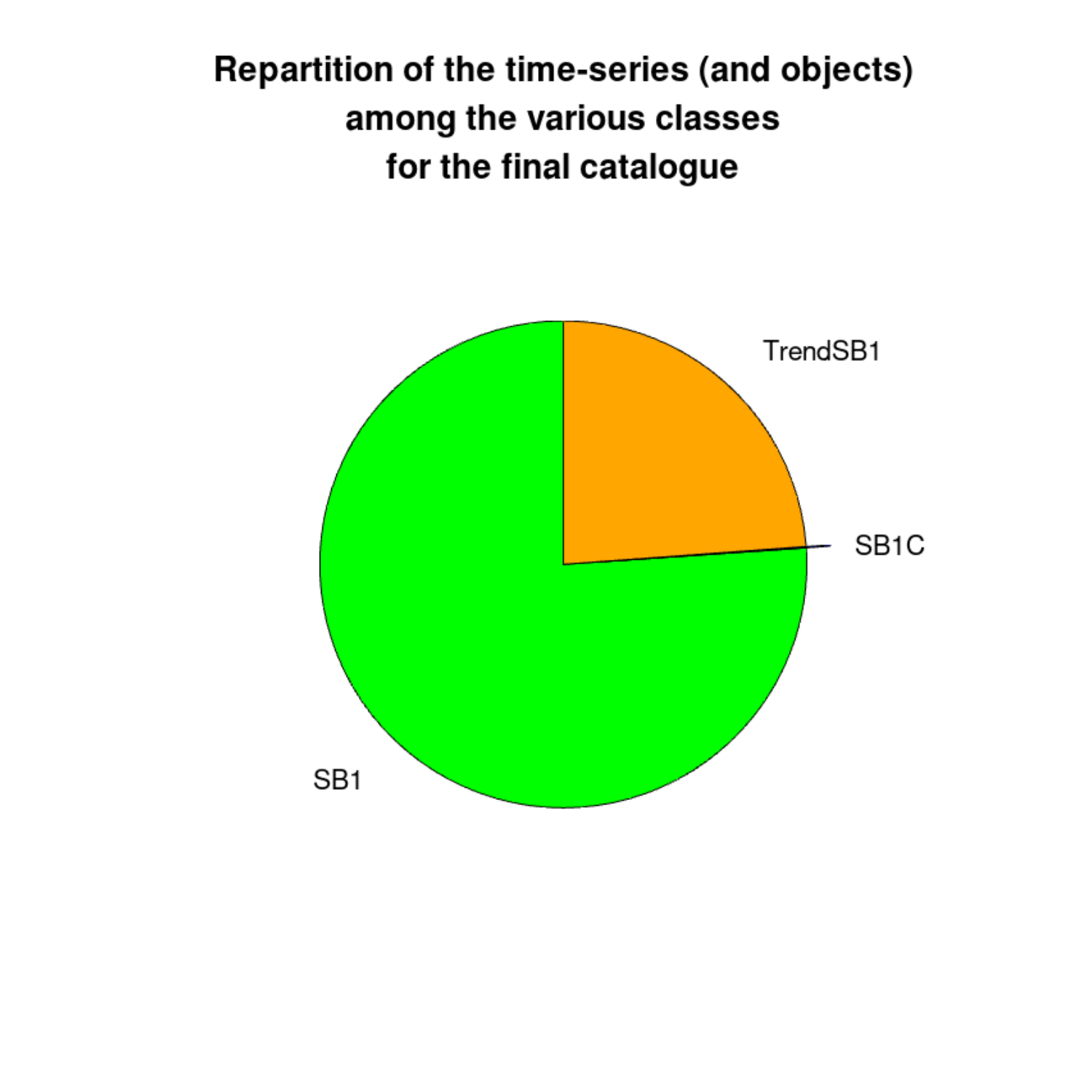}
}
\caption{Pie charts illustrating (upper row) the
repartition of the time series among 
the various solution types including and excluding the StochasticSB1
(see Table\,\ref{tab:tableresult1}) and its evolution (lower row) 
over the course of the various losses: internal filtering (left:  
Tables\,\ref{tab:tableresult2}, \ref{tab:tableresult3},
\ref{tab:tableresult4}) and combination+post-filtering 
(right: Tables\,\ref{tab:tableresult5}, \ref{tab:tableresult6},
\ref{tab:tableresult7}).
}
\label{fig:piechartA}
\end{figure*}
\begin{figure*}[!htp]
\centerline{
\includegraphics[width=0.55\textwidth]{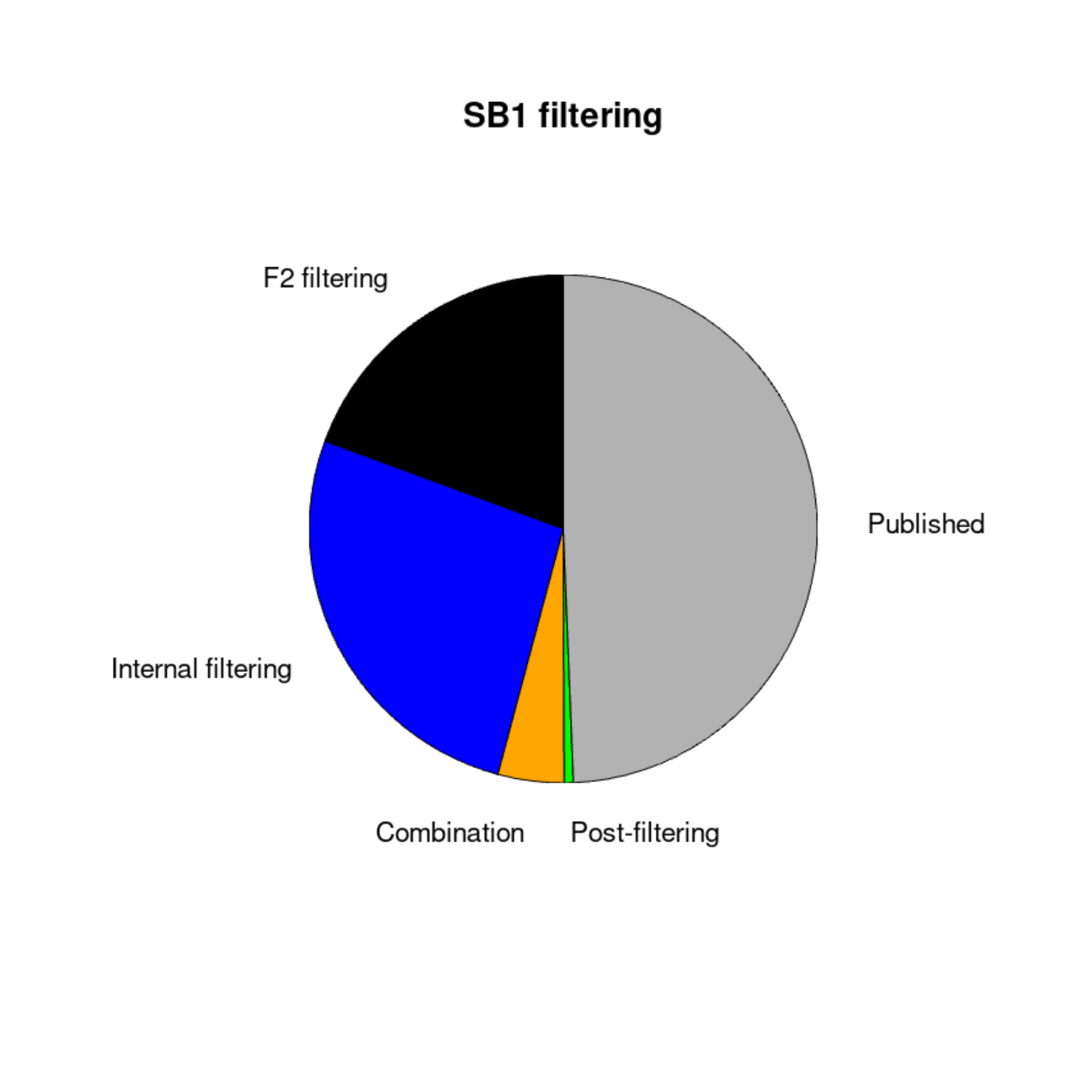}
\includegraphics[width=0.55\textwidth]{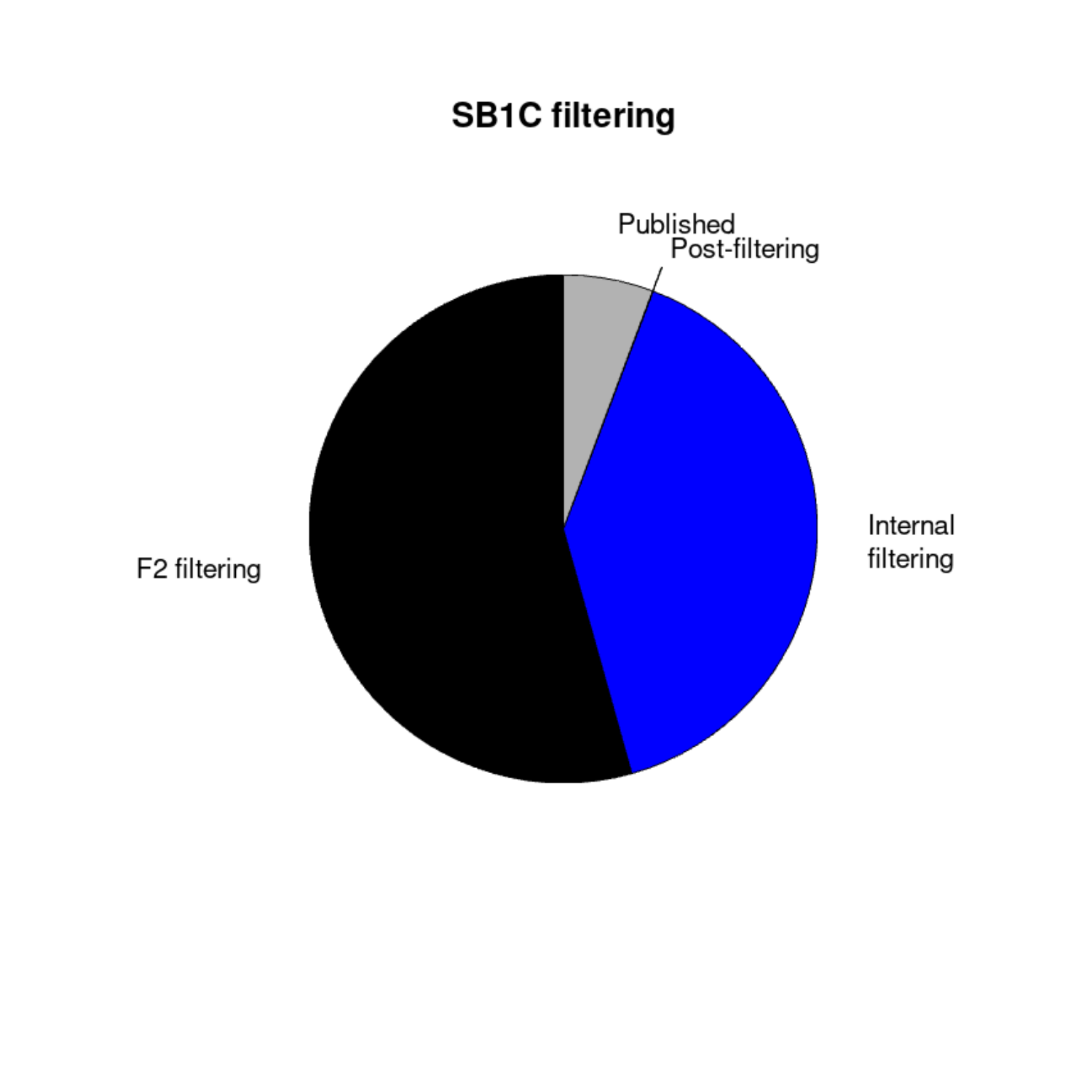}
}
\centerline{
\includegraphics[width=0.55\textwidth]{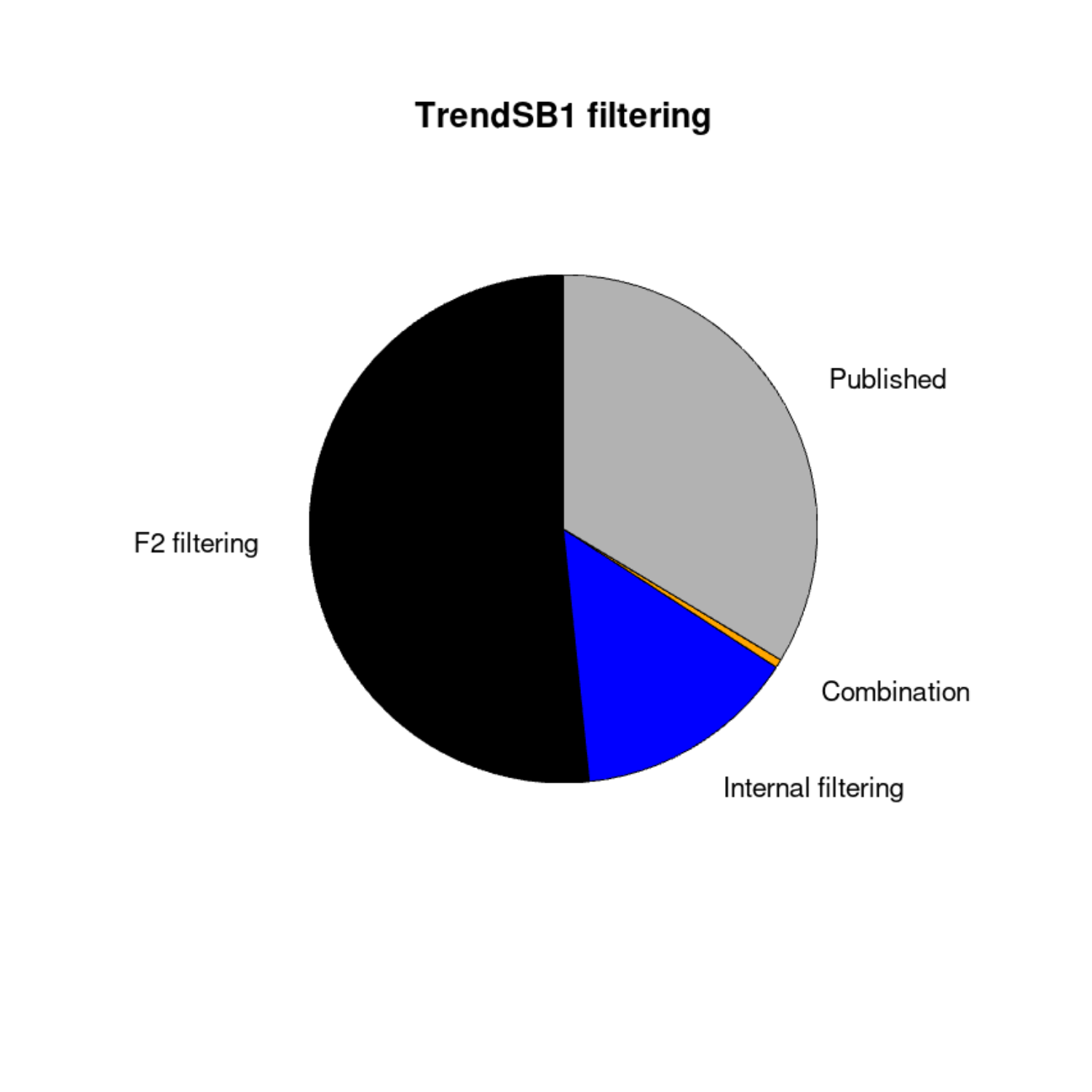}
\includegraphics[width=0.55\textwidth]{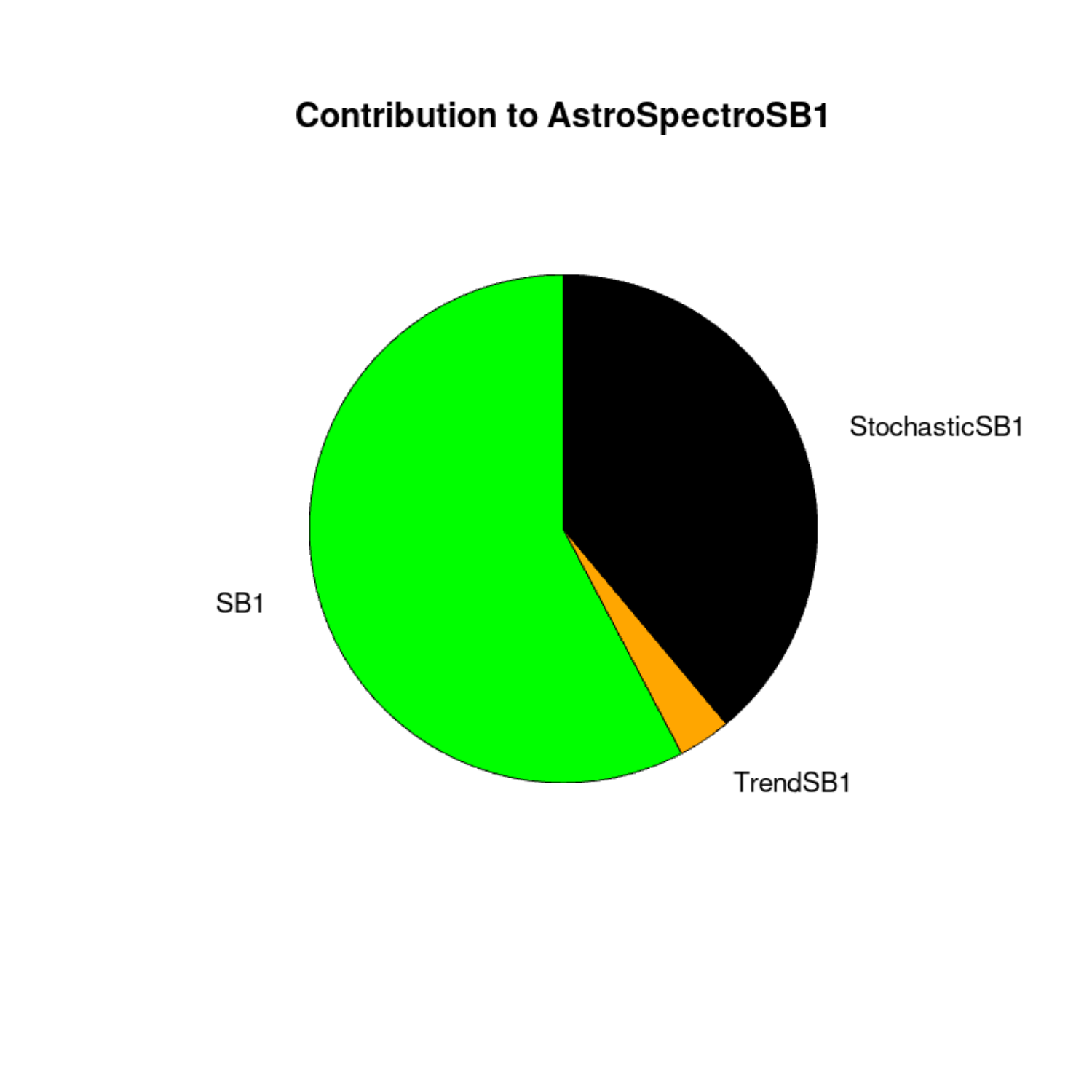}
}
\caption{Details of the percentage of discarding at the various
steps before publication for the SB1 solution type (upper left),
for the SB1C solution type (upper right) and for the TrendSB1
solution type (lower left). 
The $F_2$ filtering is the first step of the internal filtering and
its contribution is represented in black. The effect of the subsequent steps
of the internal filtering is in blue. The effect of the loss by combination
is in orange and the one of the post-filtering is in green. The gray sector
represents the proportion published.
The lower right pie chart 
gives the contribution
of the different solution types to the {\tt{AstroSpectroSB1}} 
class (see Table\,\ref{tab:tableresult8}).}
\label{fig:piechartB}
\end{figure*}
\clearpage
\section{A few examples of resulting orbital solutions}
\label{sec:appE}
\vspace*{-0.6cm} 
\begin{figure*}[!htp]
\centerline{
\includegraphics[width=0.29\textwidth]{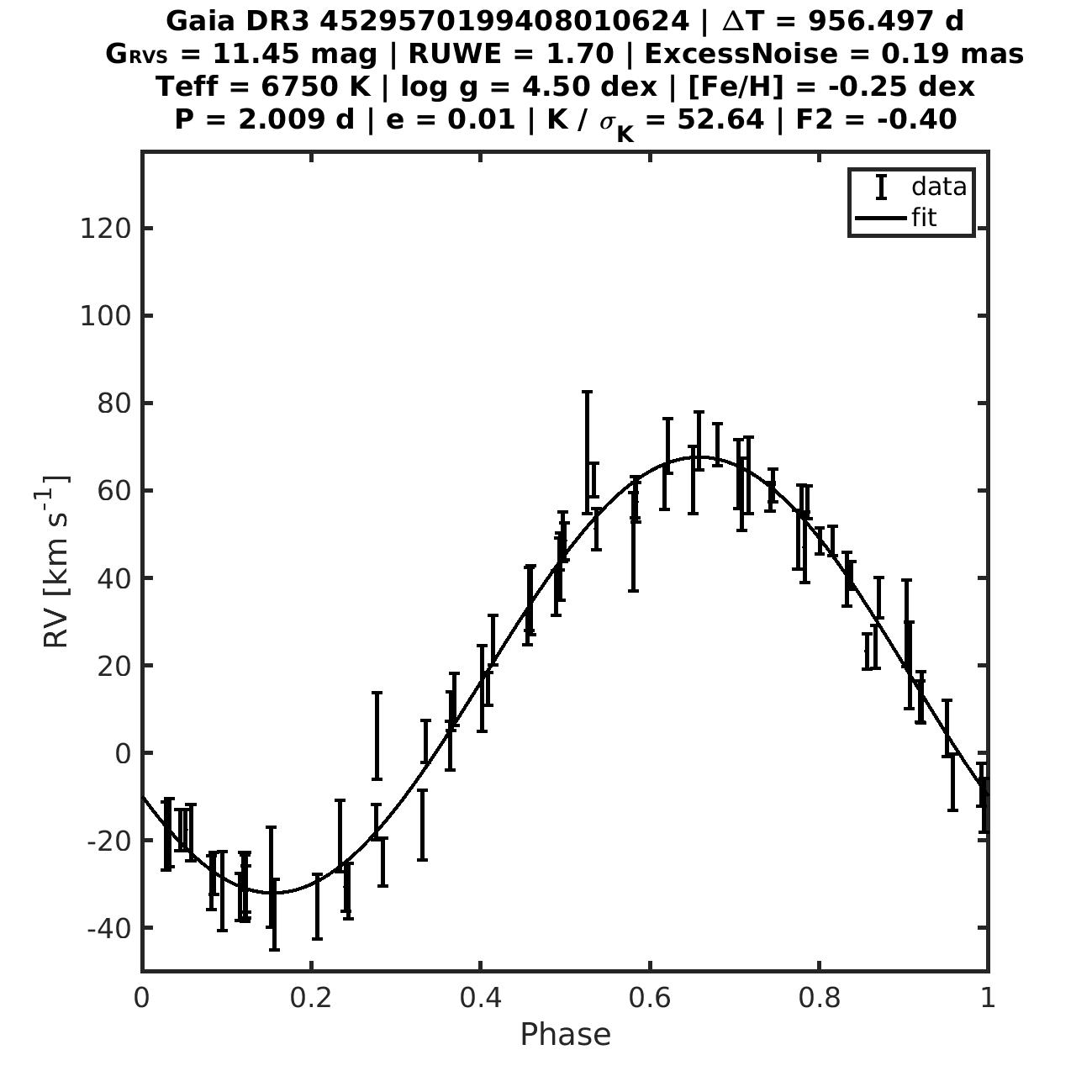}
\includegraphics[width=0.29\textwidth]{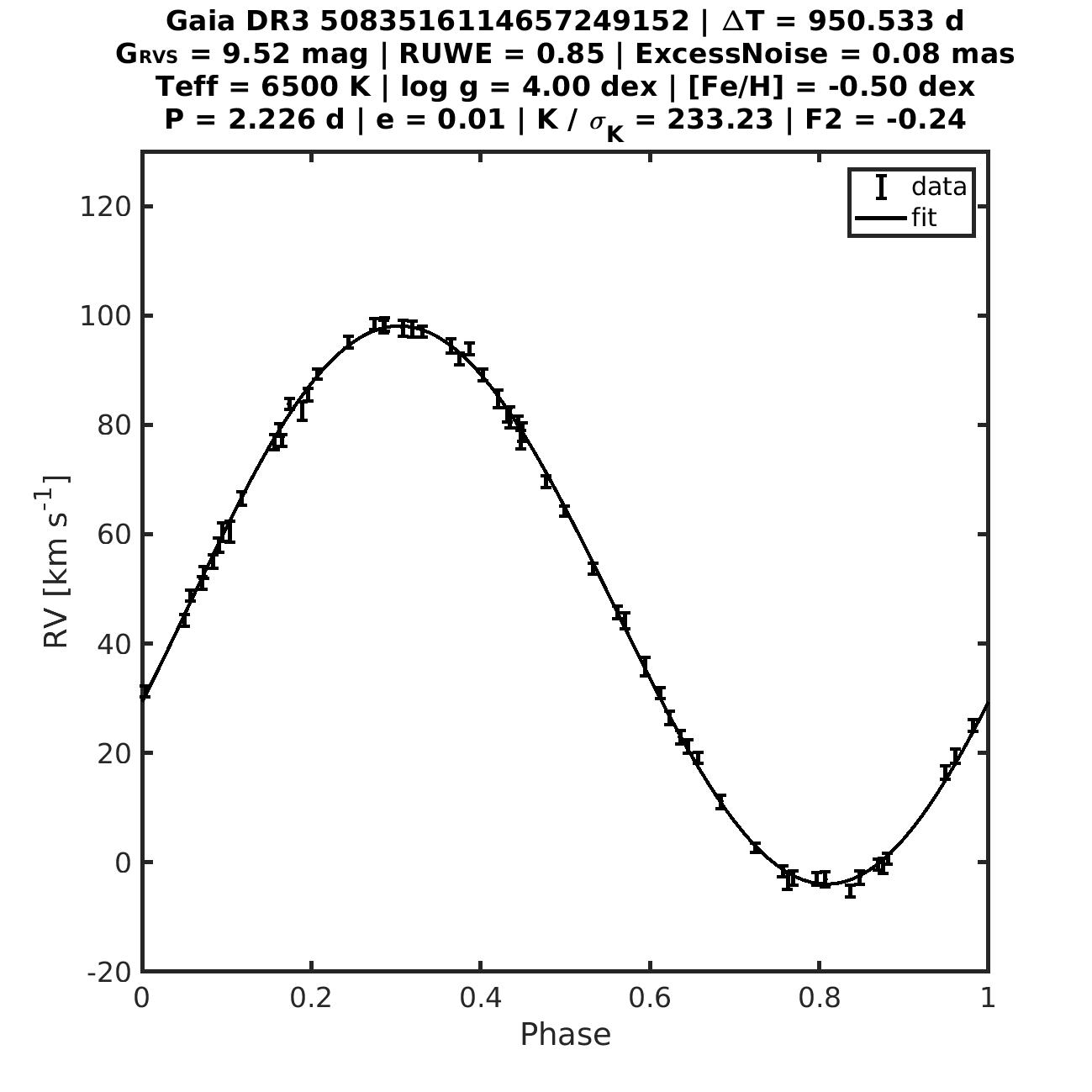} 
\includegraphics[width=0.29\textwidth]{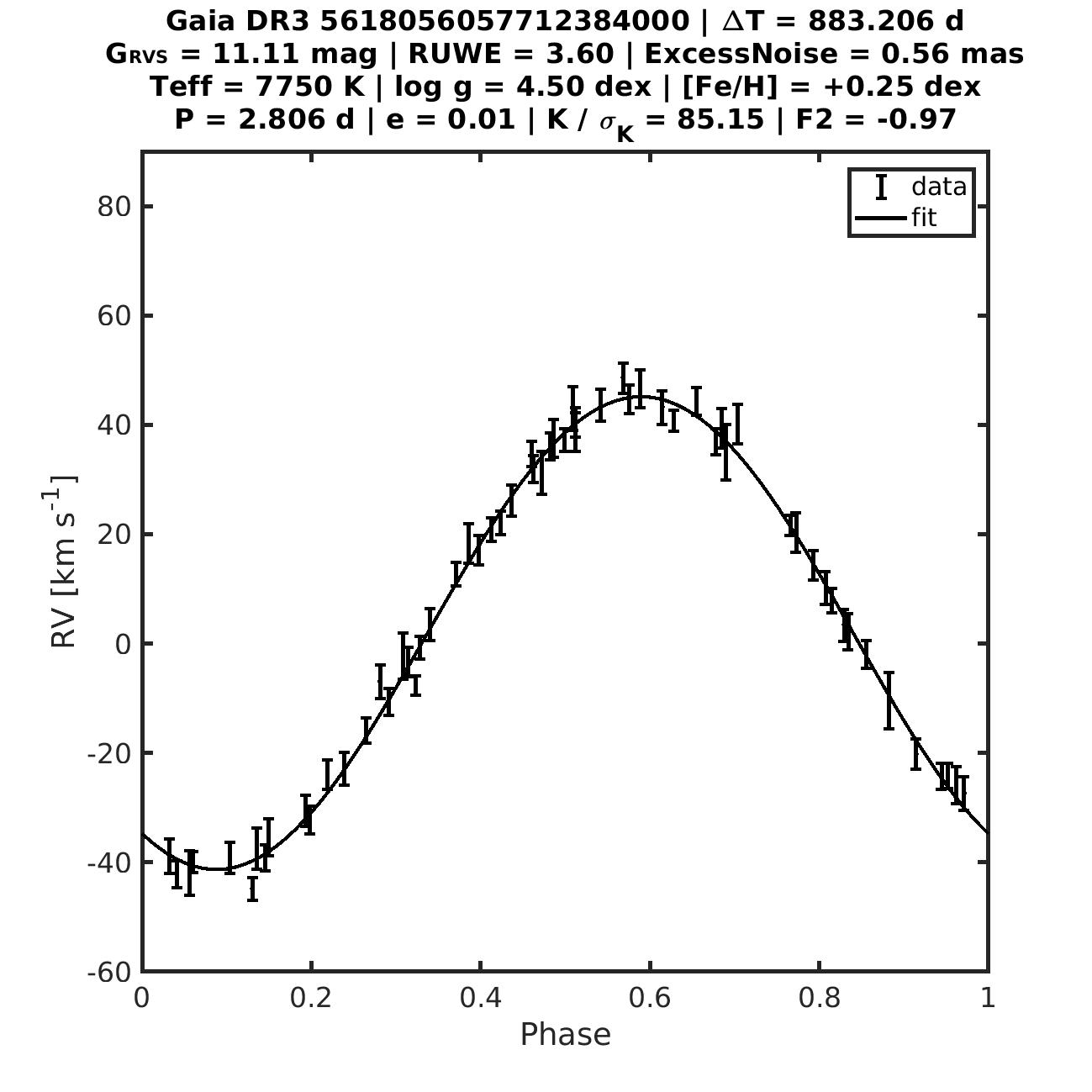} 
}
\centerline{
\includegraphics[width=0.29\textwidth]{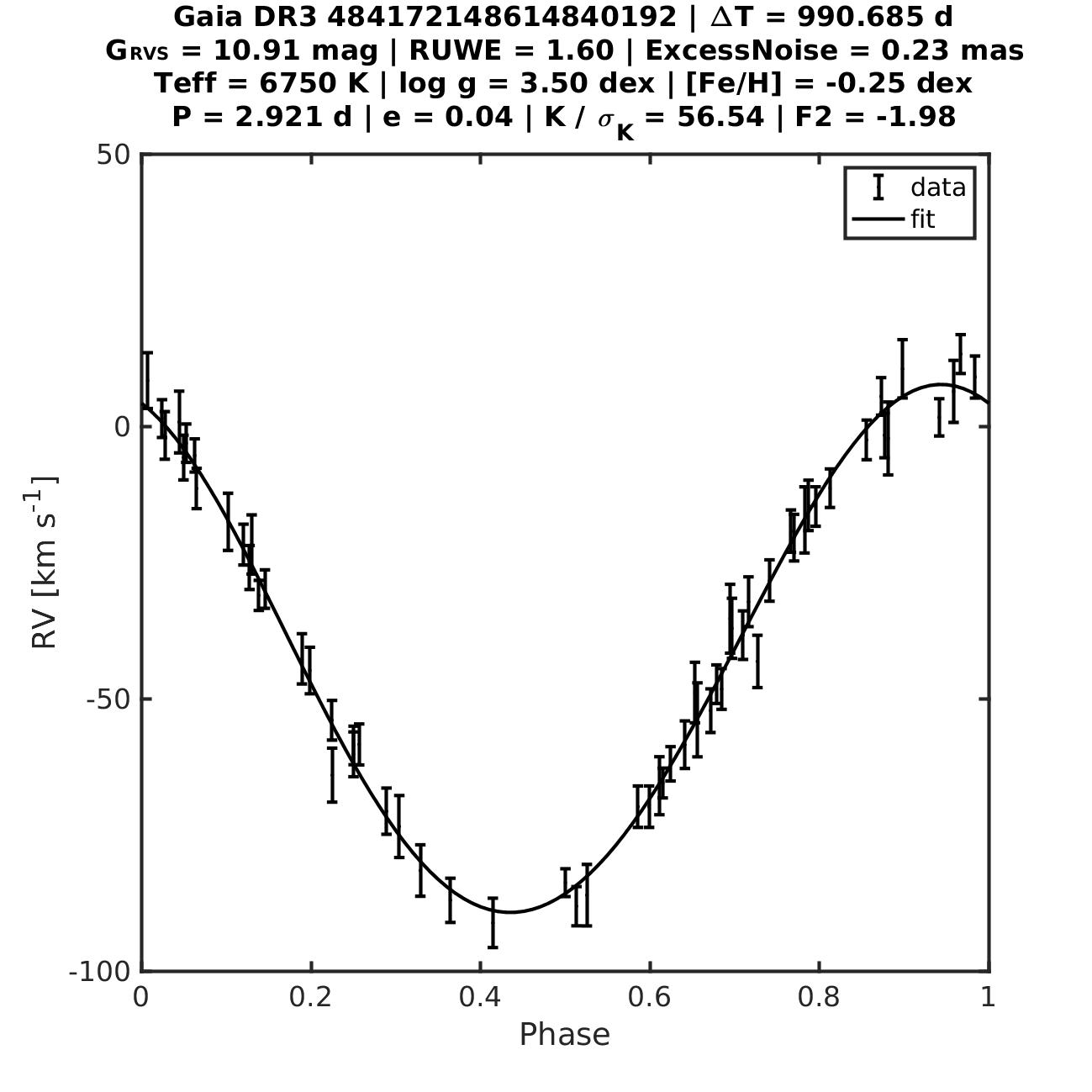}
\includegraphics[width=0.29\textwidth]{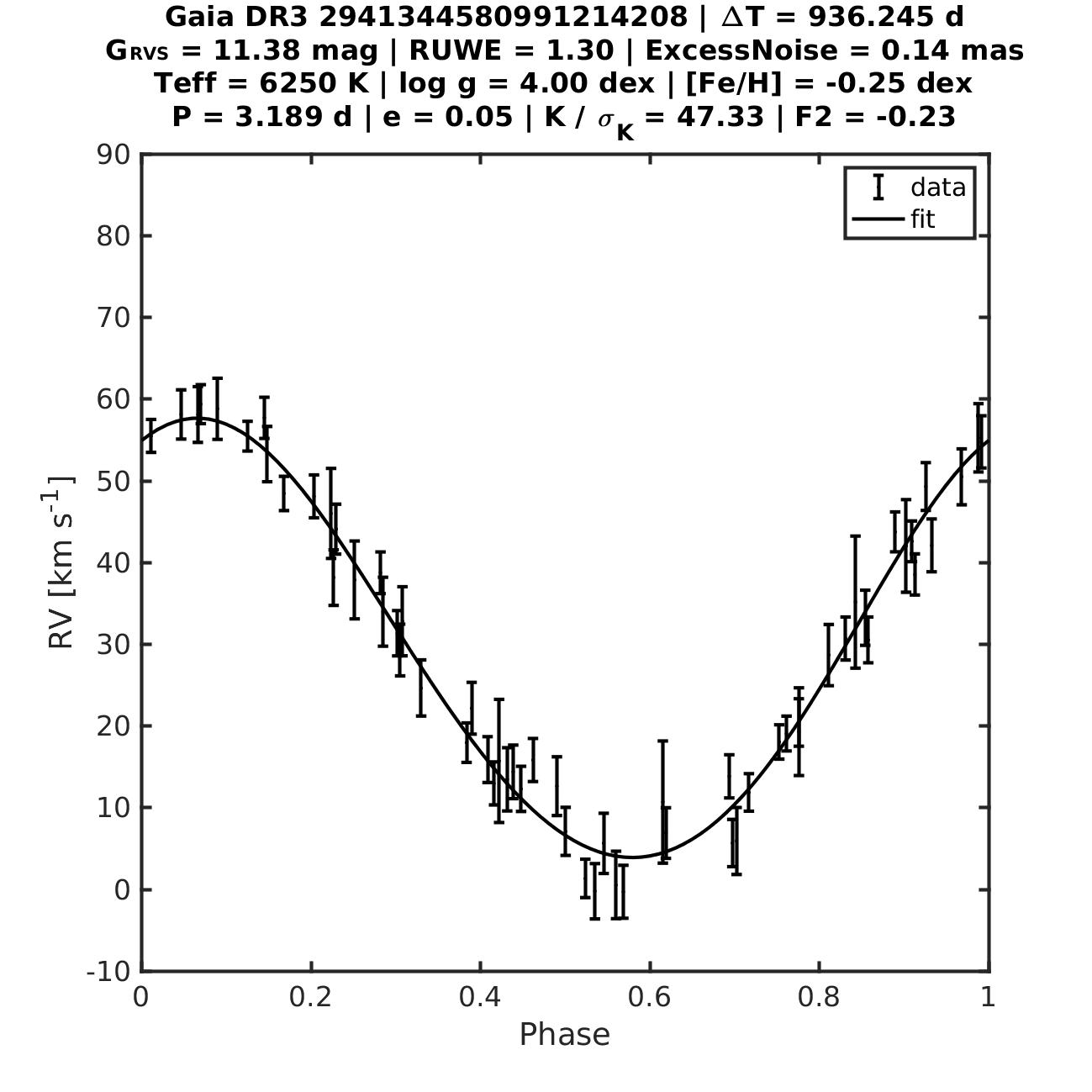}
\includegraphics[width=0.29\textwidth]{./FIGS/pp_515682154702230272_SB1.jpg}
}
\centerline{
\includegraphics[width=0.29\textwidth]{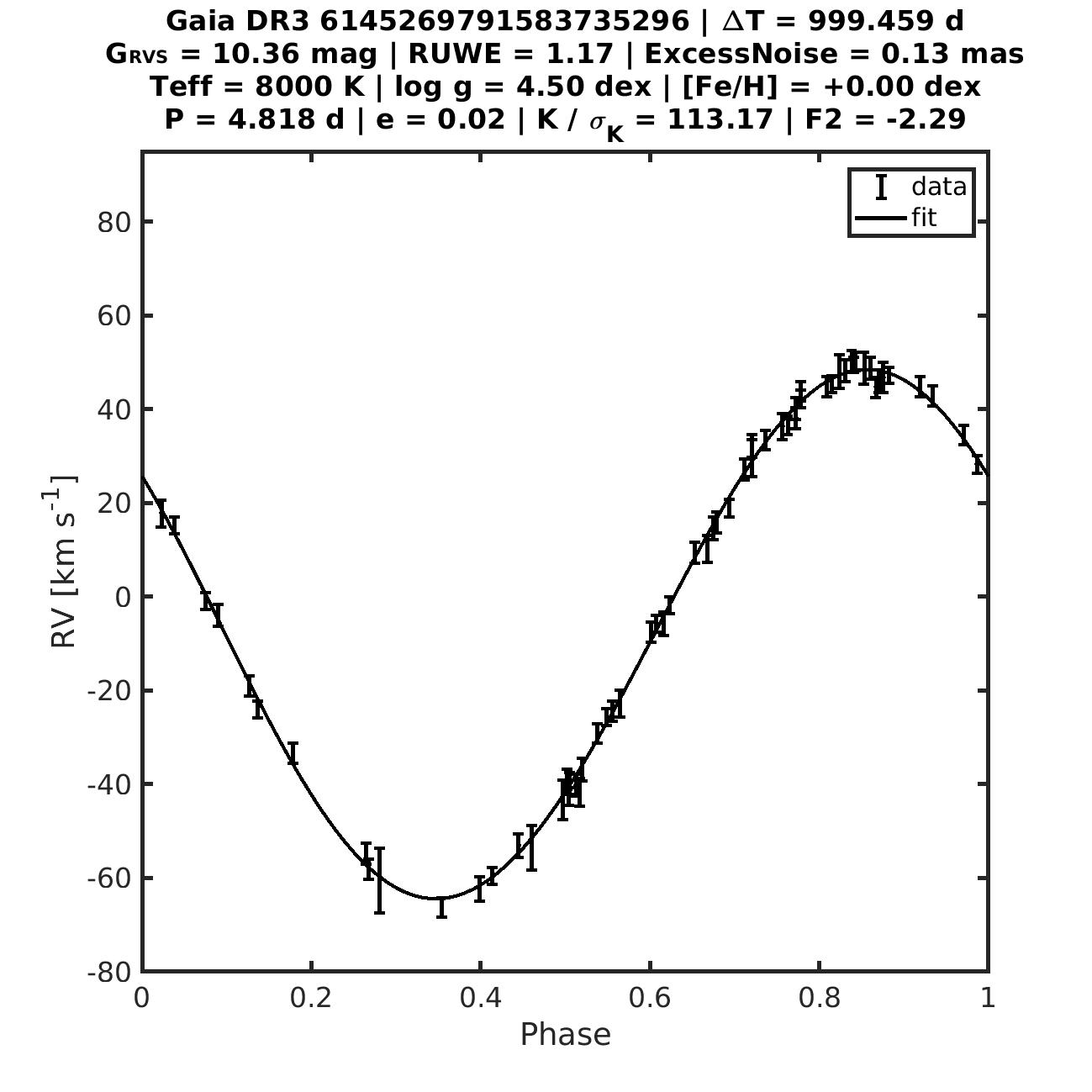}
\includegraphics[width=0.29\textwidth]{./FIGS/pp_1955225454244655872_SB1.jpg}
\includegraphics[width=0.29\textwidth]{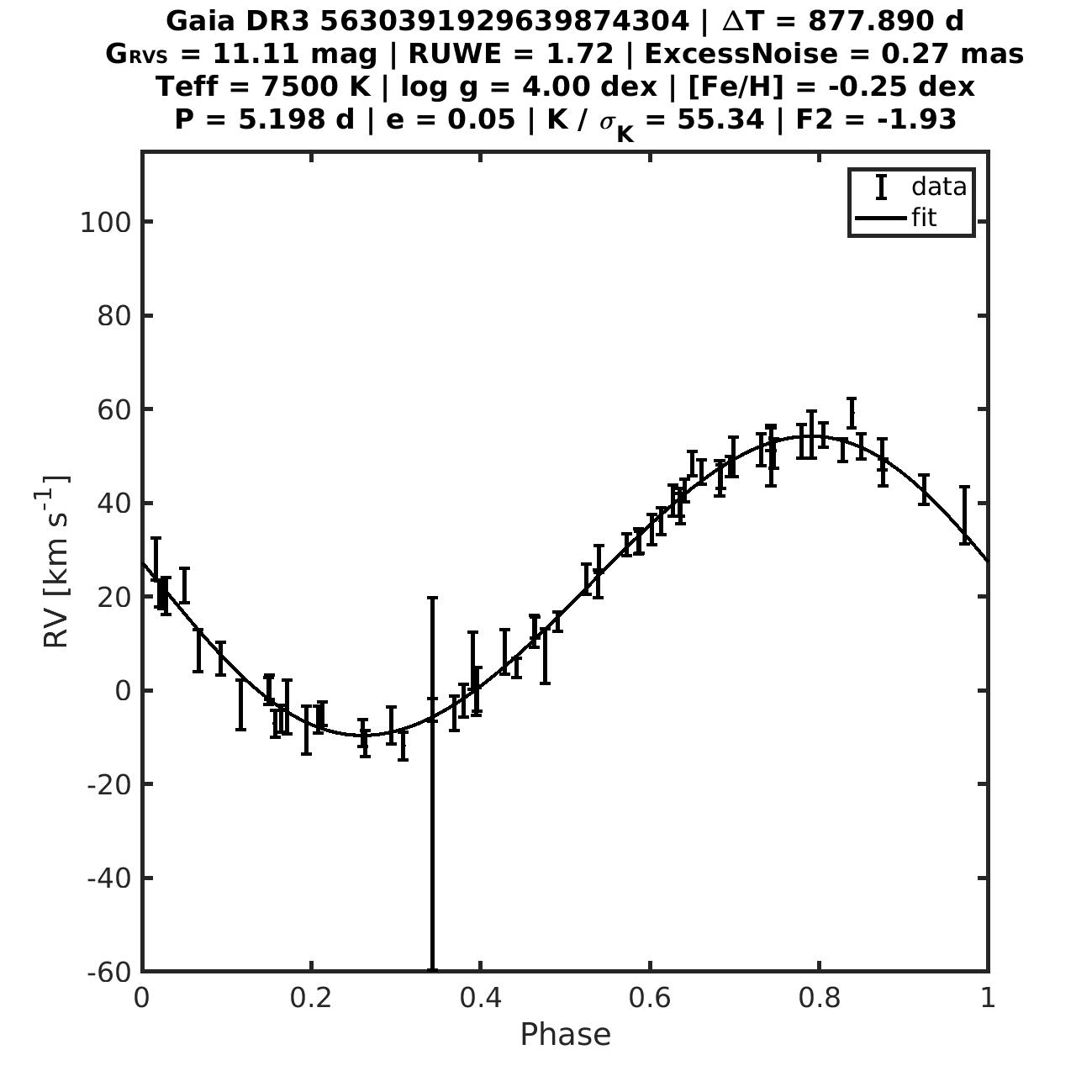}
}
\centerline{
\includegraphics[width=0.29\textwidth]{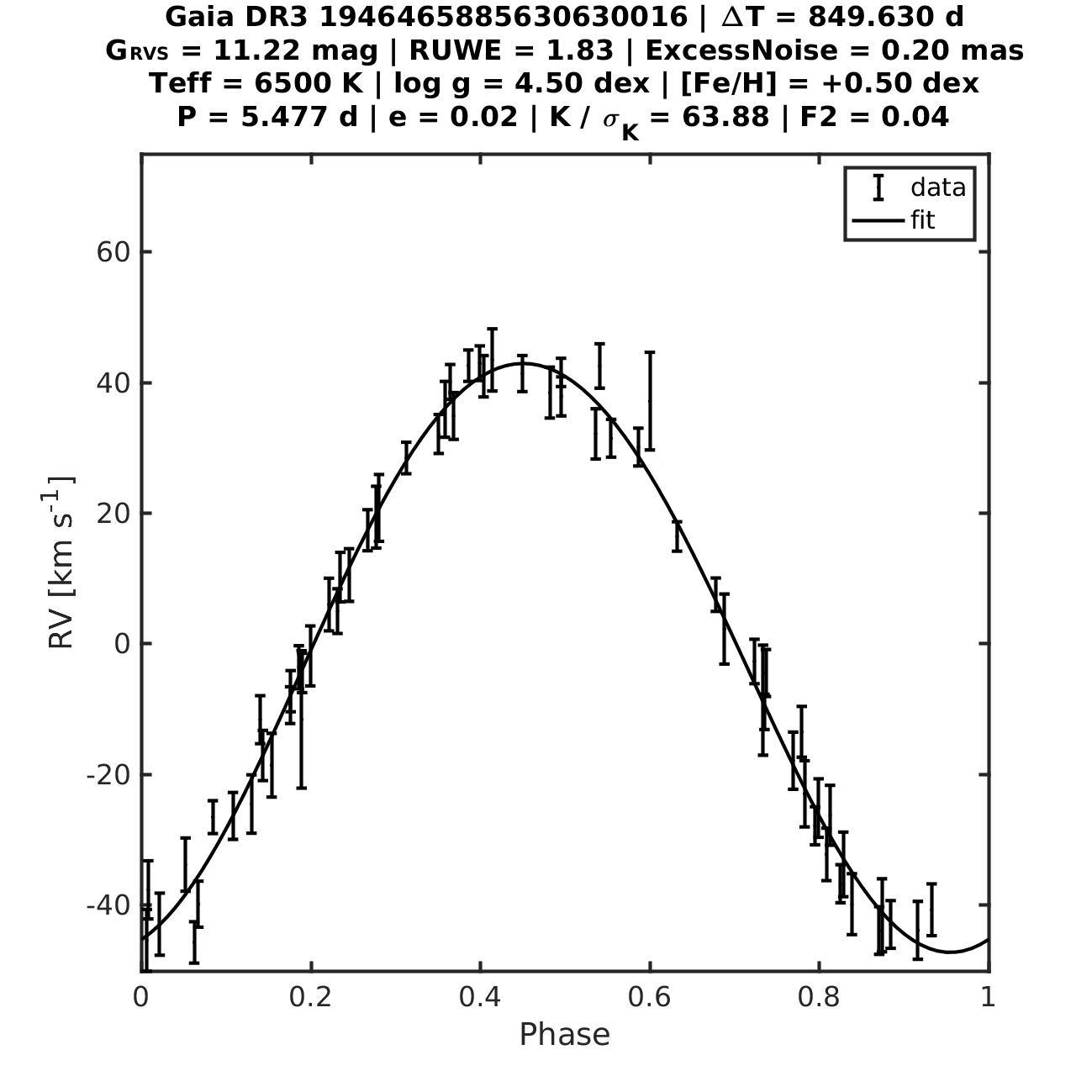}
\includegraphics[width=0.29\textwidth]{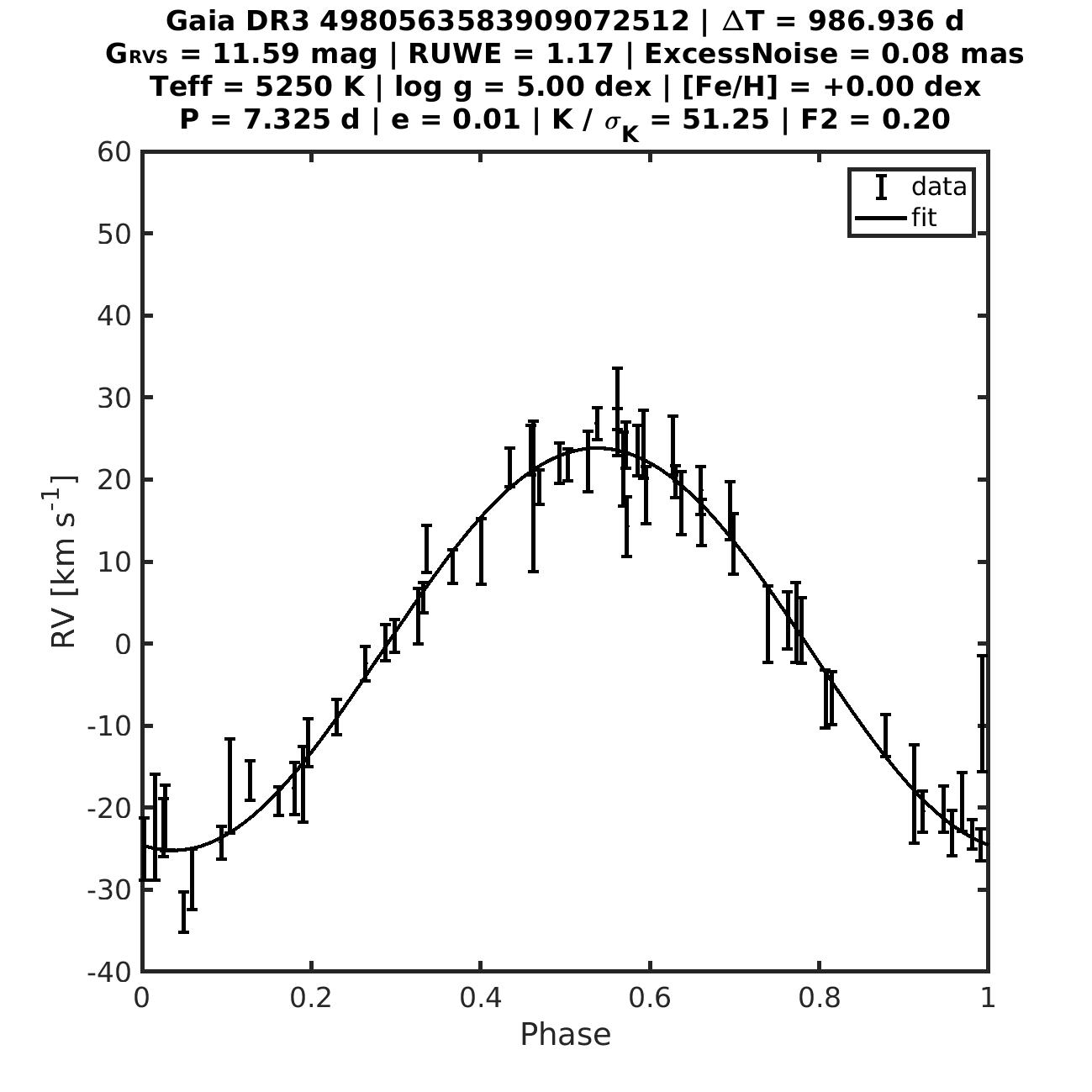}
\includegraphics[width=0.29\textwidth]{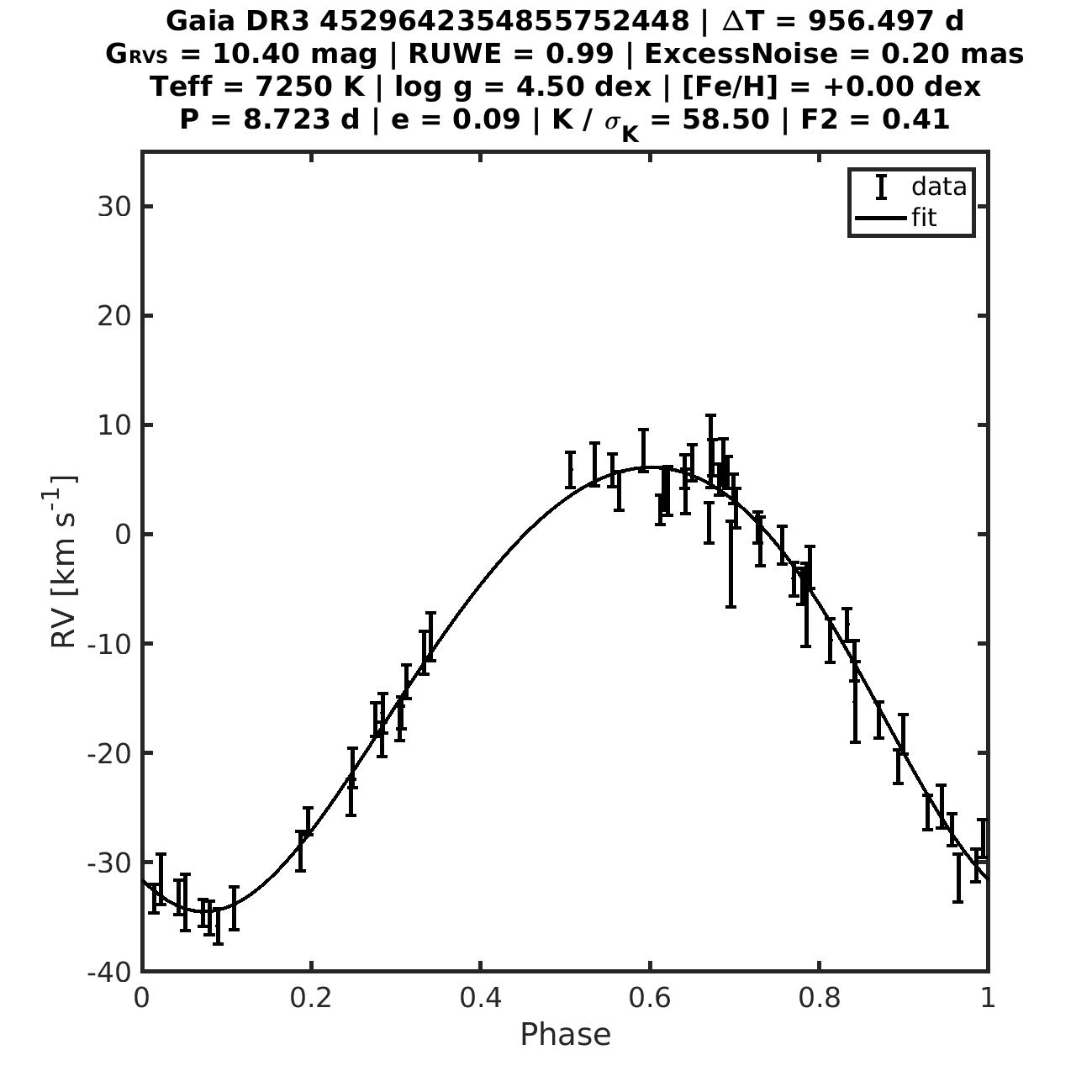}
}
\caption{Good results: a few examples illustrative of the output 
of the pipeline for SB1-type solutions. Each panel
concerns an object whose \gaia\ name is given in 
the header. Each of the panels shows the phase 
diagram containing the folded RVs (data points at 
the centre of the $\pm 1 \sigma$ error bars) 
along with the fitted orbital solution. The header
gives also the time span covered by the time series
($\Delta T$). The second line gives the 
$G_{\mathrm{RVS}}^{\mathrm{int}}$ magnitude of 
the object, the astrometric {\tt{ruwe}} and the astrometric
excess noise, whereas the third line concerns
the physical parameters identifying the template used. 
The last line indicates the period, the eccentricity, 
the significance, and the statistic
$F_2$ related to the adopted solution. The objects are
ordered by increasing period.
}
\label{fig:appgoodressb1part1}
\end{figure*}
\begin{figure*}[!htp]
\ContinuedFloat
\centerline{
\includegraphics[width=0.29\textwidth]{./FIGS/pp_4983214300285468288_SB1.jpg}
\includegraphics[width=0.29\textwidth]{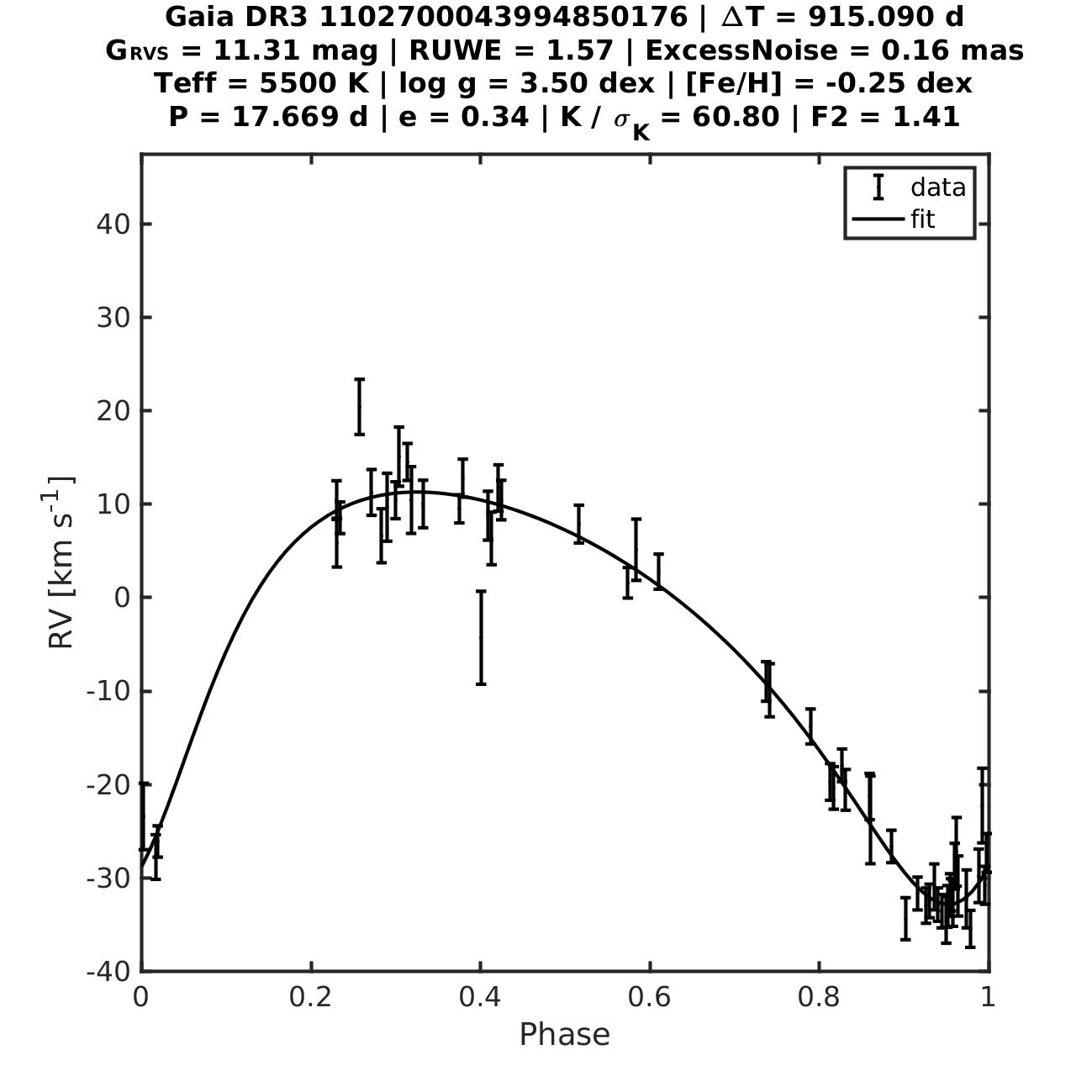} 
\includegraphics[width=0.29\textwidth]{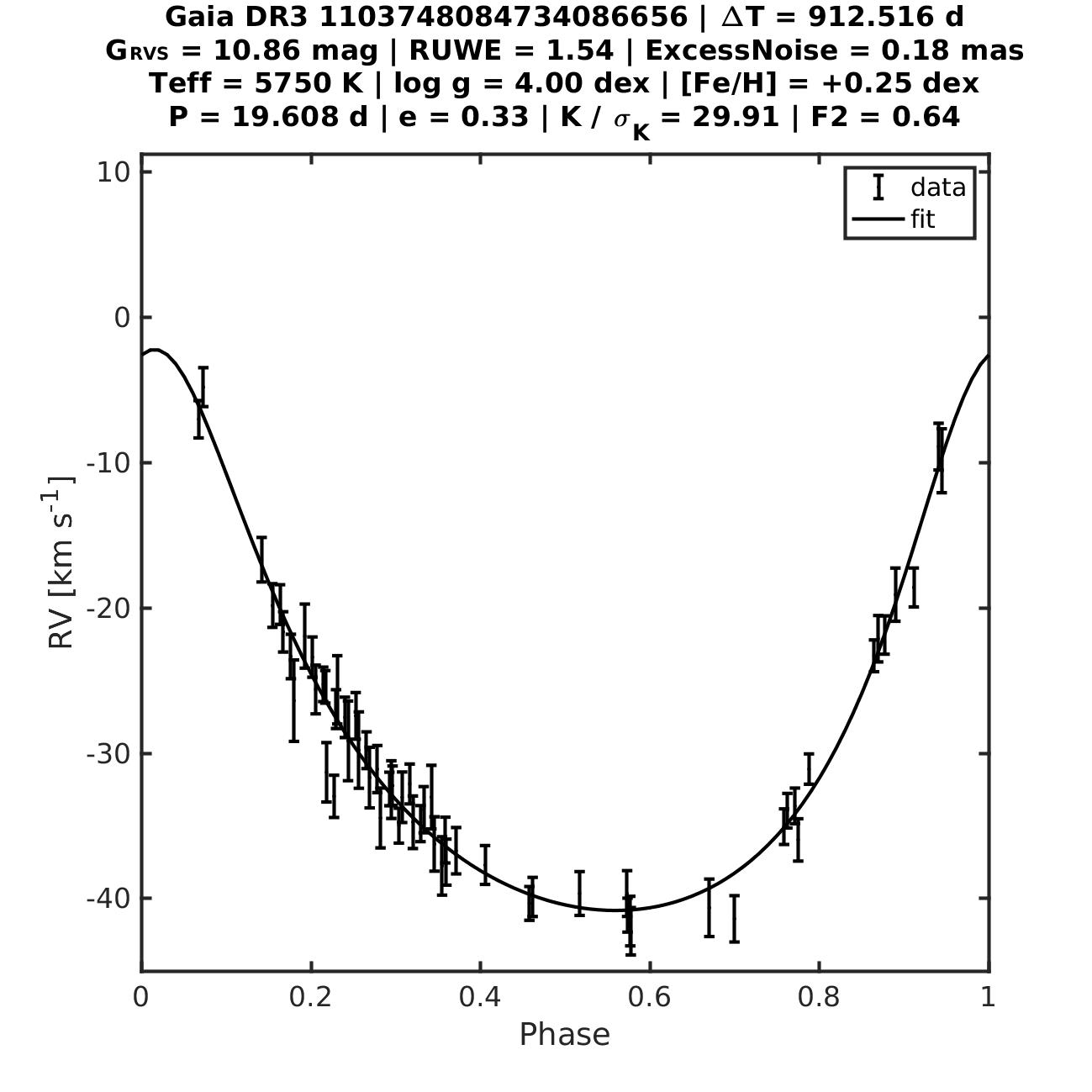} 
}
\centerline{
\includegraphics[width=0.29\textwidth]{./FIGS/pp_5059164749602128128_SB1.jpg}
\includegraphics[width=0.29\textwidth]{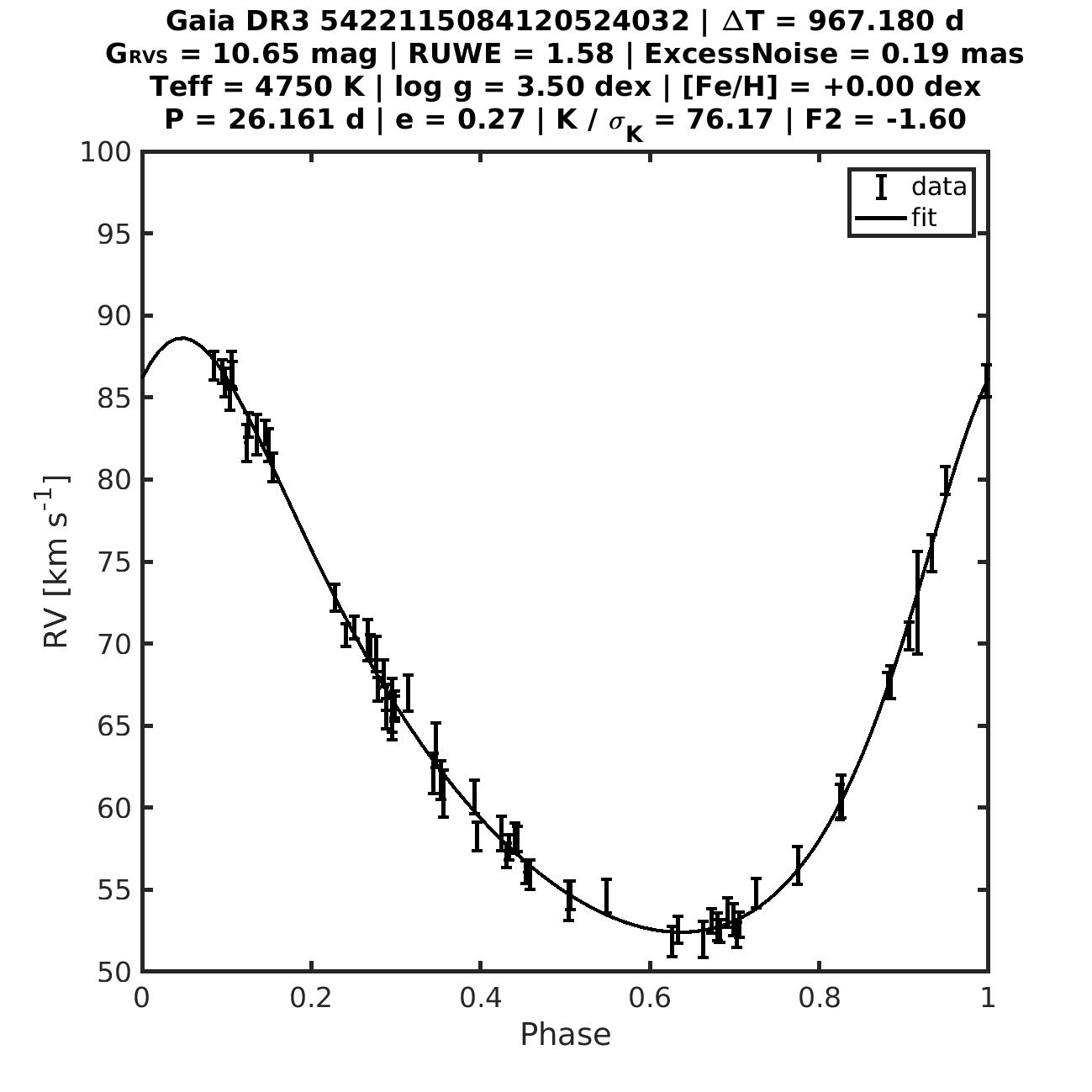}
\includegraphics[width=0.29\textwidth]{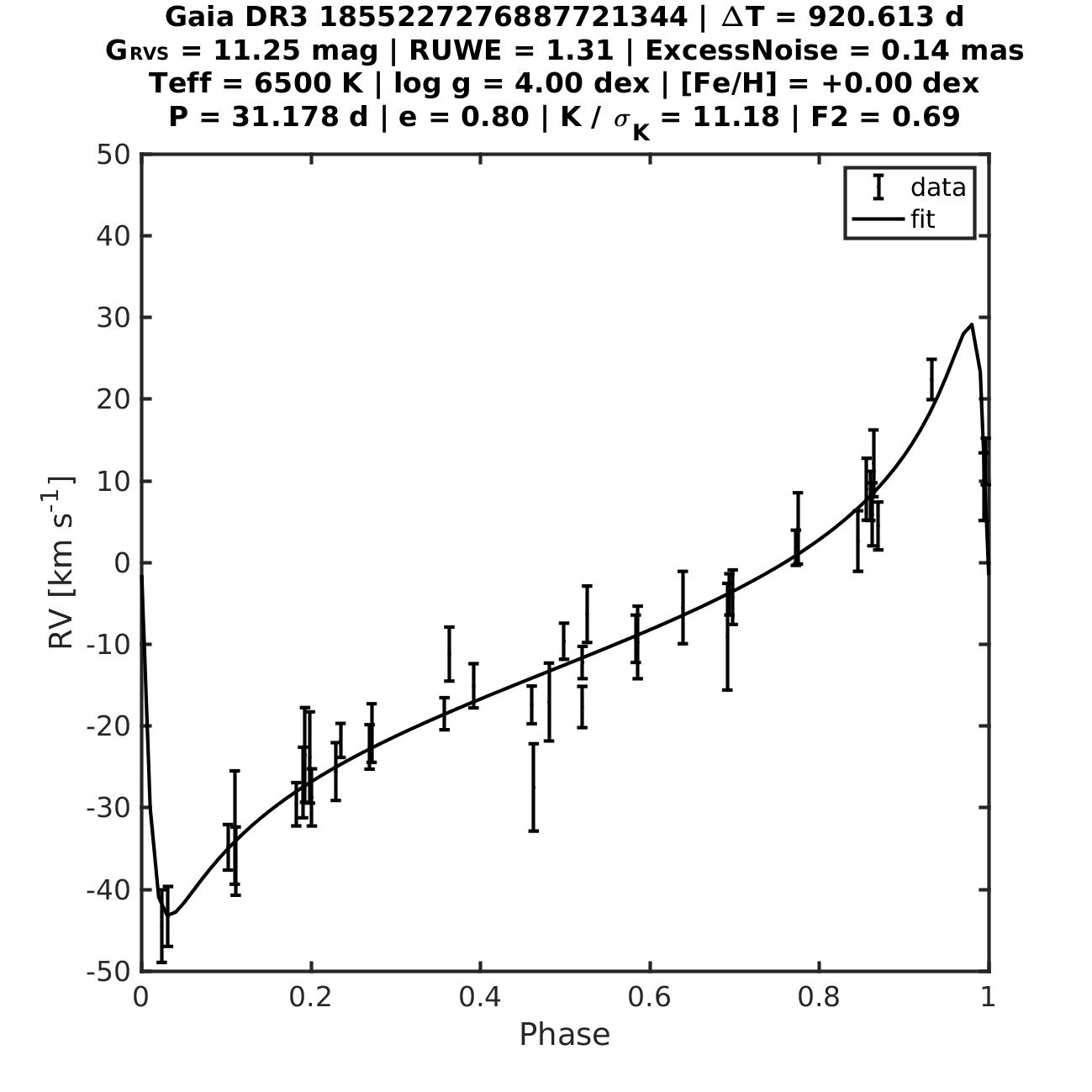}
}
\centerline{
\includegraphics[width=0.29\textwidth]{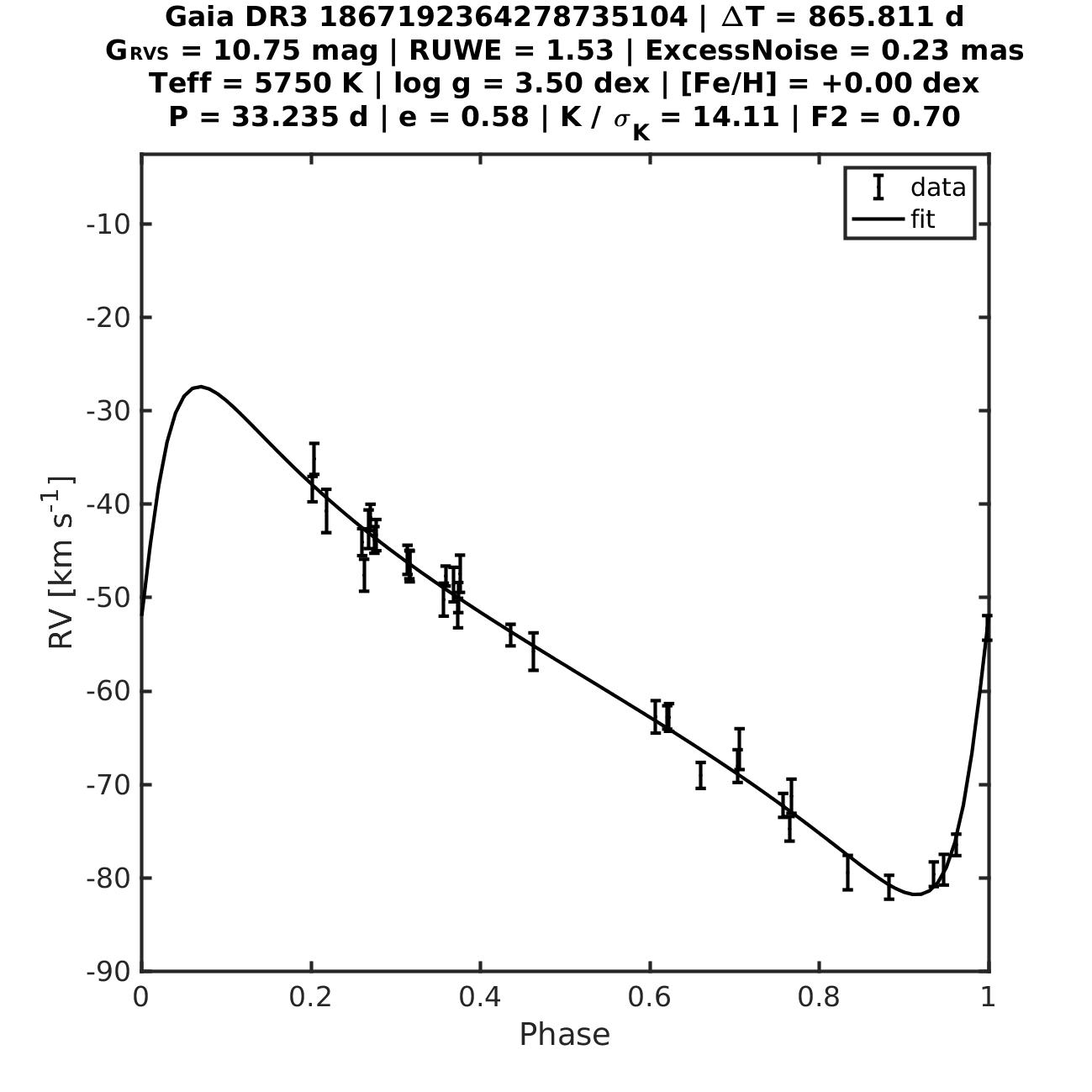}
\includegraphics[width=0.29\textwidth]{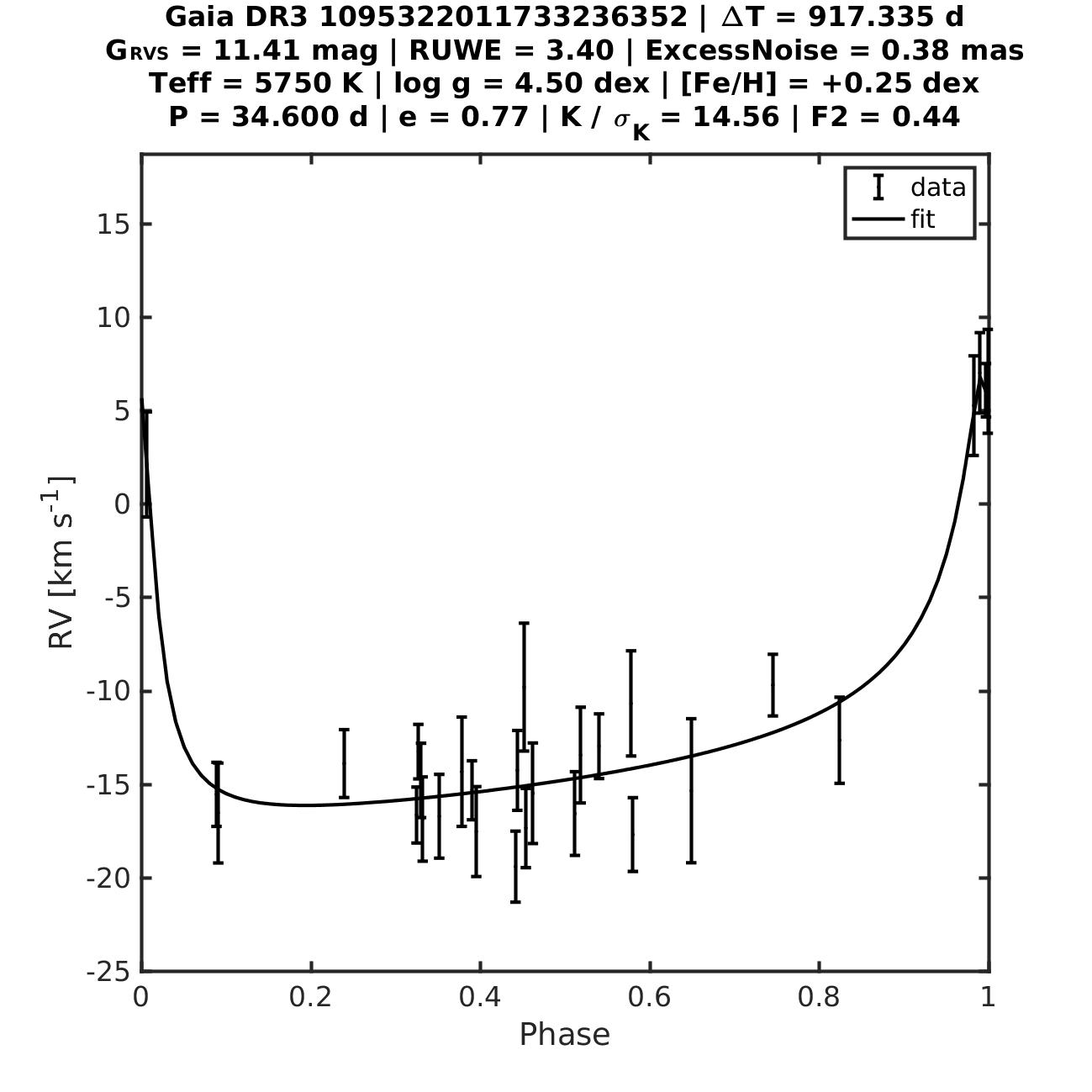}
\includegraphics[width=0.29\textwidth]{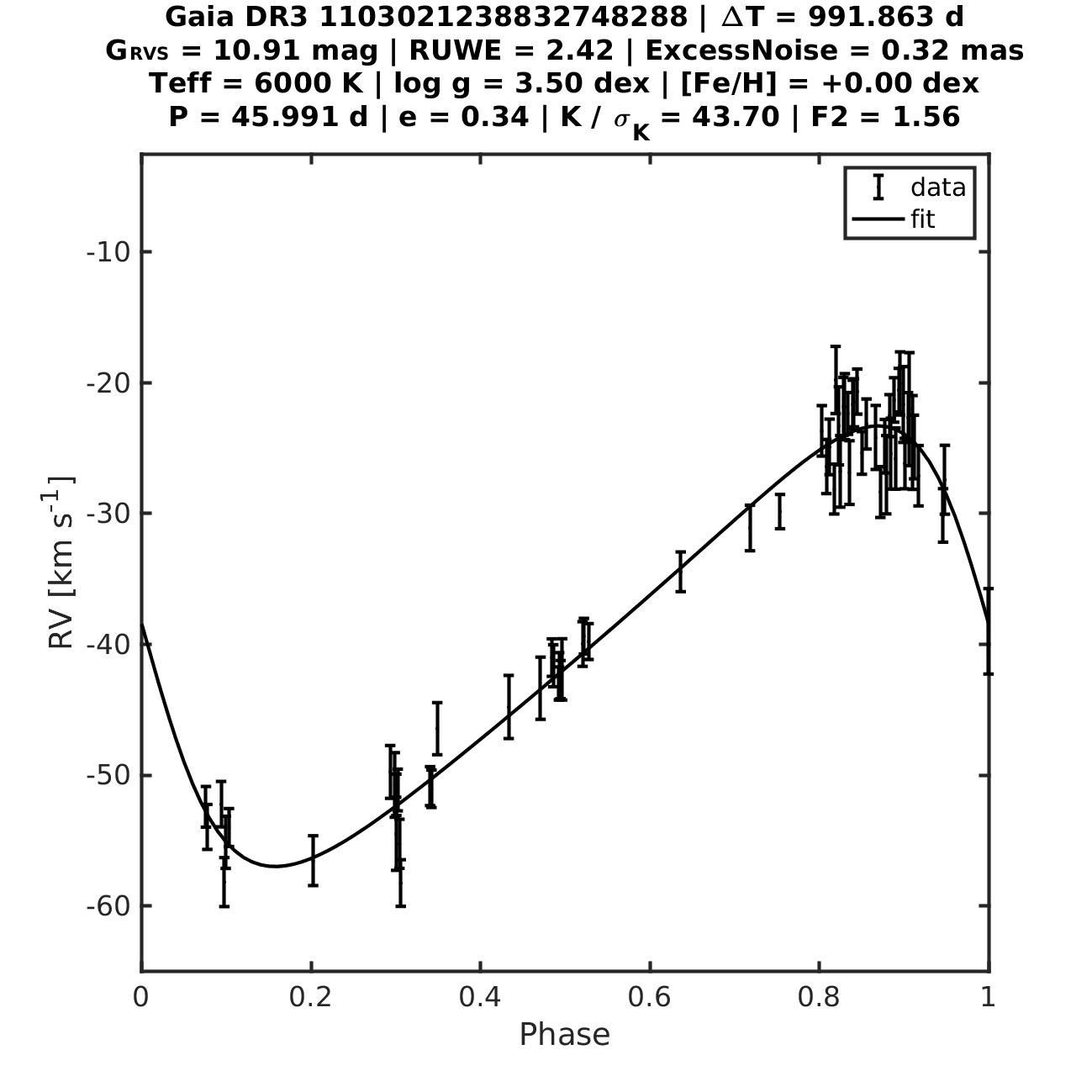}
}
\centerline{
\includegraphics[width=0.29\textwidth]{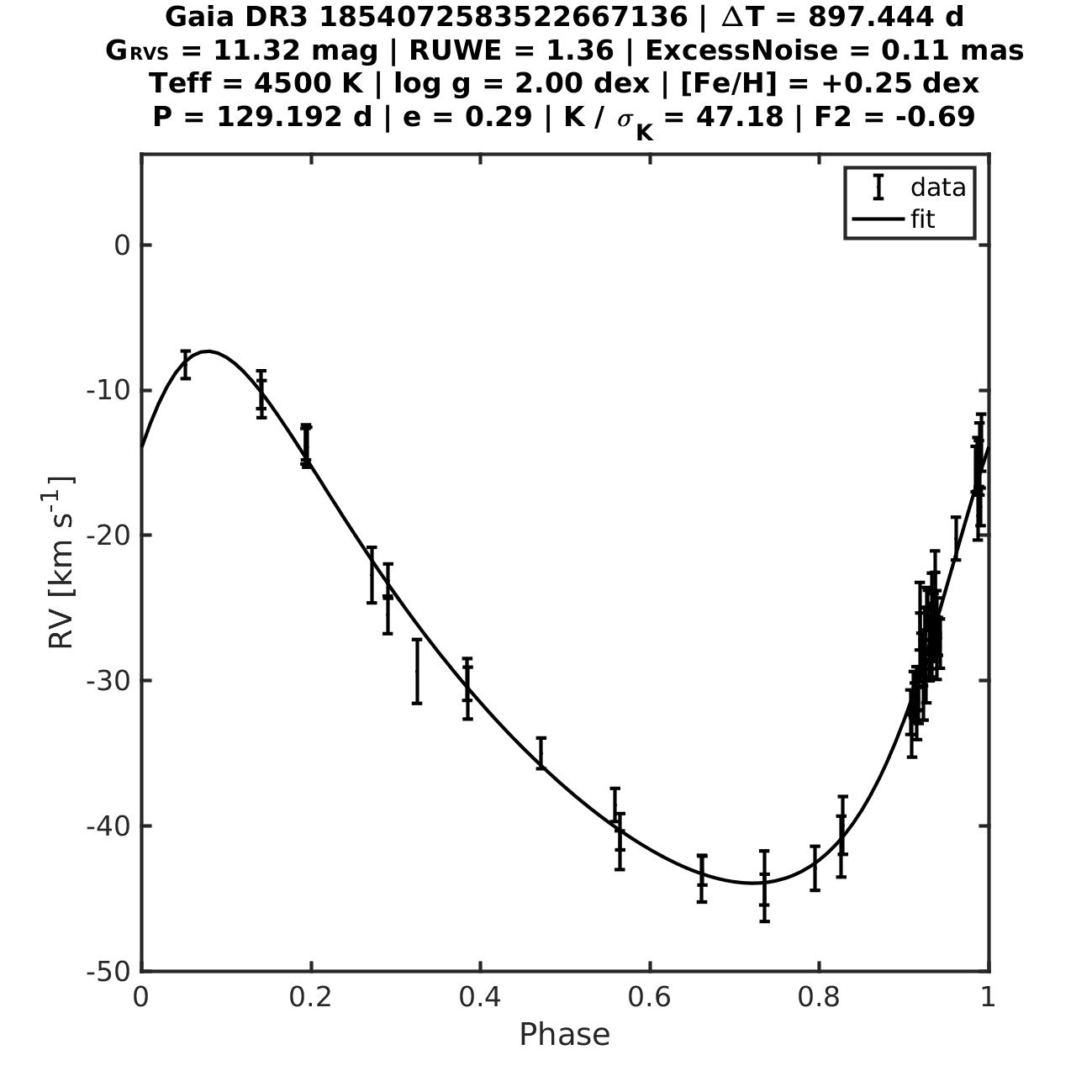}
\includegraphics[width=0.29\textwidth]{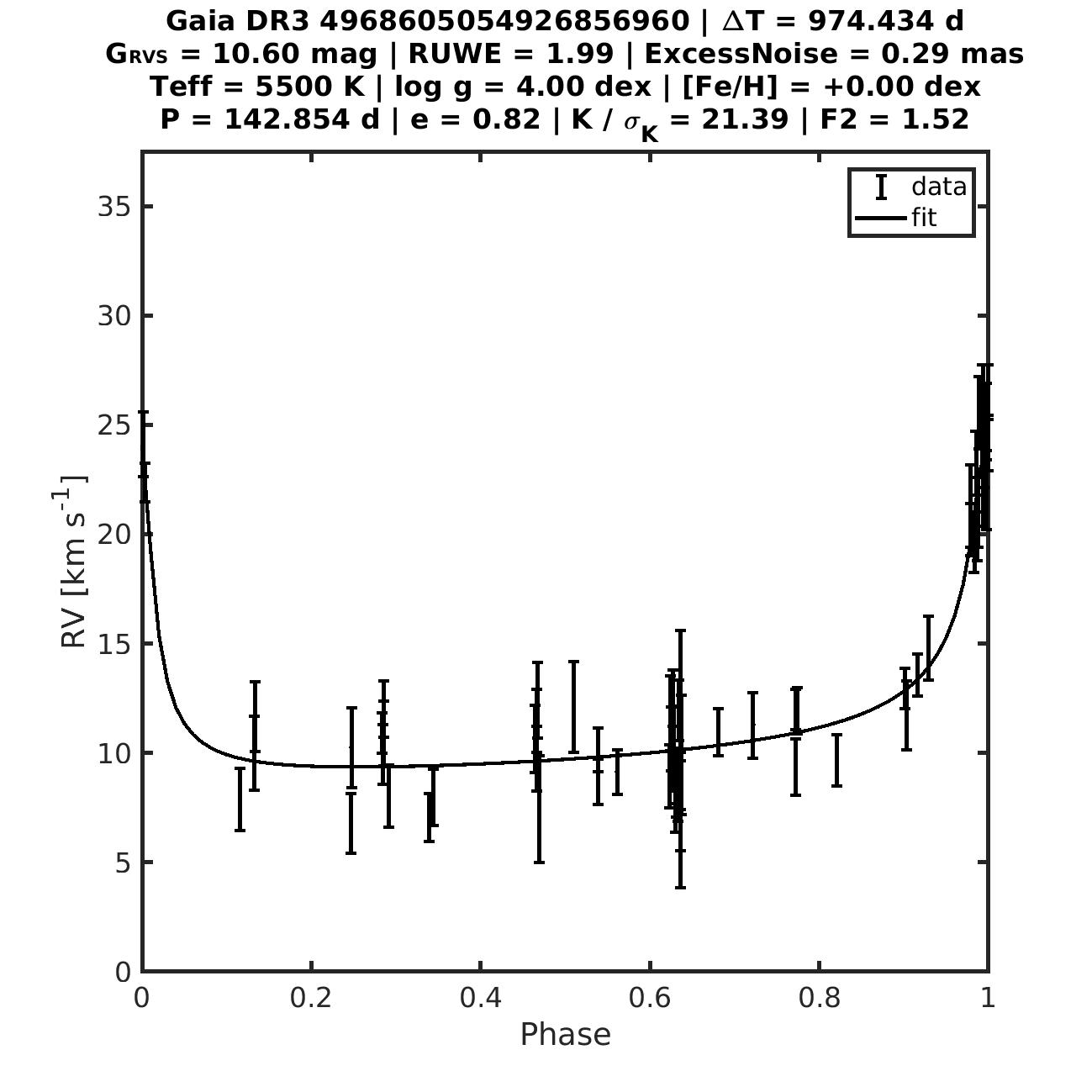}
\includegraphics[width=0.29\textwidth]{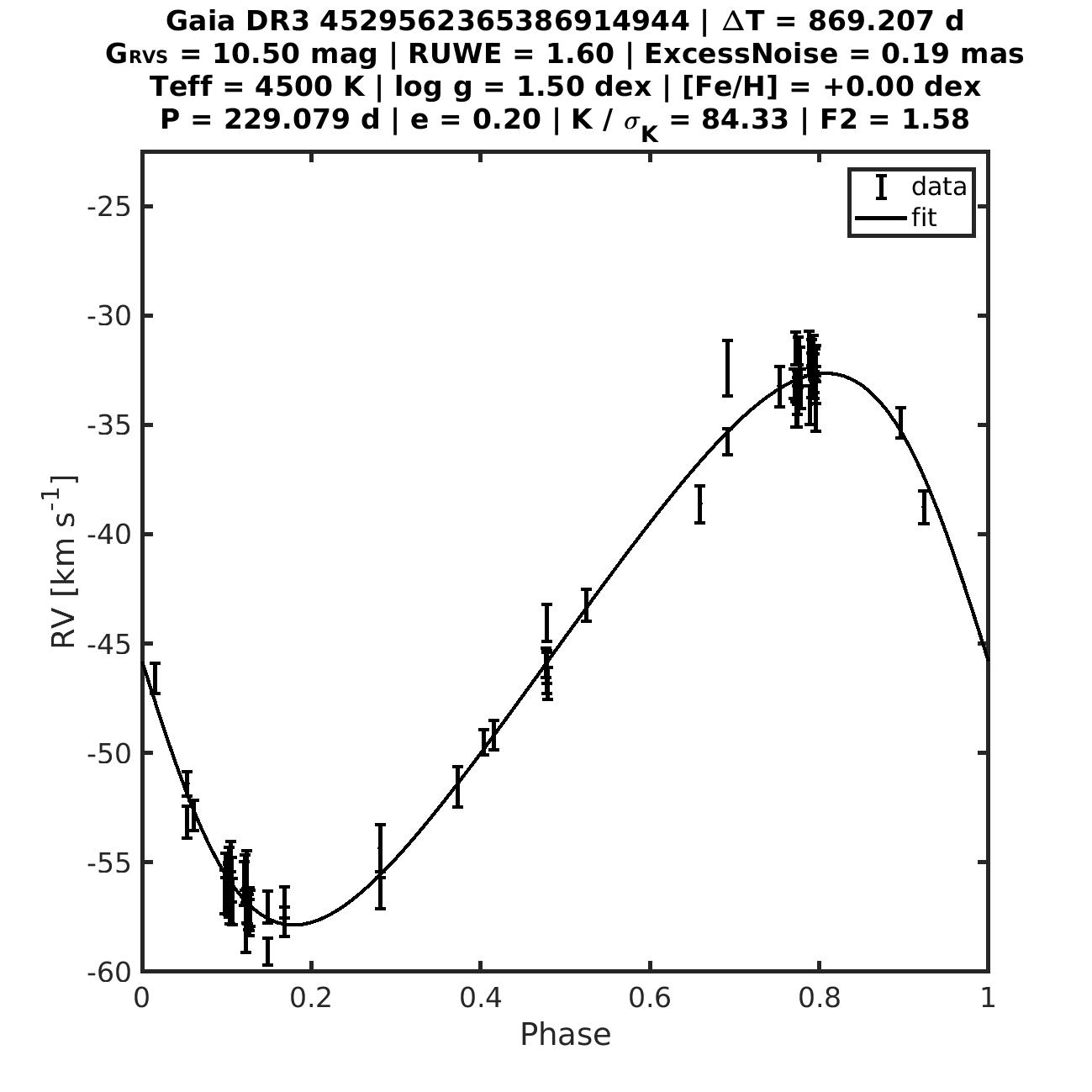}
}
\caption{continued.}
\label{fig:appgoodressb1part2}
\end{figure*}
\begin{figure*}[!htp]
\ContinuedFloat
\centerline{
\includegraphics[width=0.29\textwidth]{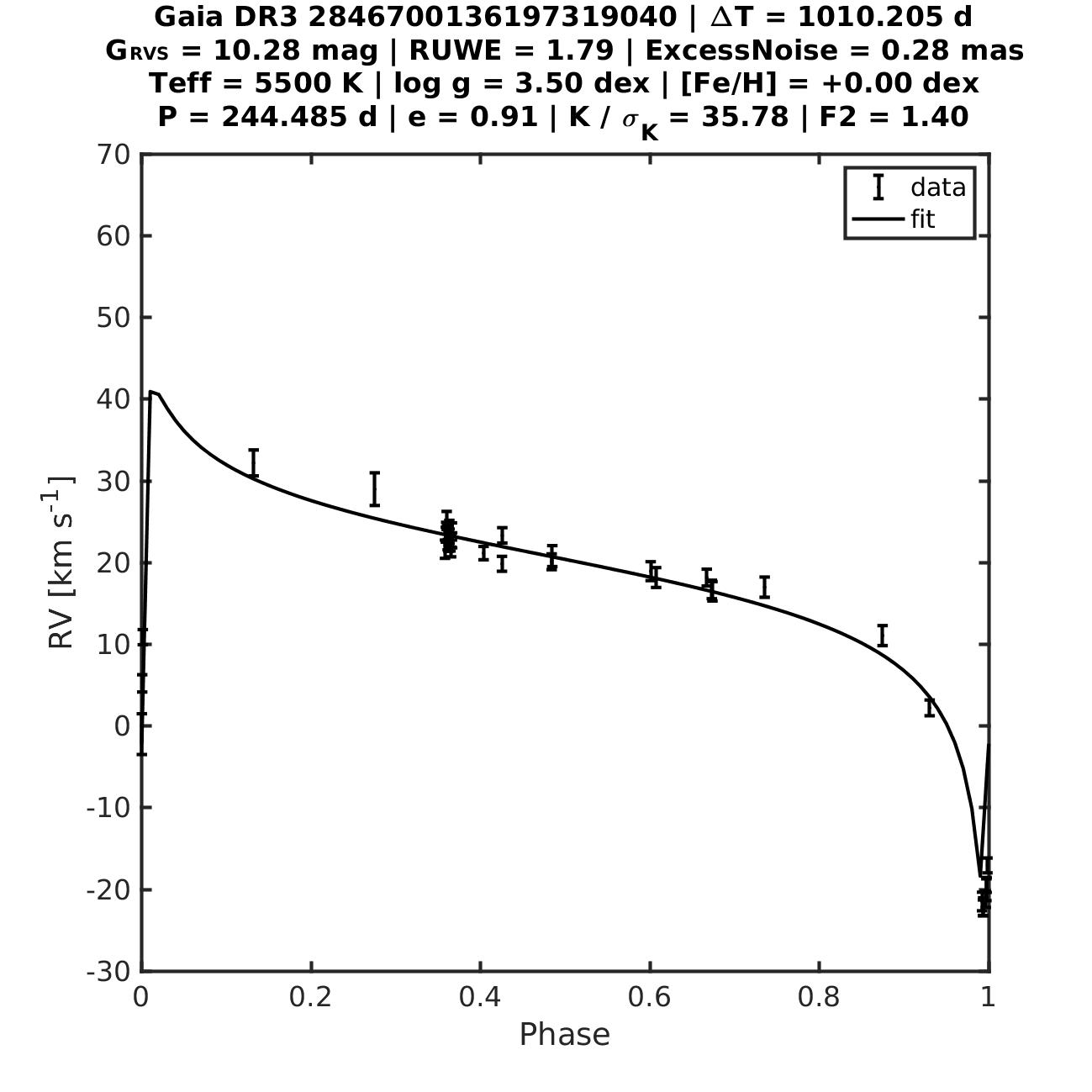}
\includegraphics[width=0.29\textwidth]{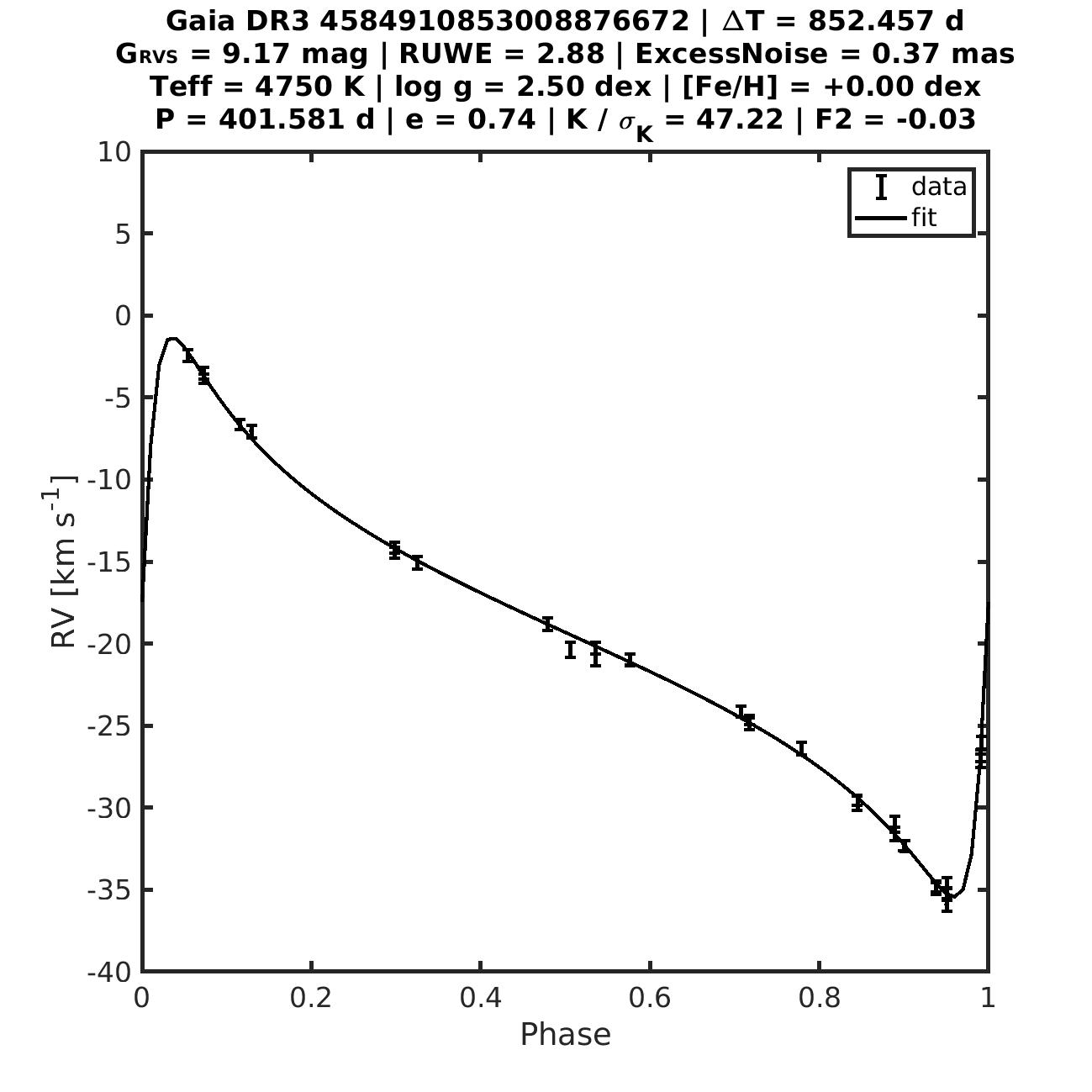} 
\includegraphics[width=0.29\textwidth]{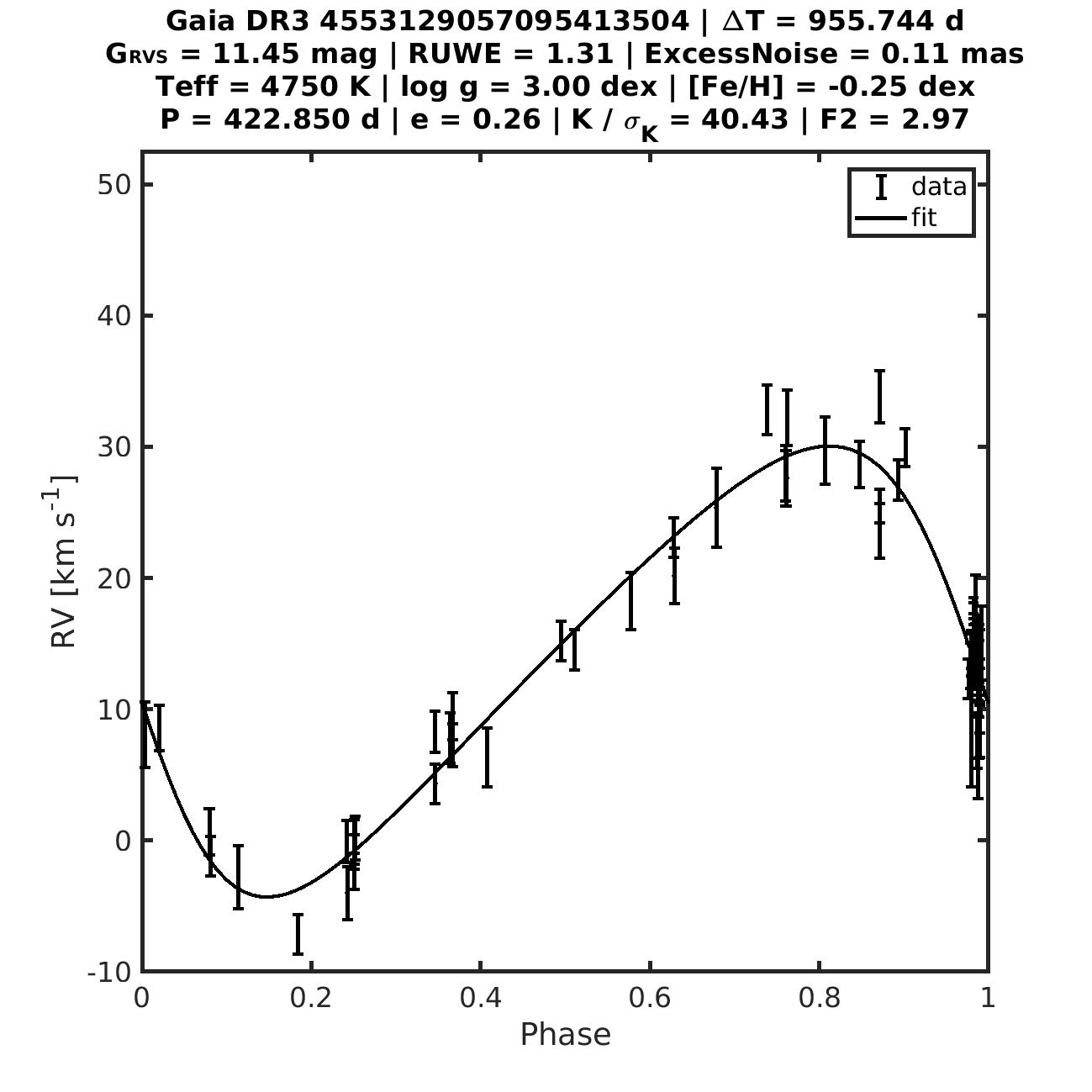} 
}
\centerline{
\includegraphics[width=0.29\textwidth]{./FIGS/pp_1956836204422469888_SB1.jpg}
\includegraphics[width=0.29\textwidth]{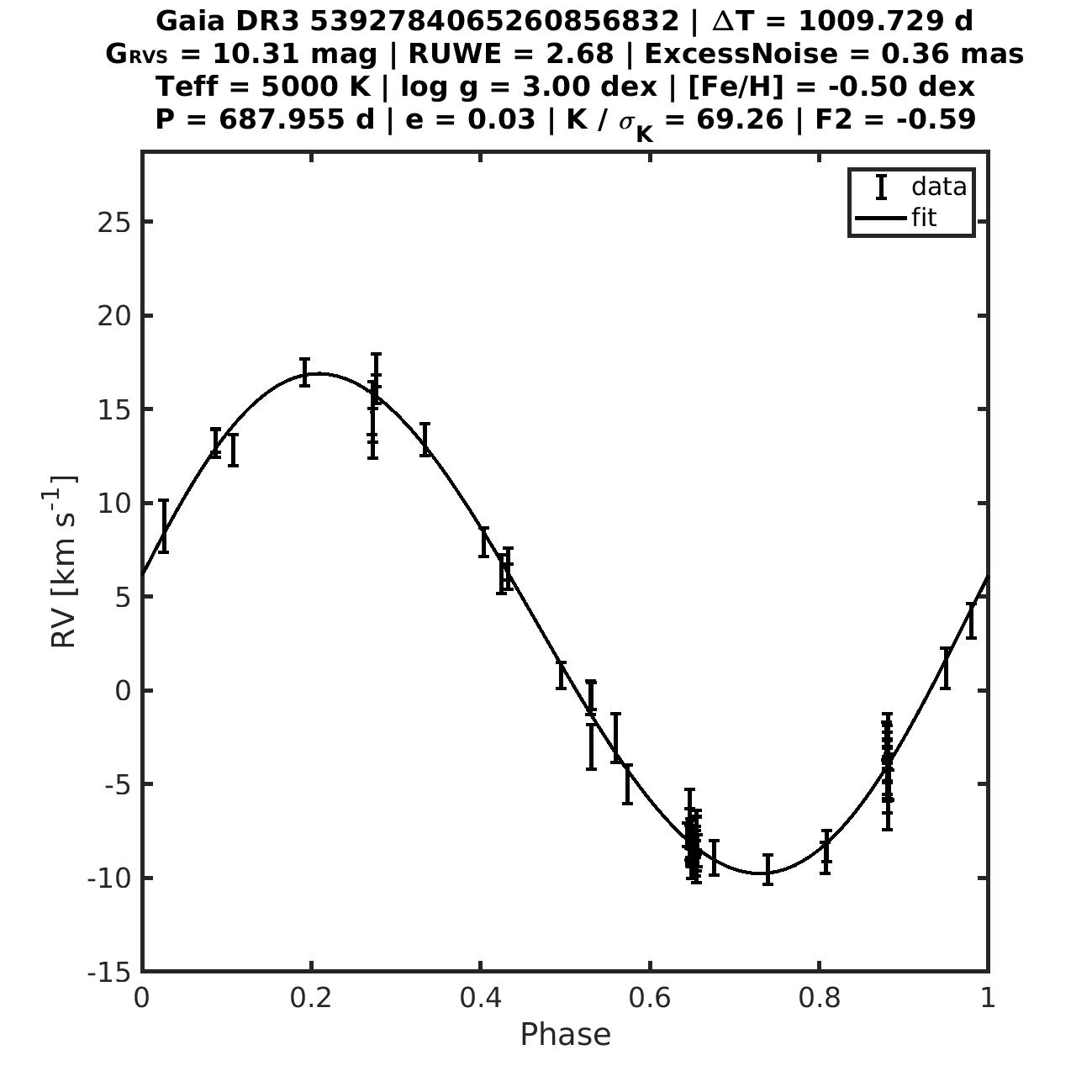}
\includegraphics[width=0.29\textwidth]{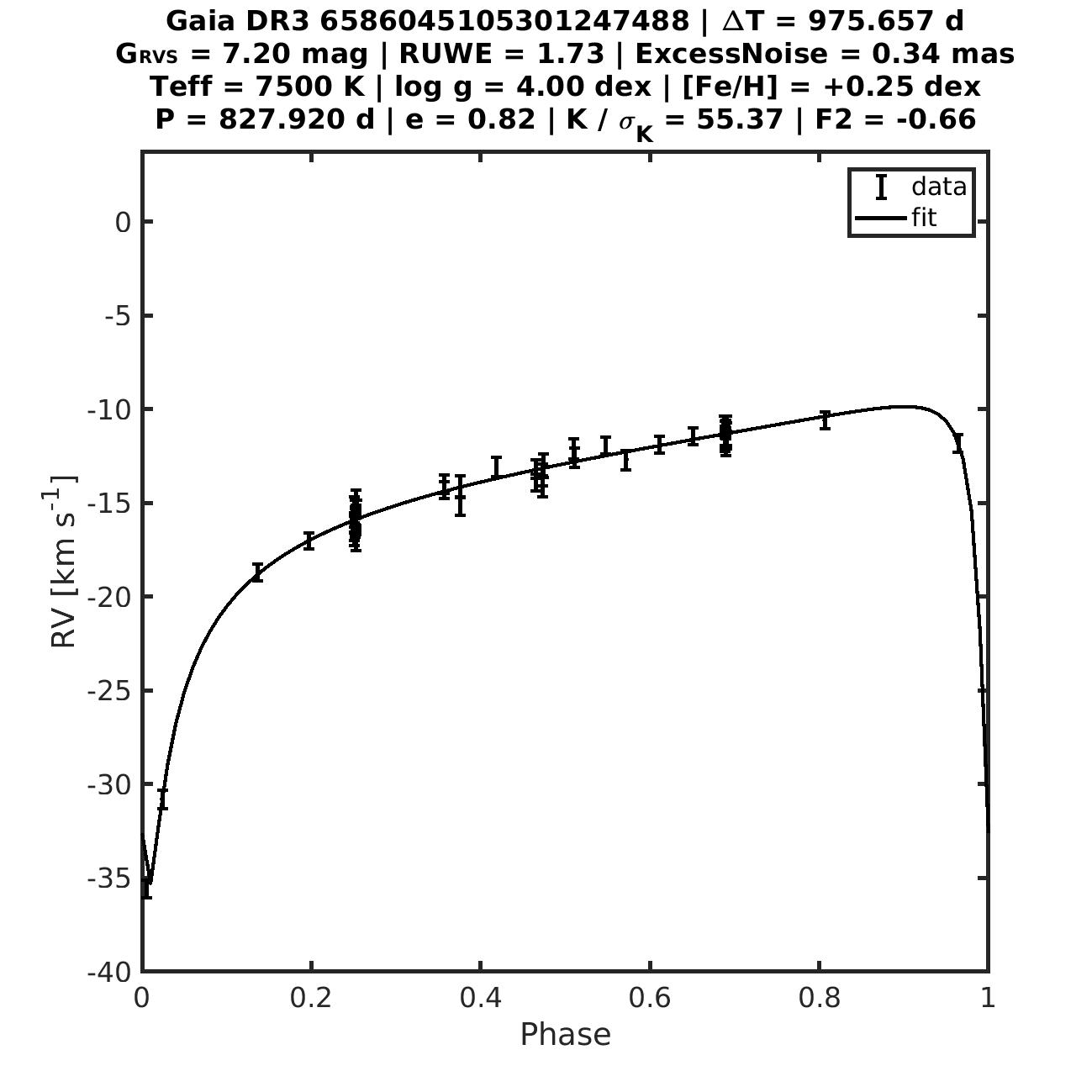}
}
\centerline{
\includegraphics[width=0.29\textwidth]{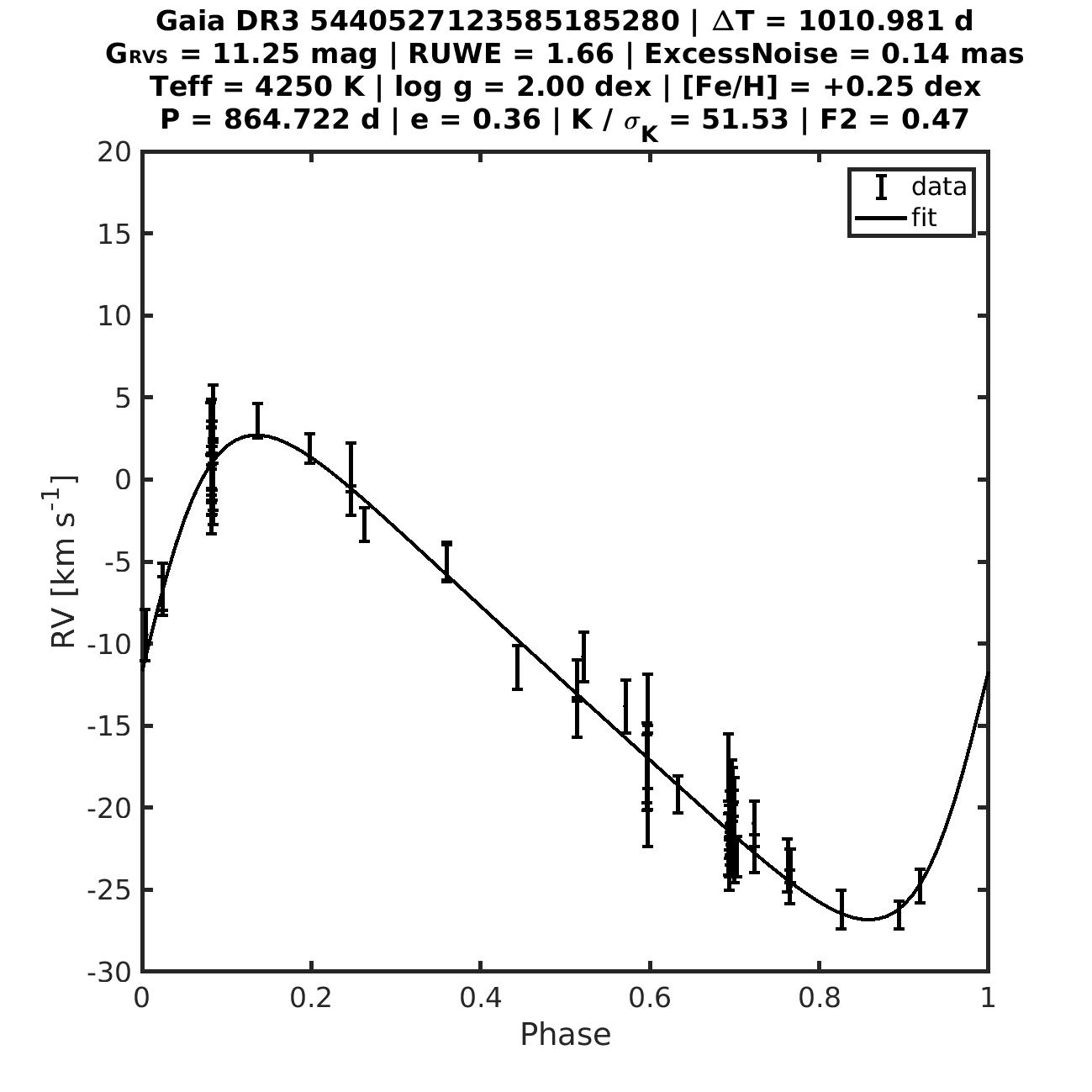}
\includegraphics[width=0.29\textwidth]{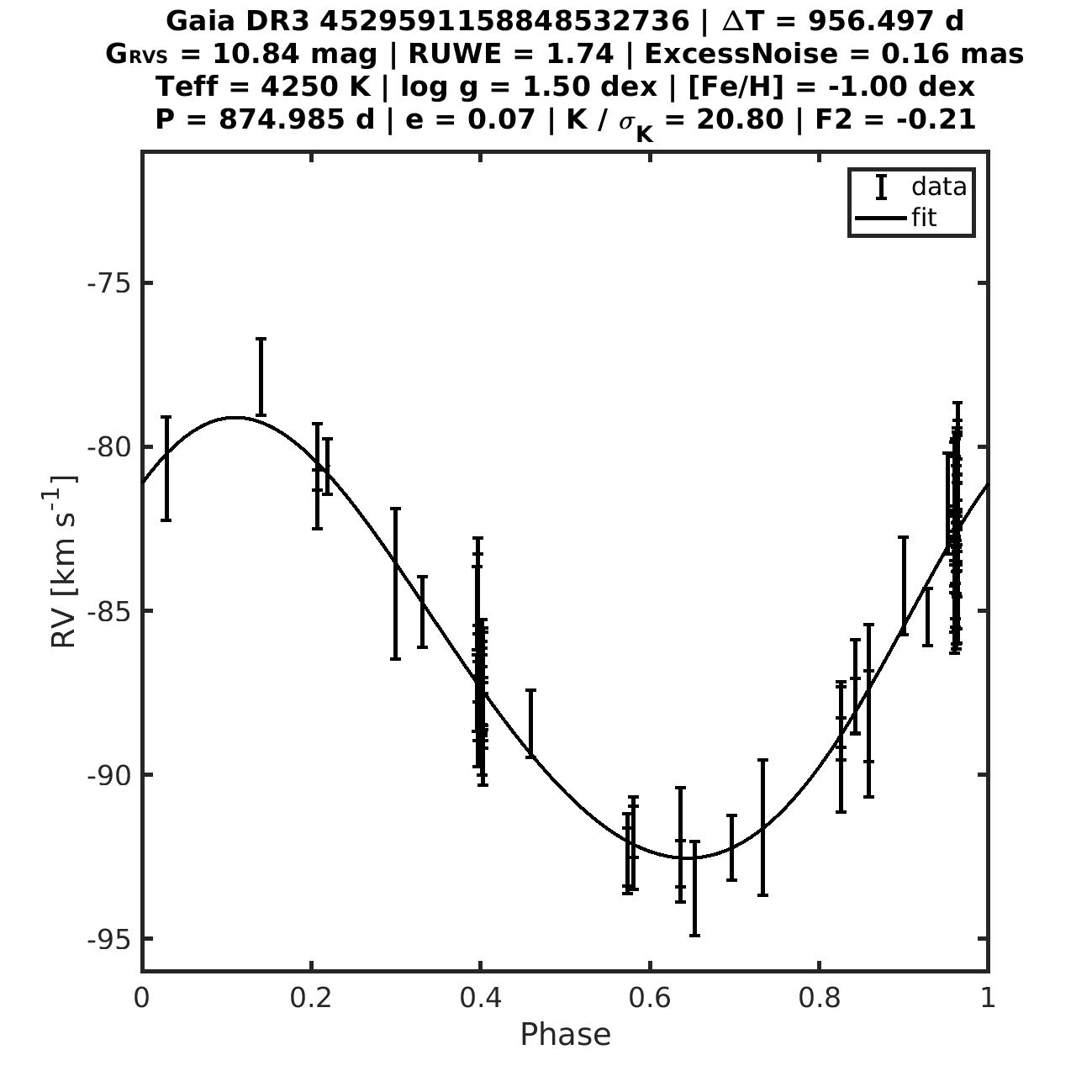}
\includegraphics[width=0.29\textwidth]{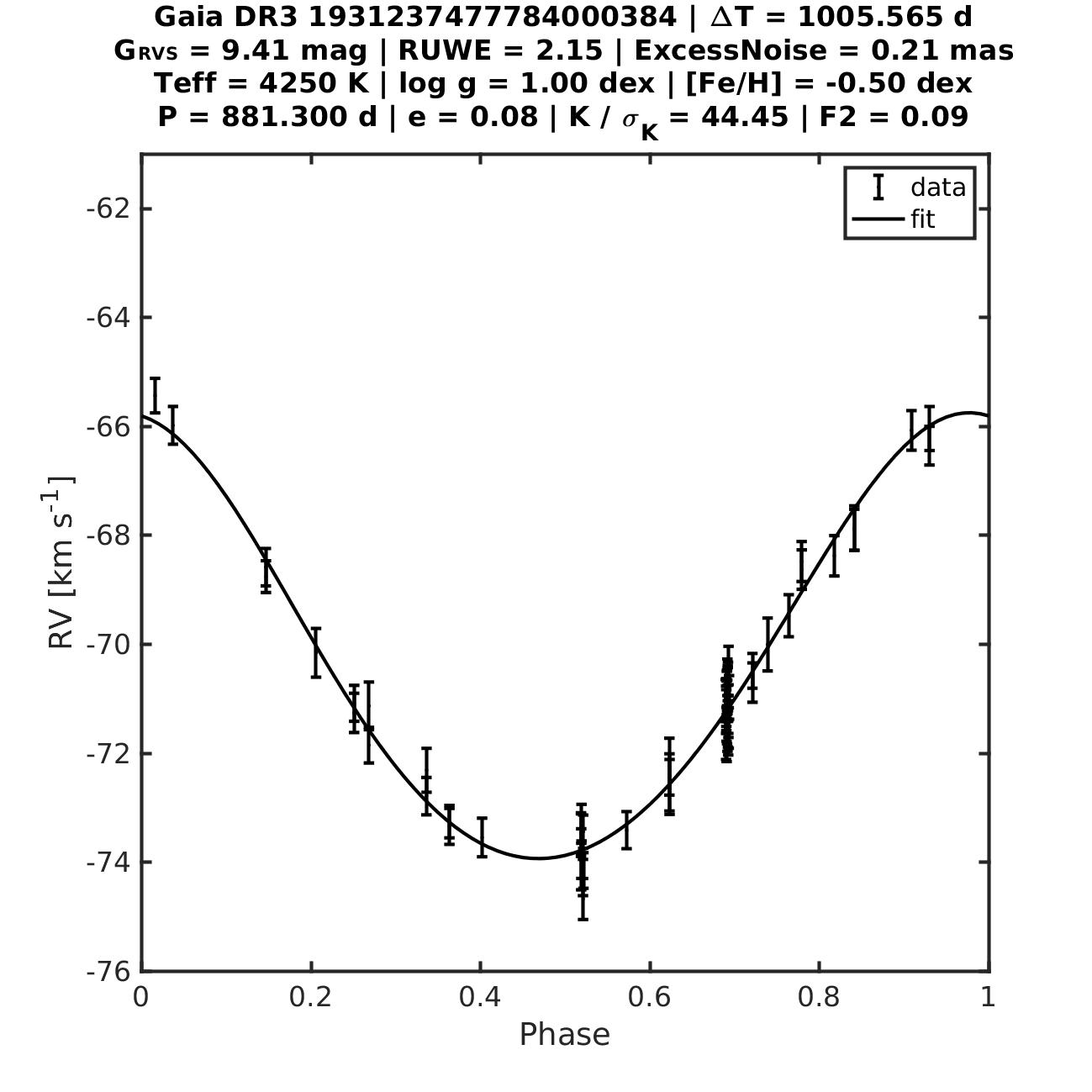}
}
\centerline{
\includegraphics[width=0.29\textwidth]{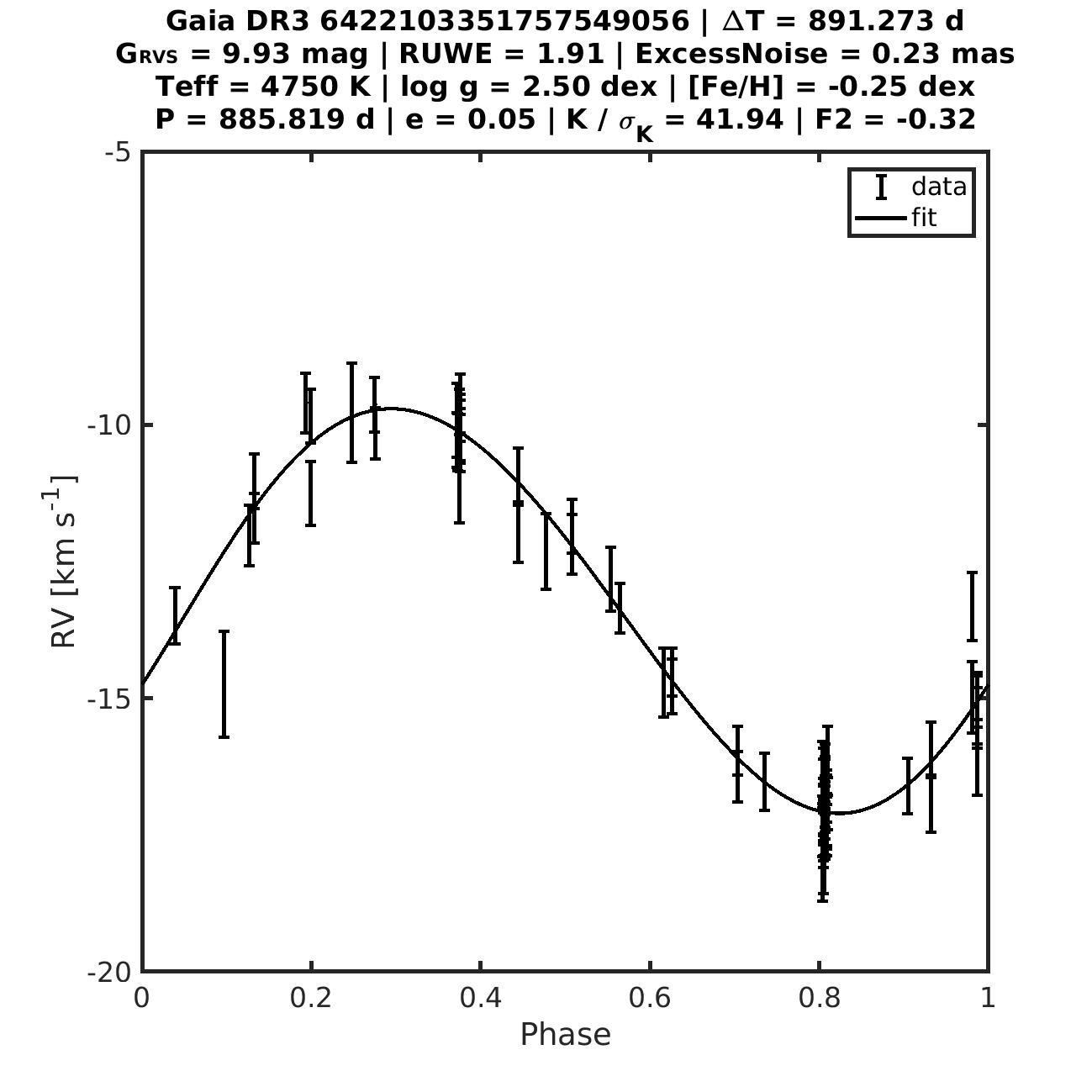}
\includegraphics[width=0.29\textwidth]{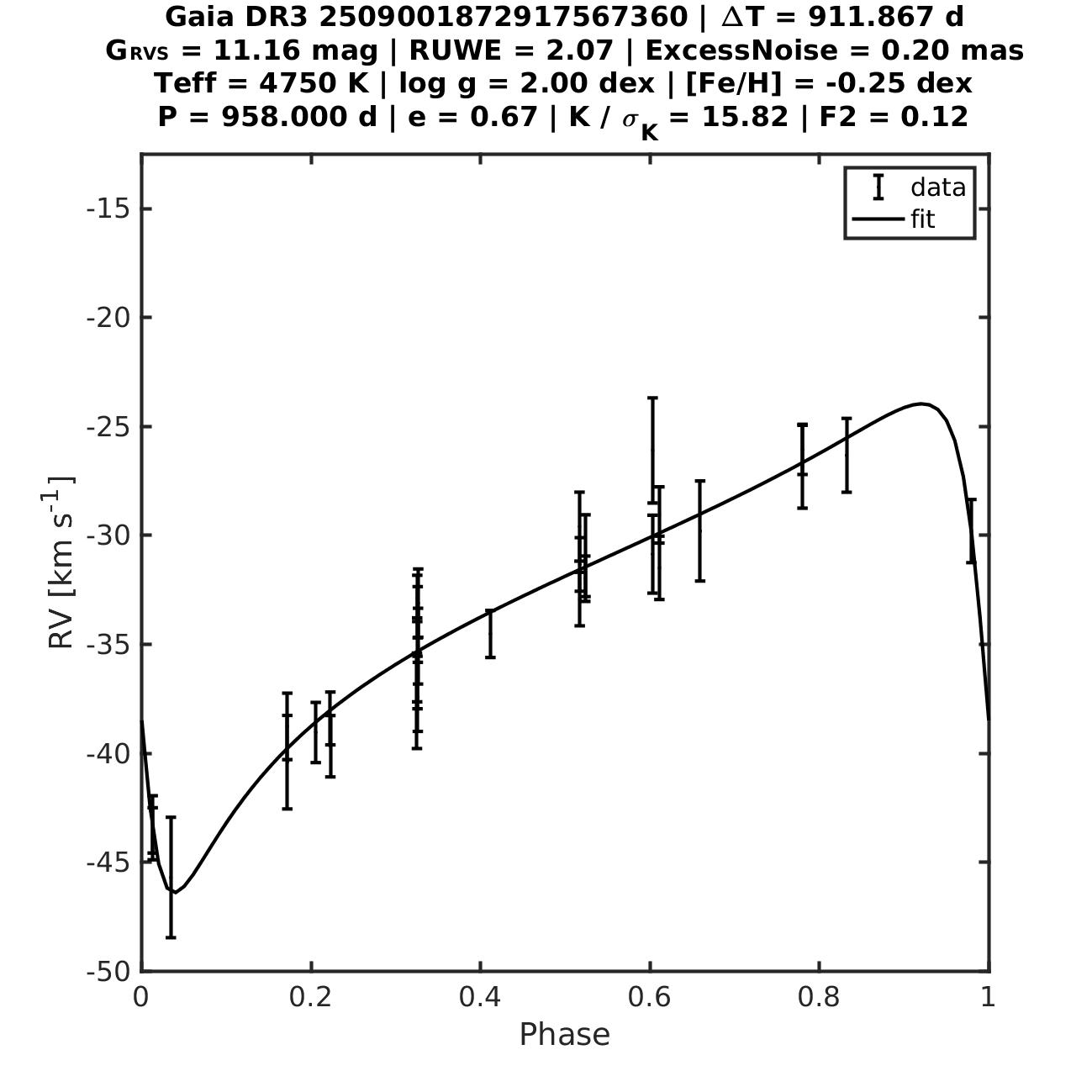}
\includegraphics[width=0.29\textwidth]{./FIGS/pp_5818903954144888320_SB1.jpg}
}
\caption{continued.}
\label{fig:appgoodressb1part3}
\end{figure*}
\begin{figure*}[!htp]
\centerline{
\includegraphics[width=0.29\textwidth]{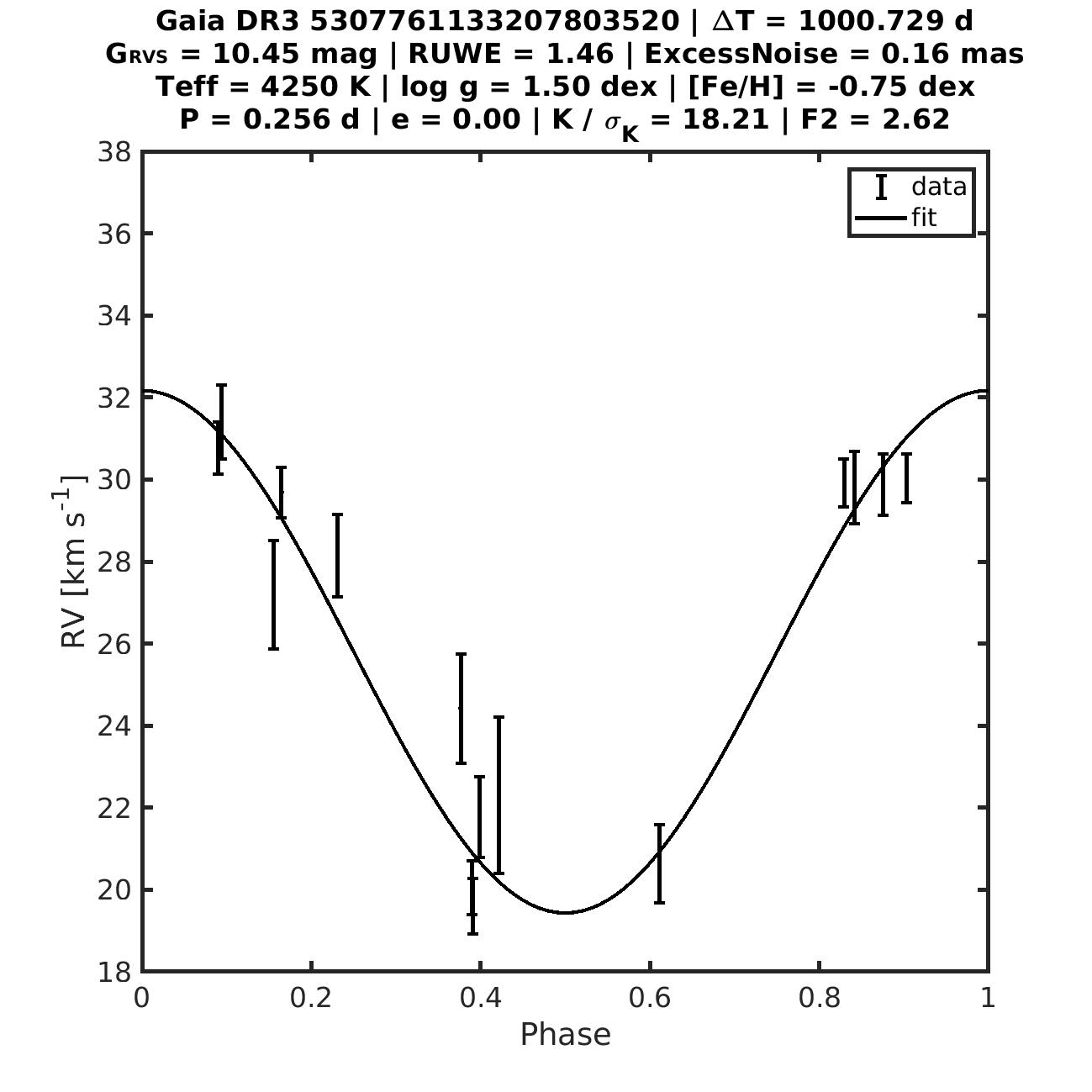}
\includegraphics[width=0.29\textwidth]{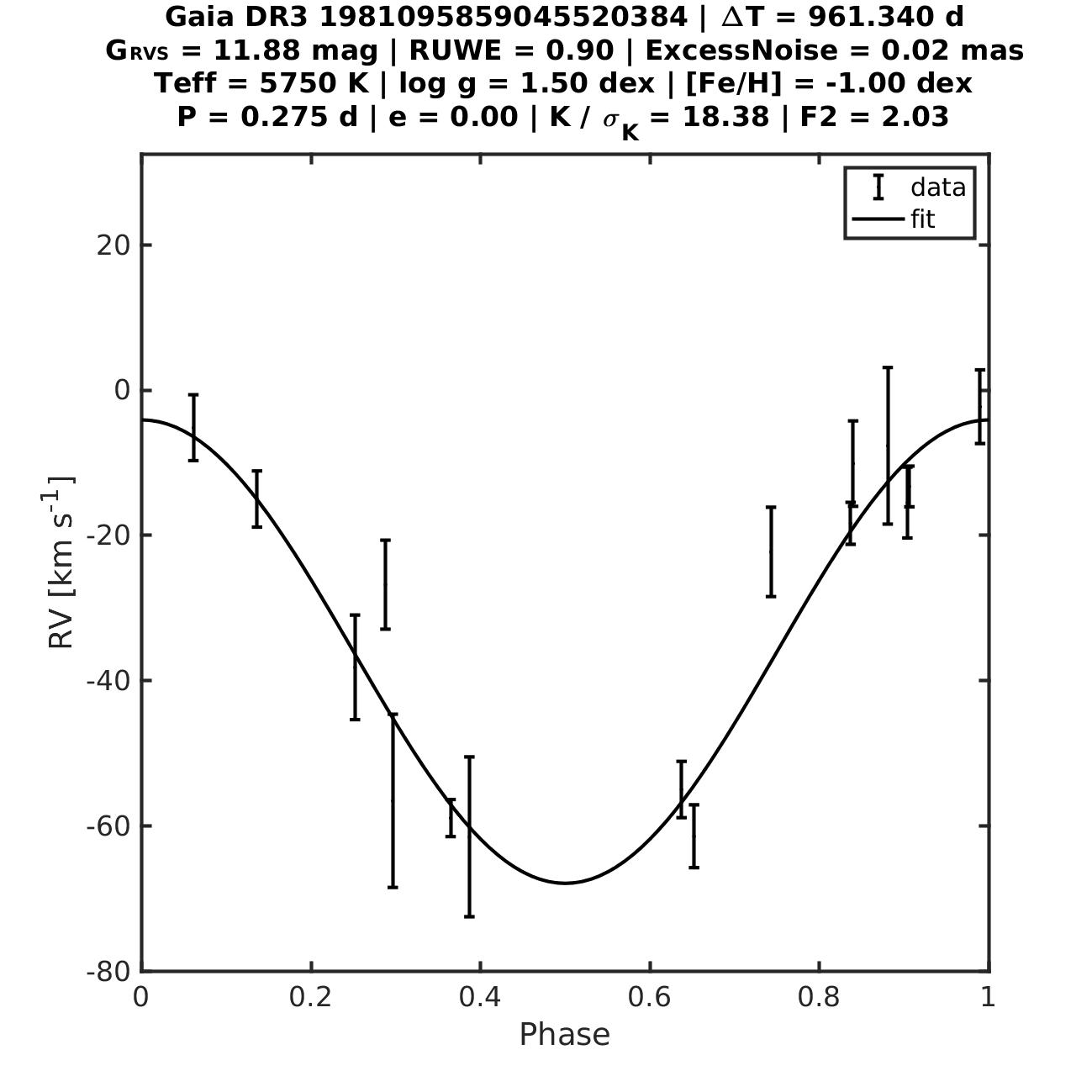} 
\includegraphics[width=0.29\textwidth]{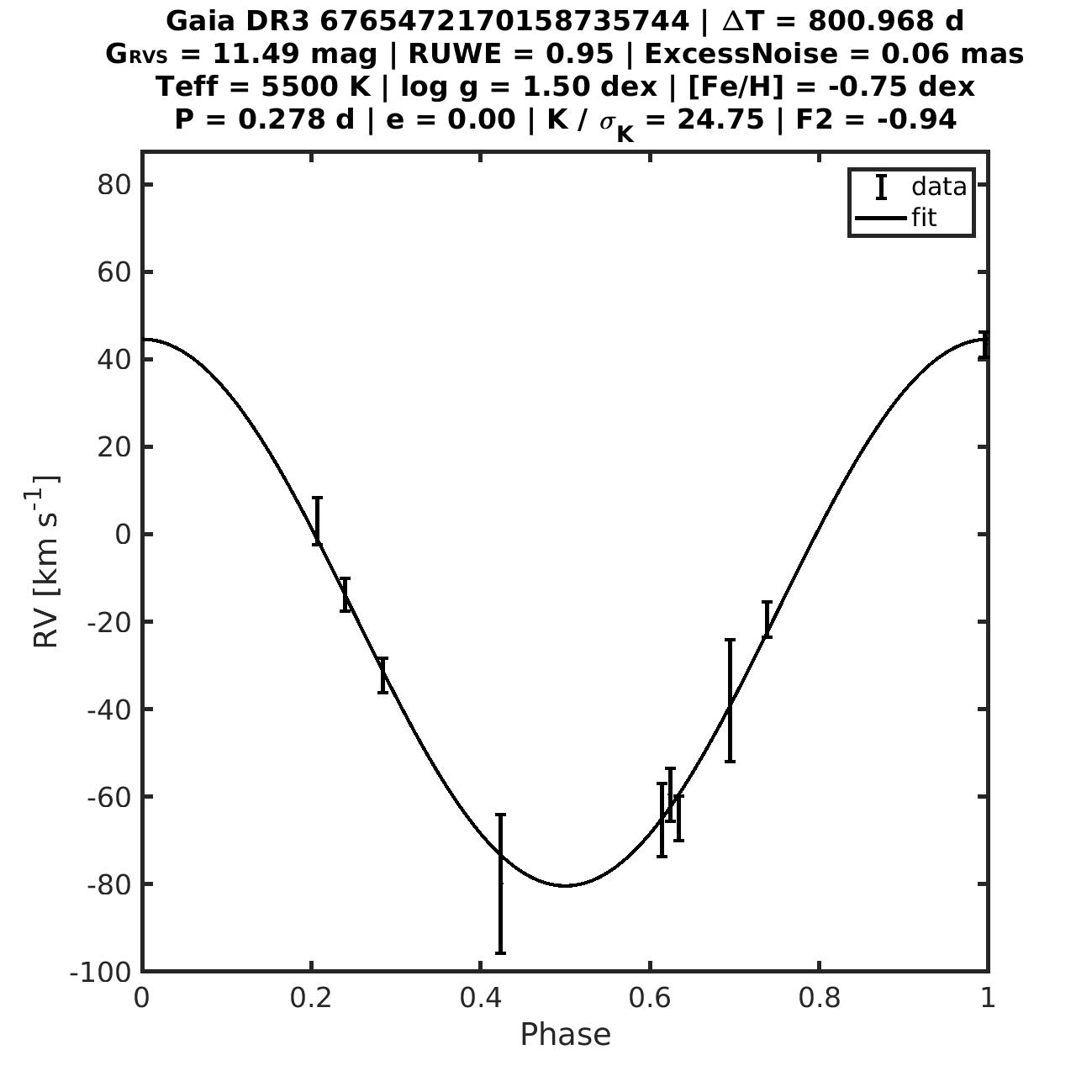} 
}
\centerline{
\includegraphics[width=0.29\textwidth]{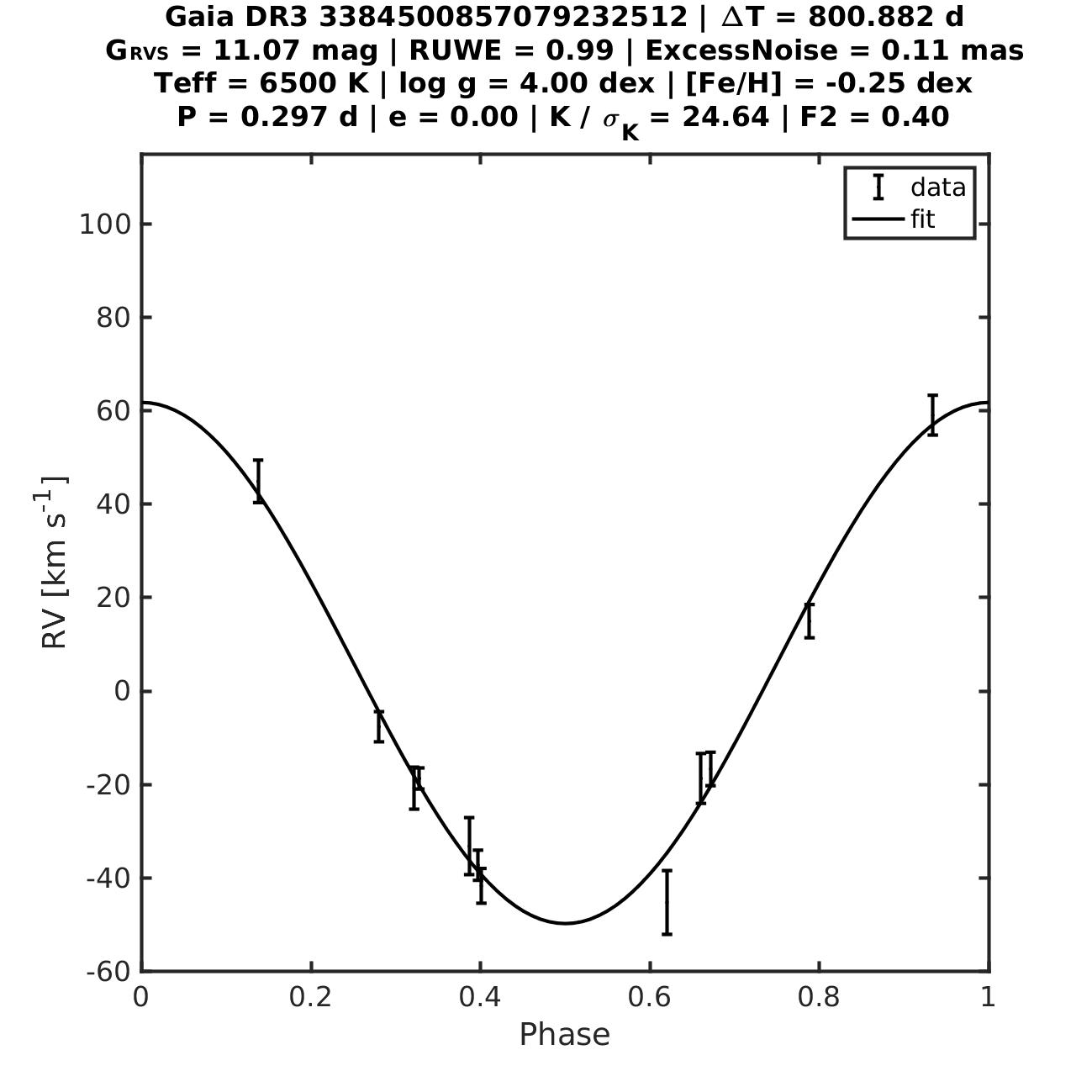}
\includegraphics[width=0.29\textwidth]{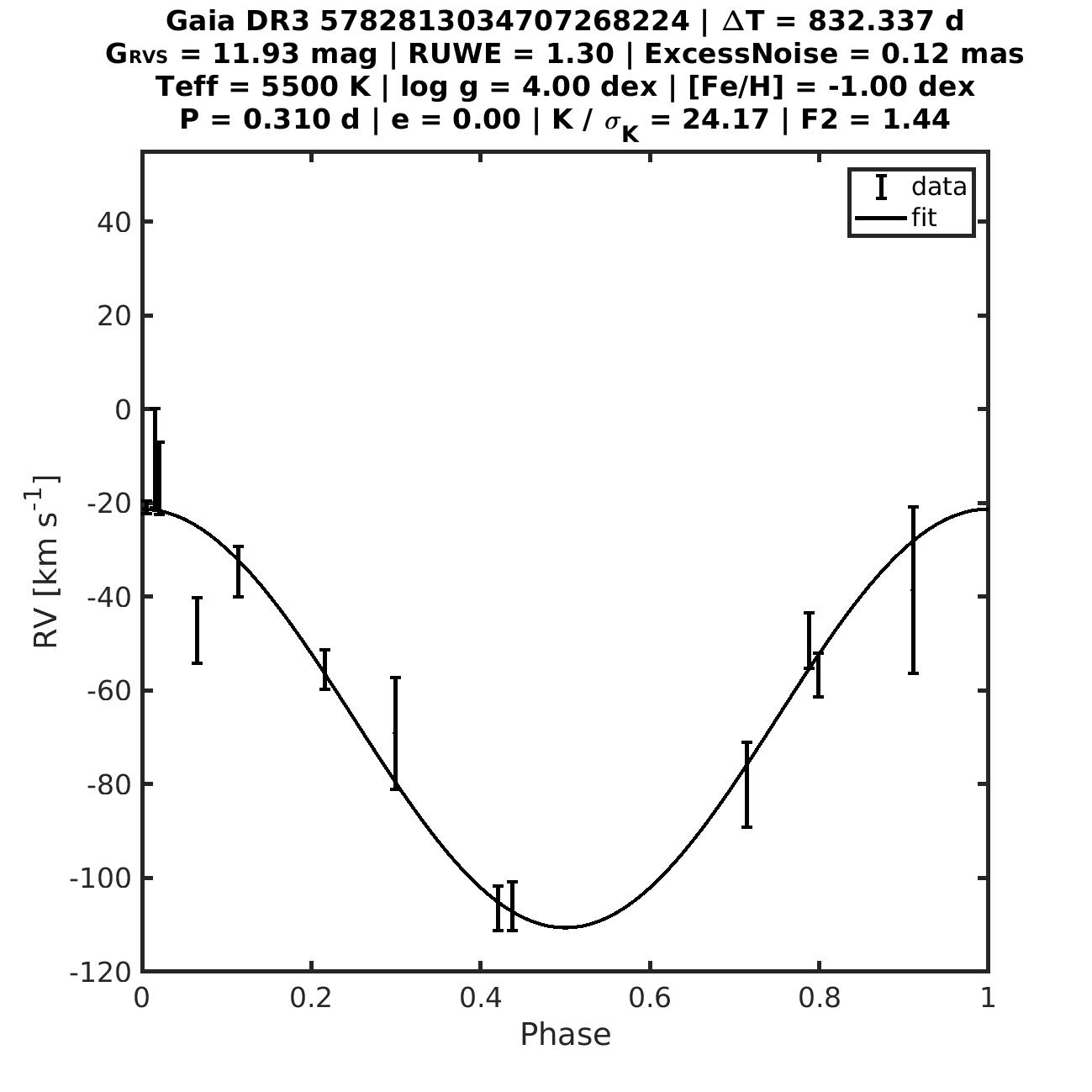}
\includegraphics[width=0.29\textwidth]{./FIGS/pp_5336447043659387136_SB1C.jpg}
}
\centerline{
\includegraphics[width=0.29\textwidth]{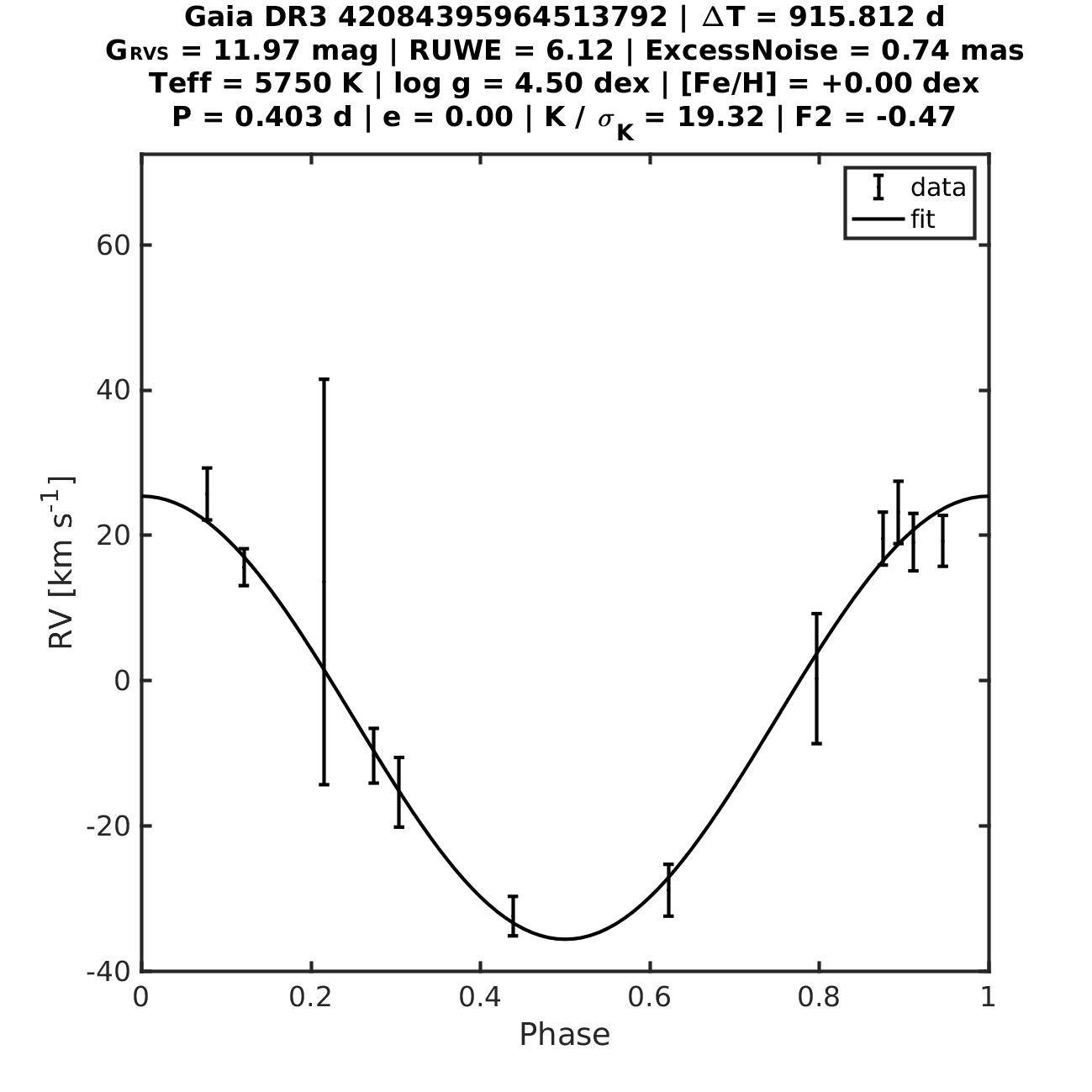}
\includegraphics[width=0.29\textwidth]{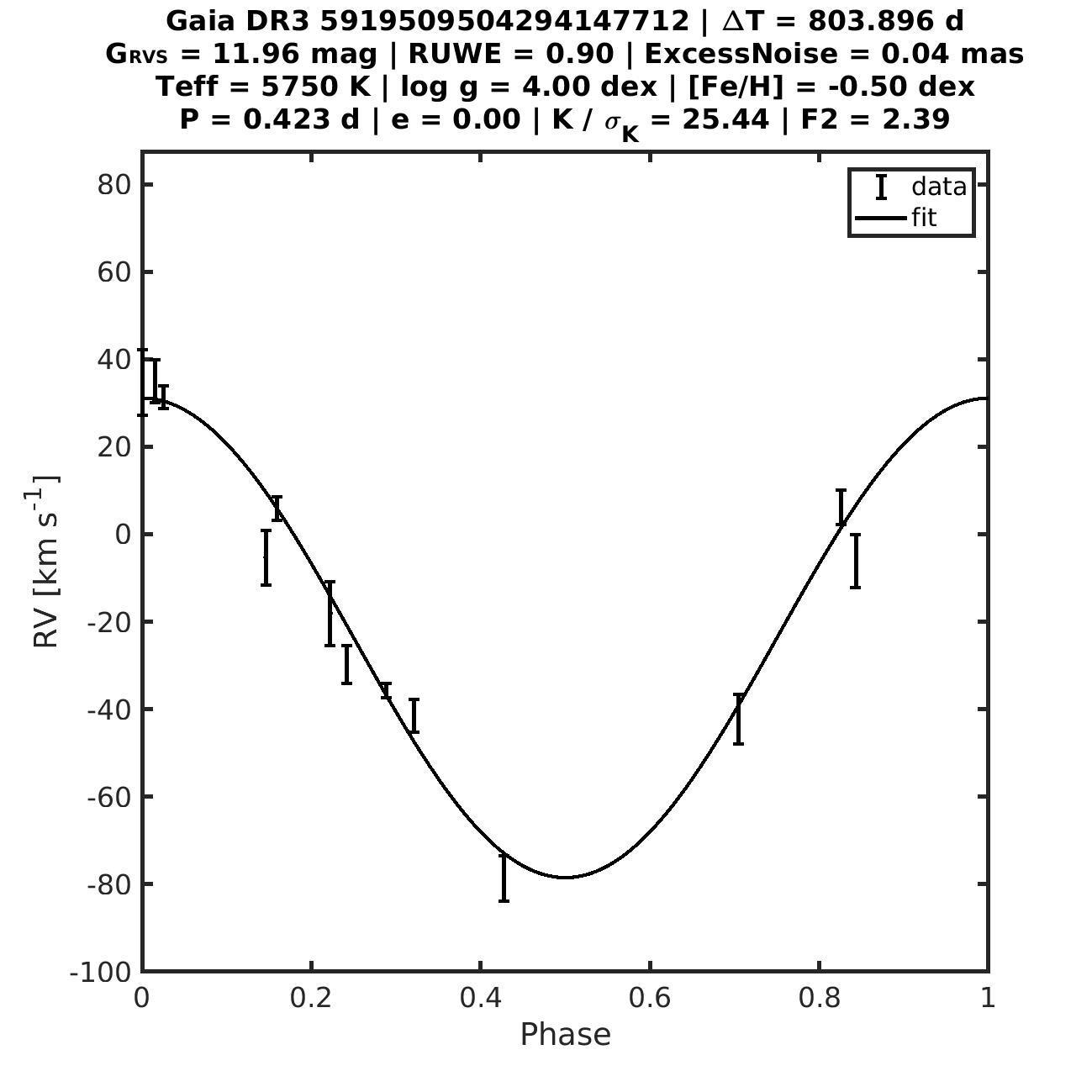}
\includegraphics[width=0.29\textwidth]{./FIGS/pp_1871154871054688384_SB1C.jpg}
}
\centerline{
\includegraphics[width=0.29\textwidth]{./FIGS/pp_1811534398483358336_SB1C.jpg}
\includegraphics[width=0.29\textwidth]{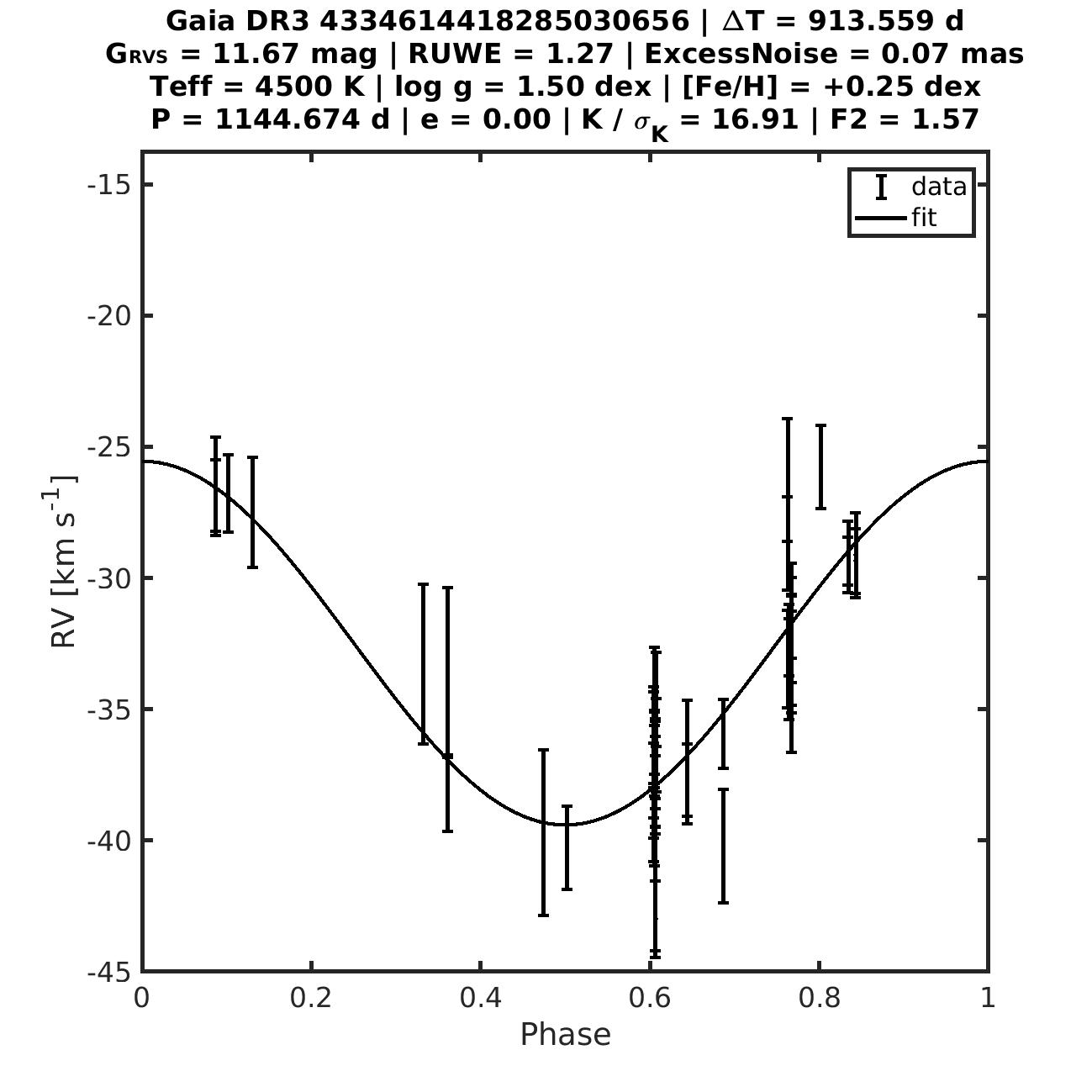}
\includegraphics[width=0.29\textwidth]{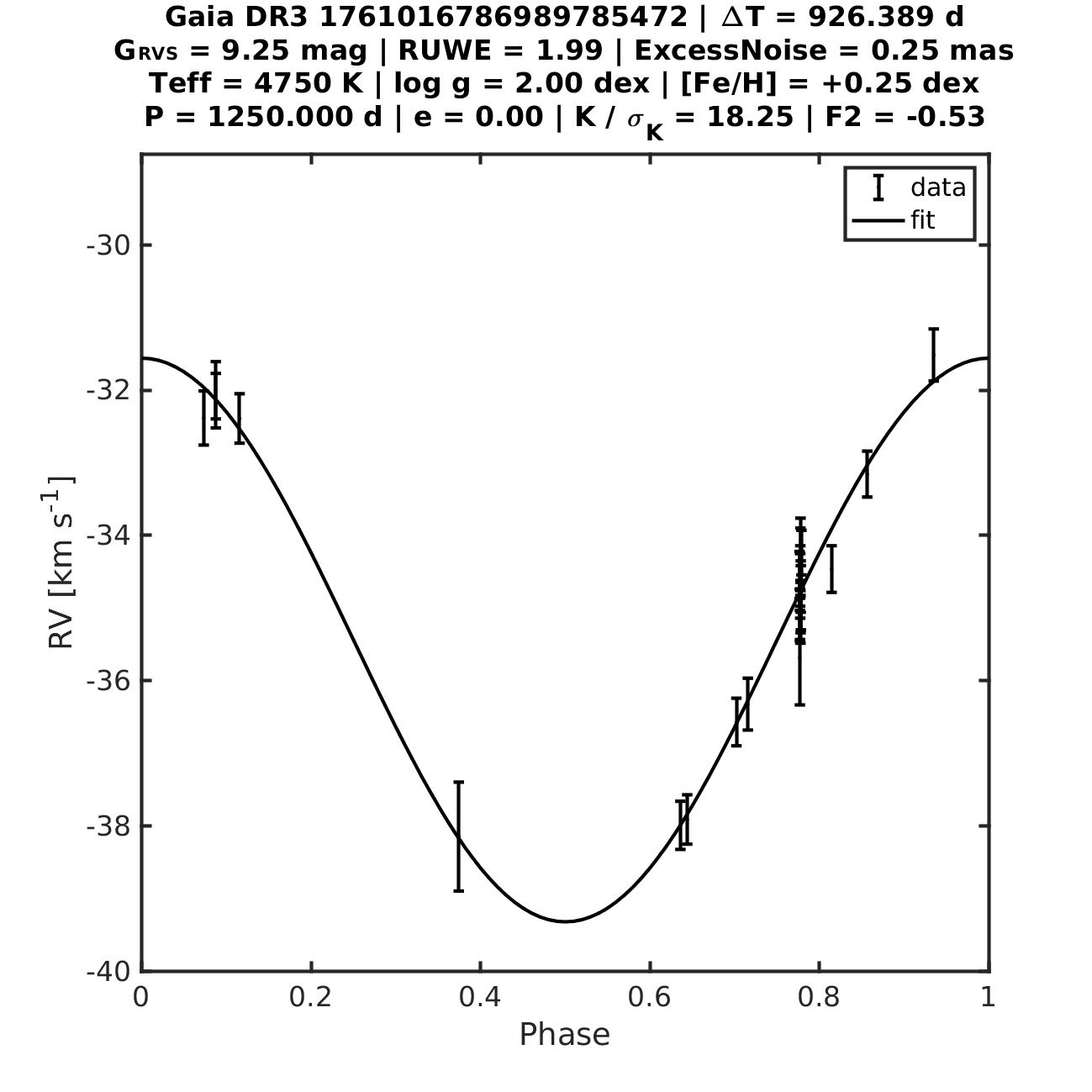}
}
\caption{Good results: same as Fig.\,\ref{fig:appgoodressb1part1} 
but concerning the SB1C-type solutions.}
\label{fig:appgoodressb1c}
\end{figure*}
\begin{figure*}[!htp]
\centerline{
\includegraphics[width=0.29\textwidth]{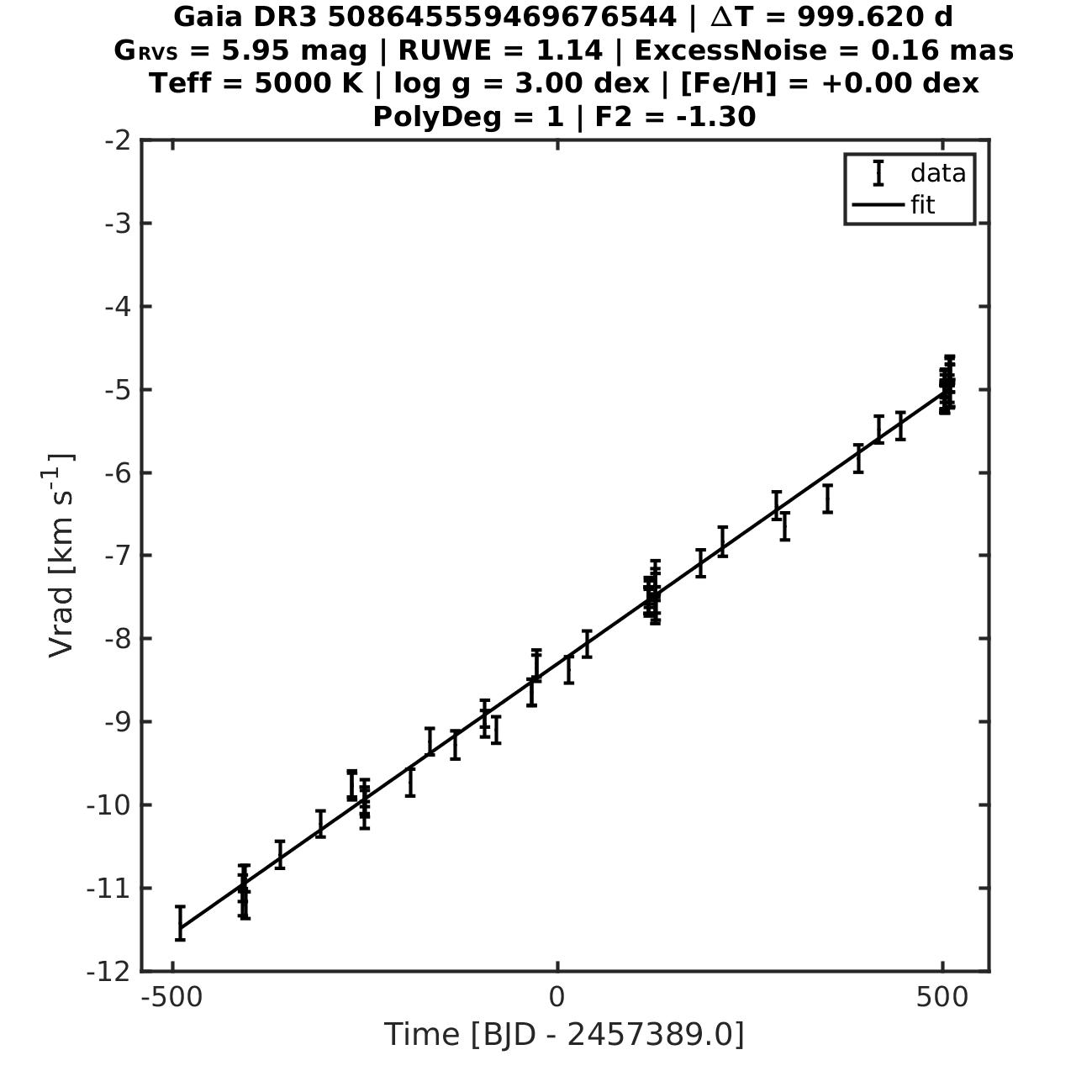}
\includegraphics[width=0.29\textwidth]{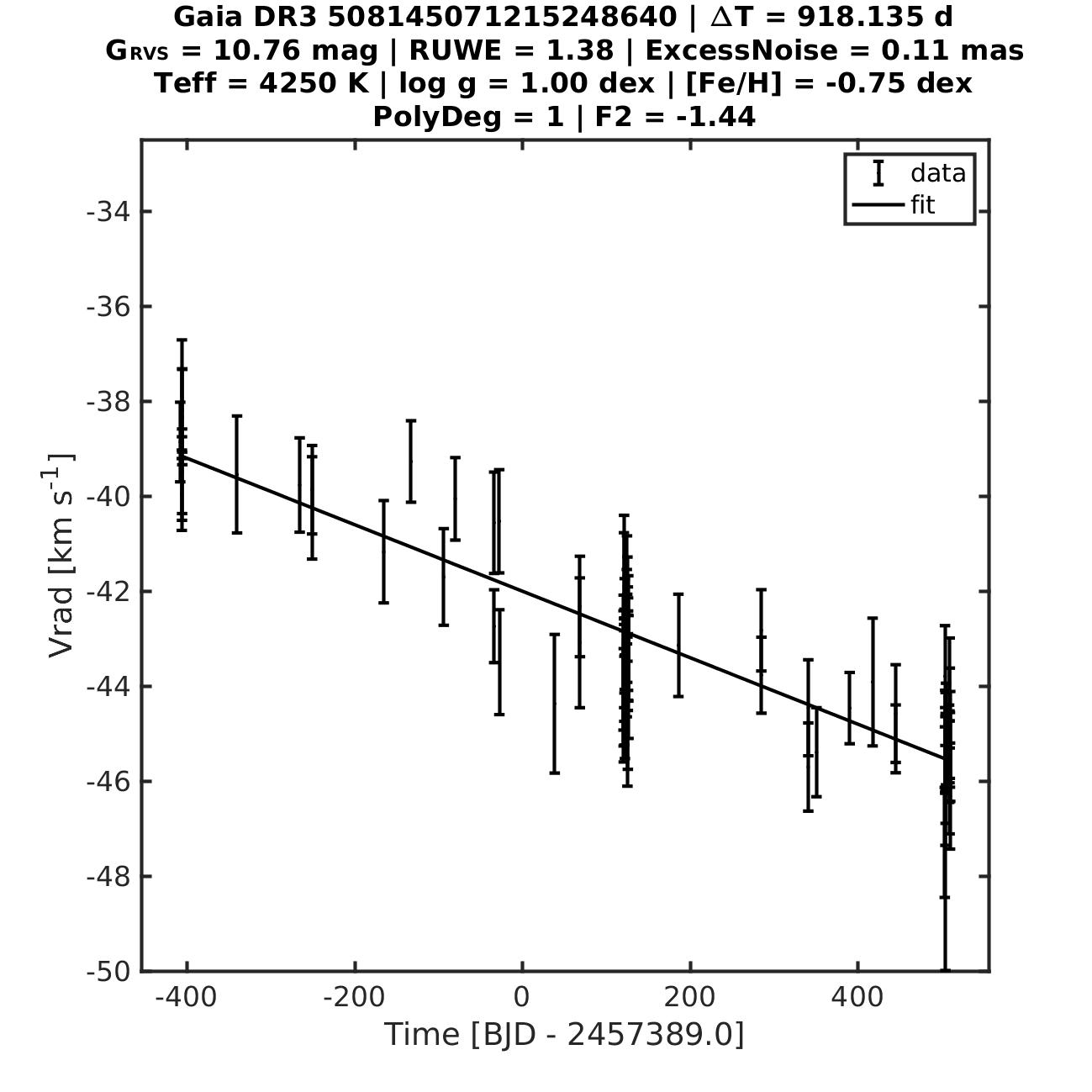}
\includegraphics[width=0.29\textwidth]{./FIGS/pp_4979506918874792576_TrendSB1.jpg}
}
\centerline{
\includegraphics[width=0.29\textwidth]{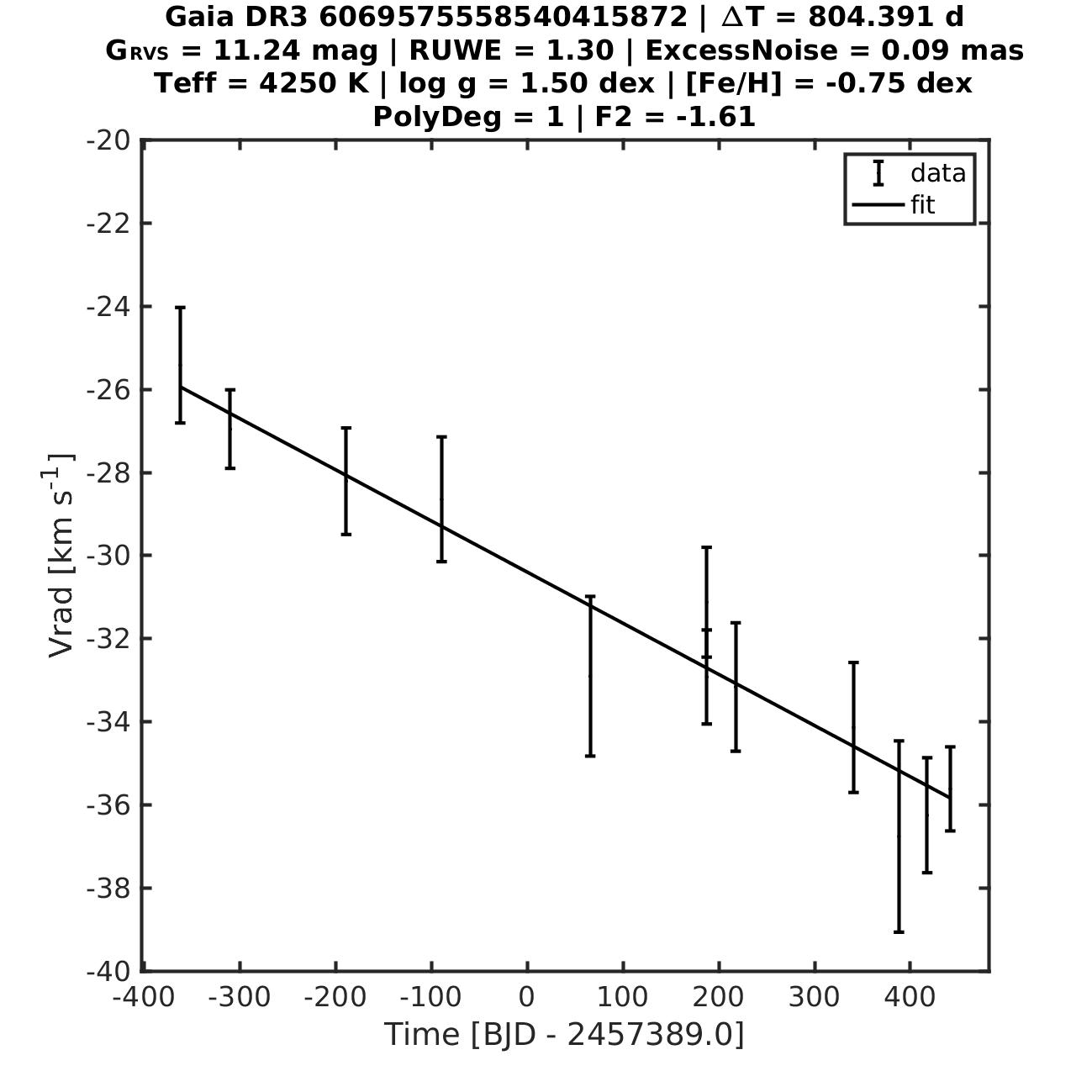}
\includegraphics[width=0.29\textwidth]{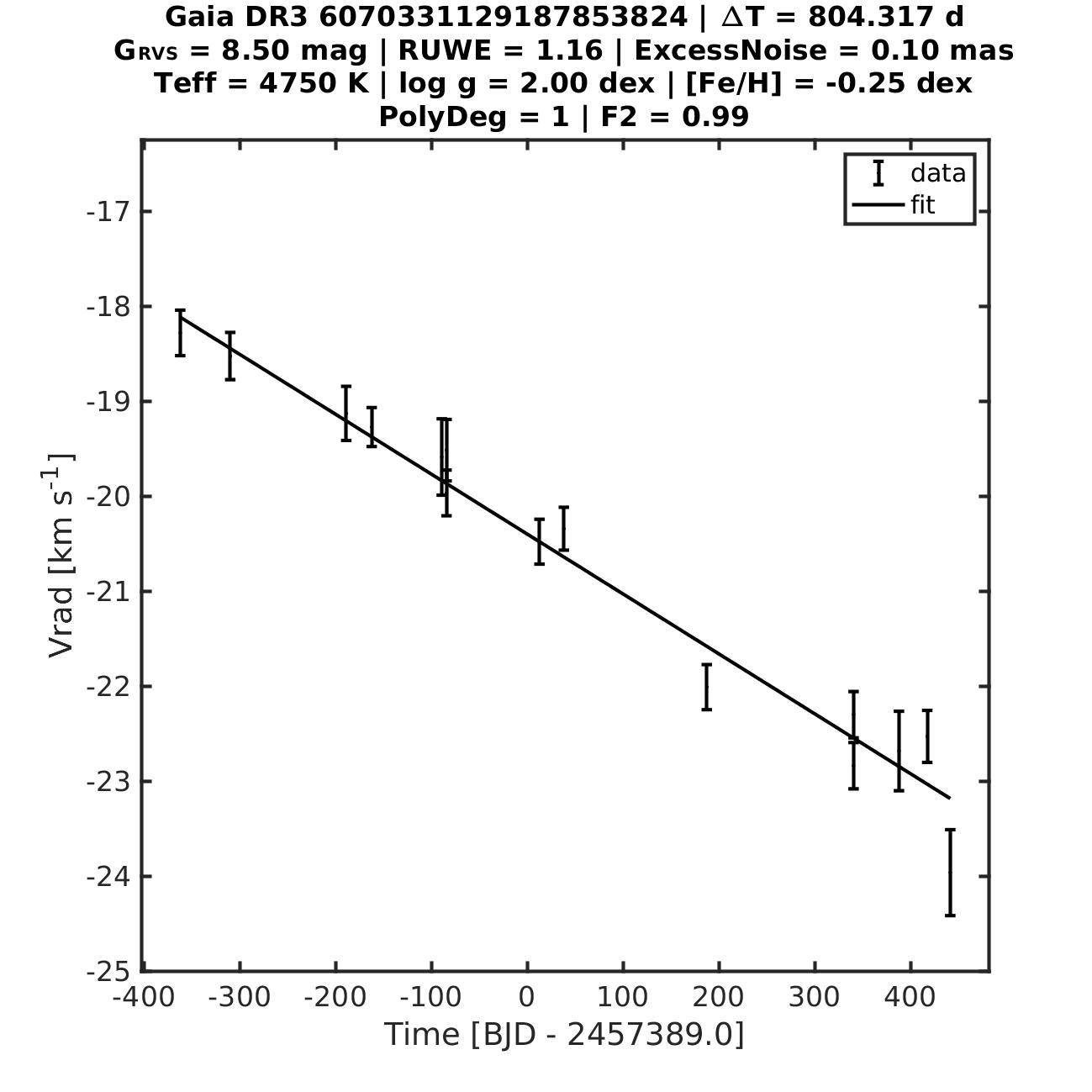}
\includegraphics[width=0.29\textwidth]{./FIGS/pp_4660348139680343424_TrendSB1.jpg}
}
\centerline{
\includegraphics[width=0.29\textwidth]{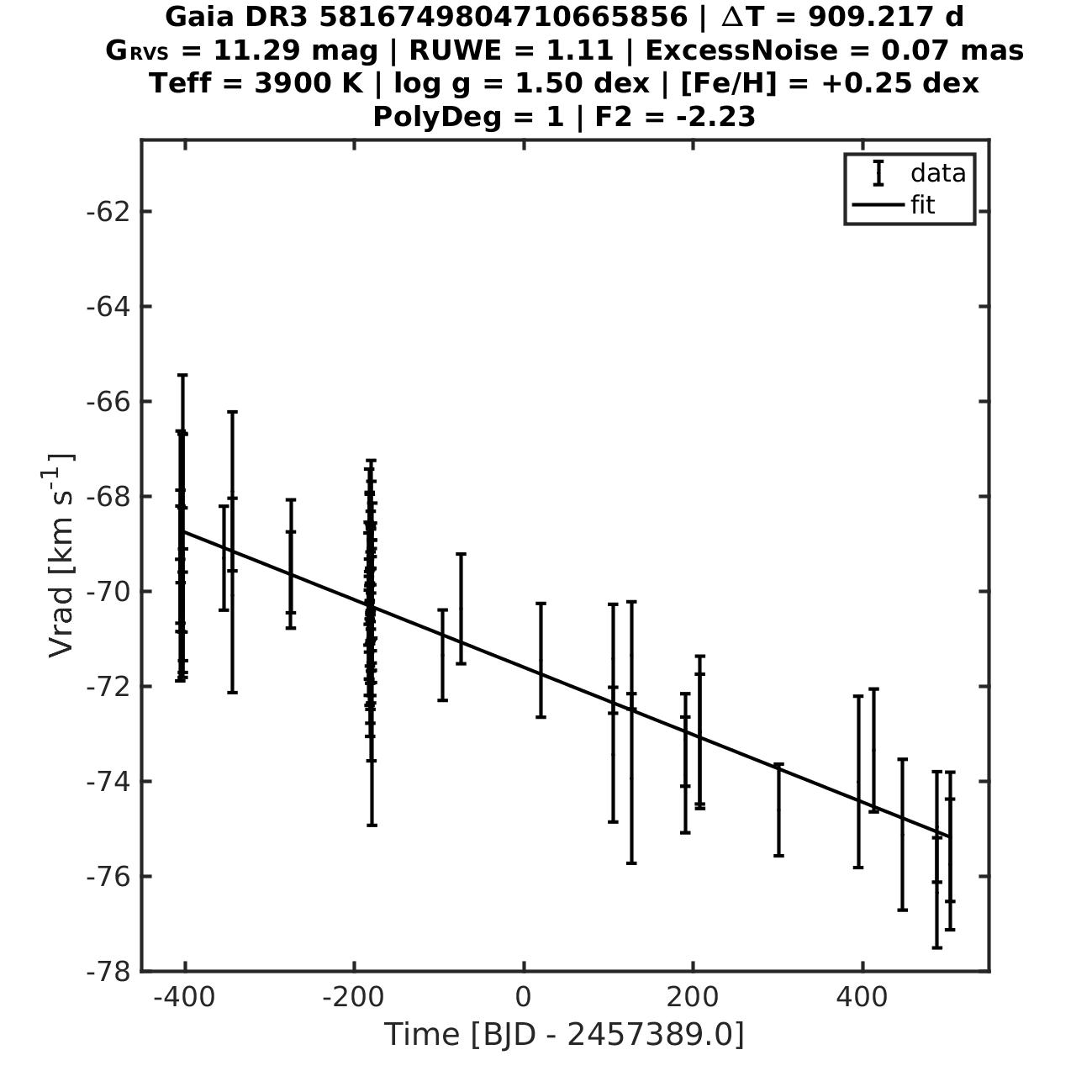}
\includegraphics[width=0.29\textwidth]{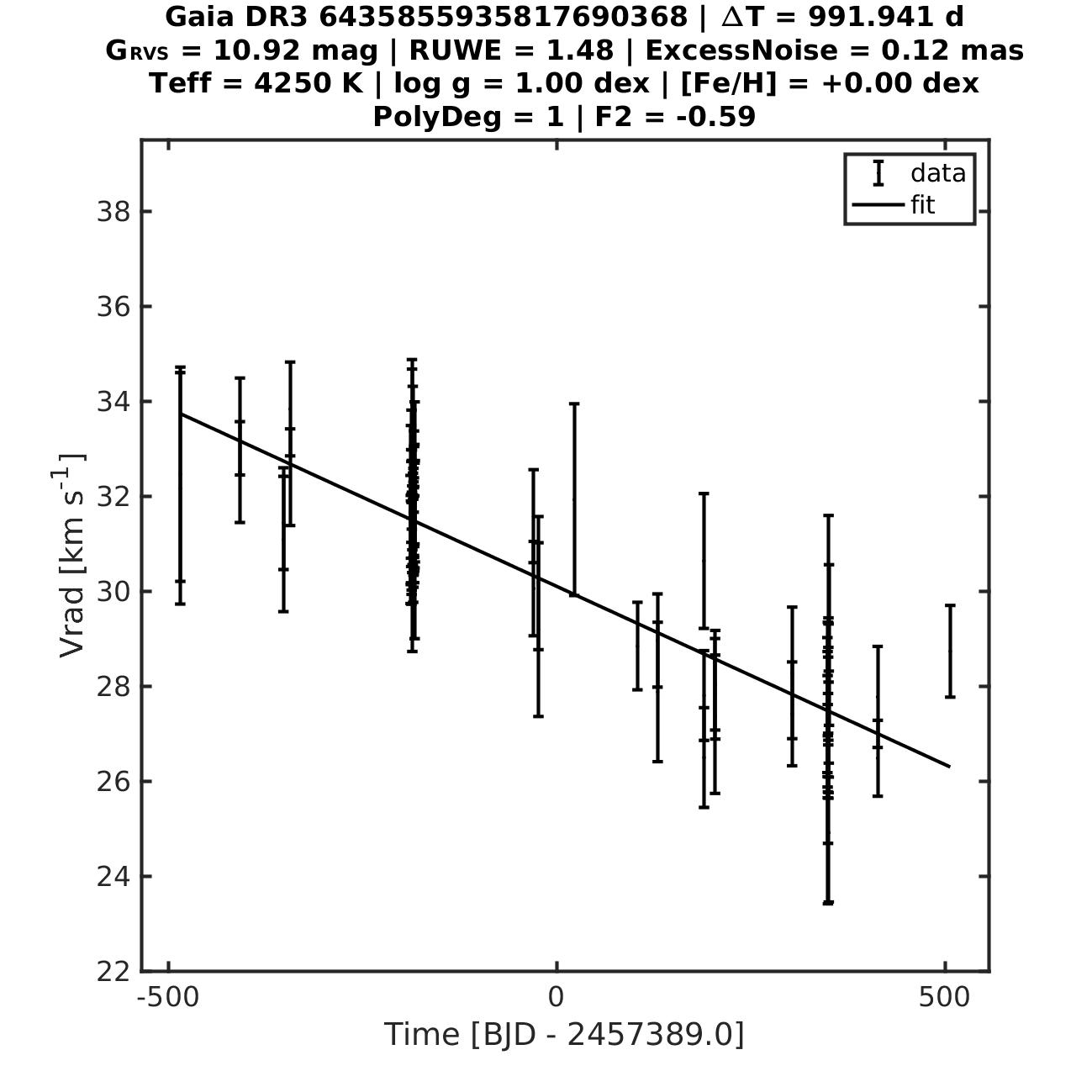}
\includegraphics[width=0.29\textwidth]{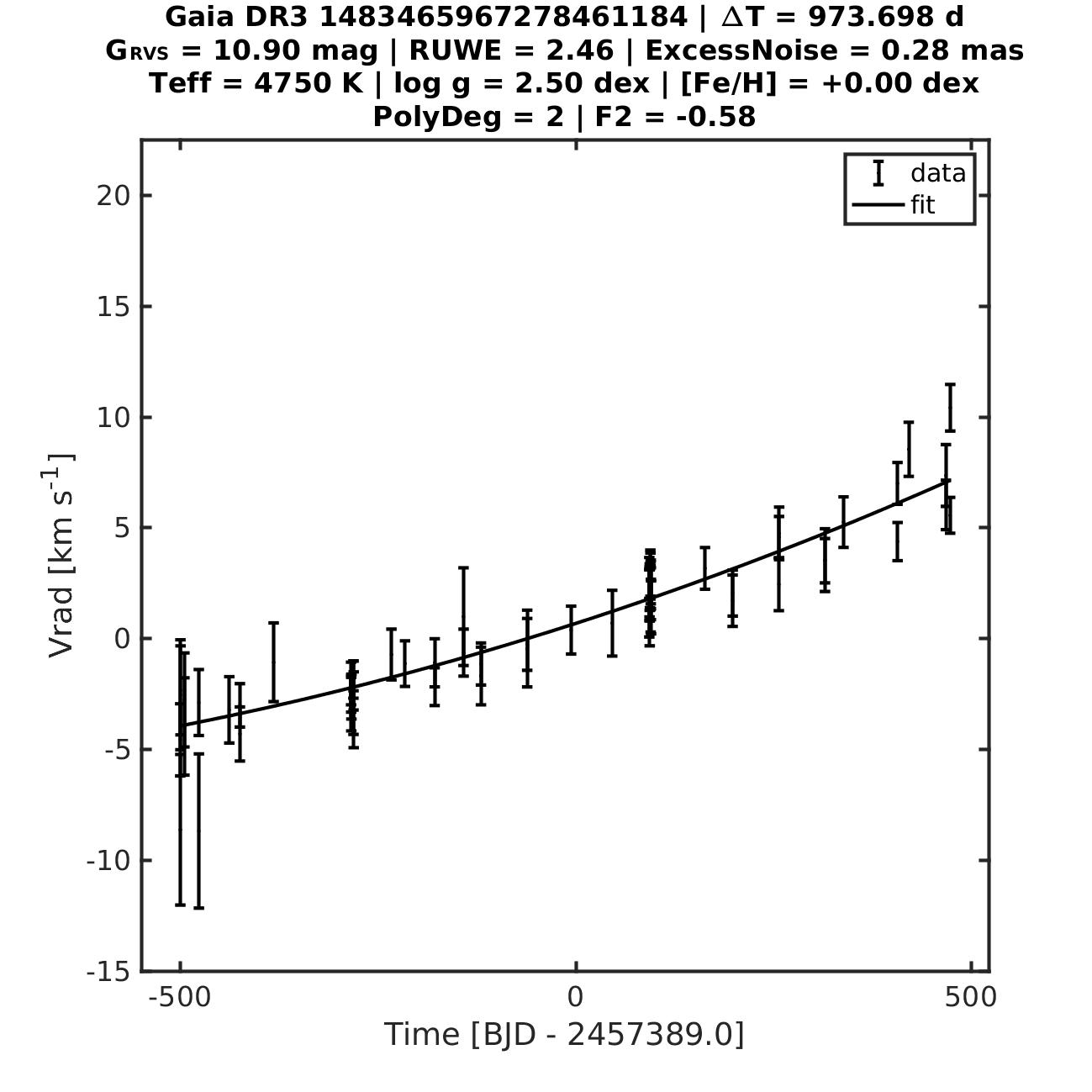}

}
\centerline{
\includegraphics[width=0.29\textwidth]{./FIGS/pp_5854429312298025728_TrendSB1.jpg}
\includegraphics[width=0.29\textwidth]{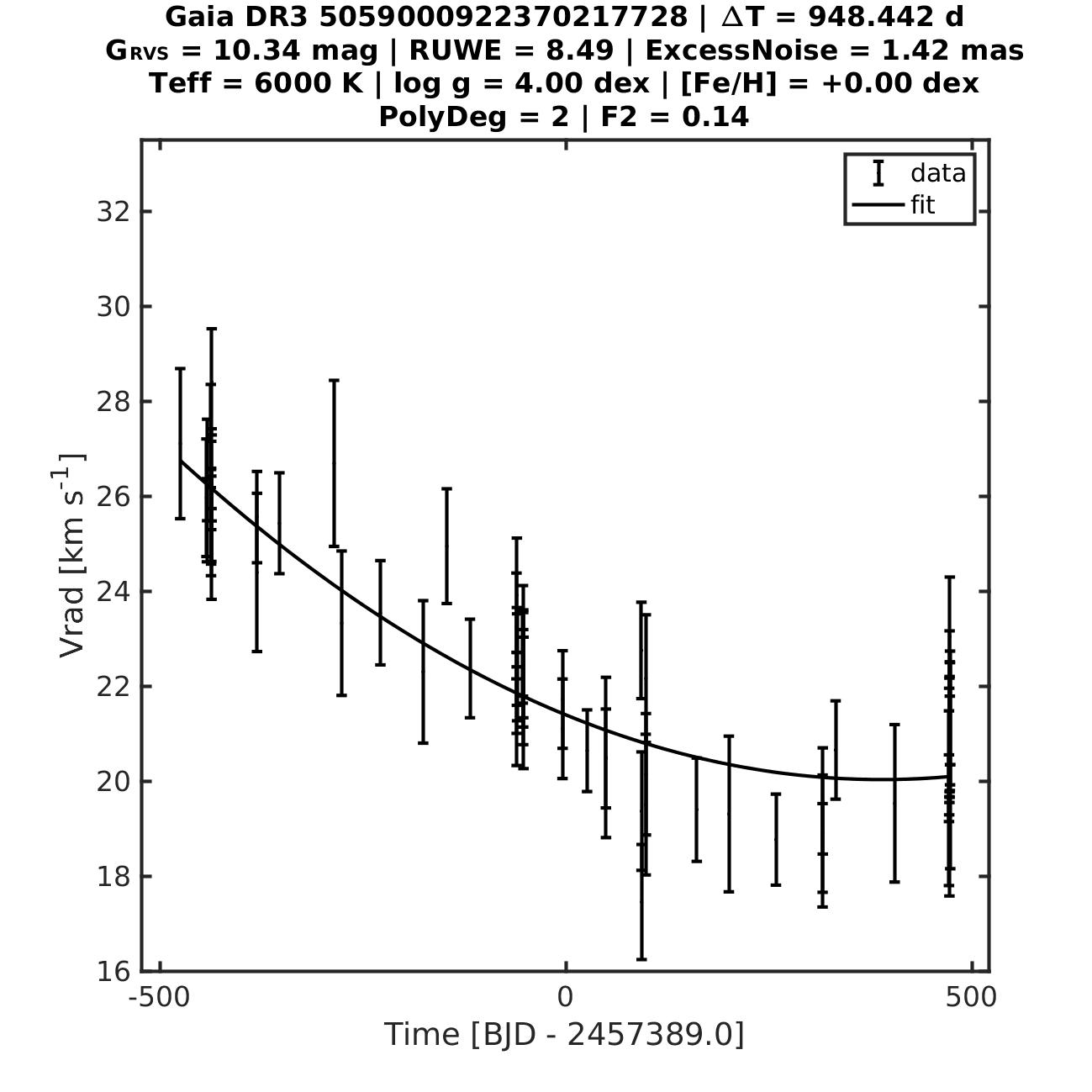}
\includegraphics[width=0.29\textwidth]{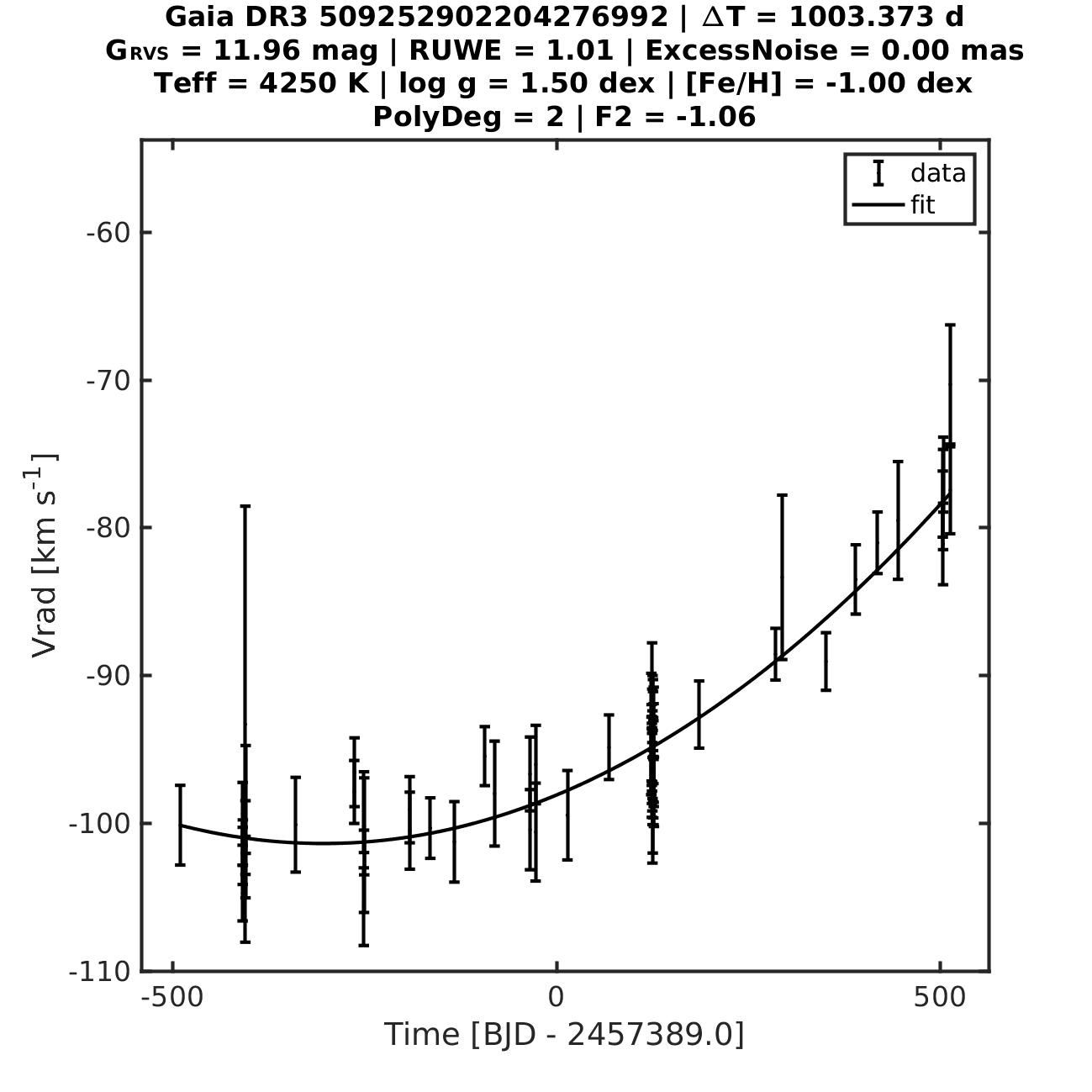}
}
\caption{Good results: same as Fig.\,\ref{fig:appgoodressb1part1} 
but concerning the TrendSB1-type solutions. 
The order of the objects is arbitrary, except that the 
second degree trends have been placed at the end.}
\label{fig:appgoodrestrend}
\end{figure*}
\begin{figure*}[!htp]
\centerline{
\includegraphics[width=0.29\textwidth]{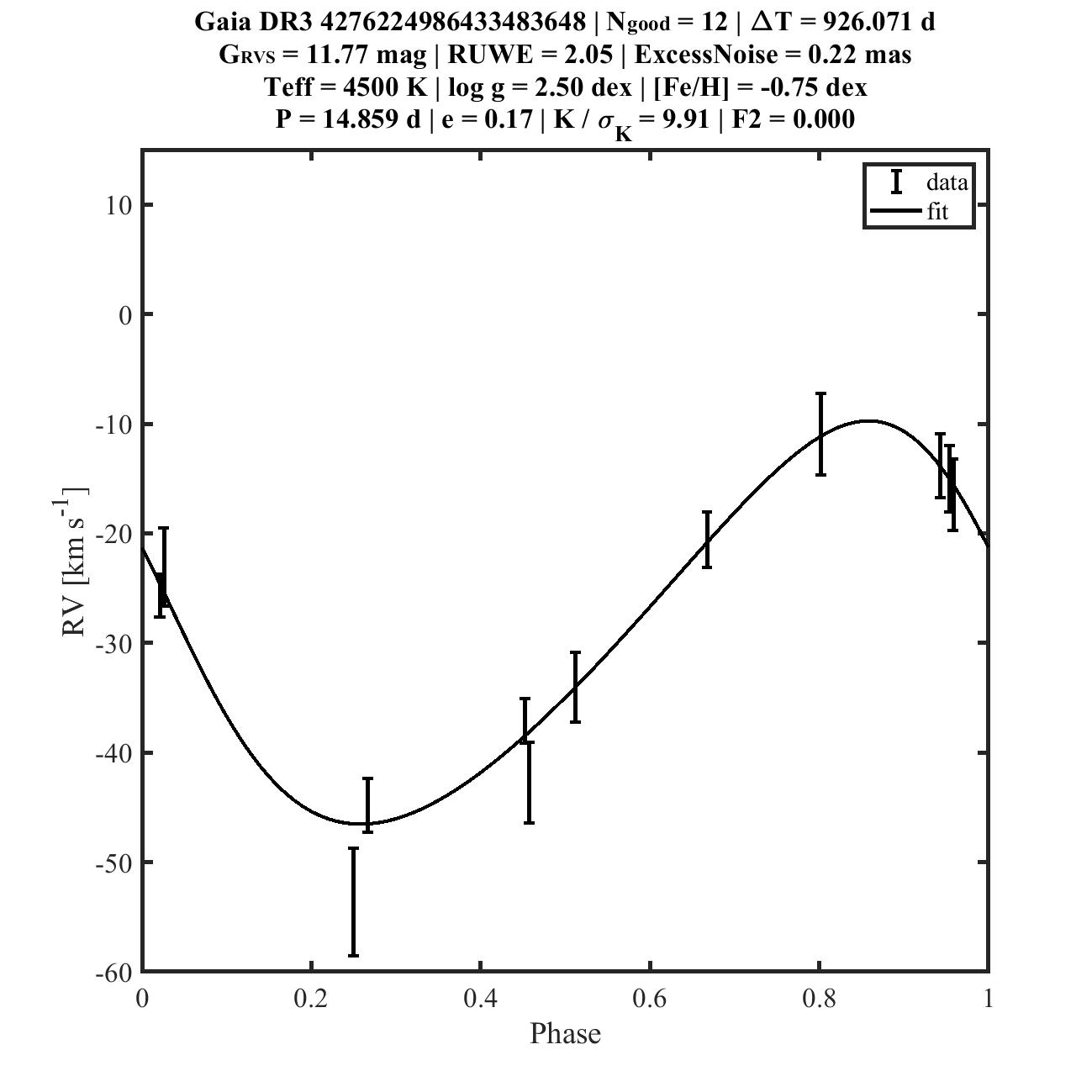}
\includegraphics[width=0.29\textwidth]{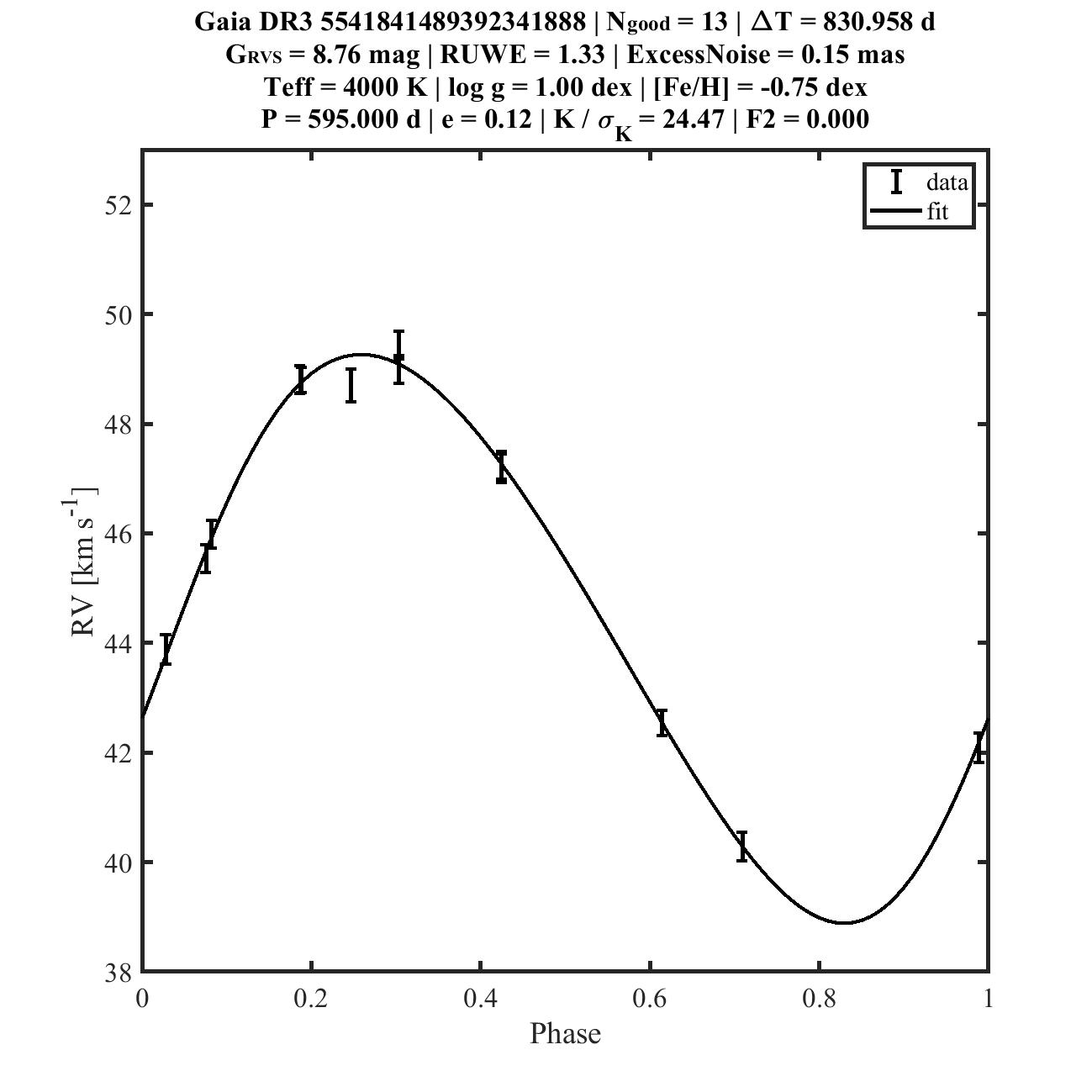} 
\includegraphics[width=0.29\textwidth]{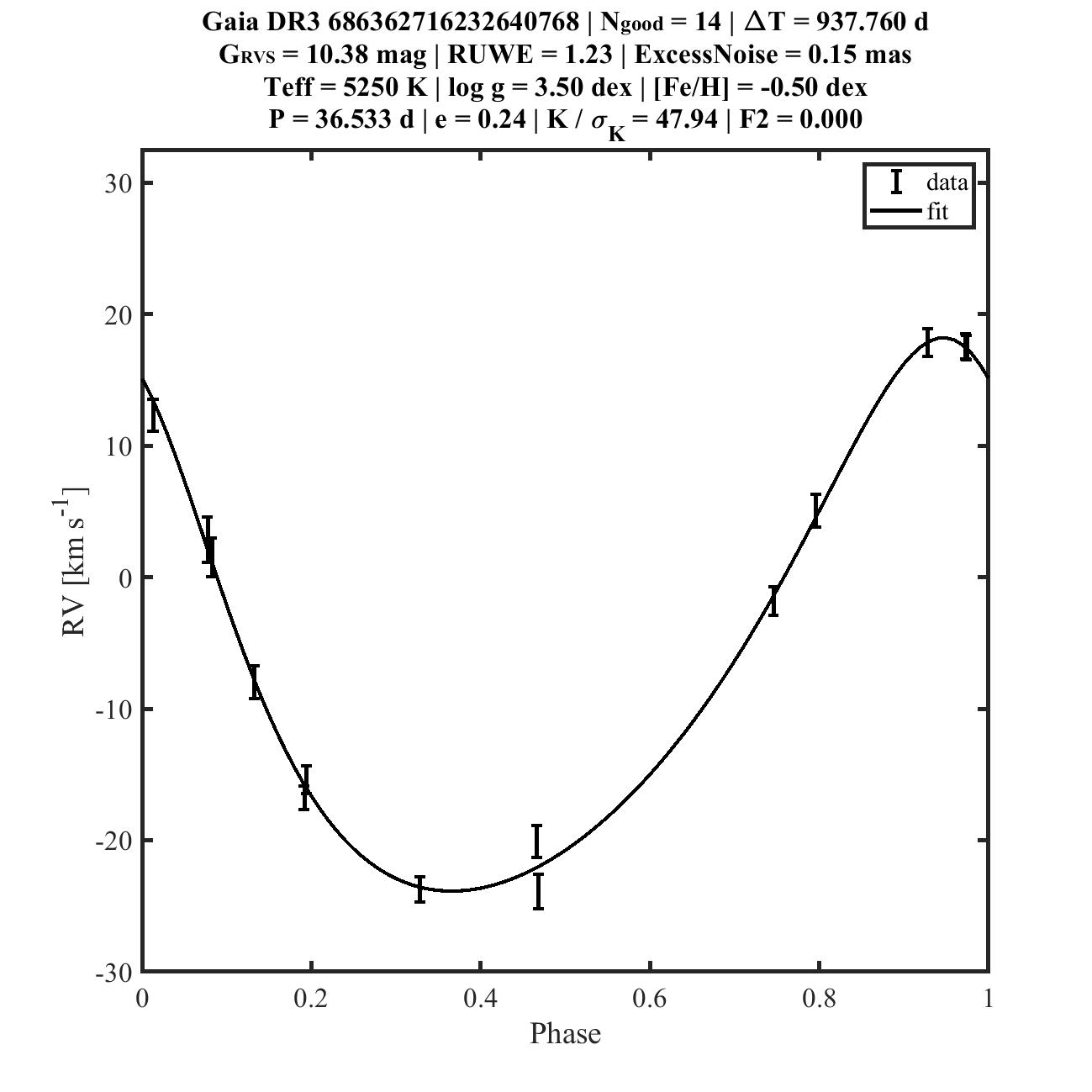} 
}
\centerline{
\includegraphics[width=0.29\textwidth]{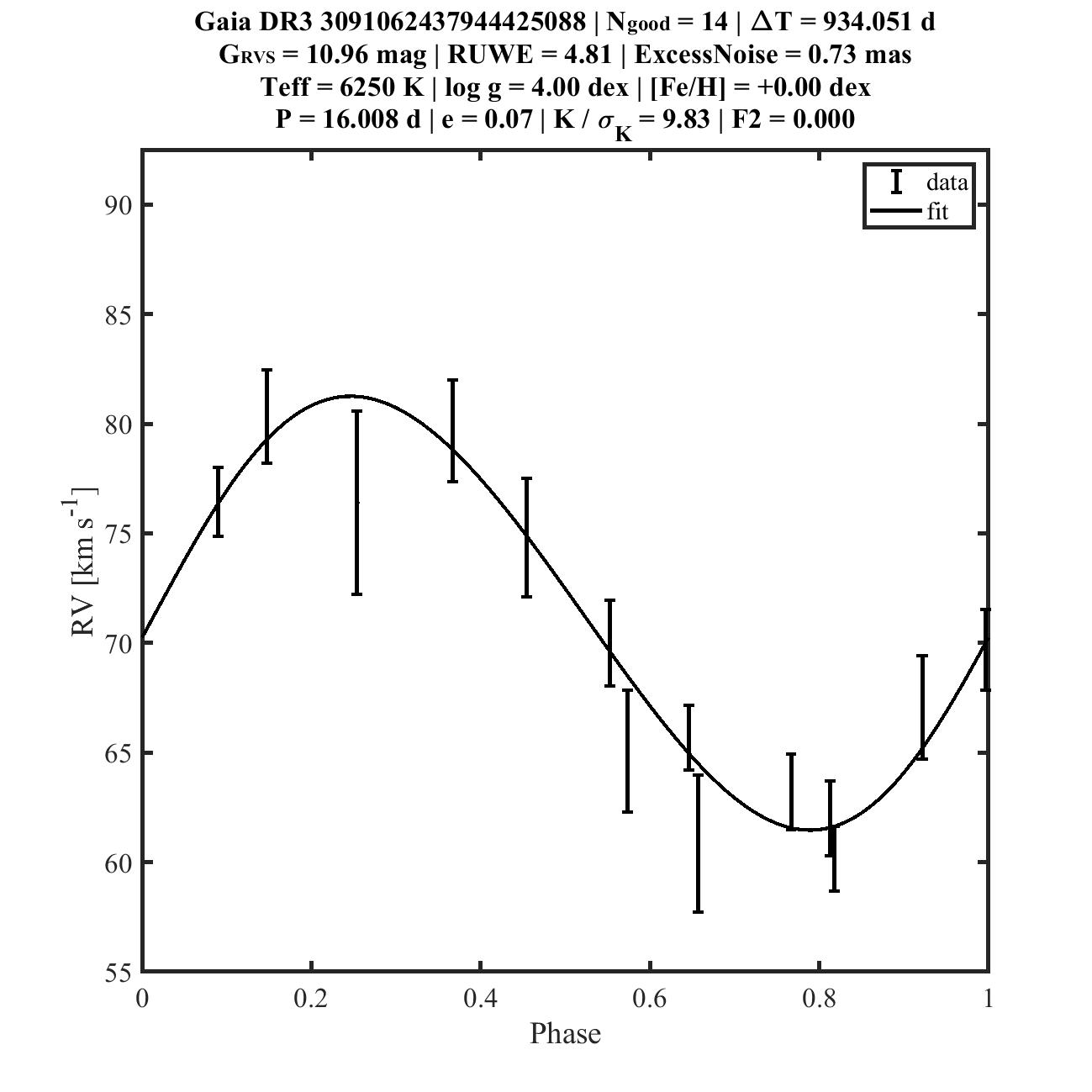}
\includegraphics[width=0.29\textwidth]{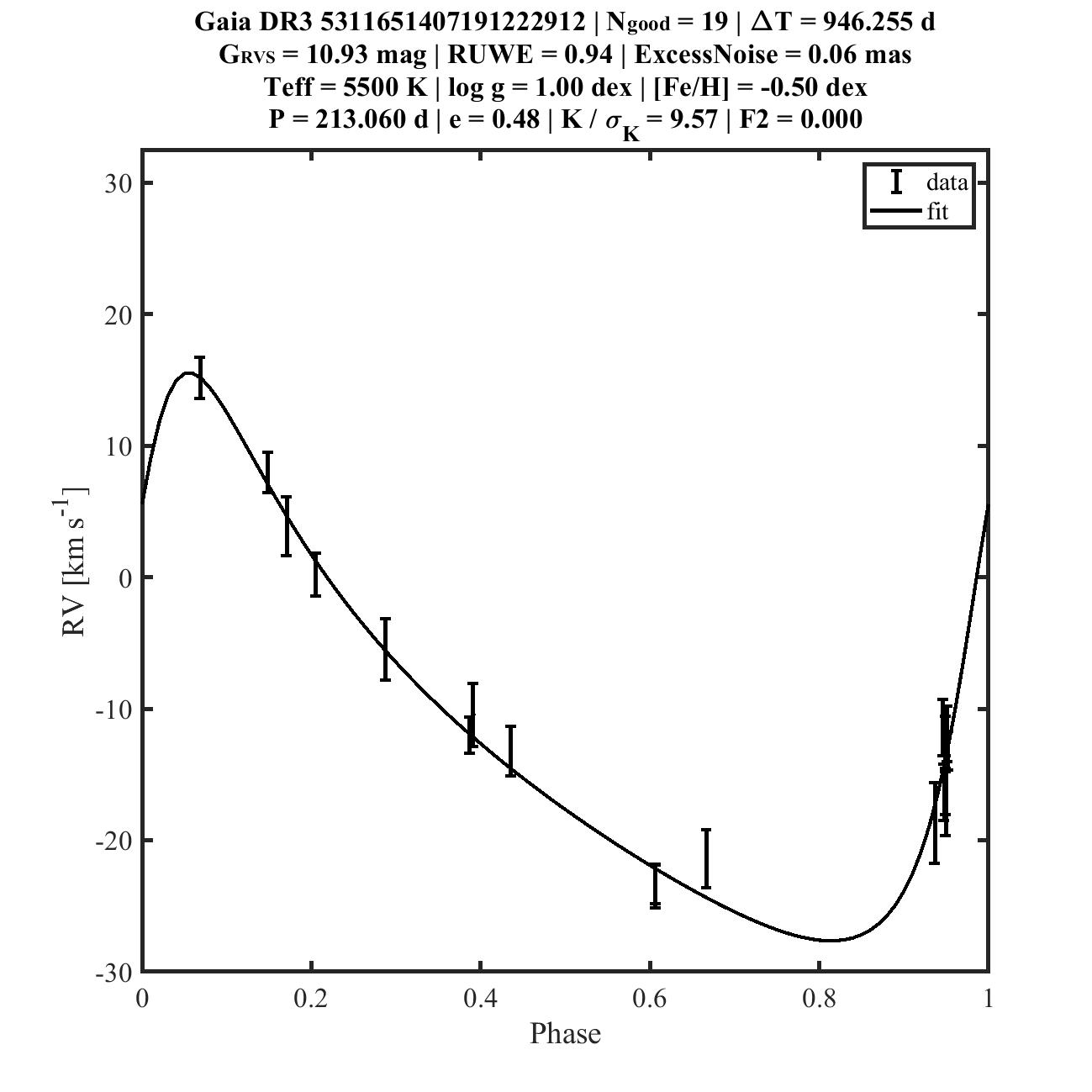}
\includegraphics[width=0.29\textwidth]{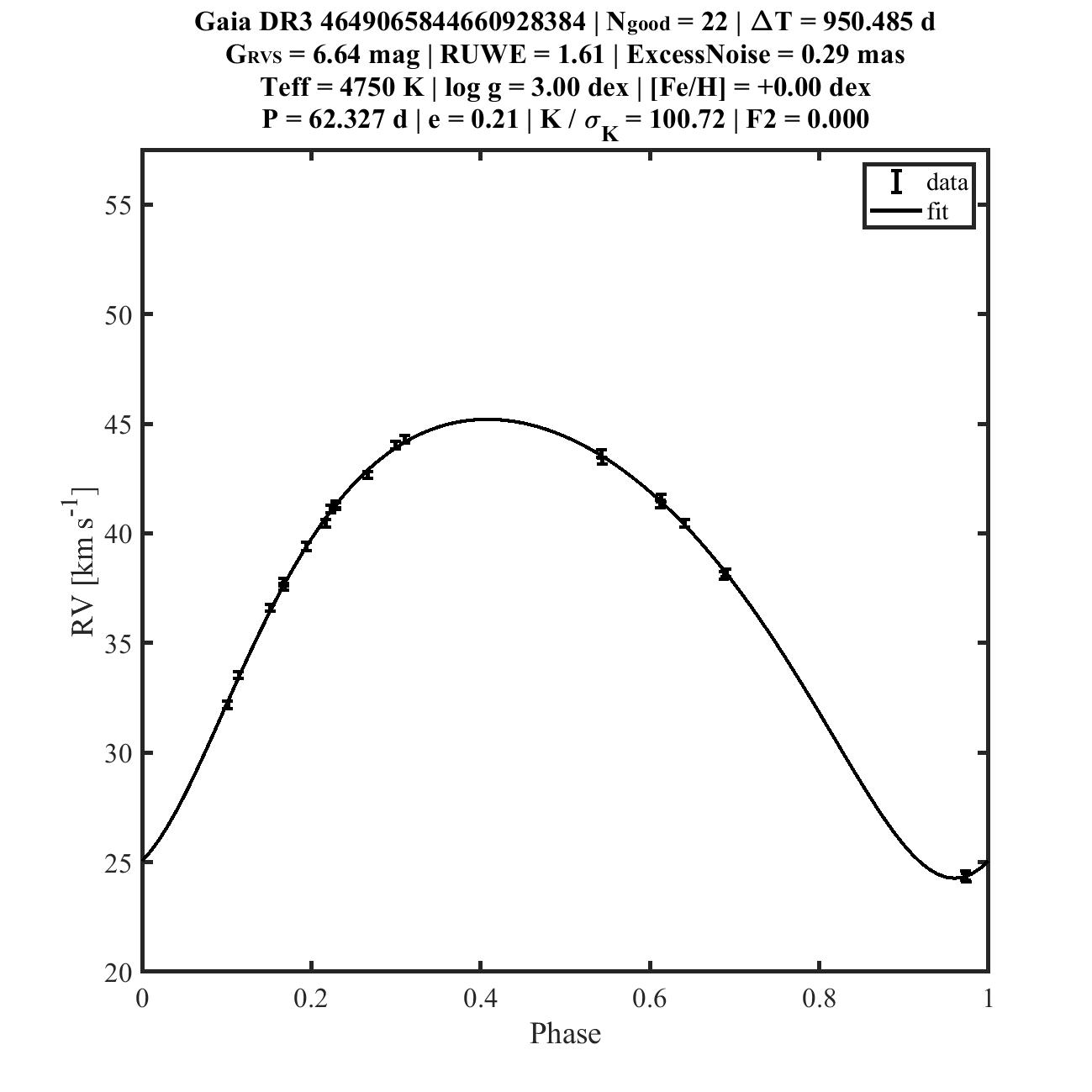}
}
\centerline{
\includegraphics[width=0.29\textwidth]{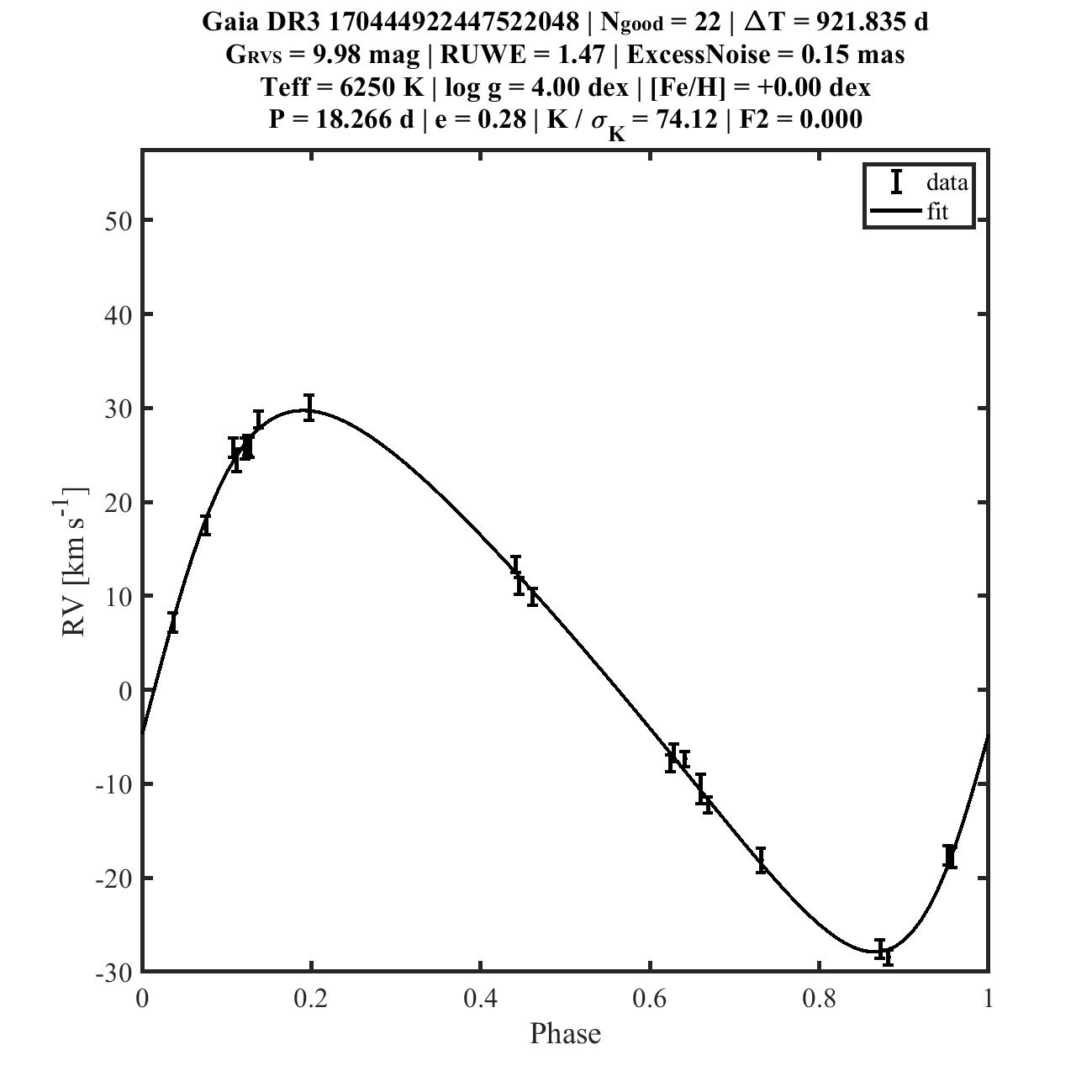}
\includegraphics[width=0.29\textwidth]{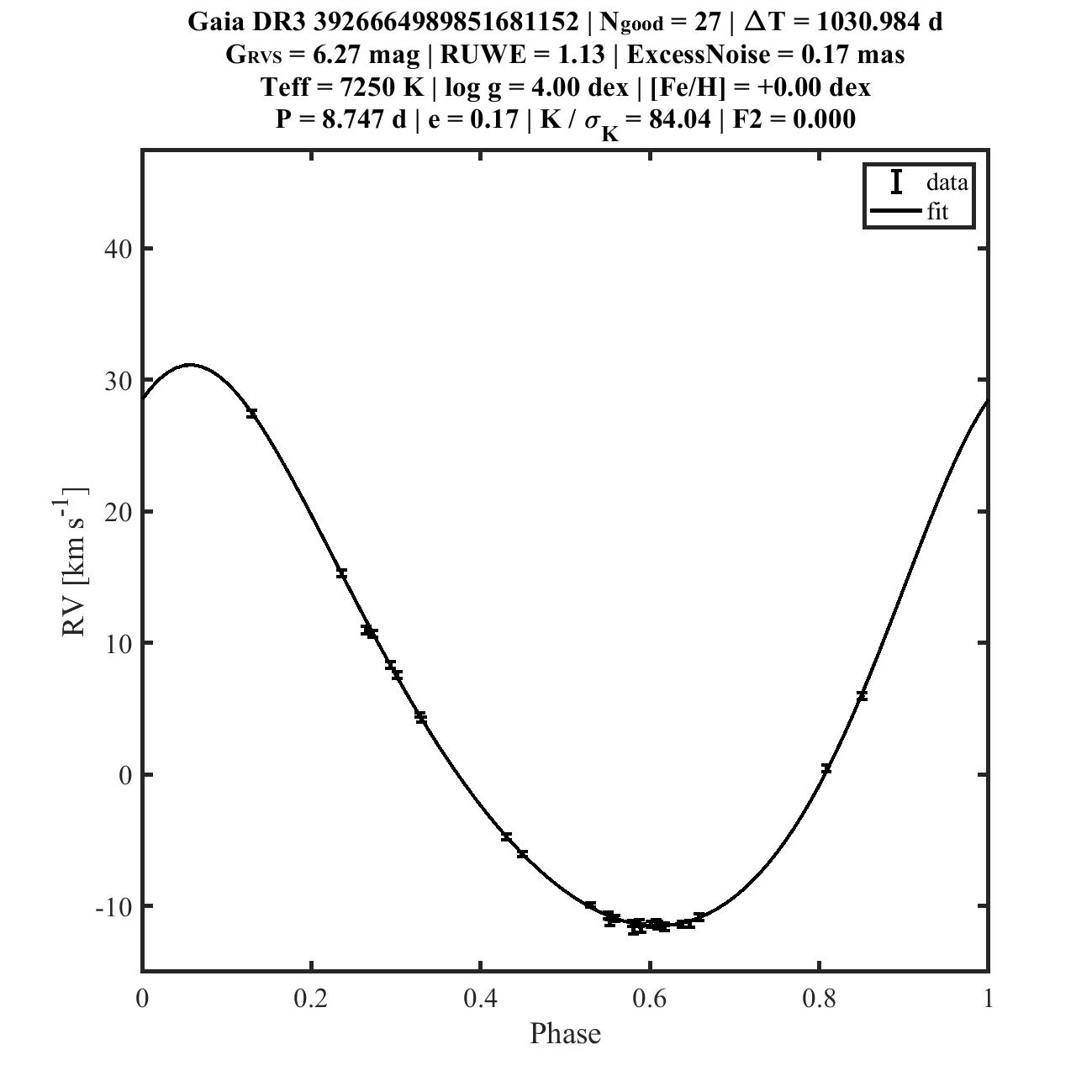}
\includegraphics[width=0.29\textwidth]{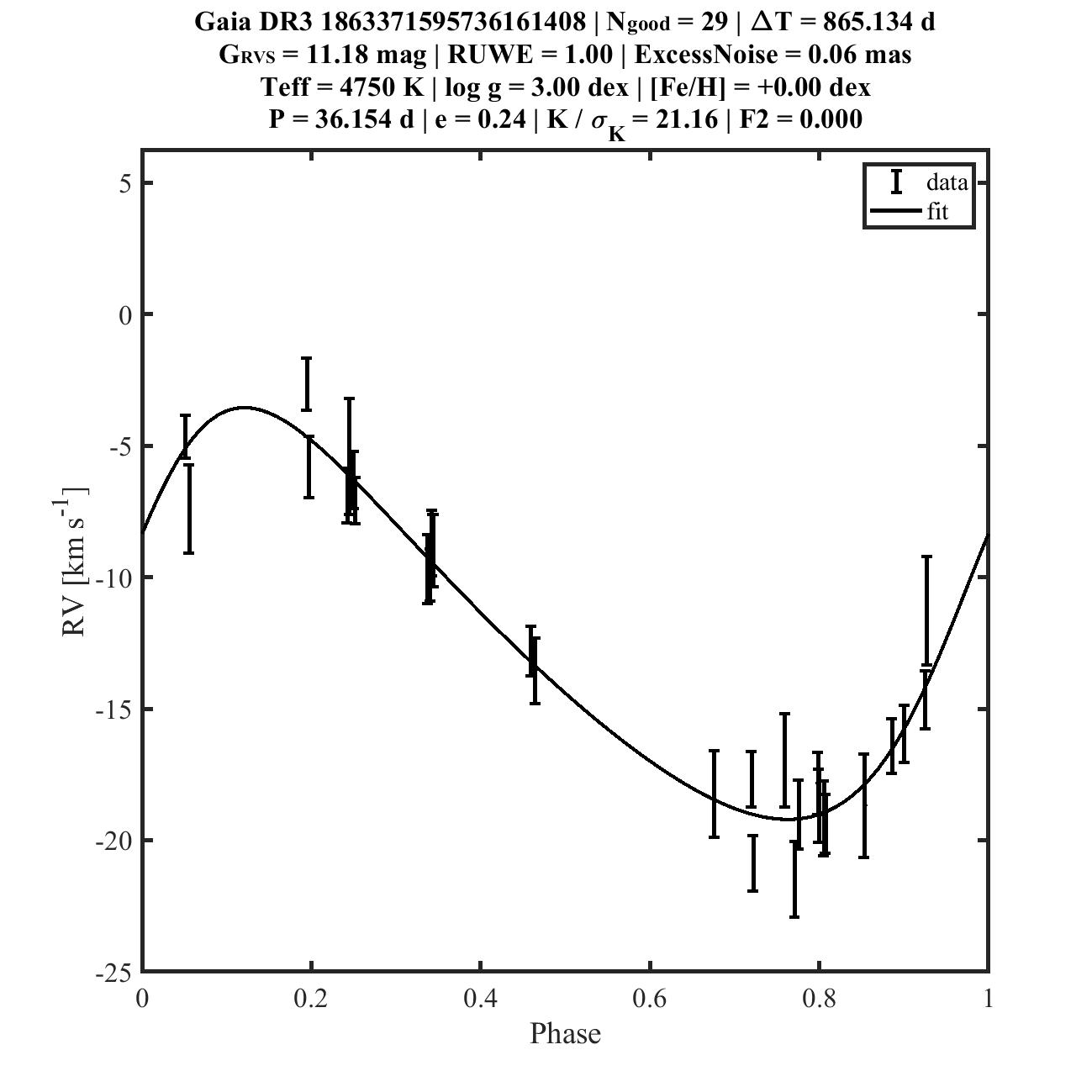}
}
\centerline{
\includegraphics[width=0.29\textwidth]{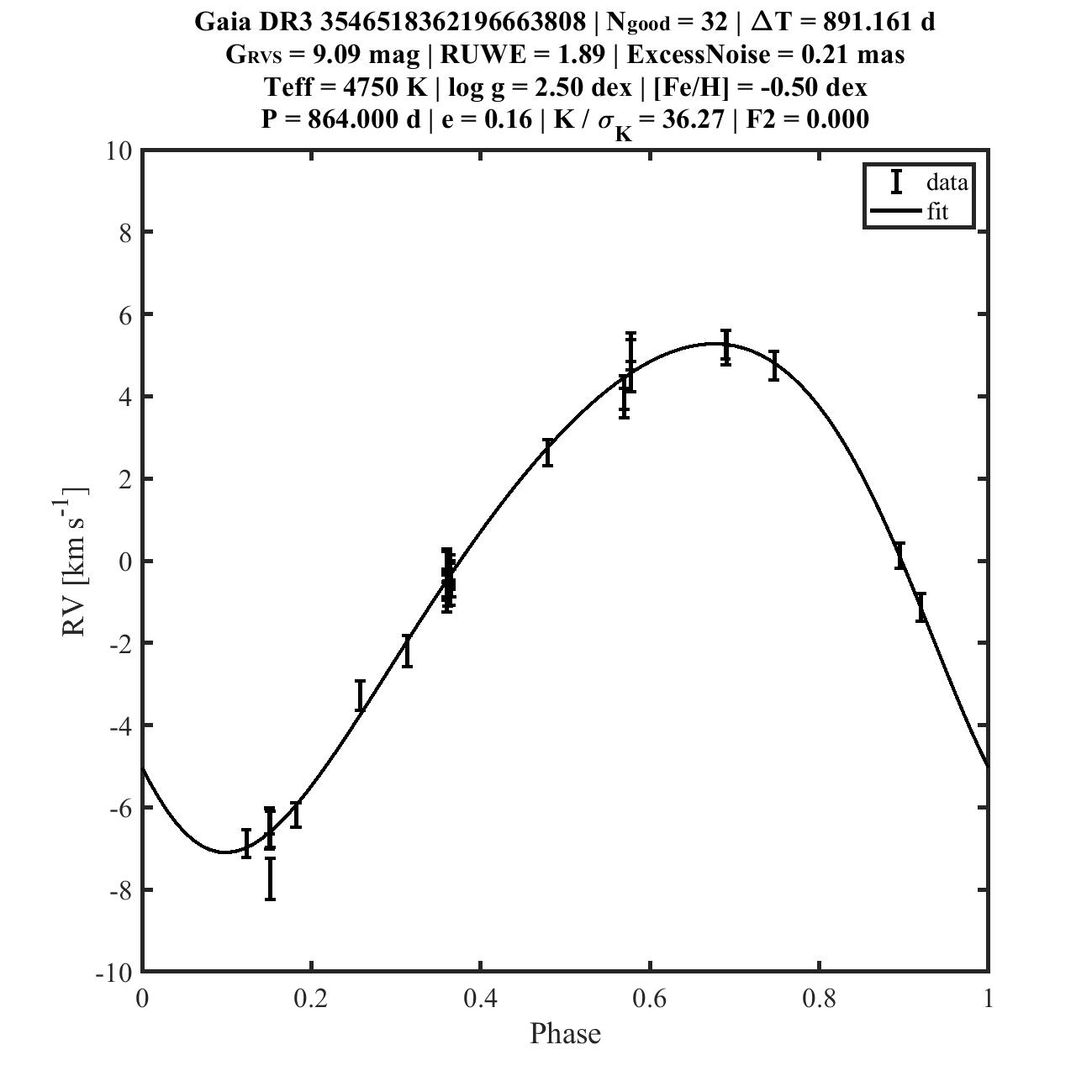}
\includegraphics[width=0.29\textwidth]{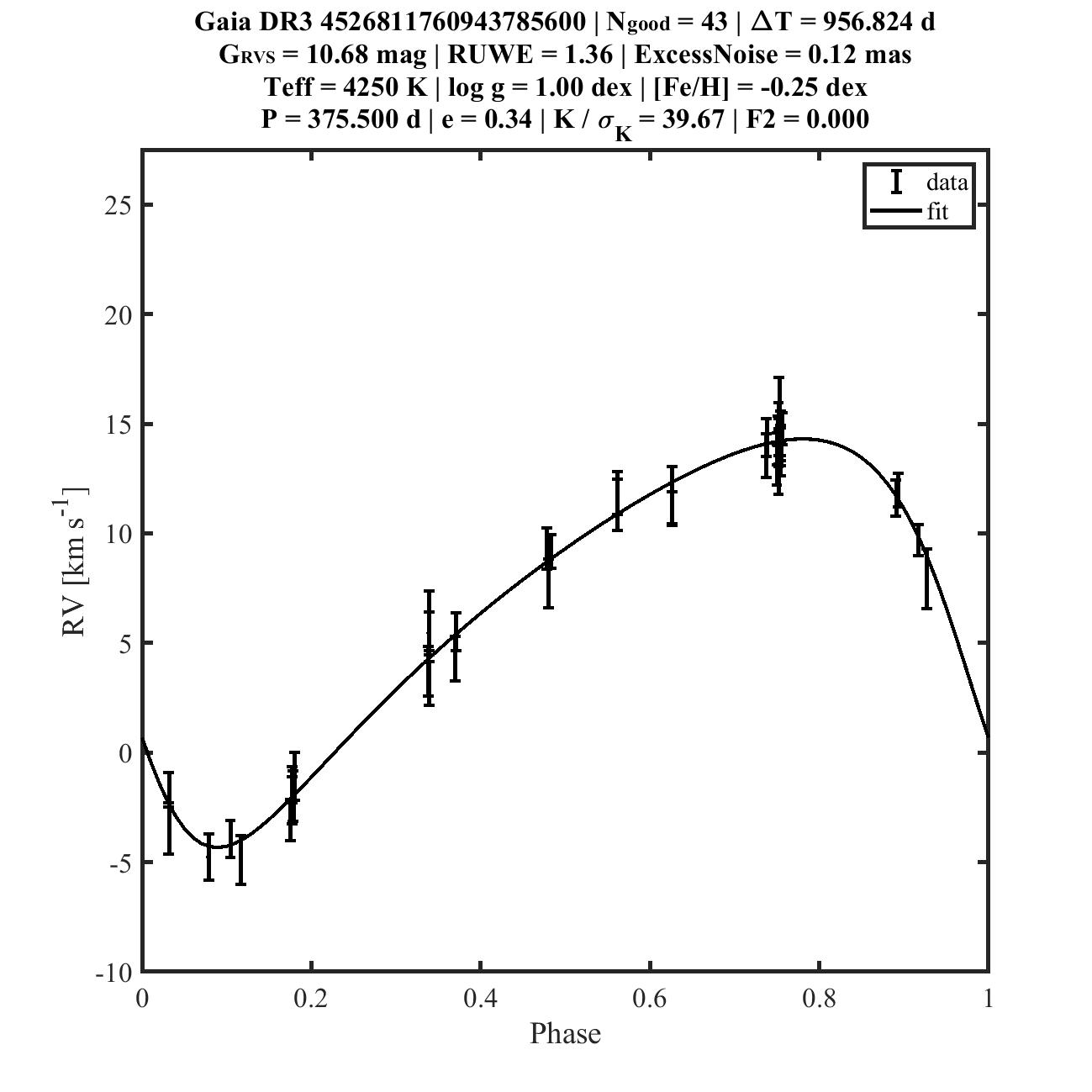}
\includegraphics[width=0.29\textwidth]{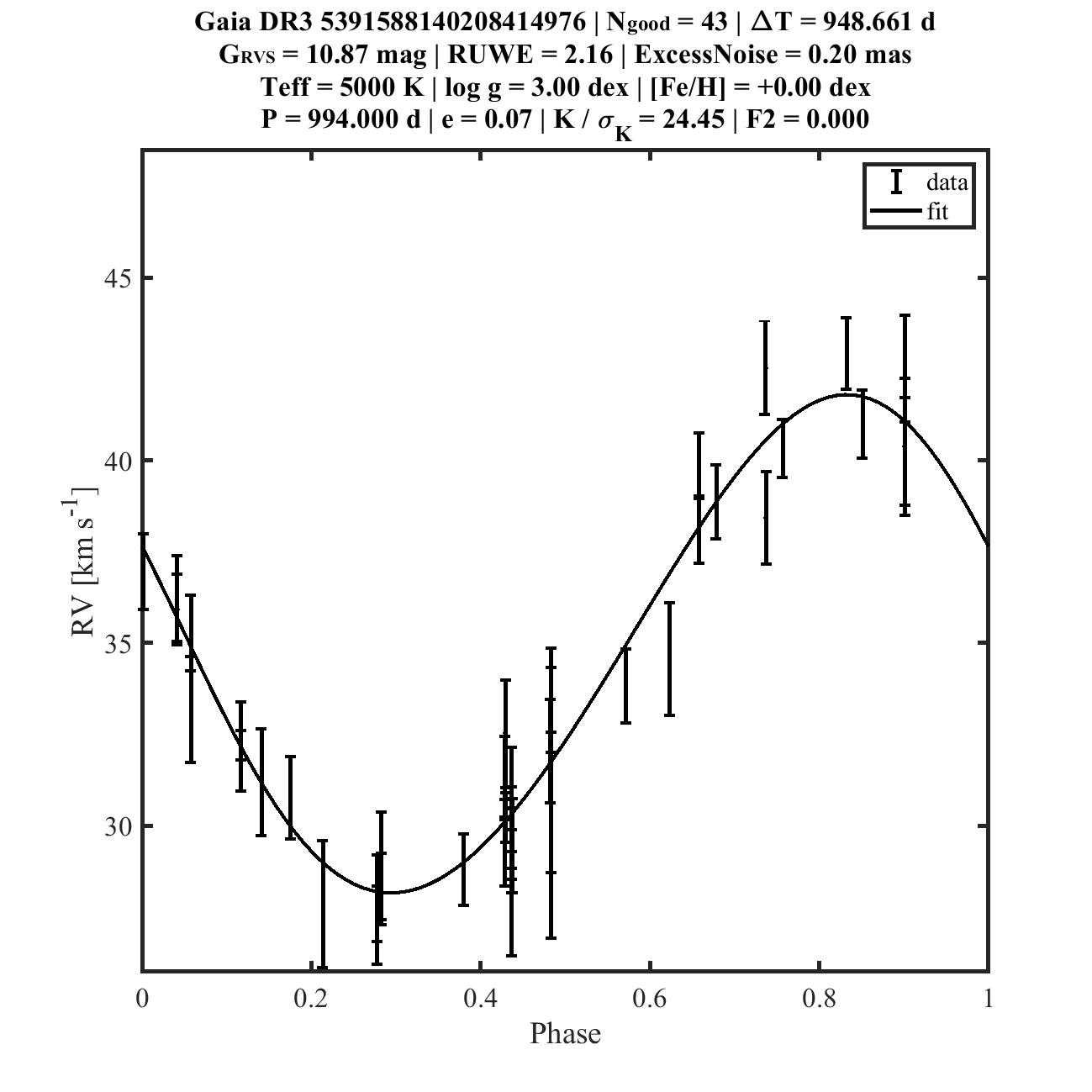}
}
\caption{This figure exhibits the RV curves and the SB1 orbital 
solutions for a random sample of objects from the catalogue. 
All the solutions have a similar $F_2$ around zero but 
have otherwise been selected at random
with $N_\mathrm{good}$ comprised between 10 and 45. 
They have been ordered by increasing $N_{\mathrm{good}}$. 
The case of larger number of data points has already been
illustrated in Fig.\,\ref{fig:appgoodressb1part1}.}
\label{fig:randomsample}
\end{figure*}
%
%
%
\clearpage
\section{Comparison of orbital parameters with respect to literature values}
\label{sec:appF}

Figures illustrating the deviations in $P$ 
(expressed in $\sigma$ units) with respect to external 
catalogues are shown in Sect.~\ref{ssec:spectroSB1_validation_otherset} 
as a function of the reference period. Here we extend such 
comparisons to other orbital parameters ($e$, $\gamma$, and $K$) 
and show the dependence with other relevant quantities
(including the $P$ from our catalogue, not the reference one).
\vspace{-0.7cm} \\
\begin{figure}[!htp]
\centerline{
\includegraphics[width=0.9\textwidth, trim= 40 385 0 130, clip]{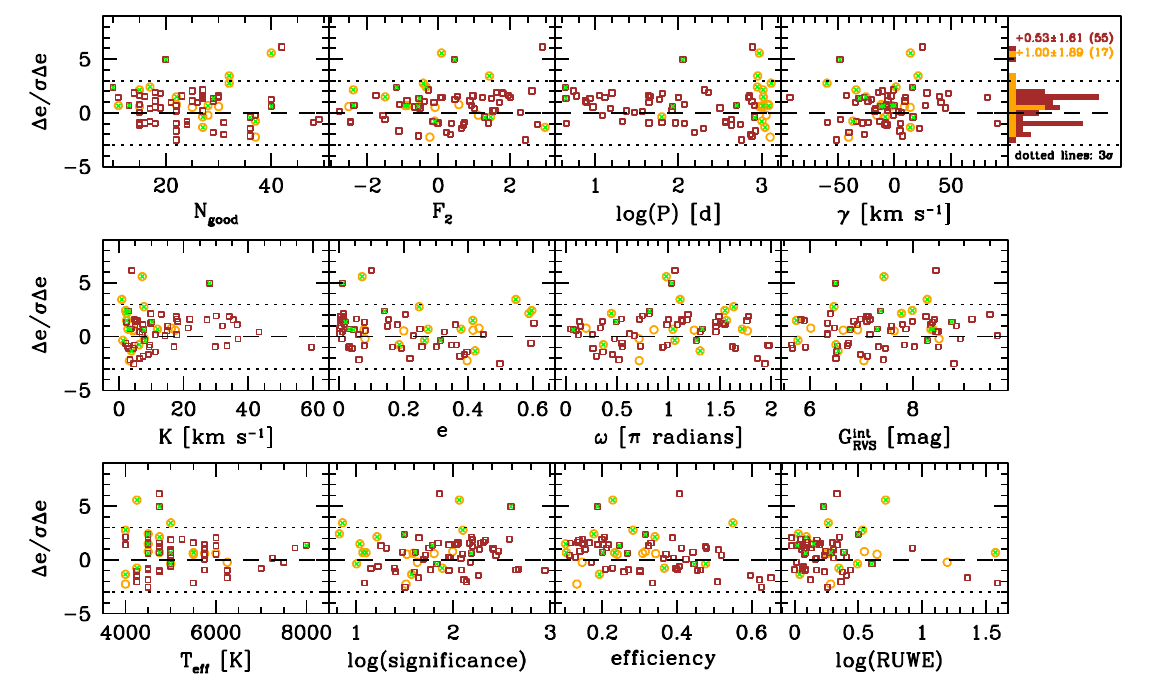}
}\caption[Comparison of the eccentricity with Griffin]
{Comparison for the eccentricity, $e$, 
between the DR3 spectroscopic values and those determined by Griffin and collaborators. 
Systems for which the period, $P$, in the literature is unlikely to be found 
(i.e. $\Delta T$ $<$ $P$) are indicated with orange circles. 
Systems for which $\Delta T$ $\geqslant$ $P$ are shown as brown squares. 
Binaries for which the reference 
period is not recovered to within 3$\sigma$ are flagged with green crosses. 
The quantity $T_\mathrm{eff}$ refers to the effective temperature of the template 
used to determine the RVs \citep{2023A&A...674A...5K}.
}
\label{fig:cu4nss_spectro_comparison_griffin_SB1_e}
\end{figure}
\vspace{-0.95cm} \\
\begin{figure}[!htp]
\centerline{
\includegraphics[width=0.9\textwidth, trim= 40 385 0 130, clip]{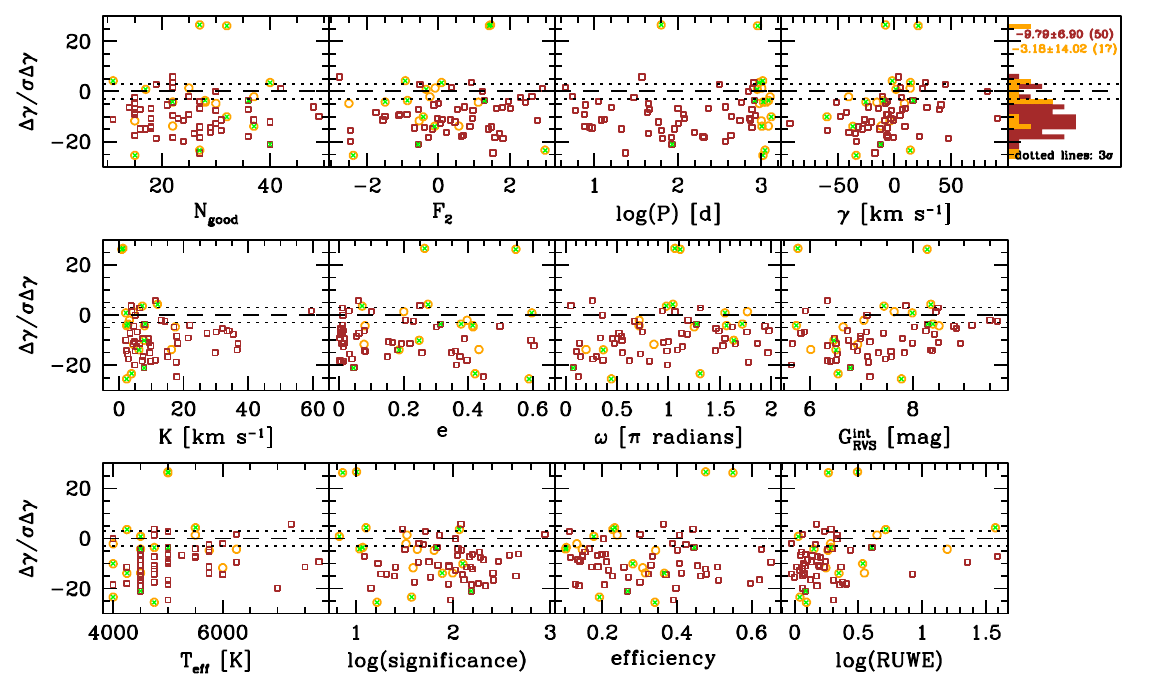}
}\caption[Comparison of systemic velocity with Griffin]{Same as 
Fig.\,\ref{fig:cu4nss_spectro_comparison_griffin_SB1_e} 
in the case of Griffin, but for $\gamma$.}
\label{fig:cu4nss_spectro_comparison_griffin_SB1_gamma0}
\end{figure}
\begin{figure}[!htp]
\centerline{
\includegraphics[width=0.9\textwidth, trim= 40 385 0 130, clip]{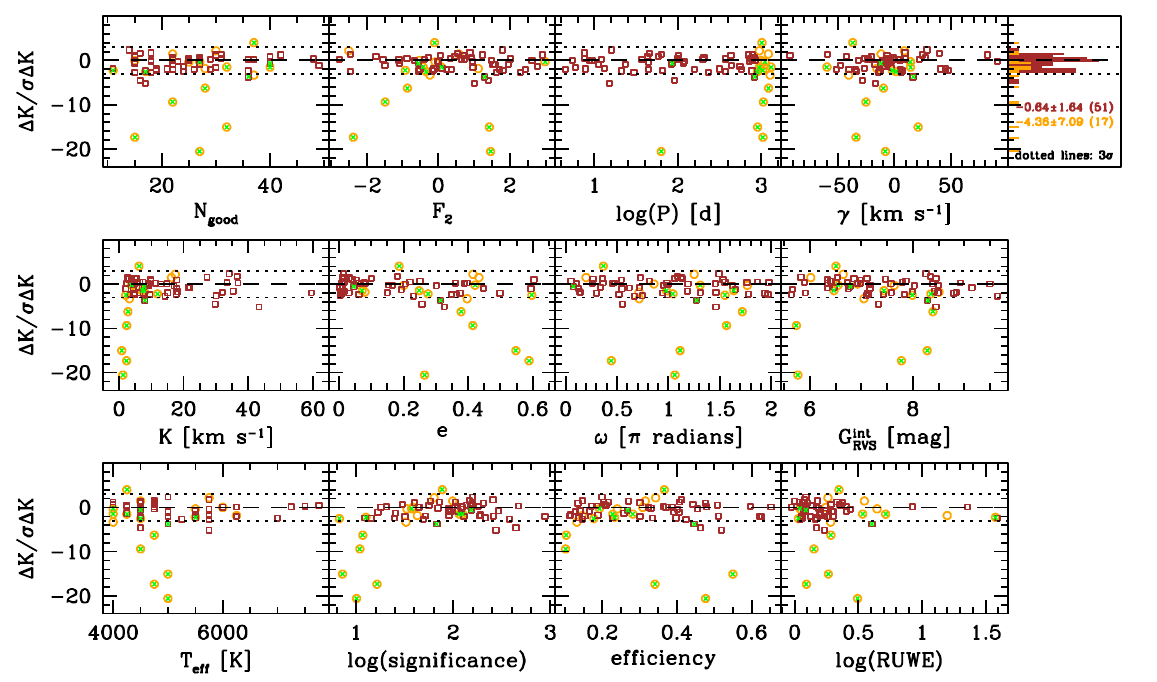}
}\caption[Comparison of semi-amplitude with Griffin]{Same as 
Fig.\,\ref{fig:cu4nss_spectro_comparison_griffin_SB1_e} 
in the case of Griffin, but for $K$.}
\label{fig:cu4nss_spectro_comparison_griffin_SB1_K}
\end{figure}
\begin{figure}[!htp]
\centerline{
\includegraphics[width=0.9\textwidth, trim= 40 385 0 130, clip]{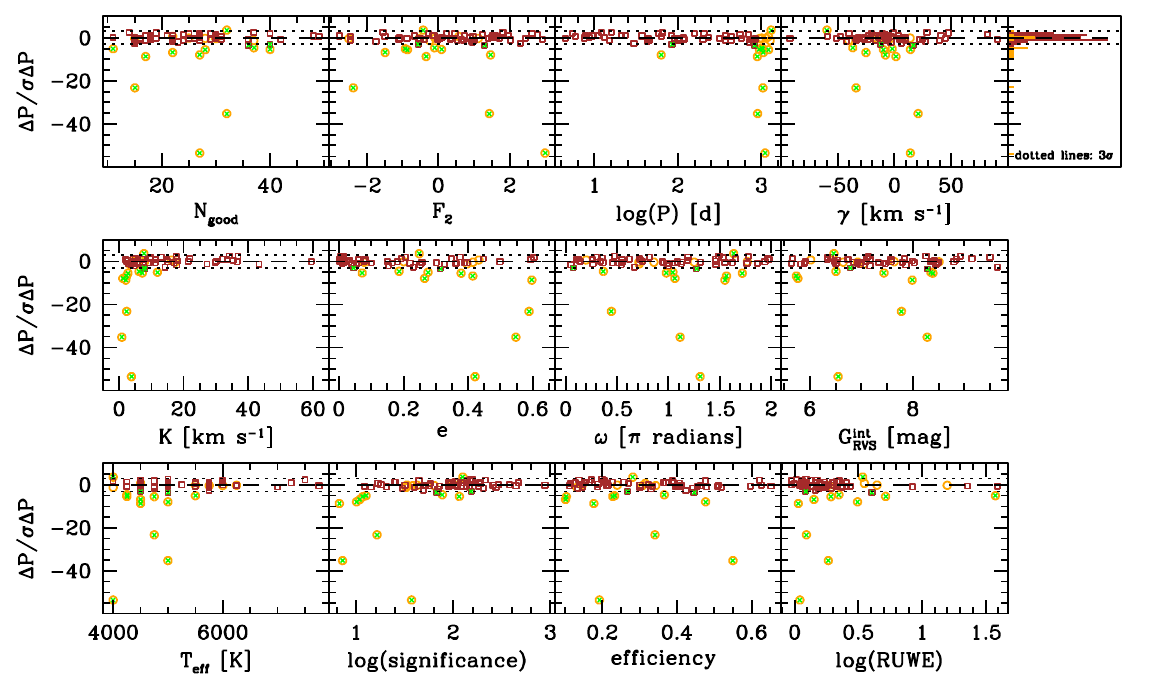}
}\caption[Comparison of period with Griffin]{Same as 
Fig.\,\ref{fig:cu4nss_spectro_comparison_griffin_SB1_e} 
in the case of Griffin, but for $P$.}
\label{fig:cu4nss_spectro_comparison_griffin_SB1_P}
\end{figure}
\begin{figure}[!htp]
\centerline{
\includegraphics[width=0.9\textwidth, trim= 40 385 0 130, clip]{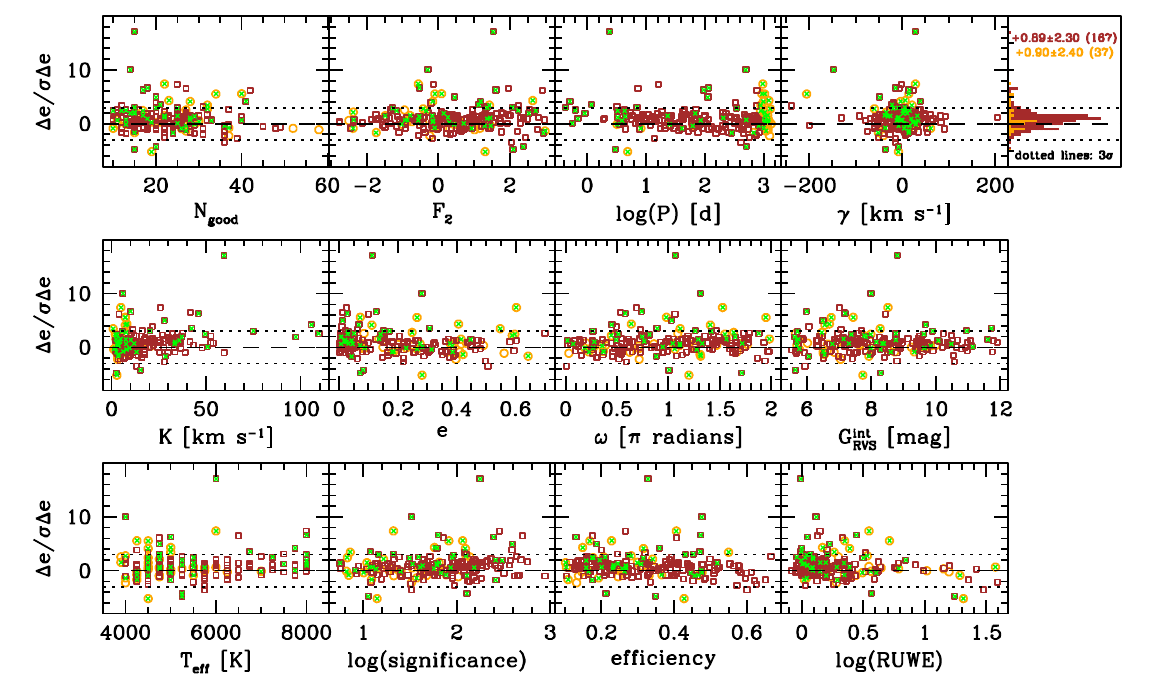}
}\caption[Comparison of the eccentricity with SB9]{Comparison for 
the eccentricity, $e$, 
between the DR3 spectroscopic values and those in the SB9. 
Systems for which the period, $P$, in the literature is unlikely to be found 
(i.e. $\Delta T$ $<$ $P$) are indicated with orange circles. 
Systems for which $\Delta T$ $\geqslant$ $P$ are shown as brown squares. 
Binaries for which the reference 
period is not recovered to within 3$\sigma$ are flagged with green crosses. 
The quantity $T_\mathrm{eff}$ refers to the effective temperature of 
the template used to determine the RVs \citep{2023A&A...674A...5K}.}
\label{fig:cu4nss_spectro_comparison_SB9_SB1_e}
\end{figure}
\begin{figure}[!htp]
\centerline{
\includegraphics[width=0.9\textwidth, trim= 40 385 0 130, clip]{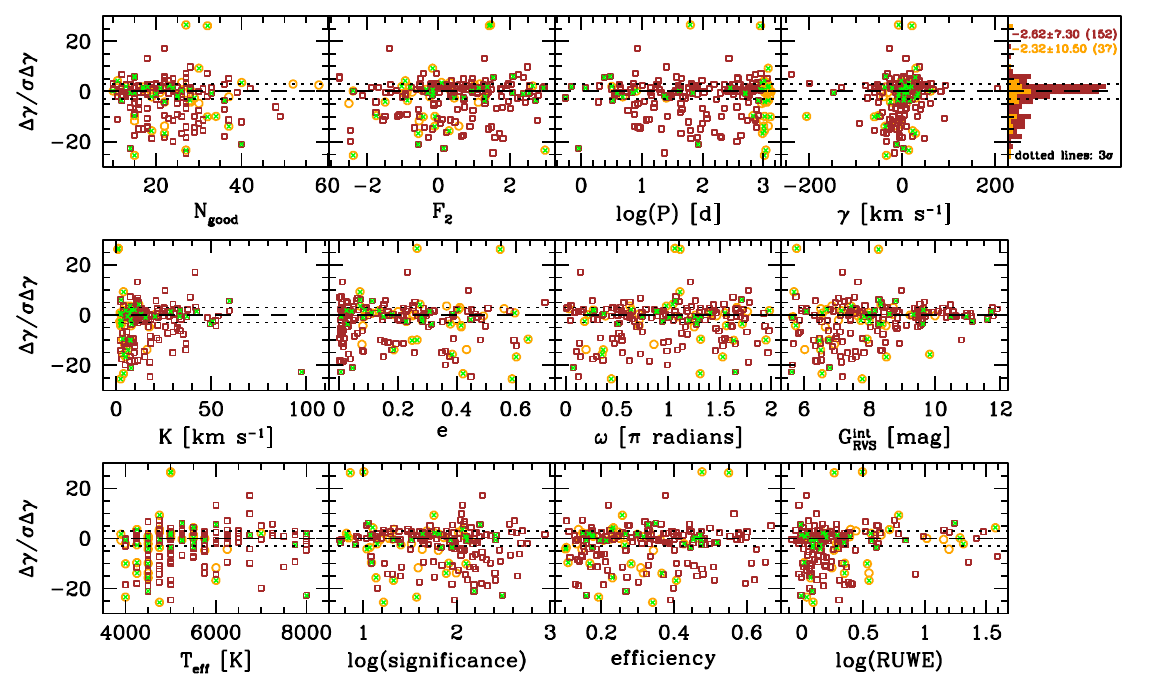}
}\caption[Comparison of systemic velocity with SB9]{Same as 
Fig.\,\ref{fig:cu4nss_spectro_comparison_SB9_SB1_e} in the case of the SB9, 
but for $\gamma$.}
\label{fig:cu4nss_spectro_comparison_SB9_SB1_gamma0}
\end{figure}
\begin{figure}[!htp]
\centerline{
\includegraphics[width=0.9\textwidth, trim= 40 385 0 130, clip]{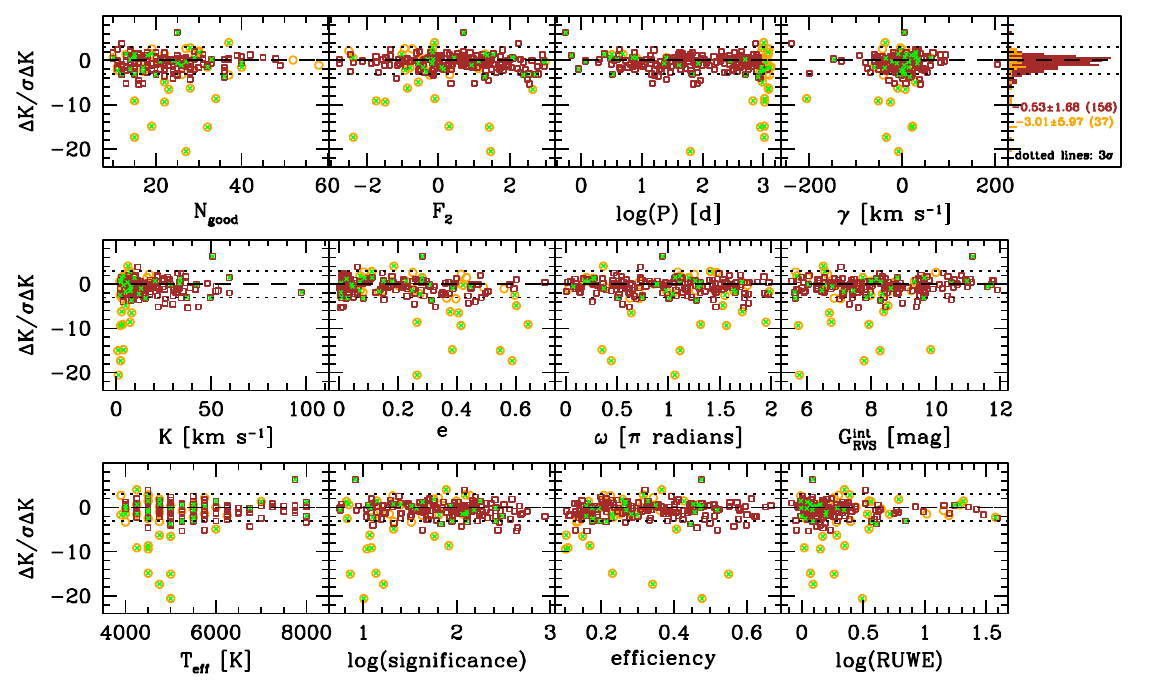}
}\caption[Comparison of semi amplitude with SB9]{Same as 
Fig.\,\ref{fig:cu4nss_spectro_comparison_SB9_SB1_e} 
in the case of the SB9, but for $K$.}
\label{fig:cu4nss_spectro_comparison_SB9_SB1_K}
\end{figure}
\begin{figure}[!htp]
\centerline{
\includegraphics[width=0.9\textwidth, trim= 40 385 0 130, clip]{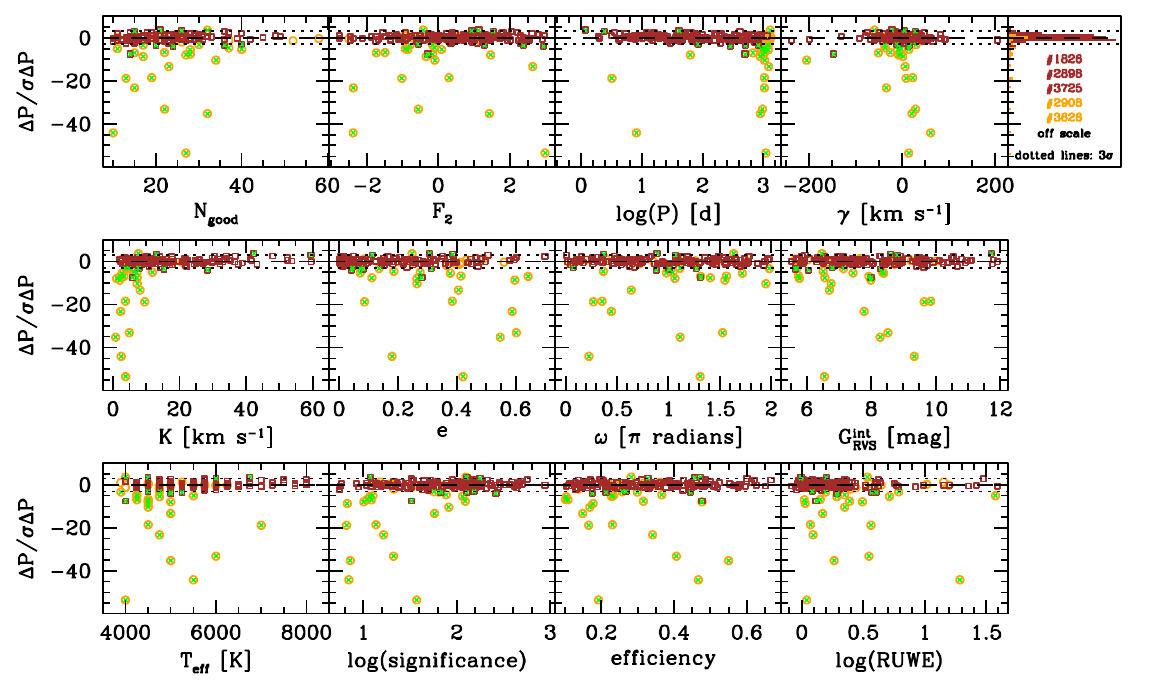}
}\caption[Comparison of period with SB9]{Same as 
Fig.\,\ref{fig:cu4nss_spectro_comparison_SB9_SB1_e} in the case of the SB9, 
but for $P$. As indicated, a few systems have 
very negative $\Delta P/\sigma\Delta P$ values and are off scale.
The SB9 IDs are given. }
\label{fig:cu4nss_spectro_comparison_SB9_SB1_P}
\end{figure}
\begin{figure}[!htp]
\centerline{
\includegraphics[width=0.9\textwidth, trim= 40 385 0 130, clip]{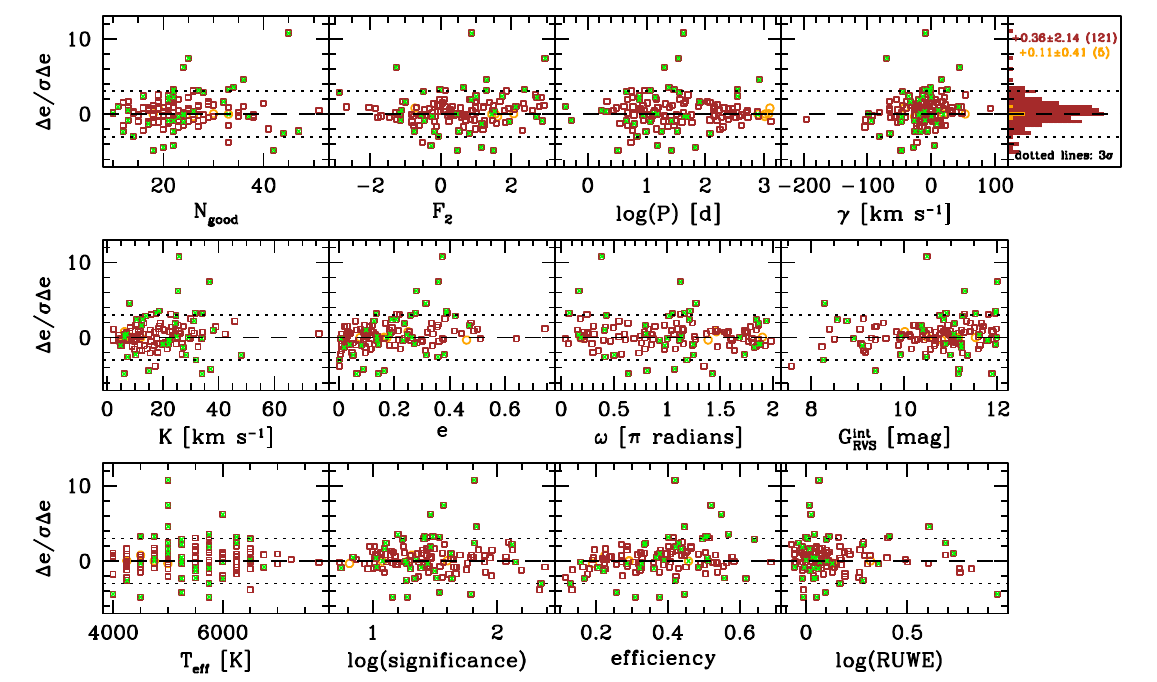}
}\caption[Comparison of the eccentricity with \citep{price_whelan20}]{Comparison 
for the eccentricity, $e$, 
between the DR3 spectroscopic values and those in PW20. 
Systems for which the period, $P$, in the literature is unlikely 
to be found (i.e. $\Delta T$ $<$ $P$) are indicated with orange circles. 
Systems for which $\Delta T$ $\geqslant$ $P$ are shown as brown squares. 
Binaries for which the reference 
period is not recovered to within 3$\sigma$ are flagged with green crosses. 
The quantity $T_\mathrm{eff}$ refers to the effective temperature of the 
template used to determine the RVs \citep{2023A&A...674A...5K}.}
\label{fig:cu4nss_spectro_comparison_PW20_SB1_e}
\end{figure}
\begin{figure}[!htp]
\centerline{
\includegraphics[width=0.9\textwidth, trim= 40 385 0 130, clip]{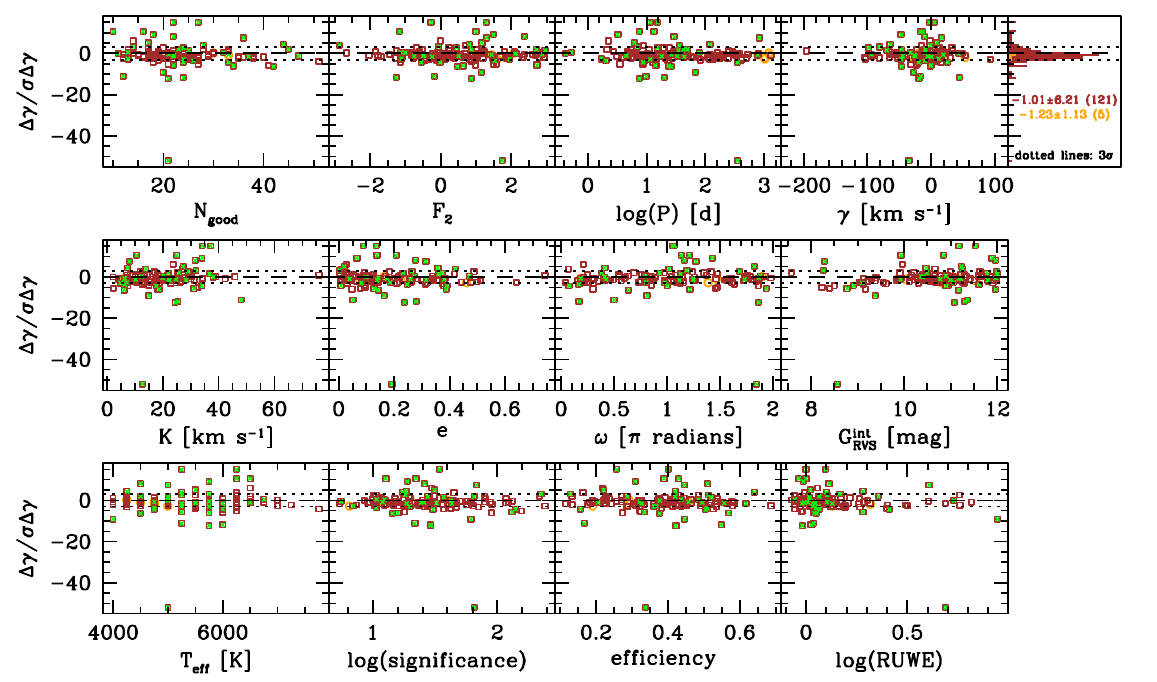}
}\caption[Comparison of systemic velocity with \citep{price_whelan20}]{Same as 
Fig.\,\ref{fig:cu4nss_spectro_comparison_PW20_SB1_e} in the case of PW20, 
but for $\gamma$.}
\label{fig:cu4nss_spectro_comparison_PW20_SB1_gamma0}
\end{figure}
\begin{figure}[!htp]
\centerline{
\includegraphics[width=0.9\textwidth, trim= 40 385 0 130, clip]{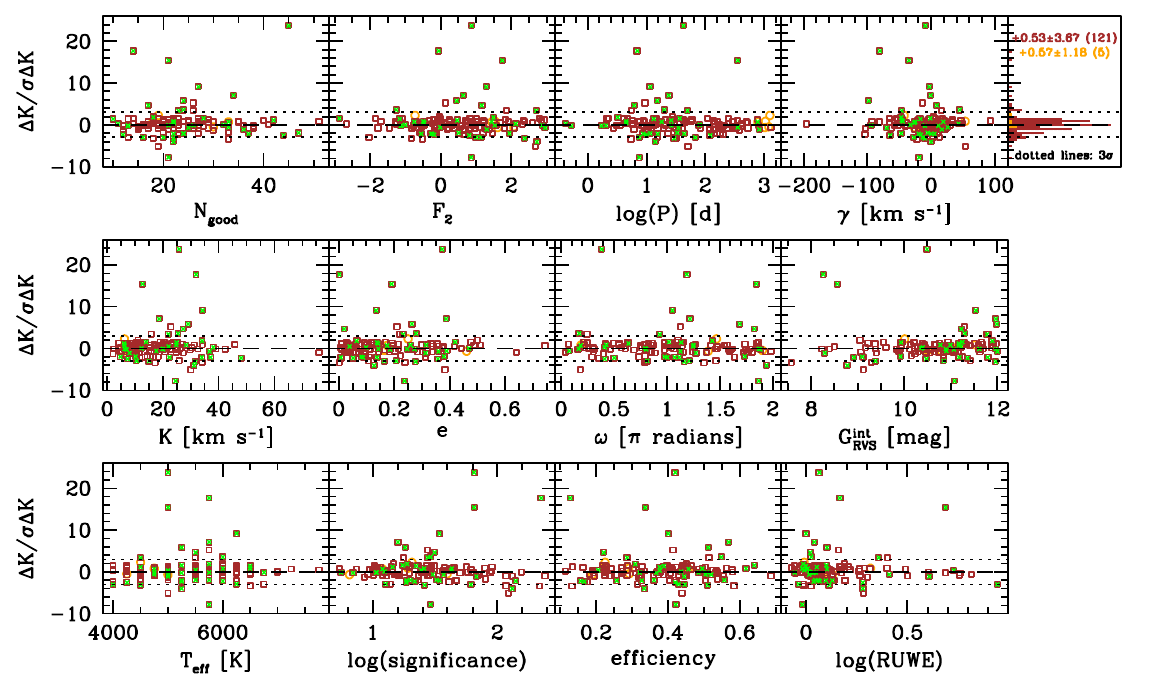}
}\caption[Comparison of semi amplitude with \citep{price_whelan20}]{Same as 
Fig.\,\ref{fig:cu4nss_spectro_comparison_PW20_SB1_e} in the case of PW20, 
but for the semi-amplitude $K$.}
\label{fig:cu4nss_spectro_comparison_PW20_SB1_K}
\end{figure}
\begin{figure}[!htp]
\centerline{
\includegraphics[width=0.9\textwidth, trim= 40 385 0 130, clip]{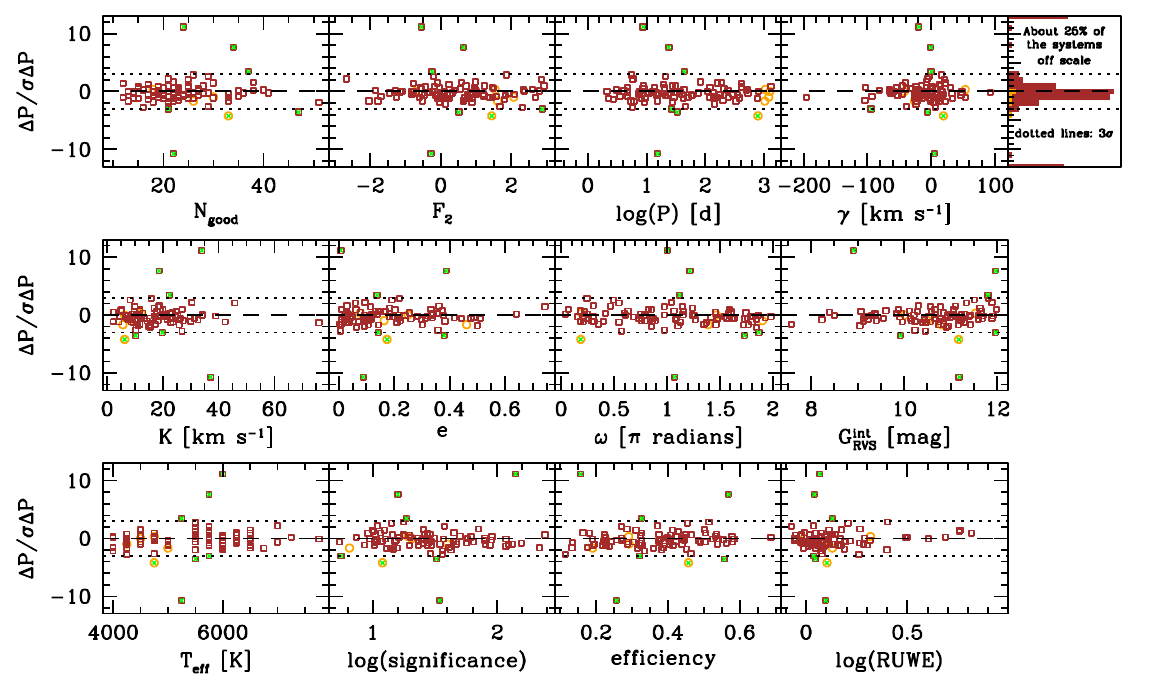}
}\caption[Comparison of period with \citep{price_whelan20}]{Same 
as Fig.\,\ref{fig:cu4nss_spectro_comparison_PW20_SB1_e} in the case of PW20, 
but for $P$. We note that a significant number of systems are off scale.}
\label{fig:cu4nss_spectro_comparison_PW20_SB1_P}
\end{figure}
\clearpage
\section{Processing steps and outcome of analysis for SB9 systems}
\label{sec:appG}
\begin{figure*}[hb]
\centerline{
\includegraphics[width=1.0\textwidth]{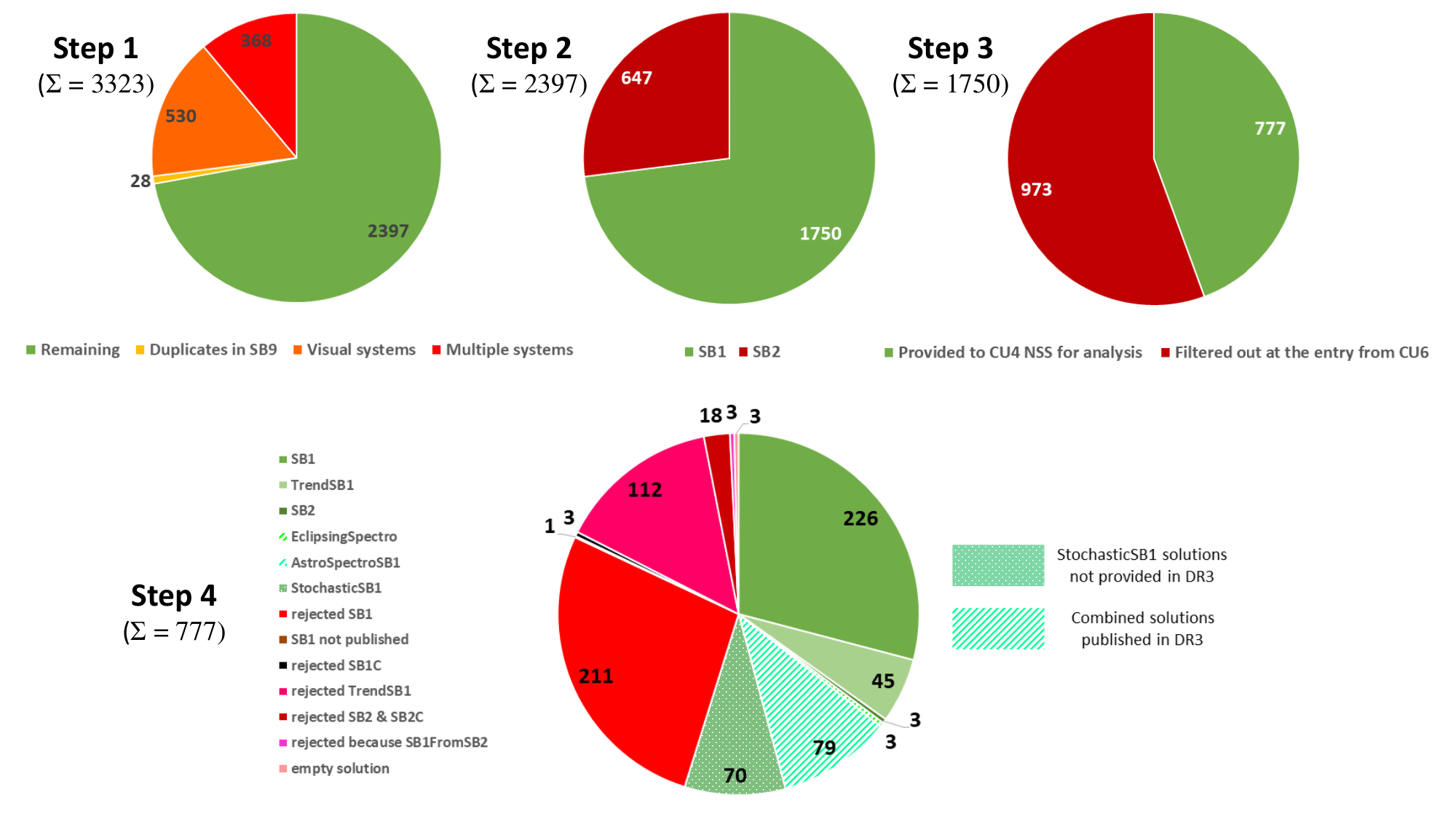}
}\caption{Overview of the cleaning of the input SB9 catalogue and processing
history of the sources selected. The bottom pie chart shows the census of 
the results at the very end of the chain. The sources that received an empty  
solution ended up with an insufficient number of valid RV measurements after 
removal of the outliers. Indicated as "not published" is 
\object{Gaia DR3 2178837257167184896}, which is among the 94 systems 
discussed in Sect.\,\ref{ssec:spectroSB1_results_numbers}, and does not 
have neither an {\tt SB1} nor a combined solution in DR3.}
\label{fig:pieCharts}
\end{figure*}

As for Griffin’s catalogue, the small number of SB9 systems 
eventually retained for validation 
(Sect.\,\ref{sssec:spectroSB1_quality_val_compSB9} and
Figs.\,\ref{fig:cu4nss_spectro_comparison_SB9_SB1_e}-
\ref{fig:cu4nss_spectro_comparison_SB9_SB1_P}) arises from the use of only a 
subset of the reference data set, along with the numerous 
selection criteria applied to the DR3 
data described in Tables\,\ref{tab:tableresult2},
\ref{tab:tableresult3}, \ref{tab:tableresult4}
and \ref{tab:appcatalogsel}. Figure\,\ref{fig:pieCharts} provides a 
summary of the various cuts that were applied, along with the number 
of SB9 sources remaining at each stage and the final outcome of the 
analysis of spectroscopic binaries by CU4/NSS.

\begin{figure*}[h]
\centerline{
\includegraphics[width=0.7\textwidth, trim= 130 130 110 130, clip]{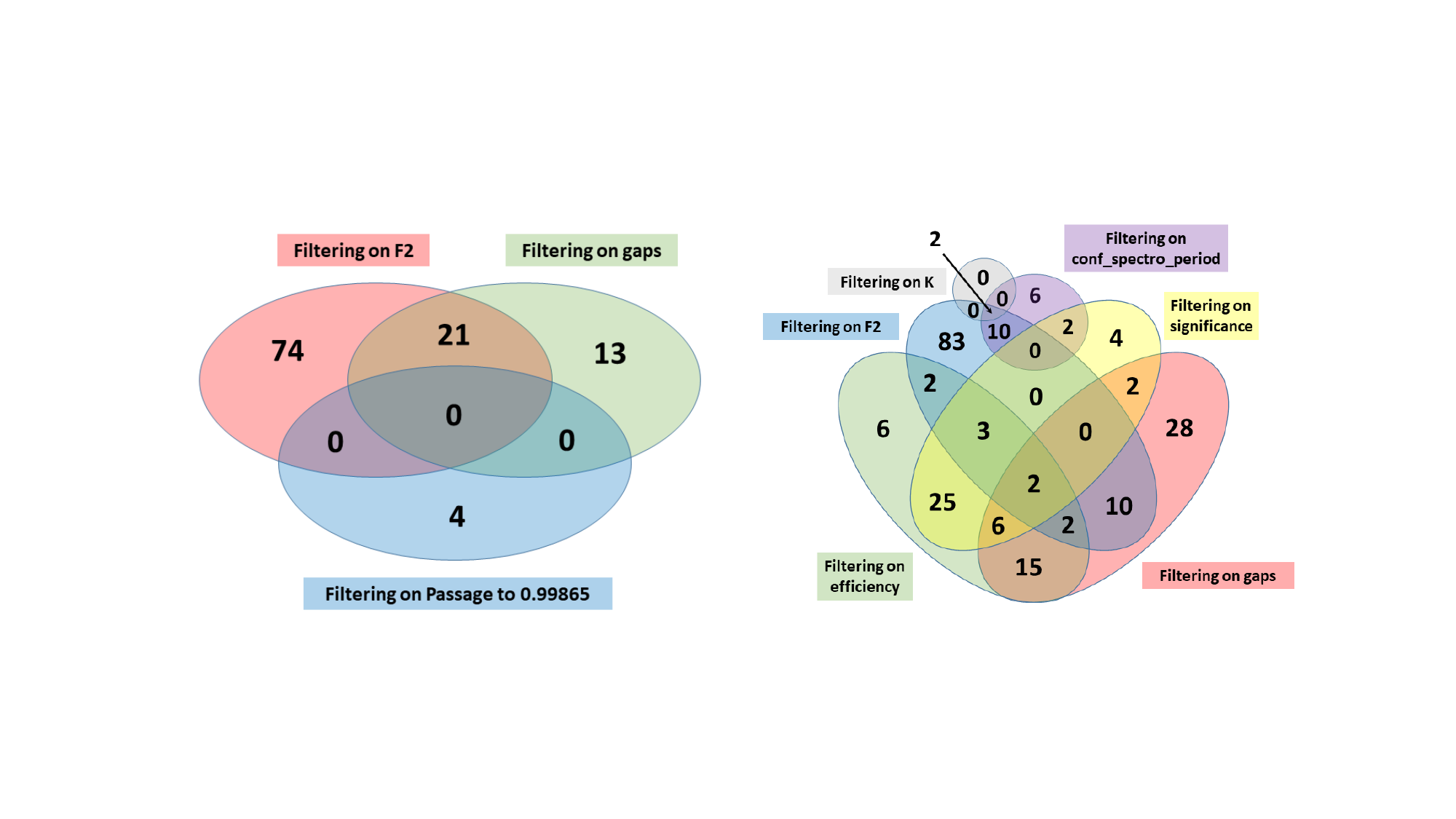}
}\caption{Venn diagrams describing the internal filtering that led to the 
rejection of 112 {\tt TrendSB1} ({left}) and 211 {\tt SB1} ({right}) 
solutions in DR3, respectively. In the latter case, three systems with 
multiple flags have been omitted for clarity.}
\label{fig:vennDiagrams}
\end{figure*}

\begin{itemize}
\item 
{\bf Step 1: selection applied to the cross-matches with SB9}\\
We start off with the 3323 cross-matches out of the 4021 binaries in the SB9. 
Among them, close to 30\% 
are visual/multiple systems, and are rejected straight away from the input 
catalogue (see beginning of Sect.\,\ref{sssec:spectroSB1_quality_val_compGriffin}). 
Furthermore, a few objects have duplicated entries in the SB9.
\\
\item 
{\bf Step 2: selection based on SB types in SB9}\\
We did not consider further the SB2 solutions reported in SB9, which 
at this stage constitute $\sim$25\% of the reference sample. 
As for the previous step, of course, rejection for validation purposes of 
a given system does not preclude it entering the DR3 subcatalogue. 
Among the 1750 SB1 objects from SB9, 1476 (=1750-226-45-3) do not 
have spectroscopic deterministic solutions (Sect.\,\ref{sssec:spectroSB1_quality_val_compSB9}).
\\
\item 
{\bf Step 3: CU6 data not feeding the present pipeline}\\
For more than half of the remaining systems, the CU6 data are either 
not transmitted to our pipeline for analysis or will not lead to any solutions. 
As discussed in Sects.\,\ref{sec:spectroSB1_input} and
\ref{sssec:spectroSB1_processing_ingest}, it is mainly because there is 
either no compelling evidence for variability in the RV time series, 
there are not enough valid RV measurements, or the source lies outside 
the $G_{\mathrm{RVS}}^{\mathrm{int}}$ or $T_\mathrm{eff}$ ranges. 
We note that $\sim$20\% of the entries in the whole SB9 contain an OBA component.
\\
\item 
{\bf Step 4: internal selection within our filtering scheme}\\
Finally, only a relatively small fraction of the remaining sources have 
either a proposed {\tt SB1}, {\tt SB1C}, or {\tt TrendSB1} solution as 
a result of the numerous quality cuts described in 
Sect.~\ref{ssec:spectroSB1_intfilt} (see also Tables\,\ref{tab:tableresult2},
\ref{tab:tableresult3}, \ref{tab:tableresult4}
and \ref{tab:appcatalogsel}). As an example, Fig.~\ref{fig:vennDiagrams} 
gives full details about the internal filtering applied to the {\tt TrendSB1} 
and {\tt SB1} solution types. In addition, we recall that the cross-matches 
with the combined and {\tt StochasticSB1} DR3 solutions are not counted in
Sect.\,\ref{sssec:spectroSB1_quality_val_compSB9}. The same holds 
for the purely astrometric orbital solutions or those for EBs only based on photometry.
\end{itemize}

\clearpage
\section{Parameters delivered to the catalogue for classes
{\tt{SB1}}, {\tt{SB1C}} and {\tt{TrendSB1}}}
\label{sec:appH}
Table\,\ref{tab:appcatalogsb1} lists the parameters computed in the
present work concerning the {\tt{SB1}} class. 
Table\,\ref{tab:appcatalogsb1c} deals with the
{\tt{SB1C}} class and Table\,\ref{tab:appcatalogtrendsb1} with the
{\tt{TrendSB1}} one. Table\,\ref{tab:appcatalogsel} summarises the selection criteria
applied on the way to the SB-subcatalogue.
\begin{table}[!ht]
\caption{List of the parameters delivered by the present work concerning 
the {\tt{SB1}} class. The first column gives the parameter name, column 2
the relevant symbol used here. The third column offers the
name of the entry in the catalogue itself. The fourth column gives
the type of the computer variable in the catalogue, and 
column 5 the physical units used. 
The last column contains any remark.}
\centering
\begin{tabular}{|l|c|l|c|c|l|}
\hline
\hline
Parameter & Symbol & Table entry & Type & Units & Remarks \\
\hline
Solution type & SB1 & {\tt{nss\_solution\_id}} & string & - & \\
Orbital period & $P$ & {\tt{period}} & double & d & \\
Uncertainty & $\sigma_P$ & {\tt{period\_error}} & float & d & \\
Systemic velocity & $\gamma$ & {\tt{center\_of\_mass\_velocity}} & double & km\,s$^{-1}$ & \\
Uncertainty & $\sigma_\gamma$ & {\tt{center\_of\_mass\_velocity\_error}} & float & km\,s$^{-1}$ & \\
Semi-amplitude & $K$ & {\tt{semi\_amplitude\_primary}} & double & km\,s$^{-1}$ & \\
Uncertainty & $\sigma_K$ & {\tt{semi\_amplitude\_primary\_error}} & float & km\,s$^{-1}$ & \\
Eccentricity & $e$ & {\tt{eccentricity}} & double & none & \\
Uncertainty & $\sigma_e$ & {\tt{eccentricity\_error}} & float & none & \\
Argument of periastron & $\omega$ & {\tt{periastron\_argument}} & double & degree & radian (this work)\\
Uncertainty & $\sigma_\omega$ & {\tt{periastron\_argument\_error}} & float & degree & \\
Periastron epoch & $T_0$ & {\tt{t\_periastron}} & double & Barycentric JD & \\
Uncertainty & $\sigma_{T_0}$ & {\tt{t\_periastron\_error}} & float & Barycentric JD & \\
Total number of points & $N_{\mathrm{tot}}$ & {\tt{rv\_n\_obs\_primary}} & integer & none & \\
Total number of valid points & $N_{\mathrm{good}}$ & {\tt{rv\_n\_good\_obs\_primary}} & integer & none & \\
Objective function & $\chi^2$ & {\tt{obj\_function}} & double & none & \\
Goodness of fit & $F_2$ & {\tt{goodness\_of\_fit}} & float & none & \\
Efficiency & - & {\tt{efficiency}} & float & none & \\
Significance & $K/\sigma_K$ & {\tt{significance}} & float & none & \\
Flags & - & {\tt{flags}}  & long & - & see appendix I \\
Confidence on period & - & {\tt{conf\_spectro\_period}} & float & - & \\
Concerned parameters & - & {\tt{bit\_index}} & long & - & $\equiv \, 127$ \\
Vector of correlations & - & {\tt{corr\_vec}} (length 15)& float & - & \\
\hline
\end{tabular}
\label{tab:appcatalogsb1}
\end{table}
\begin{table}[!ht]
\caption{Same as Table\,\ref{tab:appcatalogsb1} but for 
the {\tt{SB1C}} class.}
\centering
\begin{tabular}{|l|c|l|c|c|l|}
\hline
\hline
Parameter & Symbol & Table entry & Type & Units & Remarks \\
\hline
Solution type & SB1C & {\tt{nss\_solution\_id}} & string & - & \\
Orbital period & $P$ & {\tt{period}} & double & d & \\
Uncertainty & $\sigma_P$ & {\tt{period\_error}} & float & d & \\
Systemic velocity & $\gamma$ & {\tt{center\_of\_mass\_velocity}} & double & km\,s$^{-1}$ & \\
Uncertainty & $\sigma_\gamma$ & {\tt{center\_of\_mass\_velocity\_error}} & float & km\,s$^{-1}$ & \\
Semi-amplitude & $K$ & {\tt{semi\_amplitude\_primary}} & double & km\,s$^{-1}$ & \\
Uncertainty & $\sigma_K$ & {\tt{semi\_amplitude\_primary\_error}} & float & km\,s$^{-1}$ & \\
Maximum velocity epoch & $T_0$ & {\tt{t\_periastron}} & double & Barycentric JD & entry misleading \\
Uncertainty & $\sigma_{T_0}$ & {\tt{t\_periastron\_error}} & float & Barycentric JD & entry misleading \\
Total number of points & $N_{\mathrm{tot}}$ & {\tt{rv\_n\_obs\_primary}} & integer & none & \\
Total number of valid points & $N_{\mathrm{good}}$ & {\tt{rv\_n\_good\_obs\_primary}} & integer & none & \\
Objective function & $\chi^2$ & {\tt{obj\_function}} & double & none & \\
Goodness of fit & $F_2$ & {\tt{goodness\_of\_fit}} & float & none & \\
Efficiency & - & {\tt{efficiency}} & float & none & \\
Significance & $K/\sigma_K$ & {\tt{significance}} & float & none & \\
Flags & - & {\tt{flags}} & long & - & see appendix I \\
Confidence on period & - & {\tt{conf\_spectro\_period}} & float & - & \\
Concerned parameters & - & {\tt{bit\_index}} & long & - & $\equiv \, 31$ \\
Vector of correlations & - & {\tt{corr\_vec}} (length 6) & float & - & \\
\hline
\end{tabular}
\label{tab:appcatalogsb1c}
\end{table}
\begin{table}[!ht]
\caption{Same as Table\,\ref{tab:appcatalogsb1} but for 
the {\tt{TrendSB1}} class.}
\centering
\begin{tabular}{|l|c|l|c|c|l|}
\hline
\hline
Parameter & Symbol & Table entry & Type & Units & Remarks \\
\hline
Solution type & First DegreeTrendSB1 & {\tt{nss\_solution\_id}} & string & - & \\
& SecondDegreeTrendSB1 & & & & \\
Mean velocity & $V_0$ & {\tt{mean\_velocity}} & double & km\,s$^{-1}$ & \\
Uncertainty & $\sigma_{V_0}$ & {\tt{mean\_velocity\_error}} & float & km\,s$^{-1}$ & \\
First order derivative & $\frac{\partial V}{\partial t}$ & {\tt{first\_deriv\_velocity}} & double & km\,s$^{-1}$\,d$^{-1}$ & \\
Uncertainty & $\sigma_{\frac{\partial V}{\partial t}}$ & {\tt{first\_deriv\_velocity\_error}} & float & km\,s$^{-1}$\,d$^{-1}$ & \\
Second order derivative & $\frac{\partial^2 V}{\partial t^2}$ & {\tt{second\_deriv\_velocity}} & double & km\,s$^{-1}$\,d$^{-2}$ & \\
Uncertainty & $\sigma_{\frac{\partial^2 V}{\partial t^2}}$ & {\tt{second\_deriv\_velocity\_error}} & float & km\,s$^{-1}$\,d$^{-2}$ & \\
Total number of points & $N_{\mathrm{tot}}$ & {\tt{rv\_n\_obs\_primary}} & integer & none & \\
Total nbr of valid points & $N_{\mathrm{good}}$ & {\tt{rv\_n\_good\_obs\_primary}} & integer & none & \\
Objective function & $\chi^2$ & {\tt{obj\_function}} & float & none & \\
Goodness of fit & $F_2$ & {\tt{goodness\_of\_fit}} & float & none &  \\
Flags & - &  {\tt{flags}} & long & - & appendix I \\
Concerned parameters & - & {\tt{bit\_index}} & long & - & $\equiv \, 7$ or 15 \\
Vector of correlations & - & {\tt{corr\_vec}} (length 1 or 3) & float & - & \\
\hline
\end{tabular}
\label{tab:appcatalogtrendsb1}
\end{table}
\begin{table}[!ht]
\caption{Summary of the various selection criteria applied 
at various stages on the way to
the SB-subcatalogue (classes {\tt{SB1}} and {\tt{SB1C}}).}
\centering
\begin{tabular}{|c|c|c|}
\hline\hline
&&\\[1pt]
Stage of application & Selection criterion & Description in \\
&&\\[1pt]
\hline
&&\\[1pt]
Transfer from CU6 products & RV time series available & Katz et al. (2023) \\
& Normal stars, no spectral peculiarities & Katz et al. (2023) \\
& Object not classified as SB2 & Damerdji et al. (2024) \\
&&\\[1pt]
\hline
&&\\[1pt]
Selection of  & {\tt{rv\_renormalized\_gof}} larger than 4 & 
Sect.\,\ref{sec:spectroSB1_input} \\
objects/time series at entry & $T_{\mathrm{eff}}$ in [3875 - 8125] K  & Sect.\,\ref{sec:spectroSB1_input} \\
& $G_{\mathrm{RVS}}^{\mathrm{int}}$ mag in [5.5 - 12] & Sect.\,\ref{sec:spectroSB1_input} \\
& {\tt{rv\_chisq\_pvalue}} less than or equal to 0.01 & Sect.\,\ref{sec:spectroSB1_input} \\
& RV outlier rejected from time series & Sect.\,\ref{sssec:spectroSB1_processing_ingest}  \\
& $N_{\mathrm{good}}$ larger than or equal to 10 & Sect.\,\ref{sssec:spectroSB1_processing_ingest}  \\
&&\\[1pt]
\hline
&&\\[1pt]
Internal filtering on solutions & $ \lvert \gamma \rvert \le \, $1000 km\,s$^{-1}$ 
& Sect.\,\ref{ssec:spectroSB1_intfilt} \\
& $F_2$ less than or equal to 3 & Sect.\,\ref{ssec:spectroSB1_intfilt} \\ 
& Rejection of solutions with bad flags & Sect.\,\ref{ssec:spectroSB1_intfilt} \\
& $K$ less than or equal to 250 km\,s$^{-1}$ & Sect.\,\ref{ssec:spectroSB1_intfilt} \\
& {\tt{efficiency}} greater than or equal to 0.1 & Sect.\,\ref{ssec:spectroSB1_intfilt} \\
& $\delta \phi$ less than or equal to 0.3 & Sect.\,\ref{ssec:spectroSB1_intfilt} \\
& {\tt{conf\_spectro\_period}} 
\hspace{0.1cm}
$
\left\{ 
\hspace{0.05cm}
\parbox{5.4cm}{- $\geq$ 0.95 if $P$ $\geq$ 10\,d \\
- $\geq$ 0.995 if $P$ $\leq$ 1\,d  \\
- $\geq$ 0.995\,--\,0.045$\times \log\,P$ in between} \right\}$ & 
Sect.\,\ref{ssec:spectroSB1_intfilt} \\
& {\tt{significance}} larger than or equal to 5 & 
Sect.\,\ref{ssec:spectroSB1_intfilt} \\
& $\sigma_\omega$ less than 2$\pi$ 
\, \, (solely for SB1) & Sect.\,\ref{ssec:spectroSB1_intfilt} \\
&&\\[1pt]
\hline
&&\\[1pt]
Loss by combination & --- & Sect.\,\ref{ssec:spectroSB1_combiner} \\
&&\\[1pt]
\hline
&&\\[1pt]
Post-filtering & 380 objects rejected as intrinsic variables (Cepheids and RR Lyrae) 
& Sect.\,\ref{sssec:spectroSB1_add_cons_postfilt_intvar} \\
& 1475 objects rejected having more than 10 \% double-line transits 
& Sect.\,\ref{sssec:spectro_add_cons_postfilt_double_line_transits} \\
& 164 objects rejected suffering from a scan-angle effect 
& Sect.\,\ref{sssec:spectroSB1_add_cons_postfilt_scan angle} \\
&&\\[1pt]
\hline
\end{tabular}
\label{tab:appcatalogsel}
\end{table}
\clearpage
\section{Various flags linked to the processing}\label{sec:appI}
Table\,\ref{tab:tableh1} details the various flags that could be activated
during the spectroscopic orbital solution processing, along with the relevant 
explanations in the framework of the decision tree.
\begin{table*}[!htp]
\caption{This table is the repertory of the various flags utilised in the chain.}
\label{tab:tableh1}
\centering  
\tiny
\begin{tabular}{lcc}
\hline\hline
Flag bit nr. & Value & Logical Variable \\
\hline
8 & 256 & BAD\_UNCHECKED\_NUMBER\_OF\_TRANSITS \\
\multicolumn{3}{l}{Explanation: The number of transits is not sufficient 
to process the star (before removing bad transits)} \\[0.3cm]
9 & 512 & NO\_MORE\_VARIABLE\_AFTER\_FILTERING \\
\multicolumn{3}{l}{Explanation: The RV curve is no more variable after velocity filtering
(at the threshold 0.9)} \\[0.3cm]
10 & 1024 & BAD\_CHECKED\_NUMBER\_OF\_TRANSITS \\
\multicolumn{3}{l}{Explanation: The number of transits is not sufficient 
to process the star (after removing bad transits)} \\[0.3cm]
11 & 2048 & SB2\_REDIRECTED\_TO\_SB1\_CHAIN\_NOT\_ENOUGH\_COUPLE\_MEASURES \\
\multicolumn{3}{l}{Explanation: SB2 is redirected to the SB1 chain because there are not enough
couples of measures} \\[0.3cm]
12 & 4096 & SB2\_REDIRECTED\_TO\_SB1\_CHAIN\_PERIODS\_NOT\_COHERENT\\
\multicolumn{3}{l}{Explanation: SB2 is redirected to the SB1 chain because the
periods found by the SB1 and SB2 chains are not coherent} \\[0.3cm]
13 & 8192 & NO\_SIGNIFICANT\_PERIOD\_CAN\_BE\_FOUND \\
\multicolumn{3}{l}{Explanation: No significant period can be derived
(no period from photometric variability analysis and periodogram peaks below
the cut-off threshold)} \\[0.3cm]
14 & 16384 & REFINED\_SOLUTION\_DOES\_NOT\_CONVERGE \\
\multicolumn{3}{l}{Explanation: The refined orbital solution does not converge
(after 1000 iterations)} \\[0.3cm]
15 & 32768 & REFINED\_SOLUTION\_SINGULAR\_VARIANCE\_COVARIANCE\_MATRIX \\
\multicolumn{3}{l}{Explanation: The variance-covariance matrix can not be obtained
(singular) for the refined solution} \\[0.3cm]
16 &  65536 & CIRCULAR\_SOLUTION\_SINGULAR\_VARIANCE\_COVARIANCE\_MATRIX \\
\multicolumn{3}{l}{Explanation: The variance-covariance matrix can not be obtained
(singular) for the circular solution} \\[0.3cm]
17 & 131072 & TREND\_SOLUTION\_SINGULAR\_VARIANCE\_COVARIANCE\_MATRIX \\
\multicolumn{3}{l}{Explanation: The variance-covariance matrix can not be obtained
(singular) for the trend solution} \\[0.3cm]
18 & 262144 & REFINED\_SOLUTION\_NEGATIVE\_DIAGONAL\_OF\_VARIANCE\_COVARIANCE\_MATRIX \\
\multicolumn{3}{l}{Explanation: The diagonal of the variance-covariance matrix
is negative for the refined solution} \\[0.3cm]
19 & 524288 & CIRCULAR\_SOLUTION\_NEGATIVE\_DIAGONAL\_OF\_VARIANCE\_COVARIANCE\_MATRIX \\
\multicolumn{3}{l}{Explanation: The diagonal of the variance-covariance matrix
is negative for the circular solution} \\[0.3cm]
20 & 1048576 & TREND\_SOLUTION\_NEGATIVE\_DIAGONAL\_OF\_VARIANCE\_COVARIANCE\_MATRIX \\
\multicolumn{3}{l}{Explanation: The diagonal of the variance-covariance matrix
is negative for the trend solution} \\[0.3cm]
21 & 2097152 & CIRCULAR\_SOLUTION\_DOES\_NOT\_CONVERGE \\
\multicolumn{3}{l}{Explanation: The Lucy refined orbital solution diverges
(after 1000 iterations)} \\[0.3cm]
22 &  4194304 & LUCY\_TEST\_APPLIED \\
\multicolumn{3}{l}{Explanation: The Lucy test has been applied} \\[0.3cm]
23 & 8388608 & TREND\_SOLUTION\_NOT\_APPLIED \\
\multicolumn{3}{l}{Explanation: The trend analysis has not been applied
(case of unsorted SB2)} \\[0.3cm]
24 & 16777216 & SOLUTION\_OUTSIDE\_E\_LOGP\_ENVELOP \\
\multicolumn{3}{l}{Explanation: The orbital solution is above the
$e$ - $\log\,P$ envelop} \\[0.3cm]
25 & 33554442 & PERIOD\_FOUND\_IN\_CU7\_PERIODICITY \\
\multicolumn{3}{l}{Explanation: The period is equal to a period issued from the 
CU7 (within the quadratic sum of their uncertainties)} \\[0.3cm]
26 & 67108864 & FORTUITOUS\_SB2 \\
\multicolumn{3}{l}{Explanation: The RV1 and RV2 seem to be uncorrelated
(contrary to the case of any binary system)} \\[0.3cm]
\hline 
\end{tabular} 
\end{table*}
\clearpage
\section{Examples of retrieving a gold sample}
\label{sec:appJ}
\begin{itemize}
\item 
For an {\tt{SB1}} gold sample, we suggest limiting the efficiency
to values larger than or equal to 0.2. To secure the
determination of the periods and the subsequent fits,
we advise adopting $N_{\mathrm{good}}$ larger or equal to
20. The most important selection is related to the significance
that must remain larger or equal to 40. This condition 
eliminates dubious solutions, notably with small periods
and large eccentricity.
Solutions with periods larger than 800\,d will not be
secure, and are known over 1000\,d to present biases and
a small efficiency. Periods between 2 and 10 \,d can be
kept thanks to the choice on the significance. 
An alternative possibility would be to relax the condition 
on the significance to 25-30 and to discard periods below
10\,d.
Below 2\,d,
we advise being cautious, although some positive validation
took place. The following ADQL command will deliver some
15\,314 objects: \\
\\
{\tt{SELECT * FROM gaiadr3.nss\_two\_body\_orbit  \\
WHERE nss\_solution\_type = 'SB1'  \\
AND efficiency >= 0.2  \\
AND rv\_n\_good\_obs\_primary >= 20  \\
AND significance >=40  \\
AND period <= 800  \\
AND period >= 2 }} .\\

The recovery rate of the period for this gold sample reaches 93\,\%.\\
\\
\item 
Since objects characterised by a {\tt{ruwe}} above 1.4
have thus a dubious position and consequently 
possible problems with the RVs, we recommend an additional
selection on the astrometric {\tt{ruwe}}. This selection
implies the use of two tables and restricts the sample
to 9578 objects: \\
\\
{\tt{SELECT * FROM gaiadr3.nss\_two\_body\_orbit  \\
INNER JOIN gaiadr3.gaia\_source\_lite USING (source\_id) \\
WHERE nss\_solution\_type = 'SB1'  \\
AND efficiency >= 0.2  \\
AND rv\_n\_good\_obs\_primary >= 20  \\
AND significance >=40  \\
AND period <= 800  \\
AND period >= 2  \\
AND ruwe < 1.4 }}  .\\
\\
\item 
Finally, we also present a gold selection
on the linear trends that delivers 12\,650 objects: \\
\\
{\tt{SELECT * FROM gaiadr3.nss\_non\_linear\_spectro  \\
WHERE nss\_solution\_type = 'FirstDegreeTrendSB1'  \\
AND rv\_n\_obs\_good\_primary >= 20 }} .\\
\end{itemize}
\end{appendix}
\end{document}